\documentclass[11pt,a4paper]{article}
\usepackage{jheppub}

\usepackage{graphicx}		

\newtheorem{theorem}{Theorem}

\usepackage[font=small,labelfont=bf]{caption}

\usepackage[font=footnotesize]{subcaption}

\usepackage[nameinlink,noabbrev ]{cleveref}	

\usepackage{placeins} 	

\usepackage{csquotes} 	

\usepackage{tipa} 		

\usepackage{tabularray}	

\newcommand{\abs}[1]{\lvert #1 \rvert}
\renewcommand{\d}{\textnormal{d}}
\newcommand{\ren} {{  \mathcal R}}
\newcommand{\period}{\mathcal P}
\newcommand{\Aut}{\operatorname{Aut}}
\newcommand{\diam}{\operatorname{diam}}

\renewcommand{\|}{\rule[-0.4ex]{0.2ex}{1.2em}}

\title{Statistics of Feynman amplitudes in $\phi^4$-theory}

\author[1]{Paul-Hermann Balduf
\note{ Most of the numerical computations of this work were done while the author was affiliated with Humboldt-Universität zu Berlin. The author thanks Karen Yeats, David Broadhurst, Gerald Dunne, Erik Panzer, Michael Borinsky, and Dirk Kreimer for discussion and comments.}}

\affiliation{Department of Combinatorics \& Optimization, University of Waterloo, \\
	200 University Avenue West, Waterloo, Ontario, Canada N2L 3G1.
}

\emailAdd{pbalduf@uwaterloo.ca}

\arxivnumber{2305.13506}

\keywords{Large-Order Behaviour of Perturbation Theory, Renormalization Group, 1/N Expansion, Feynman period}

\abstract{
The amplitude of subdivergence-free logarithmically divergent Feynman graphs in $\phi^4$-theory in 4 spacetime dimensions is given by a single number, the Feynman period. We numerically compute the periods of 1.3 million completed graphs, this represents more than 33 million graphs contributing to the beta function. Our data set includes all primitive graphs up to 13 loops, and non-complete samples up to 18 loops, with an accuracy of ca. 4 significant digits.

We implement all known symmetries of the period in a new computer program and count them up to 14 loops. Combining the symmetries, we discover relations between periods that had been overlooked earlier. All expected symmetries are respected by the numerical values of periods.

We examine the distribution of the numerically computed Feynman periods. We confirm the leading asymptotic growth of the average period with growing loop order, up to a factor of 2. At high loop order, a limiting distribution is reached for the amplitudes near the mean.  A small class of graphs, most notably the zigzags, grows significantly faster than the mean and causes the limiting distribution to have divergent moments even when normalized to unit mean. 
We examine the relation between the period and various properties of the underlying graphs. We confirm the strong correlation with the Hepp bound, the Martin invariant, and the number of 6-edge cuts. We find that, on average, the amplitude of planar graphs is significantly larger than that of non-planar graphs, irrespective of $O(N)$ symmetry.
  
We estimate the primitive contribution to the 18-loop beta function of the $O(N)$-symmetric theory. We find that primitive graphs  constitute a large part of the  beta function in MS for $L\rightarrow \infty$ loops.  The relative contribution of planar graphs increases with growing $N$ and decreases with growing loop order $L$.

}

\begin{document}

\maketitle



\section{Introduction}\label{sec:introduction}

	\subsection{Motivation}\label{sec:motivation}
	
	One of the most widely used frameworks to compute observables in quantum field theory is perturbation theory: 
	An approximation to the sought-after function is constructed as a power series in a coupling parameter, where the series coefficients are given by Feynman integrals, which in turn can be indexed by Feynman graphs. The first coefficients of the perturbation series often give a decent approximate result, comparable with current experimental precision, but there are various reasons to believe that the perturbation series is ultimately divergent. This does not necessarily invalidate perturbation theory, because it might be possible to infer non-perturbative finite results from a divergent perturbation series via resummation. More broadly, the properties of divergent series can be examined within the framework of \emph{resurgence} theory, but such a study needs as input the asymptotic growth of series coefficients at high order. Promising steps in this direction have been taken in the last years, see for example \cite{dunne_resurgence_2012,borinsky_nonperturbative_2020,bellon_resurgent_2021,borinsky_semiclassical_2021,clavier_borelecalle_2021,borinsky_resonant_2022}, however, they are often restricted to cases  where extra information, such as a differential equation  or an exact solution of the involved Feynman integrals, are available to efficiently generate high-order series coefficients.
	
	The number of non-isomorphic Feynman graphs per loop order grows factorially with the loop order, see \cref{sec:asymptotics}. Considering only a structurally simple subclass is not a reliable approximation for the amplitude of \enquote{typical} Feynman graphs of the theory at high loop order.  The goal of the present work is to provide such a broad picture, by numerically computing as many graphs as possible, albeit only to limited precision.  This allows us to not only examine the average of those amplitudes, but also their distribution and various correlations,  which is vital for estimating the reliability of non-complete samples or extrapolations. 
	
	Concretely, we examine all  subdivergence-free Feynman integrals contributing to the 4-point function (the \emph{periods}, see \cref{sec:period}) in $O(N)$-symmetric $\phi^4$-theory up to, and including, 13 loops, and non-complete samples of such graphs up to 18 loops. Details are explained in \cref{sec:samples}.

	\subsection{The period}\label{sec:period}
	The renormalized perturbative 4-point function (amplitude) of $\phi^4$-theory is given by a sum over all vertex graphs of $\phi^4$-theory, that is, graphs which are 4-regular except for having exactly 4 external edges, 
	\begin{align}\label{perturbation_series}
		G_\ren^{(4)}=\sum_{G \text{ is $\phi^4$ vertex graph}} \frac{4!}{\abs{\Aut(G)}} \mathcal F_\ren(G), 
	\end{align}
	Here $\mathcal F_\ren(G)$ denotes the renormalized Feynman integral associated to the graph $G$. 
	In full generality, a renormalized Feynman integral is a function of masses, external momenta, coupling constants, the renormalization scheme and -scale, and the dimension and signature (Euclidean or Minkowski) of spacetime. 
	
	In the present work, we consider  Euclidean $\phi^4$-theory with interaction $\frac{\lambda}{4!}\phi^4$ in  $D=4-2\epsilon$ spacetime dimensions. The vertex graphs of this theory have $\abs{E_G}=2 L_G$ internal edges and $\abs{V_G}= L_G+1$ vertices, where $L_G$ is the loop number. Their superficial degree of convergence   is $\omega_G=\abs{E_G}-  L_G \frac D 2=L_G\cdot \epsilon$, making them superficially log-divergent. 
	The un-renormalized Feynman integral can be expressed as a parametric integral according to  
	\begin{align}\label{feynman_integral_parametric2}
		\mathcal F (G) &=   \frac{ (- \lambda )^{\abs{V_G}}}{(4\pi)^{\abs{L_G} \frac D 2}}    \Gamma(\omega_G)  \prod_{e\in E_G} \int \limits_0^\infty  \d a_e   \, \delta \left( 1-\sum_{e=1}^{\abs{E_G}} a_e \right) \frac{\psi_G^{\omega_G-\frac D 2}}{\phi_G^{\omega_G}} .
	\end{align}
	See \cite{nakanishi_graph_1971,panzer_feynman_2015} for details and derivations.
	In \cref{feynman_integral_parametric2}, the structure of the graph is encoded in the first and second Symanzik polynomials $\psi_G$ and $\phi_G$ \cite{bogner_feynman_2010}. These polynomials are homogeneous functions of the Schwinger parameters $\left \lbrace a_e \right \rbrace   $, and $\phi_G$ is additionally a function of masses and momenta. Using dimensional regularization, $D=4-2\epsilon$, the divergence of the integral \cref{feynman_integral_parametric2} takes the form of poles in $\epsilon$.

	Furthermore, we restrict ourselves primitive graphs, this means that no proper subgraph of $G$ has zero or negative degree of convergence in the limit $\epsilon \rightarrow 0$. In that case, the only ultraviolet divergence in \cref{feynman_integral_parametric2} is the gamma function $\Gamma(\omega_G)=\frac{1}{L \epsilon}+ \mathcal O \left( \epsilon^0 \right) $, and any potential divergence of the integral itself is    of kinematic or infrared nature.
	
	Let $p$ be a suitable linear combination of external momenta, then $s \propto p^2$ defines an overall \emph{scale}, and all masses and momenta can be expressed relative to $s$ by  scale-free ratios, called \emph{angles}. For dimensional reasons, we also need  a reference scale $s_0$, which is arbitrary but fixed.
	Expanding \cref{feynman_integral_parametric2} in $\epsilon$, we then find the general structure of the integral of a primitive log-divergent graph:
	\begin{align}\label{amplitude_period} 
		\mathcal F (G) &=\Lambda \left(  \frac{  \period (G)}{  L_G  } \frac{1}{\epsilon}- \period(G) \ln \left( \frac s {s_0}  \right)   +C_G\left( \left \lbrace \theta \right \rbrace   \right)   + \mathcal  O \left( \epsilon \right) \right).
	\end{align}
	Here,  $\Lambda = 	  (4\pi)^{ L_G  (- 2 )}    (- \lambda )^{\abs{V_G}} $,   and $C_G$ is a finite quantity which might depend on the angles $\left \lbrace \theta \right \rbrace  $, but not on the scale $s$. $\period(G)$ is the \emph{period}
	\begin{align}\label{def:period}
		\period (G)&=  \left( \prod_{e\in E_G} \int \limits_0^\infty \d a_e  \right)   \; \delta \left( 1-\sum_{e\in E_G}  a_e \right) \frac{1} {\psi_G^{ 2}}.
	\end{align}
	The period is free of UV-divergences and it does not depend on the kinematics of the graph.
	For a more rigorous discussion of the convergence of   integrals and the precise dependence on angle variables, as well as alternative integral representations for the period, see \cite{bloch_motives_2006,schnetz_quantum_2010,brown_angles_2013}. The fact that the same coefficient describes the $\ln(s)$-dependence and the $\epsilon^{-1}$-pole is expected from the renormalization group \cite{wilson_renormalization_1974}. In $\phi^4$-theory, the period is known analytically for the infinite family of zigzag-graphs (\cref{zigzag_amplitude}), and hundreds of other graphs  \cite{broadhurst_knots_1995,schnetz_quantum_2010,brown_singlevalued_2015,panzer_analytic_2013,panzer_galois_2017,schnetz_numbers_2018,schnetz_hyperlogprocedures_2023}.
	
	For a  primitive  graph, renormalization merely amounts to subtraction of the pole term and --- depending on the scheme --- potentially finite terms of the unrenormalized amplitude. Crucially, renormalization can not alter the $s$-dependent part of the primitive amplitude because that would require an $s$-dependent, hence non-local, counterterm. Therefore, the renormalized integral of a primitive graph \cref{amplitude_period} has the form 
	\begin{align}\label{amplitude_period_renormalized} 
		\mathcal F_\ren (G) &=\Lambda \left(   - \period (G)\ln \left( \frac s {s_0}  \right)   +\text{finite terms depending on $\theta$, but not on $s$}   + \mathcal  O \left( \epsilon \right) \right).
	\end{align}
	In particular, the period encodes the $s$-dependence at $s_0$ of \cref{amplitude_period_renormalized},
	\begin{align}\label{period_derivative}
		\frac{\partial}{\partial \ln (s)} \mathcal F_\ren (G) \Big|_{s=s_0} = -\Lambda \cdot \period(G).
	\end{align}

	The central physical motivation to consider the period is its direct contribution to the beta function via \cref{period_derivative}, as will be discussed in \cref{sec:beta}. However, the period is also interesting from a conceptual perspective because it is structurally simpler than  Feynman integrals with subdivergences, and is being studied with respect to   the algebraic structures or the number content of Feynman integrals. Thirdly, a period in the pure mathematics sense is any number which is defined by a rational integral with rational boundaries, and these periods constitute a  class of numbers   between algebraic and complex numbers. Consequently,  periods have received considerable interest in the past two decades, an incomplete list of relevant publications is  \cite{kontsevich_periods_2001,belkale_periods_2003,brown_multiple_2006,schnetz_quantum_2010,brown_periods_2010,kreimer_quantum_2015,todorov_polylogarithms_2014,crump_period_2016,nasrollahpoursamami_periods_2016,crump_properties_2017,hu_further_2022,laradji_results_2021,borinsky_graphical_2022,borinsky_recursive_2022}.

\subsection{Samples}\label{sec:samples}

	The periods contributing to the 4-point function of $\phi^4$-theory   are invariant under  completion, that is, joining the 4 external edges of the graph $G$ to a new vertex, see \cref{sec:automorphism}. The resulting graph is a primitive (subdivergence-free)  vacuum graph  of $\phi^4$-theory, called a \emph{completion} of $G$. Owing to the completion symmetry of the period, the present paper is largely organized in terms of completions, not in terms of the underlying vertex graphs called \emph{decompletions}.
	
	We generated all completions and decompletions up to $L=16$ loops, counts will be discussed in \cref{sec:number_of_graphs}. 

	For numerical integration, we use the algorithm recently developed by Borinsky \cite{borinsky_tropical_2023a}.
	Up to including $L=13$ loops, we computed the periods of all primitive completions --- even of  those which are known to have equal periods, see \cref{sec:symmetries}. For higher loop order, we drew   random samples, generated by multiple runs of the \texttt{genrang} procedure supplied with \texttt{nauty} version 2.7 \cite{mckay_practical_2014}, and corrected for non-uniform sampling according to  \cref{sec:uniform_sampling}. The concrete numbers of graphs (prior to correction) are reported in \cref{tab:samples}, where the non-complete samples are indicated by appending \enquote{s} to the loop number. 
	To assess the reliability of the sampling, we generated, using \texttt{genrang}, a random sample of 10000 13-loop graphs, processed it analogously to the other samples and compared results to the full data set of all 13-loop graphs.

\begin{table}[h]
	\centering
	\begin{tblr}{  vlines, 
			hline{1}={solid},
			hline{2,Z}={solid},
			rowsep=0pt,
			columns={halign=r},
			column{1}={halign=c,mode=math},
			column{4}={halign=c},
			row{1}   = {  halign=c,valign=m,font = \fontsize{10pt}{12pt}\selectfont  }
		}
		L         &  {computed\\ completions} &        proportion & {relative \\ accuracy} & {computed \\ decompl.} & {planar \\ decompl.} & {inferred \\ compl.} &{inferred \\ decompl.} \\
		 
		1		  &      1 &                   1 &   4 /   4 &        1 &     1 &      1 &        1 \\
		3         &      1 &                   1 &   3 /   3 &        1 &     1 &      1 &        1 \\
		4         &      1 &                   1 &   3 /   3 &        1 &     1 &      1 &        1 \\
		5         &      2 &                   1 &   6 /   8 &        3 &     2 &      2 &        3 \\
		6         &      5 &                   1 &   5 /   6 &       10 &     5 &      5 &       10 \\
		7         &     14 &                   1 &   7 /   9 &       44 &    19 &     14 &       44 \\
		8         &     49 &                   1 &   9 /  15 &      248 &    58 &     49 &      248 \\
		9         &    227 &                   1 &  24 /  33 &     1688 &   235 &    227 &     1688 \\
		10        &   1354 &                   1 &  68 /  90 &    13094 &   880 &   1354 &    13094 \\
		11        &   9722 &                   1 &  36 /  90 &   114016 &  3623 &   9722 &   114016 \\
		12        &  81305 &                   1 &  56 /  90 &  1081529 & 14596 &  81305 &  1081529 \\
		13        & 755643 &                   1 & 217 / 289 & 11048898 & 60172 & 755643 & 11048898 \\
		13\text{s}       &  10000 & $1.4 \cdot 10^{-2}$ & 143 / 284 &   148389 &   849 &  26329 &   364065 \\
		 13\star  &  10000 & $1.4 \cdot 10^{-2}$ & 130 / 284 &   146084 &   843 &  32476 &   412477 \\
		14\text{s}        & 216190 & $2.9 \cdot 10^{-2}$ & 246 / 289 &  3436145 &  7314 & 489993 &  7690996 \\
		14 \star  &  109999 & $1.4 \cdot 10^{-2}$ & 217 / 240 &  1734993 &  3649 & 302704 &  4677617 \\
		15\text{s}        & 101207 & $1.2 \cdot 10^{-3}$ & 220 / 288 &  1713395 &  1283 & 258802 &  4320337 \\
		15 \star  &  67048 & $8.1 \cdot 10^{-4}$ & 212 / 260 &  1129937 &   879 & 211718 &  3475930 \\
		16\text{s}        &  10196 & $1.1 \cdot 10^{-5}$ & 217 / 274 &   182990 &    39 &  23329 &   414073 \\
		17\text{s}        &   5178 & $4.6\cdot 10^{-7}$  & 172 / 262 &    98180 &     7 &  11644 &   217002 \\
		18\text{s}        &    912 & $6.2\cdot 10^{-9}$  & 108 / 221 &    18206 &     0 &   1743 &    34442 \\
	 	\end{tblr}
	\caption{Overview of the graphs computed in this work. Loop number of decompletion, with s indicating non-complete samples and $\star$ indicating a sample drawn from the complete set of graphs. For \enquote{s}, the numbers refer to the samples prior to correction (\cref{sec:uniform_sampling})  \| Number of completed primitives computed \| proportion computed relative to total number (\cref{tab:periods_count}) \| average / maximum relative accuracy, in ppm, average taken over computed completions  \|  decompletions (= non-isomorphic 4-valent Feynnman graphs) of the computed completions  \| planar decompletions \|  completions which can be inferred from the computed graphs by symmetries, including the computed ones.  \| decompletions which can be inferred from the computed graphs. The latter is the total number of non-isomorphic graphs whose period is known numerically. }
	\label{tab:samples}
\end{table}

	For 13, 14 and 15 loops, we drew uniform random samples out of the set of \emph{all} completions using \texttt{Mathematica} version 13.1, they are denoted $13\star, 14\star$, and $15 \star$ in \cref{tab:samples}. Those samples serve as a check for potential bias in  the random graph generation algorithm of \texttt{genrang}. The overlap between the samples $14s$ and $14\star$, and between $15s$ and $15\star$, is very small.

In total, we computed the period of more than 1.3 million different completions, they correspond to more than 16.5 million different decompletions (=non-isomorphic Feynman graphs of the 4-point function \cref{perturbation_series}). The periods of over 33 million different decompletions can be inferred from this data by exploiting known symmetries of the period, see the last colum of \cref{tab:samples}.

We express the accuracy of our results  as the standard deviation, relative to the absolute value. For $L \leq 12$ loops, all numerical periods have $\leq 90$ppm relative accuracy, that is, slightly more than 4 significant digits. For all other graphs, we reach $\leq 290$ppm accuracy, see column 4 in \cref{tab:samples}. 

\FloatBarrier
	
\subsection{Content and results}\label{sec:results}

A large-scale statistical analysis necessarily involves many tables and plots to show the various coefficients, distributions, and correlations. To make this work more accessible to the reader, the present section is a survey of all central outcomes. References point to the concrete subsections and tables in the main text which support our claims empirically. 
\bigskip

\noindent
\Cref{sec:number_of_graphs} concerns the number of graphs, frequency of subclasses, growth rate, and symmetry factors.
\begin{itemize}
	\item \Cref{sec:automorphism} is a review of  completions, decompletions, and symmetry factors. 
	\item A well-understood model in graph theory, which contains the completions of $\phi^4$-theory as a subset, is the class of random 4-regular simple graphs $\mathcal{G}_{n,4}$  (\cref{thm:random_graphs}).
	\item If multiedges are excluded, most of the vertex-type graphs of $\phi^4$-theory are primitive, in accordance with the prediction from $\mathcal{G}_{n,4}$ (\cref{sec:asymptotics}).
	\item The proportion of planar graphs at high loop order is vanishingly small (\cref{tab:periods_count}).
	\item With growing loop order, automorphisms and symmetry factors become increasingly irrelevant (\cref{sec:asymptotics_symmetry}), at 15 loops, the average symmetry factor is $0.95$, more than 90\% of all completions have symmetry factor unity and more than 99\% of all decompletions are non-isomorphic (\cref{tab:aut}). This trend is expected from   $\mathcal{G}_{n,4}$. The \enquote{asymptotic} regime starts rather abruptly at $L_\text{crit}\approx 9$ loops (\cref{fig:ratio_correction,fig:Cj_growth}).
	\item In terms of symmetry factors, planar graphs behave similarly to the full data set (\cref{tab:aut_planar}), and there is no significant correlation between the symmetry factor of a completion and how many of its decompletions are planar (\cref{sec:planar_graphs}). 
\end{itemize}
	
\noindent
In \cref{sec:symmetries}, we discuss the symmetries of the period (not to be confused with the symmetry factors of \cref{sec:number_of_graphs}), and how they are used for the numerical integration.
\begin{itemize}
	\item In \cref{sec:period_symmetries}, we review the known symmetries and the Hepp bound, and comment on our programs to compute each of them. 
	\item The number of symmetric graphs is reported in \cref{tab:symmetries}.  The bulk of the symmetries can be interpreted as twist and/or Fourier split. Pure Fourier identities are very rare.
	\item  Our result agrees with earlier publications at lower loop order, but we find slightly more graphs related by symmetries at higher loop order than previously assumed. This is because we use more combinations of the known symmetries  (\cref{sec:symmetries_count}). 
	\item Still, not all cases where the Hepp bound agrees can be explained by known symmetries. These cases are rare  compared to the total number of graphs (\cref{tab:symmetries}).
	\item The tropical Feynman quadrature algorithm  \cite{borinsky_tropical_2023a} works as anticipated. The process has been fully automated in order to handle  large sample sizes (\cref{sec:numerical_integration}).
	\item Numerical results agree whenever we expect the graphs to be related by a period symmetry. We use this to improve the accuracy of the numerical results.
\end{itemize}

\FloatBarrier

\begin{figure} 
	\centering
	\includegraphics[width=.9\linewidth]{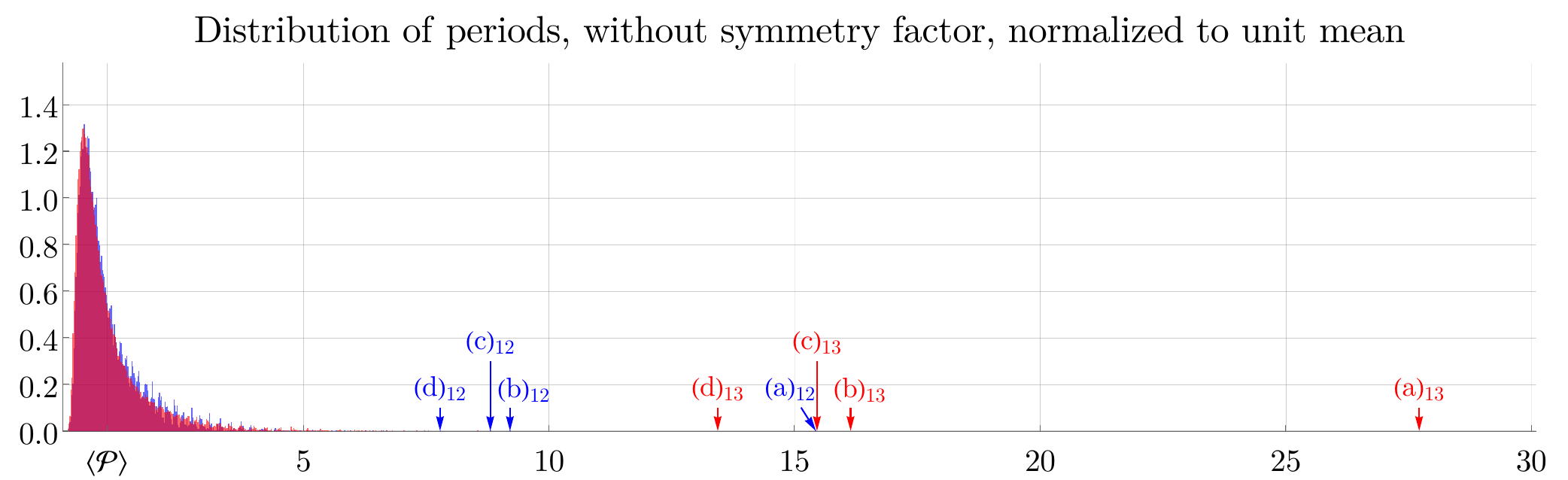}
	\caption{Distribution of periods for 12 loops (blue) and 13 loops (red), normalized with respect to their mean $\left \langle \period \right \rangle $. Labels $(a)_L, (b)_L, \ldots$ refer to the largest, 2\textsuperscript{nd} largest, \ldots period of the $L$-loop sample, see \cref{sec:largest}. $(a)_L$ is the $L$-loop zigzag graph. Note that the \enquote{continuous} part around $\left \langle \period \right \rangle $  (shown in \cref{fig:fit_comparison}) is almost identical for the two distributions, whereas the \enquote{outliers} of the 13-loop sample  are larger than those of the 12-loop sample by a constant factor $\approx  2$. }
	\label{fig:distribution}
\end{figure}
	
\begin{figure}[h!]
	\centering
	\begin{subfigure}[b]{.49 \textwidth}
		\includegraphics[width=\linewidth]{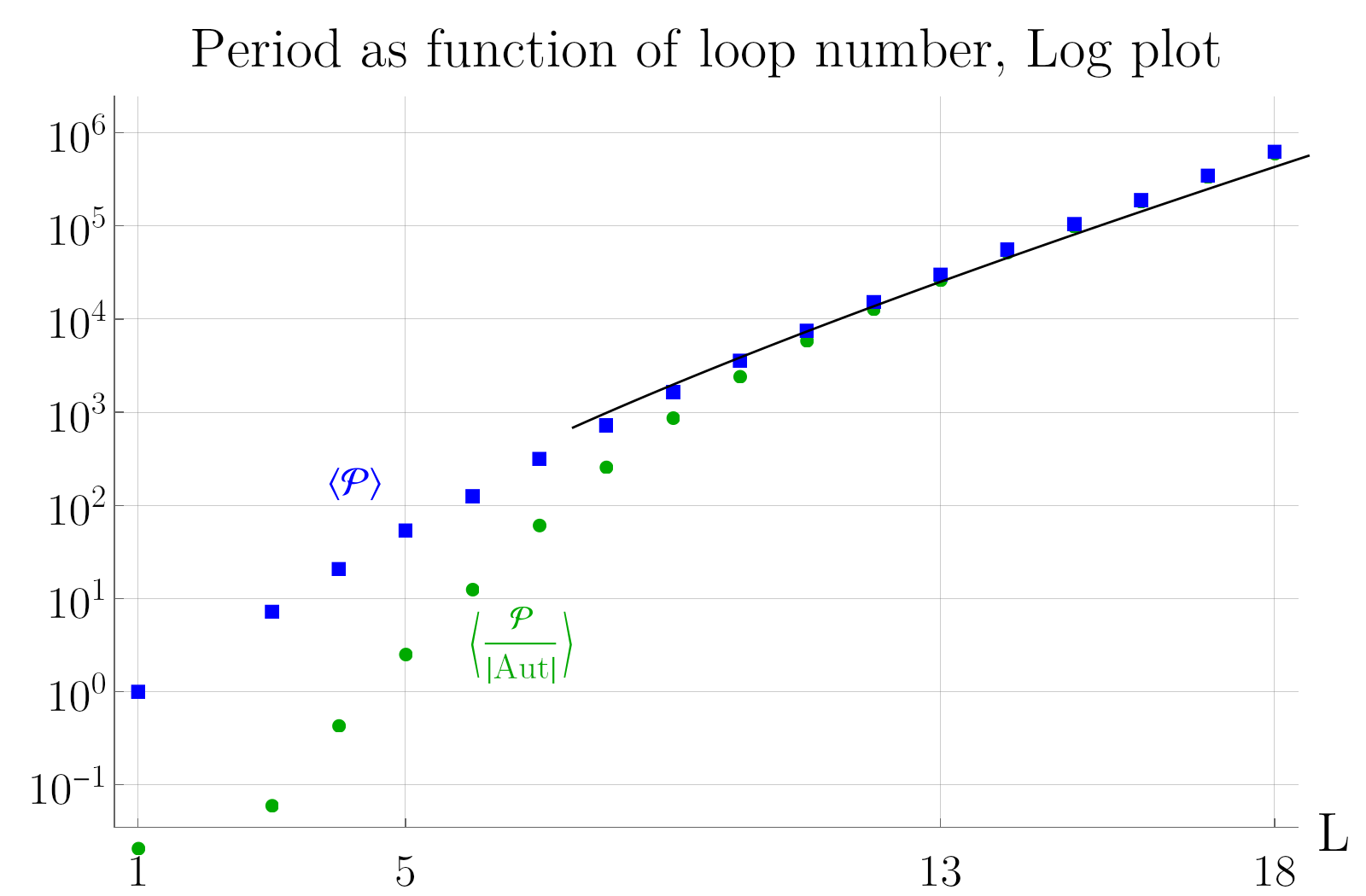}
		\subcaption{}
		\label{fig:period_mean}
	\end{subfigure}
	\begin{subfigure}[b]{.49 \textwidth}
		\includegraphics[width=\linewidth]{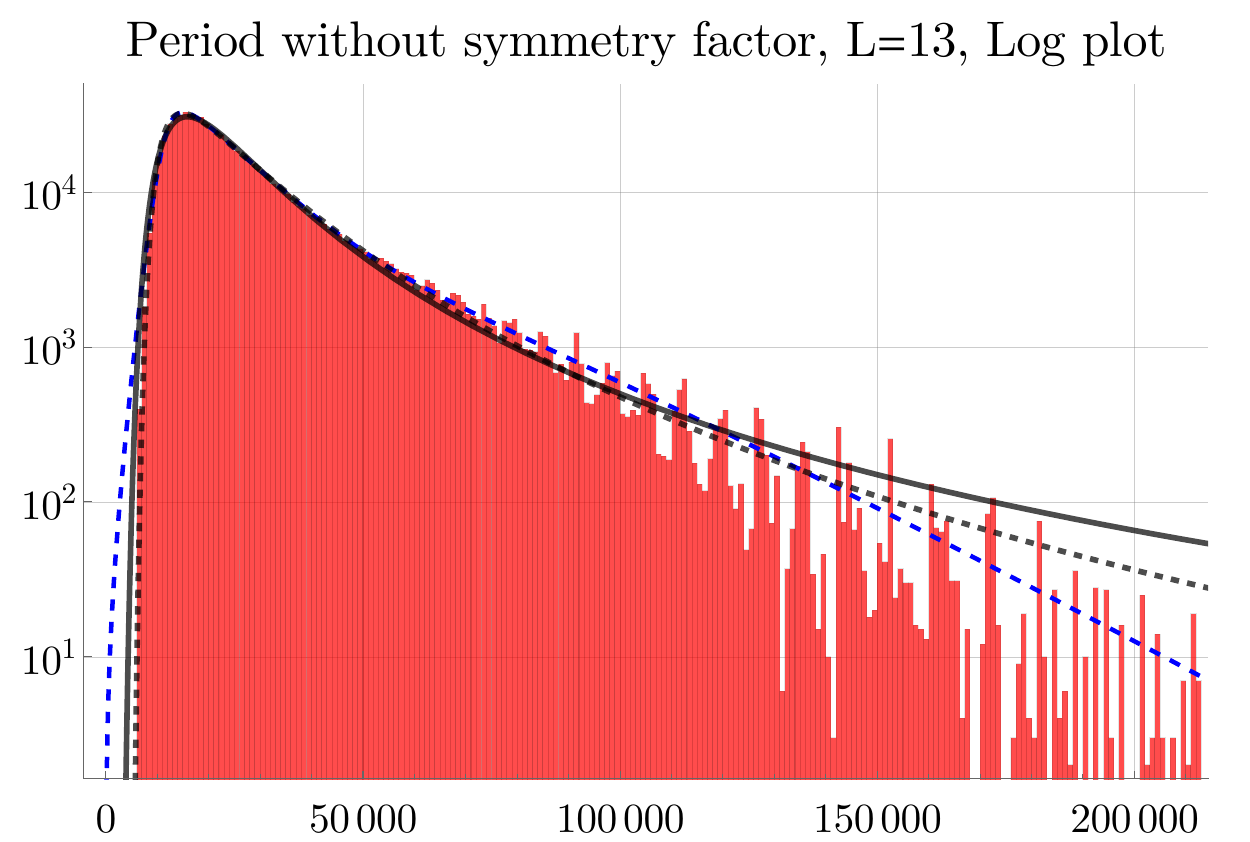}
		\subcaption{}
		\label{fig:fit_comparison}
	\end{subfigure}
	
	\caption{(a) Mean period of completed primitive graphs, where $L$ is the loop number of decompletions. Blue squares: without symmetry factor (\cref{def:period_mean}), green circles: with symmetry factor (\cref{def:period_mean_aut}). Black line: supposed asymptotics \cref{P_mean_asymptotics}.
		(b) Histogram of periods, without symmetry factor, absolute count with bin size 1000. The horizontal plot range corresponds to $[0,7]$ in \cref{fig:distribution}. Black line: Quadratic model function \cref{fitN}, Black dashed line: Quartic model \cref{fitQ}. Blue long dashed line: Empirical model \cref{P_distribution}.  Note that the quartic model accurately captures the falloff on the left, while the empirical model is superior for the decay on the right.
	}
\end{figure}

\noindent
\Cref{sec:distribution} is about the empirical distribution of periods as observed from our samples.
\begin{itemize}
	 \item For $L \leq 13$ loops, we determine the mean of the period to 6 significant figures with (\cref{tab:means_Aut}) and without (\cref{tab:means}) symmetry factor. For $14 \leq L \leq 18$ loops, the uncertainty of the mean is, by far, dominated by the sampling uncertainty (\cref{sec:moments}). The mean grows by a factor of $\frac 3 2$ per loop order (\cref{fig:period_mean}, \cref{sec:asymptotics_mean}).
	  \item The average period of decompletions is similar to that of completions. The average period of planar graphs is significantly larger than that of all graphs  (\cref{sec:mean_decompletions}).
	 \item The distribution of periods consists of two qualitatively different regions: A continuous distribution around the mean, and a small set of very large outliers (\cref{fig:distribution}).  
	 \item Even when the distribution is normalized to unit mean, the variance and higher moments   grow monotonically with loop order and are likely infinite for the limiting distribution (\cref{fig:Cj_growth}). Even large non-complete samples of periods have significant sampling uncertainty, which can be estimated as discussed in \cref{sec:sampling_uncertainty,sec:sampling_example}.
	 \item The divergence of moments is caused by the outliers  which grow faster than the average period and lie farther off the mean the higher the loop order. For each loop number, we find the largest graph to be the $(1,2)$-circulant (\enquote{zigzag}) of that loop number. The next-largest graphs can be described as \enquote{almost zigzags}, where larger distortion from the zigzag shape results in smaller periods. (\cref{sec:largest}).
	 \item Besides the $(1,2)$-circulant being the largest period, the other circulants don't show remarkable features, see \cref{tab:circulants}. For $L \leq 11$ loops, the smallest period is a circulant.
	 \item The continuous part of the distribution, roughly the periods between $0.5$ and $4$ times the mean, can be  described well by an empirical distribution function (\cref{sec:distribution_central}). In this regime, the probability for a period $x \cdot \left \langle \period \right \rangle $ falls off like $e^{-1.2 x}$.
	 The parameters of the distribution function change  only slowly with loop order and likely assume finite limiting values as $L\rightarrow \infty$. 
	  \item The inverse square root of the period has an almost symmetric distribution, which can be well approximated by a normal distribution (\cref{sec:nonlinear}) This gives rise to a second class of functions to model the distribution of periods. Either of the two models is reliable in a certain region, see \cref{fig:fit_comparison}.
	 \item The logarithm of the period behaves much better in terms of outliers, it is possible, but not certain, that its limiting distribution has finite moments. Conversely, the logarithm is affected by symmetry factors much more than the  period itself (\cref{sec:logarithm}).
\end{itemize}

\noindent
In \cref{sec:beta}, we examine the consequences of our data for  the beta function of a $O(N)$-symmetric $\phi^4$-theory.
\begin{itemize}
	\item In \cref{sec:beta_background}, we review the computation  of the  beta function and the conjecture that primitive graphs dominate the large-order asymptotics of the beta function in MS. 
	\item Coefficients of the primitive beta function for $L \leq 18$ are listed in \cref{tab:beta}. They agree with previous 7-loop analytic results and with an 11-loop extrapolation based on the Hepp bound. 
	\item Regarding the asymptotics, our results are ambivalent: At $L \approx 18$, the primitive coefficients reach, and even exceed, the supposed leading order asymptotics. This supports the conjecture that the primitive coefficients dominate the beta function in MS, but the numerical values come out significantly too large to indicate convergence to the limit (\cref{sec:beta_numerical}) . One possible interpretation is that the leading asymptotics is missing a factor of two. An alternative interpretation is that the asymptotic regime will only be reached at $L \gg 18$ loops. 
	\item We argue, and find empirically, that the primitive contribution to the $O(N)$-symmetric beta function always exceeds the leading asymptotic value for any fixed $L$ if $N$ is chosen large enough. This implies that our data, even at $L=18$, is not yet close to the asymptotics if $N$ is significantly larger than unity (\cref{sec:beta_N_dependence}). 
	\item The relative contribution of planar graphs to the beta function grows with growing $N$, but becomes relevant only at $N \gg 10$. Extrapolating the planar graphs to all graphs gives a too large result even at $N=1$ since planar graphs have above average periods (\cref{tab:means_dec}).
	\item The supposed asymptotic growth of the coefficients of the beta function implies a leading asymptotic growth of the mean period. Our data is consistent with the expected leading growth rate $\left( \frac{2}{3} \right) ^L L^{\frac 5 2}$. Again, introducing a factor of two into the absolute value of the asymptotics significantly improves the convergence, see \cref{sec:asymptotics_mean}.
\end{itemize}

\noindent
Finally, in \cref{sec:relations}, we examine the statistical correlations between the period and various properties of the graph.
\begin{itemize}
	\item The period of a graph is only weakly correlated with its symmetry factor or the number of non-isomorphic decompletions (\cref{sec:relations_symmetry_factor}). However, there is a strong, almost linear correlation between the period and the number of \emph{planar} decompletions of a graph, which gets more pronounced with growing loop order (\cref{fig:P_Drelpla}).
	\item The diameter or the girth of a graph turn out to be not useful for the loop orders under consideration since they only assume very few distinct values (\cref{thm:random_graphs,sec:small_circuits}).
	\item  There is a significant correlation between the mean distance between two vertices in a graph, and its period (\cref{sec:diameter}). This correlation is consistent across loop orders \emph{without} any normalization, that is, graphs with similar mean distance tend to have similar periods (in absolute value), even if their loop order differs (\cref{fig:period_distance}).
	\item The period is correlated with the number of triangles, and the pattern is regular once the period is scaled to its leading asymptotic growth (\cref{fig:period_triangles}). We find an empirical lower bound for the periods in our samples (\cref{period_triangle_bound}). The period is only weakly correlated with the number of 4-cycles, and anticorrelated with the number of 5-cycles. 
	\item The correlation between the period and the number of 6-edge cuts is so strong that it allows to predict the period with an average error of less than 15\% (\cref{sec:relations_6cuts}).
	\item The Hepp bound grows faster with loop order than the period,  its higher moments are divergent similarly to the ones of the period, but its overall distribution is more concentrated around the mean (\cref{tab:hepp}).
	\item We confirm the known strong correlation between period and Hepp bound, and determine new fit coefficients (\cref{sec:hepp}). This allows to predict, using a 5-parameter fit function for each loop order, the value of every period in our samples with a maximum error below 5\%, and an average error below 1\% (\cref{tab:hepp_fit2}). 
	\item The residues of the prediction by the Hepp bound fall into distinct strips, all graphs within a given strip share the same number of 6-edge cuts (\cref{fig:residue_hepp}). This will be used in a future publication to predict the period.
	\item The period is strongly correlated with the Martin invariant (\cref{sec:martin}), but not quite as strong as with the Hepp bound.
\end{itemize}

\subsection{Discussion}\label{sec:discussion}

All in all, our data supports the conjecture that primitive graphs represent a large contribution to the asymptotic beta function in MS at high loop order, see \cref{sec:beta_numerical}. In particular, we confirm the expected growth per loop order both for the average period and for the coefficients of the beta function.   
Interestingly, this growth ratio quickly approaches its asymptotic value starting from a loop order $L \approx 9$ (see \cref{fig:period_growth_correction}), which is very similar to the critical loop order observed for the   purely graph-theoretic properties, such as symmetry factors (\cref{sec:asymptotics_symmetry}). 

However, we can not clearly confirm the convergence of primitive graphs towards the supposed asymptotics of the MS beta function in absolute value. At $L \approx 18$, our data for primitive graphs is larger than the supposed asymptotics by a factor $\approx 2$. If one introduces a correction factor 2 into the asymptotics, it becomes possible to extract a finite subleading correction coefficient to from our data  for loop orders $L \geq 10$, see \cref{fig:period_mean_correction}. This is no proof for the behavior $L \rightarrow \infty$, but at least it suggests that with a factor of 2, the asymptotics becomes a useful description for $10 \leq L \leq 18$. Interestingly, in that scenario the transition to the \enquote{asymptotic} regime occurs at the same loop order $L_\text{crit}\approx 9$ for the period values as it does for the pure graph-theoretical properties such as symmetry factors (\cref{sec:asymptotics_symmetry}). In any case, our findings show that the asymptotic approximation is inaccurate below $L_\text{crit}\approx 9$ loops.

Another remarkable outcome concerns the  role of planar graphs. Firstly, for large loop orders, the proportion of planar graphs is very small (\cref{thm:random_graphs}) and their contribution to the beta function is negligible. If we introduce $O(N)$ symmetry then also non-planar graphs contribute to the leading order in $N$ (\cref{sec:beta_N_dependence}).  Nevertheless, for very large values of $N$, the contribution of planar graphs to the beta function becomes significant (\cref{fig:beta_planar_relative}). This effect is noticeable, but relatively small,  with $N \gg 1000$ required already at $L \approx 15$. This shows that for an $O(N)$ symmetric theory, one should not expect to obtain a good approximation to the beta function by leaving out non-planar graphs, at least for reasonably small values of $N$. 

Secondly, planar graphs on average have considerably higher periods than non-planar ones (\cref{tab:means_dec}). Consequently, planar graphs can not simply be \enquote{extrapolated}  without introducing significant error. In view of both effects, great care is required when planar graphs are used as an approximation to the beta function in $\phi^4$-theory. 

We stress that this question is different from a situation where, by some other mechanism, the theory contains \emph{only} planar graphs to begin with. In that case, our  data is compatible with the assumption that such a planar beta function is a convergent power series in the coupling, at least for fixed small $N$.

The present work impressively demonstrates the power of the tropical Feynman quadrature algorithm \cite{borinsky_tropical_2023a,borinsky_tropical_2023}. We can conveniently evaluate the periods up to   $L\approx 15$ loops. The limiting factors for higher loop orders are the non-parallelized preprocessing (several hours per graph) and the memory requirement (increasing by a factor of 4 per loop order).

Most likely, all higher moments of the normalized period distribution diverge in the limit $L \rightarrow \infty$. 
This has sobering consequences for the reliability of samples. We have seen that even a naive sample of far more than $10^5$ completions, corresponding to over 1 million non-isomorphic 4-valent graphs, still barely produces 3 significant digits for the beta function (compare 14s and $14\star$ in \cref{tab:means} or the experiments in \cref{sec:sampling_example}). The zigzag graphs, which are the only analytically known infinite class of $\phi^4$-periods, are a very poor proxy for a \enquote{typical} period, as they are by far the largest graphs of a given loop order (\cref{sec:largest}).

Regardless of the concrete numerical values for $\phi^4$ periods, some qualitative conclusions from the present work can likely be extended to other classes of Feynman amplitudes, for example that the distribution parameters change monotonically and predictably with growing loop order (\cref{cumulant_growth}) or that \enquote{special} classes of graphs tend to be a poor proxy for the \enquote{generic} case.

A different pathway to higher loop orders is to use the correlations between the period and various graph properties (\cref{sec:relations}) to predict amplitudes without  explicit numerical integrations. Such an extrapolation, based on the Hepp bound, had been given in \cite{kompaniets_minimally_2017} up to $L=11$ loops  and we found it confirmed (\cref{tab:beta}). At higher loop orders, the number of graphs grows factorially and the accuracy of predicting individual periods becomes largely irrelevant as long as the prediction is correct on average.  It seems well possible that an approximation of all 83 million completions at 15 loops from those correlations produces a more accurate estimate of the beta function than numerically computing $10^5$ of them as in  the present work. A more systematic study of that approach is currently in preparation.

\newpage

\section{Number of graphs and automorphisms}\label{sec:number_of_graphs}

\subsection{Graph automorphisms and completions}\label{sec:automorphism}

We consider Feynman graphs of $\phi^4$-theory, this means the graphs contain one type of particle (all edges are of the same type) and a 4-point elementary interaction (all internal vertices are 4-valent). We restrict ourselves to graphs which have either 4, or zero, external edges. The \emph{loops} of a Feynman graph are a basis of cycle space, not to be confused with the term loop in graph theory. A connected Feynman graph $G$ with 4 external edges and $L_G$ loops has 
\begin{align}\label{vertex_count}
	\abs{V_G}=L_G+1
\end{align}
vertices. 
As always for Feynman graphs in physics, by \enquote{graph} we mean an  isomorphism class of labeled graphs which do not need to be simple. Two graphs are called \emph{isomorphic}, and are being identified, if they differ only by a different labeling of their internal edges or vertices. An \emph{automorphism} is a permutation of labels which transforms the labeled graph into itself,  $\Aut(G)$ is the group of automorphisms of a graph $G$. In the perturbation series (\cref{perturbation_series}), the graph is weighted  with a \emph{symmetry factor} $\abs{\Aut(G)}^{-1}$.

\begin{figure}[htb]
	\begin{subfigure}[b]{.24 \textwidth}
		\includegraphics[width=\linewidth]{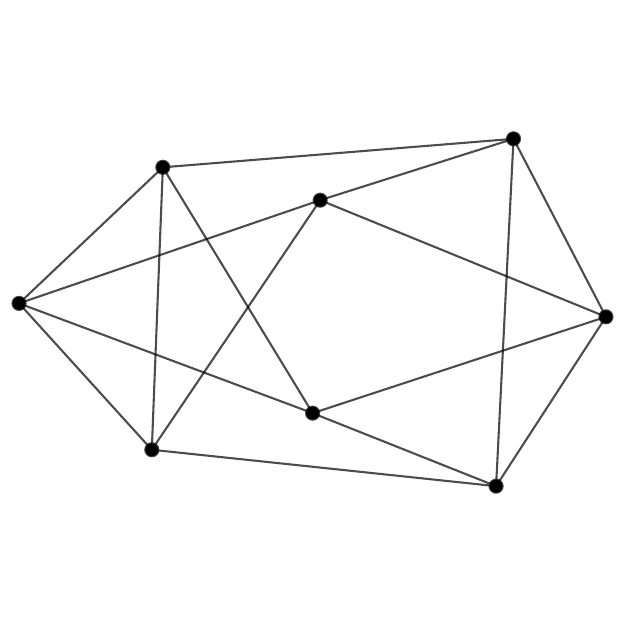}
		\subcaption{}
	\end{subfigure}
	\begin{subfigure}[b]{.24 \textwidth}
		\includegraphics[width=\linewidth]{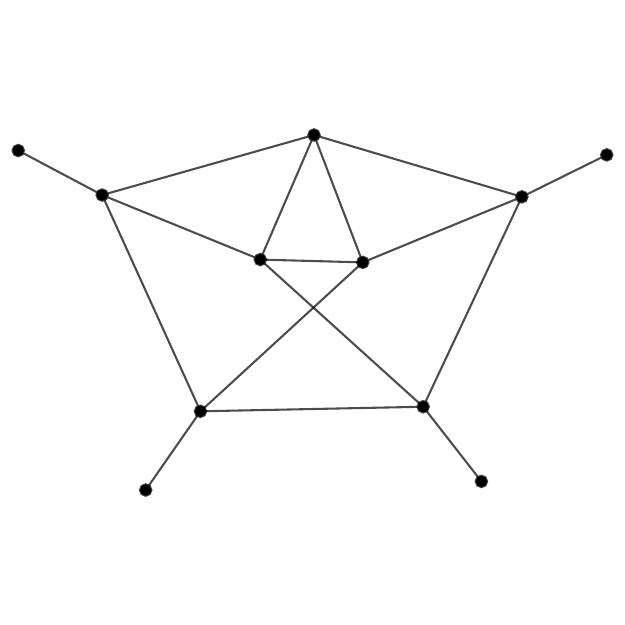}
		\subcaption{}
	\end{subfigure}
	\begin{subfigure}[b]{.24 \textwidth}
		\includegraphics[width=\linewidth]{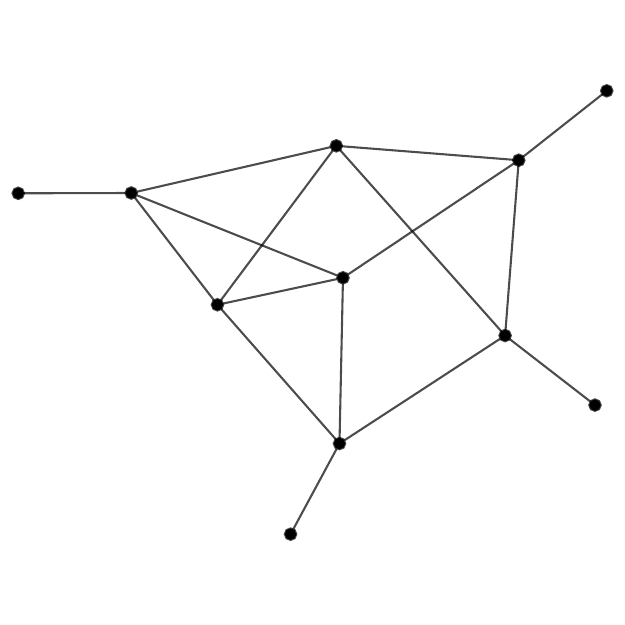}
		\subcaption{}
	\end{subfigure}
	\begin{subfigure}[b]{.24 \textwidth}
		\includegraphics[width=\linewidth]{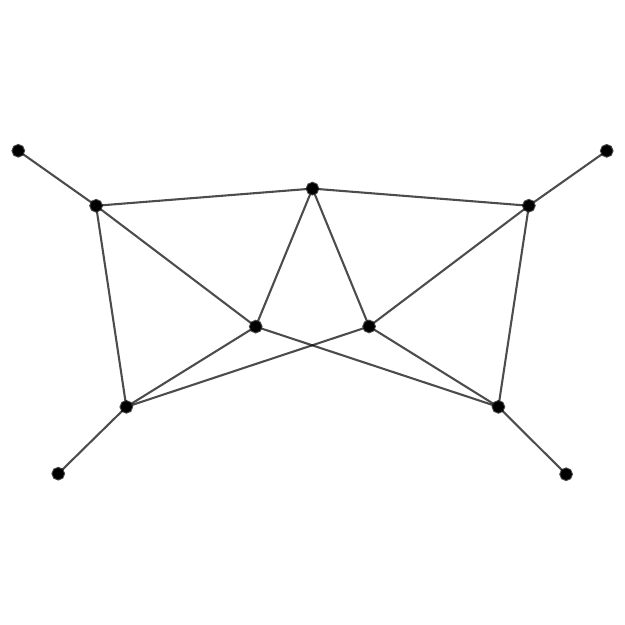}
		\subcaption{}
	\end{subfigure}
	
	\caption{(a) a 6-loop completion, and its 3  decompletions (b),(c),(d). External edges are highlighted by adding a 1-valent vertex. Even if (a) has 8 vertices, each of the 8 possible decompletions is isomorphic to either (b) or  (c) or (d). Note that (a) and (b) are not planar, but (c) and (d) are.}
	\label{fig:completion_decompletion}
\end{figure}

If a graph $G$  has 4 external edges, we can introduce one new vertex $v$ and join the 4 external edges of $G$ to $v$. This way, we obtain a 4-regular graph (that is, every vertex is 4-valent, not just the internal ones) without external edges. In physics terminology, this new graph $G\cup v$ is a vacuum graph of $\phi^4$-theory. We call $G\cup v$ the \emph{completion} of $G$, or conversely, we call $G$   a \emph{decompletion} of $G\cup v$. In the following, \emph{decompletion} will always mean \enquote{graph of $\phi^4$-theoy with 4 external edges}. Technically, the completion has 3 more loops than its decompletion, however, we define that that the \emph{loop number of a completion} coincides with the number of loops of its decompletion.   By \cref{vertex_count}, all decompletions of a given completion have the same loop number. 
As an example, \cref{fig:completion_decompletion} shows a 6-loop completion (i.e. the decompletions have 6 loops), and its three non-isomorphic decompletions.

The symmetry factors of completions and decompletions are related by the identity
\begin{align}\label{aut_decompletion}
	\left( L_G+2 \right) \frac{1}{\abs{\Aut(G)}} &=\sum_{ \substack{g \text{ decompletion}\\\text{of } G}} \frac{1}{\abs{\Aut(g)}}   ,
\end{align}
where the automorphisms do not have fixed vertices. Note that this convention is different from the usual physics one  where external vertices are demanded to be fixed by automorphisms. In our convention, a decompletion contributes to a 4-point amplitude (\cref{perturbation_series}) with all $4!$ different orientations of external edges (unless two external edges are incident to the same vertex, but, except for the 1-loop graph, our data set does not include such graphs as they are not primitive, see below), whereas in the usual physics convention, only the non-isomorphic orientations contribute.

We use the following notation to identify graphs: All graphs are transformed to the \enquote{canonical form for multigraphs} as given by \texttt{nauty} version 2.7 \cite{mckay_practical_2014}. Note that this is different from \texttt{nauty}'s canonical form for simple graphs.  The so-obtained graph has vertices numbered from $0$ to $\abs{V_G}-1$; we transform these labels to single-digit integers and letters, $0,1,\ldots,8,9,a,b,c,\ldots$. We then express the edges of the graph by the  \emph{Nickel index}\cite{nickel_compilation_1977}, that is, a list of vertex adjacencies according to
\begin{align}\label{def:nickel}
	/ \text{ vertices adjacent to 0 }/\text{ vertices adjacent to 1 }/\text{ vertices adjacent to 2 }/\ldots/.
\end{align}
We leave out any adjacency that has been specified earlier, consequently, our Nickel index contains $\abs{E_G}$ numbers, not $2\cdot \abs{E_G}$.
For example, a 1-loop multiedge has Nickel index $/11/$ and  the graph (a)  in \cref{fig:completion_decompletion} has  /4567/2367/357/46/57/6///.

In $\phi^4$-theory in 4 dimensions, a graph is superficially divergent if and only if it has 4 or fewer external edges. Consequently, a graph is \emph{primitive} (free of subdivergences) if and only if it is internally 6-edge connected, that is, no non-trivial subgraph is connected to the remainder by fewer than 6 edges.
A decompletion is primitive if and only if its completion is primitive \cite{schnetz_quantum_2010}.
For a fixed loop number $L$, we define the following counts of primitive graphs in $\phi^4$-theory:
\begin{align}\label{def:NL}
	N^{(D)}_L &:= \text{Number of primitive decompletions with $L$ loops}\\
	N^{(C)}_L &:= \text{Number of primitive  completions whose decompletions have $L$ loops}.\nonumber
\end{align}
The completion of a given decompletion is unique by construction. A given completion generally has multiple non-isomorphic decompletions, therefore $N^{(D)}_L \geq N^{(C)}_L$. By \cref{vertex_count}, a completion with $L_G$ loops  has $L_G+2$ vertices and removing any of them potentially produces a decompletion, therefore, $N^{(D)}_L \leq (L_G+2)N^{(C)}_L$. The number of decompletions is lower than this upper bound  if some of the decompletions are isomorphic, as happens in the example \cref{fig:completion_decompletion}.

\begin{table}[htbp]
	\centering
\begin{tblr}{  vlines, 
		hline{1}={solid},
		hline{2,Z}={solid},
		rowsep=0pt,
		columns={halign=r},
		row{1}   = {  halign=c,valign=m,font = \fontsize{10pt}{12pt}\selectfont  }
	}
	
		$L$  &    {1PI \\ completions} & { 4-edge conn. \\ completions} & $N^{(C)}_L$&  {pla. \\ com.} &  {3-vtx irr. \\ compl.}  &$N^{(D)}_L$  &  {planar \\ decompl. }   \\
		3  &          1 &         1 &         1 &         0 &           1 &           1 &        1 \\
		4  &          1 &         1 &         1 &         1 &           1 &           1 &        1 \\
		5  &          2 &         2 &         2 &         0 &           1 &           3 &        2 \\
		6  &          6 &         6 &         5 &         1 &           4 &          10 &        5 \\
		7  &         16 &        16 &        14 &         1 &          11 &          44 &       19 \\
		8  &         59 &        58 &        49 &         2 &          41 &         248 &       58 \\
		9  &        264 &       262 &       227 &         2 &         190 &        1688 &      235 \\
		10 &       1542 &      1532 &      1354 &         9 &        1182 &       13094 &      880 \\
		11 &      10768 &     10726 &      9722 &        11 &        8687 &      114016 &     3623 \\
		12 &      88126 &     87915 &     81305 &        37 &       74204 &     1081529 &    14596 \\
		13 &     805281 &    804108 &    755643 &        79 &      700242 &    11048898 &    60172 \\
		14 &    8036251 &   8028714 &   7635677 &       249 &     7160643 &   120451435 &   246573 \\
		15 &   86214189 &  86159508 &  82698184 &       671 &    78270390 &  1393614379 &  1015339 \\
		16 &  985816860 & 985373593 & 952538165 &           &             &  17041643034           &    4169613      \\
	\end{tblr}
	\caption{Counts of  4-regular-graphs. Loop order of the decompletion \| Number of\ldots ~ 2-vertex connected 1PI completed \emph{simple} graphs (i.e. no multiedges, no tadpoles) \| 4-edge connected simple completions \| primitive completions \| planar completions \| 4-vertex-connected (i.e. 3-vertex irreducible) completions \| primitive decompletions \| primitive planar decompletions.}
	\label{tab:periods_count}
\end{table}

\subsection{Asymptotics of the numbers of graphs}\label{sec:asymptotics}

We generated, using \texttt{nauty}'s \texttt{geng} routine \cite{mckay_practical_2014}, all 2-vertex connected 4-regular simple graphs of a fixed loop order $L \leq 16$. In physics terminology, this amounts to those 1-particle irreducible completed graphs of $\phi^4$-theory which have neither multiedges   nor  tadpoles as subgraphs, and which can not be disconnected by removing a single vertex. The count of these graphs is reported in \cref{tab:periods_count} as \enquote{1PI compl.}. For $L\leq 8$, the number coincides with the number of 4-regular connected  simple graphs \cite[A006820]{oeis}, for $L \geq 9$, they differ as not every connected graph is 2-vertex connected. 

In the next step, we removed from the above sets  those graphs which are not 4-edge connected. From a physics perspective, this step amounts to removing all remaining propagator corrections (recall that multiedges and tadpoles had been excluded before). The resulting number is \enquote{4-e compl.} in \cref{tab:periods_count}.

Lastly, we removed the graphs which are not internally 6-edge connected, that is, we remove all subgraphs which are quantum corrections to the 4-valent vertex. The remaining graphs are the primitive completed graphs  (\cref{def:NL}), their number is \enquote{$N^{(C)}_L$} in \cref{tab:periods_count}. Automatically, those graphs are at least 3-vertex connected.

\Cref{tab:periods_count} additionally shows the number of 4-vertex-connected primitive completions, and of planar graphs. Those numbers will be relevant for the symmetries of the period to be discussed in \cref{sec:symmetries}. Our results for $N^{(C)}_L$ and the 4-vertex connected 1PI graphs agree with the numbers listed, for $L\leq 14$, in \cite{schnetz_quantum_2010}.
The number of planar completions is consistent (i.e. lower or equal than) the number of planar 4-valent (but not necessarily primitive) graphs generated by \texttt{genreg} \cite{meringer_fast_1999} and reported on the website \cite{meringer_regular_2009}.

For a qualitative understanding of the results of the present work, we recall some facts from the theory of large random graphs. Let $\mathcal{G}_{n,d}$ be the probability space of $d$-regular  simple graphs  with $n$ labeled vertices, where all graph occurs with equal probability \cite{wormald_models_1999,bollobas_random_2001,bollobas_modern_1998}. The graphs in $\mathcal{G}_{n,4}$ are not necessarily 1PI, or even connected,  but they are still useful for our setting in the limit $n \rightarrow \infty$ due to the following properties:

\begin{theorem}\label{thm:random_graphs}
	Consider a random 4-regular simple graph  $G \in \mathcal{G}_{n,4}$. 
	Then the  following statements  hold in the limit where $G$ becomes infinitely large, $n\rightarrow \infty$:
\begin{enumerate}
	\item (Wormald  \cite{wormald_asymptotic_1981,bollobas_random_2001})  $G$ is almost surely 4-edge connected, and the probability of being not 4-edge connected scales like $n^{-2}$.
	\item (mentioned in \cite{wormald_models_1999} as known) If $F$ is a graph with more edges than vertices, then   $G$ almost surely does not contain a subgraph isomorphic to $F$. 
	\item (Bollobás, Wormald  \cite{bollobas_probabilistic_1980,wormald_asymptotic_1981a}) Let $X_{k}$ be the number of cycles of length $k\geq 3$ in  $G$. Then  the $X_k$ are independent Poisson distributed random variables with means $\frac{3^k}{2k}$.
	\item (Bender\&Canfield, Bollobás \cite{bender_asymptotic_1978,bollobas_probabilistic_1980,bollobas_asymptotic_1982}) The number  of  unlabeled 4-regular simple graphs with $n=L-1$ vertices  grows like
	\begin{align*}
	 \frac{1}{4! (L+2)}\frac{\Gamma(L+3) e^{-\frac{15}{4}}}{\sqrt 2 \pi }\left( \frac 2 3 \right) ^{L+3} \left( 36+ \mathcal O \left( \frac 1 L \right)   \right) . 
	\end{align*}
	\item (Bonichon et. al. \cite{bonichon_planar_2006}) The set of unlabeled planar graphs with $n$ vertices, which contains as a subset the 4-regular planar graphs with $n$ vertices, grows asymptotically slower than $31^n$.
\end{enumerate}
Many more results of this type can be found in \cite{wormald_models_1999}.
\end{theorem}

\noindent
\Cref{thm:random_graphs} helps to interpret the numbers  reported in \cref{tab:periods_count} for larger loop order:
\begin{enumerate}
	\item Most of the 1PI simple completions are 4-edge connected, in accordance with \cref{thm:random_graphs}(1).
	\item Most of the 1PI simple completions are primitive. Having excluded multiedges and tadpoles, this can be understood from \cref{thm:random_graphs} (2): Each possibly divergent subgraph has asymptotically zero probability of being present, and by \cref{thm:random_graphs} (4), the number of such subgraphs is relatively small since subgraphs have lower  $L$ than the original graph. See also the explicit results in \cite{cvitanovic_number_1978,borinsky_renormalized_2017}.
	\item For the completions, and even more strikingly for the decompletions in \cref{tab:periods_count}, the proportion of planar graphs is small at higher loop order. This is plausible  from   \cref{thm:random_graphs} (5), namely that the number of planar graphs grows only exponentially, not factorially.
\end{enumerate}

\subsection{Asymptotics of symmetry factors}\label{sec:asymptotics_symmetry}

Symmetry factors, or equivalently graph automorphisms, play a crucial role in the  QFT perturbation series \cref{perturbation_series} at low loop orders. 
 We define the sum of all symmetry factors of a loop order, taking into account the $4!$ different orientations of a decompletion, by
\begin{align}\label{def:NLAut}
	N^{(\Aut)}_L &:= \sum_{\substack{g \text{ decompletion}\\{L \text{ loops}}}} \frac{4!}{\abs{\Aut g}}= \sum_{ \substack{G \text{ completion}\\  L \text{ loops}}} \frac{4! \left( L_G+2 \right)  }{\abs{\Aut G}}.
\end{align}
Here, the sums are over non-isomorphic graphs and the automorphisms do not have any fixed vertices. Recall  that the loop order of a completion is defined to be the number of loops of its decompletions   and that, by \cref{vertex_count}, a $L$-loop completion has $n=L+2$ vertices.

The great advantage of $N^{(\Aut)}_L$ over $N^{(C)}_L$ and $N^{(D)}_L$ (\cref{def:NL}) is that $N^{(\Aut)}_L$ can be obtained directly from a manipulation of the various generating functions of graphs  \cite{cvitanovic_number_1978,borinsky_renormalized_2017}, without explicitly generating the graphs themselves. Consequently, the values $N^{(\Aut)}_L$ and their asymptotic behavior, including subleading corrections, can be calculated to almost arbitrary order.

\begin{theorem}\label{thm:asymptotic_symmetry_factor}
	The following statements hold asymptotically in the limit $L\rightarrow \infty$ for completions, where $L$ is the loop number of the decompletion:
	\begin{enumerate}
		\item (Bollobás, McKay \& Wormald \cite{bollobas_asymptotic_1982,mckay_automorphisms_1984}) A graph $G\in \mathcal{G}_{n,4}$ almost surely has no non-trivial automorphism, $\abs{\Aut(G)}\rightarrow 1$.
		\item (Borinsky \cite{borinsky_renormalized_2017}) The sum of symmetry factors of  1PI primitive decompletions, \cref{def:NLAut},  grows like $\bar N^{(\Aut)}_L \left( 1+ \mathcal{O}\left( L^{-2}  \right)   \right)$, where
		\begin{align*} 
		  \bar N^{(\Aut)}_L &:=\frac{\Gamma(L+3) e^{-\frac{15}{4}}}{\sqrt 2 \; \pi} \left( \frac 2 3  \right) ^{L+3} \left( 36 - \frac 32 \frac{243}{2}\frac{1}{L+2} \right) .
		\end{align*}
	\end{enumerate}
\end{theorem}

We used the methods of \cite{cvitanovic_number_1978,borinsky_renormalized_2017} to determine $N^{(\Aut)}_L$ (\cref{def:NLAut}) for $L\leq 18$, see  \cref{tab:aut}. For loop orders $L\geq 9$, $ N^{(\Aut)}_L$ differs from its asymptotics $\bar N^{(\Aut)}_L$ (\cref{thm:asymptotic_symmetry_factor}) by no more than 1\%.

When symmetry factors become trivial, the sum of symmetry factors  $N^{(\Aut)}_L$ approaches the sum of graphs  $N^{(C)}_L$ (\cref{def:NL}), as reflected by the coincidence of their leading asymptotics \cref{thm:random_graphs} (4) and \cref{thm:asymptotic_symmetry_factor} (2). As  $N^{(\Aut)}_L$ is much easier to compute, this leads to the question: How good is the approximation of $N^{(C)}_L$ by $N^{(\Aut)}_L$ at finite loop order? The corresponding ratio, by \cref{def:NLAut}, can equivalently be interpreted as the average symmetry factor,
\begin{align}\label{def:average_symfactor}
	\frac{N^{(\Aut)}_L}{4! (L+2)N^{(C)}_L} &=  \frac{1}{N^{(C)}_L}\sum_{ \substack{G \text{ completion}\\  L \text{ loops}}} \frac{1 }{\abs{\Aut (G)}} = \left \langle \frac 1 {\abs{\Aut}} \right \rangle .
\end{align}
 From \cref{thm:asymptotic_symmetry_factor} (1), this quantity is expected to converge to unity.

We used \texttt{nauty} \cite{mckay_practical_2014} to compute the automorphism groups of all $1.04$ billion primitive completions up to $L=16$ loops.  We verified that the sum of symmetry factors for each loop order coincides, according to \cref{def:NLAut},  with the value of $N^{(\Aut)}_L$ determined earlier. 
The ratio \cref{def:average_symfactor} is reported in \cref{tab:aut}, we see that it indeed converges to unity, but considerably slower than   the ratio between $N^{(\Aut)}_L$ and its asymptotics $\bar N^{(\Aut)}_L$, the relative difference being $\geq 5\%$ as high as $L \leq 15$. 

For future reference, it will also be useful to know the averages
\begin{align*}
	\left \langle \abs{\Aut} \right \rangle  := \frac{1}{N^{(C)}_L}\sum_{\substack{G \text{ completion}\\ L \text{ loops} }} \abs{\Aut(G)}, \qquad 
	\left \langle \ln \abs{\Aut} \right \rangle   :=  \frac{1}{N^{(C)}_L}\sum_{ \substack{G \text{ completion}\\ L \text{ loops}}}  \ln \abs{\Aut(G)}.
\end{align*}
Note that, since $x^{-1}$ and $\ln x$ are non-linear functions of $x$, the relationship between the three averages of symmetry factors is non-trivial, as can be seen in \cref{tab:aut}.

Apart from the average size of automorphism groups, it is also interesting to know how many graphs have a trivial automorphism group, or symmetry factor unity. The number is reported in \cref{tab:aut}, normalized to the total number of graphs. No  graphs have trivial symmetry factor for $L\leq 7$, but around half of the graphs have trivial symmetry factor for $L=10$, and more than 90\%  for $L \geq 15$.

\begin{table}[htbp]
	\centering
	\begin{tblr}{  vlines, 
			hline{1}={solid},
			hline{2,Z}={solid},
			cells = {  font= \fontsize{11pt}{12pt}\selectfont  },
			rowsep=0pt,
			columns={halign=r}, 
			row{1}   = {  halign=c,valign=m,font = \fontsize{10pt}{12pt}\selectfont  ,mode=math}
		}
		L    &     N^{(\Aut)}_L   &  \frac{N^{(\Aut)}_L}{\bar N^{(\Aut)}_L}   &  \left \langle \frac{1}{\abs{\Aut}} \right \rangle     &  \left \langle \abs{\Aut } \right \rangle   &     \left \langle \ln \abs{\Aut } \right \rangle    &  \frac{ \text{\#}(\abs{\Aut}=1)}{N^{(C)}_L}    & \left \langle D^\text{rel} \right \rangle  \\
		5  & $\scriptstyle \frac{31}{2}$                 & 1.494 & 0.04613095 & 31.000000 & 3.25512917 & 0        & 0.214286  \\
		6  & $\scriptstyle \frac{529}{6}$                & 1.201 & 0.09184027 & 240.00000 & 3.29312666 & 0        & 0.250000 \\
		7  & $\scriptstyle \frac{2277}{4}$               & 1.085 & 0.18824405 & 15.000000 & 2.16260435 & 0        & 0.349206  \\
		8  & $\scriptstyle \frac{16281}{4}$              & 1.031 & 0.34610969 & 17.551020 & 1.52008078 & 0.081633 & 0.506123  \\
		9  & $\scriptstyle \frac{254633}{8}$             & 1.006 & 0.53112276 & 4.0088106 & 0.89829287 & 0.281938 & 0.676011  \\
		10 & $\scriptstyle \frac{2157349}{8}$            & 0.995 & 0.69154313 & 3.3589365 & 0.56087794 & 0.497784 & 0.805884 \\
		11 & $\scriptstyle \frac{39327755}{16}$          & 0.990 & 0.81034314 & 1.7121991 & 0.31139120 & 0.670438 & 0.902125 \\
		12 & $\scriptstyle \frac{383531565}{16}$         & 0.990 & 0.87745449 & 1.4215362 & 0.19321715 & 0.780272 & 0.950152  \\
		13 & $\scriptstyle \frac{7968073183}{32}$        & 0.991 & 0.91534365 & 1.2316028 & 0.12832106 & 0.843603 & 0.974790  \\
		14 & $\scriptstyle \frac{87837901499}{32}$       & 0.992 & 0.93616672 & 1.1601109 & 0.09456207 & 0.879872 & 0.985926  \\
		15 & $\scriptstyle \frac{6145866829559}{192}$    & 0.994 & 0.94869319 & 1.1209582 & 0.07484757 & 0.902197 & 0.991283  \\
		16 & $\scriptstyle \frac{75602802752227}{192}$   & 0.995 & 0.95690877 & 1.0982303 & 0.06226253 & 0.917166 & 0.993932  \\
		17 & $\scriptstyle \frac{1957218888255863}{384}$ & 0.997 &       & & & & \\
		18 & $\scriptstyle\frac{26597510458983139}{384}$ & 0.998 &       & & & & \\
	\end{tblr}
	\caption{ Loop order \| Sum of symmetry factors \| Ratio with the asymptotics \cref{thm:asymptotic_symmetry_factor}, including the $\frac 1 {L+2}$ correction \| Average symmetry factor   \|  Average size of the automorphism group \| Average logarithm of the size of the automorphism group \|  Proportion of graphs with symmetry factor unity \| Average relative number of non-isomorphic decompletions.  }
	\label{tab:aut}
\end{table}

As discussed in \cref{sec:automorphism}, a completion can have between 1 and $(L+2)$ decompletions. Similarly to \cref{def:average_symfactor}, the average number of decompletions per completion can equivalently be expressed as a ratio of the counts of decompletions and completions, 
\begin{align}\label{def:Drel}
	\left \langle D^\text{rel}\right \rangle   := \left \langle \frac{\#( \text{decompletions of }G)}{L_G + 2} \right \rangle  =\frac{N^{(D)}_L}{(L+2)N^{(C)}_L}.
\end{align}
For growing loop number, this ratio is expected to approach unity from below, expressing that for large loop orders, a completion with $L+2$ vertices on average gives rise to almost $(L+2)$ non-isomorphic decompletions, or equivalently, almost all $(L+2)$ decompletions are non-isomorphic. 

\begin{figure}[htb]
	\centering
	\begin{subfigure}[b]{.49 \textwidth}
		\includegraphics[width=\textwidth]{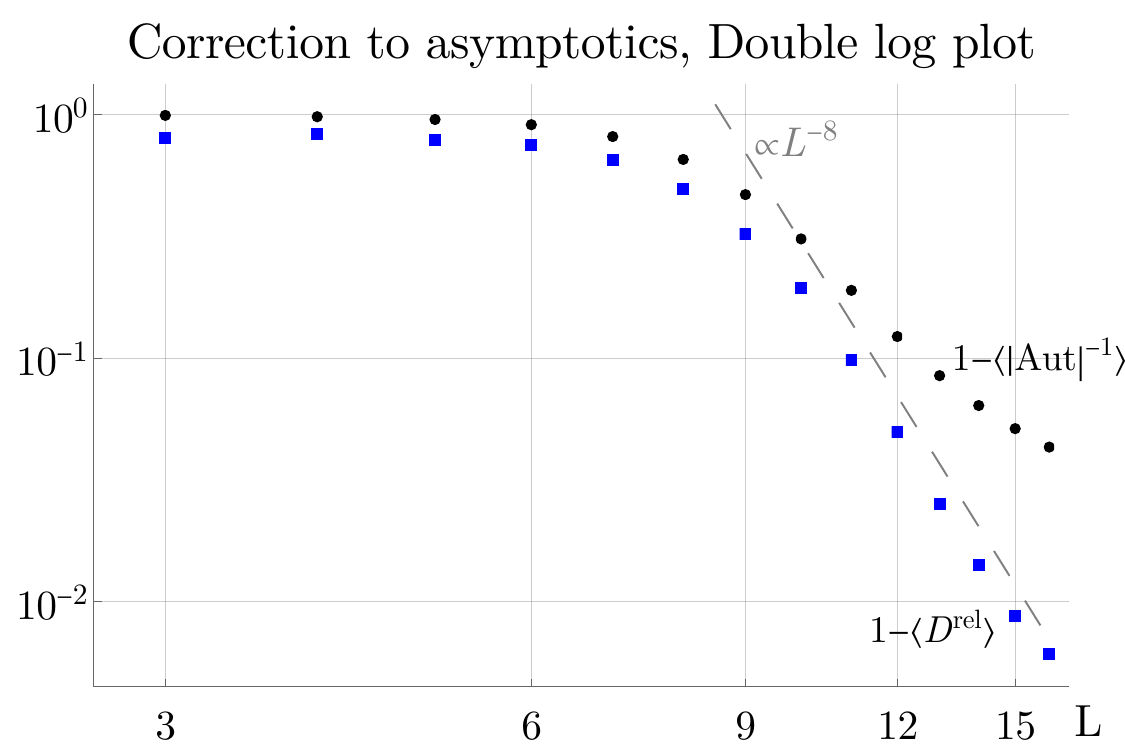}
		\subcaption{}
		\label{fig:ratio_correction}
	\end{subfigure}
	\begin{subfigure}[b]{.49 \textwidth}
		\includegraphics[width=\textwidth]{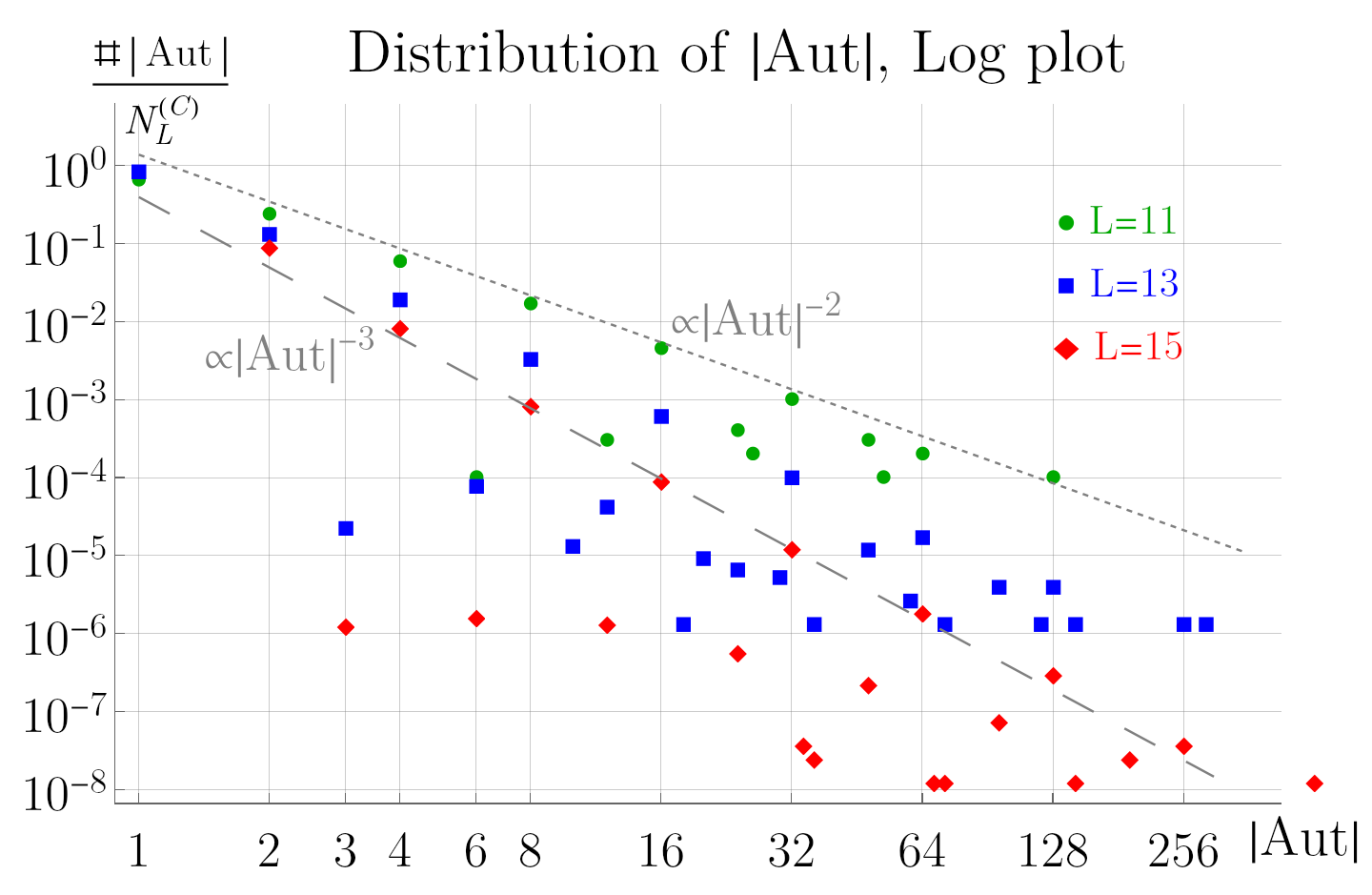}
		\subcaption{}
		\label{fig:aut_distribution}
	\end{subfigure}
	\caption{(a) Deviations of $ \left \langle D^\text{rel}\right \rangle $ and $\left \langle \frac{1}{\abs{\Aut}} \right \rangle  $ from their asymptotic value unity. Note the abrupt change in slope at $L_\text{crit} \approx 9$ loops. 
	(b) Distribution of the size of the automorphism group. The relative prevalence of large groups decays polynomially with growing $\abs{\Aut}$, and the decay becomes steeper as the loop number increases, from $\sim \abs{\Aut}^{-2}$ at $L=11$ to $\sim \abs{\Aut}^{-3}$ at $L=15$. }
\end{figure}

The data in \cref{tab:aut} shows that the convergence of $	\left \langle D^\text{rel}\right \rangle$  (\cref{def:Drel}) is significantly faster than the convergence of the average symmetry factor \cref{def:average_symfactor}. 
For both ratios, we examined the deviation from their asymptotic value unity, see \cref{fig:ratio_correction}. One might expect a $\frac 1 L$ correction, therefore we choose a double logarithmic plot. Interestingly, we do not find such a $\frac 1 L$ scaling. Instead, the correction is large below some \enquote{critical} loop number $L_\text{crit} \approx 9$, and falls off sharply for $L > L_\text{crit}$, and the decay is much faster than $ \frac 1 L$, perhaps even  exponential.

Having examined various averages, we also look at the concrete distribution of symmetry factors.
In \cref{fig:aut_distribution}, we show the count of completed graphs for each size of the automorphism group, i.e. the distribution of $\abs{\Aut(G)}$. The larger automorphism groups are less frequent than the smaller ones, the prevalence falls off roughly following a power law, where the power is smaller (the decay steeper) for higher loop orders. Nevertheless, in absolute numbers, there are more graphs with large symmetry group at higher loop order. For example, there are 24 graphs with $\abs{\Aut(G)}=128$ at 15 loops, but only three at 13 loops.

\begin{figure}[htb]
	\centering
	\begin{subfigure}[b]{.49 \textwidth}
		\includegraphics[width=\textwidth]{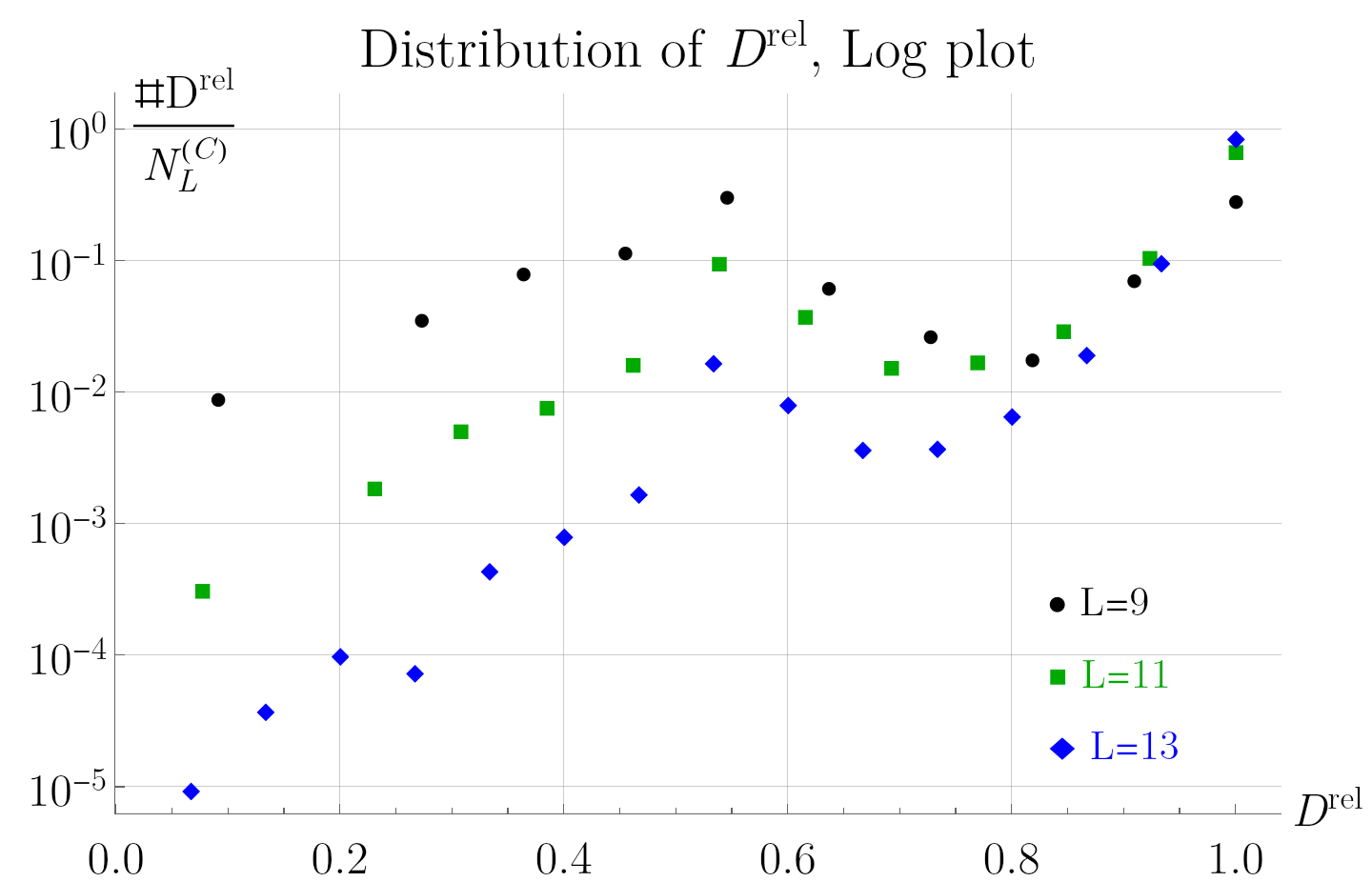}
		\subcaption{}
		\label{fig:Drel_distribution}
	\end{subfigure}
	\begin{subfigure}[b]{.49 \textwidth}
		\includegraphics[width=\textwidth]{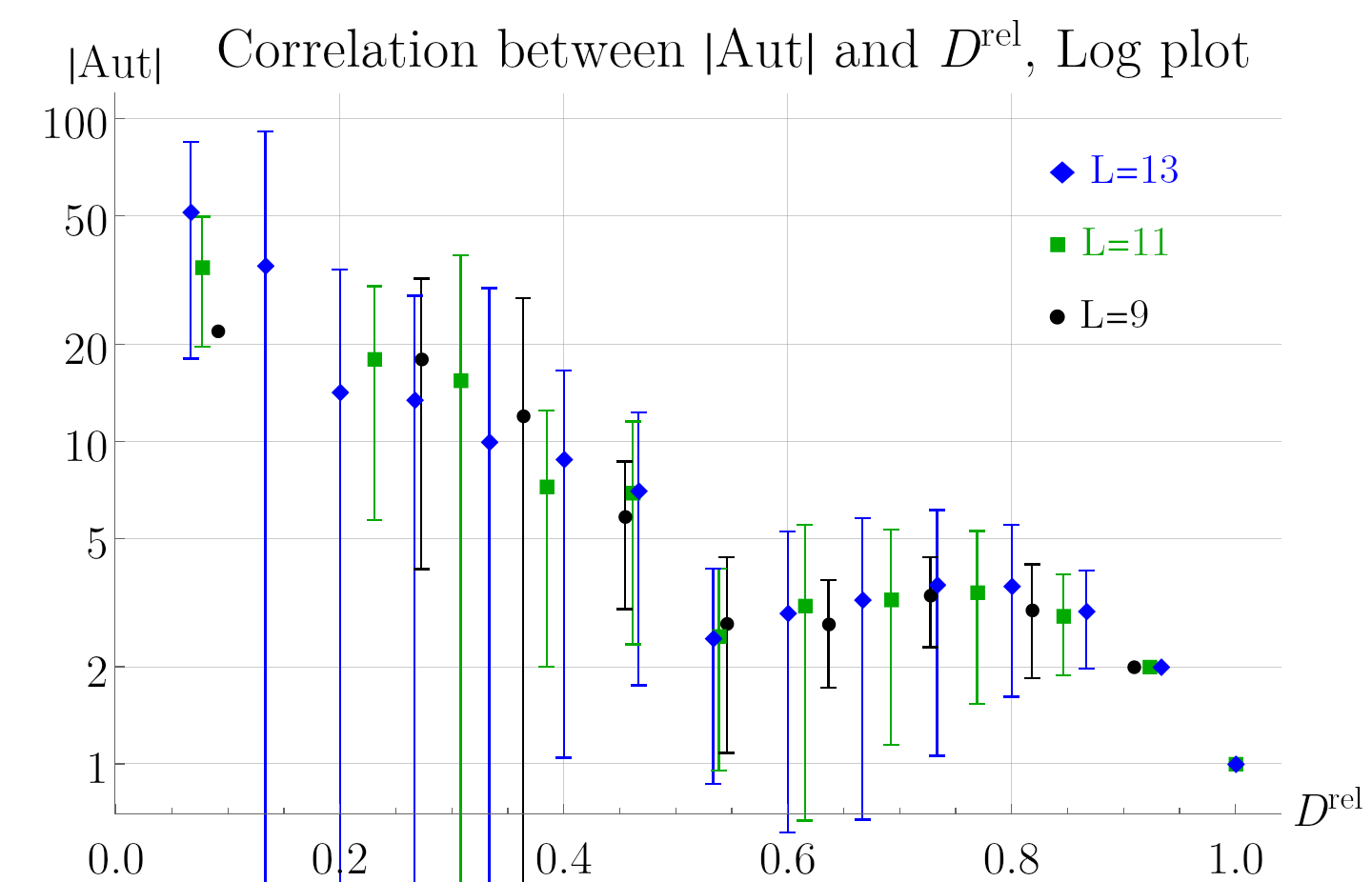}
		\subcaption{}
		\label{fig:Drel_correlation}
	\end{subfigure}
	\caption{(a) Number of graphs for a given $D^\text{rel}$, scaled to the  total number of graphs of that loop order. Most graphs have $D^\text{rel}(G)=1$ as expected, but note the second local maximum at $D^\text{rel}=\frac 12$.   \\
	(b) Correlation between $D^\text{rel}(G)$ and $\abs{\Aut(G)}$. Error bars correspond to the standard deviation. The relation is approximately exponential (linear in the log plot), with notable fluctuations. }
	\label{fig:Drel}
\end{figure}

An analogous behavior is observed for the distribution of $D^\text{rel}$, see \cref{fig:Drel_distribution}, but all slopes are inverted since $D^\text{rel} \leq 1$ and the limit is $\abs{D^\text{rel}} \rightarrow 1$. Remarkably, the distribution shows a local maximum at $D^\text{rel}=\frac 12$, apart from the global maximum at $D^\text{rel}=1$. Moreover, the overall tendency is linear in the log plot \cref{fig:Drel_distribution}, hinting at an exponential decay instead of the polynomial one in \cref{fig:aut_distribution}. A possible explanation for this is that $\abs{\Aut}$ is not bounded, i.e. the slopes in \cref{fig:aut_distribution} would be different if we were to normalize $\abs{\Aut}$ to the unit interval for each $L$.

Finally, the relation between $\abs{\Aut(G)}$ and $D^\text{rel}(G)$ is shown in \cref{fig:Drel_correlation}. We expect that if a graph has large automorphism group, then many of its decompletions will be isomorphic, so $D^\text{rel}(G)$ will be small. This is indeed what happens on average, again the functional dependence is roughly exponential with a notable deviation at $D^\text{rel}=\frac 12$ and significant fluctuations.

\FloatBarrier

\subsection{Planar graphs}\label{sec:planar_graphs}

While the set of primitive 1PI Feynman graphs can be reasonably well approximated by the set of random 4-regular graphs, the mathematical literature on the set of \emph{planar} 4-regular graphs is sparse. 

Planar 2-connected graphs are known to have exponentially large automorphism groups in the asymptotic limit  \cite{bender_number_2002}. Conversely, our empirical  data  in \cref{tab:aut_planar} shows that for $\phi^4$-theory, the average symmetry factor of planar decompletions  is not significantly different from the one of the full set of decompletions; in fact, it is slightly closer to unity than in the non-planar case. The set of all 2-connected planar graphs is therefore not a good model for our case since planar decompletions of $\phi^4$-theory constitute only a very small subset.

The number of decompletions per completion depends on the particular graph, consequenly, the averages taken over   completions, reported in \cref{tab:aut}, do not necessarily coincide with the averages taken over decompletions. With $\left \langle  \cdots \right \rangle _\text{dec}$, we indicate that an average is taken over decompletions,  and with $\left \langle \cdots \right \rangle _{\text{pla}}$  over planar decompletions, as in 
\begin{align} \label{def:mean_dec}
	\left \langle \abs{\Aut}    \right \rangle _\text{dec} &:= \frac{1}{  N^{(D)}_L  } \sum_{\substack{ \text{decompletions }g \\ L \text{ loops} }}  \abs{\Aut(g)} ,  \\
	\left \langle  \abs{\Aut } \right \rangle _\text{pla} &:= \frac{1}{  \# \text{planar decompletions} } \sum_{\substack{ \text{planar dec. }g \\ L \text{ loops} }}  \abs{\Aut (g)}. \nonumber
\end{align}
Recall that even if a completion is non-planar, some or all of its decompletions can be planar, as happens for the example in  \cref{fig:completion_decompletion}. 
Averages over decompletions are reported in \cref{tab:aut_planar}, note that their concrete numerical values differ from \cref{tab:aut} in particular for low loop orders, but the qualitative behavior for large $L$ is similar to that of the averages over completions.

\begin{table}[htb]
	\centering
	\begin{tblr}{  vlines,  
			vline{5}={1}{-}{solid},
			vline{5}={2}{-}{solid},
			hline{1}={solid},
			hline{2,Z}={solid},
			rowsep=0pt,
			cells={font=\fontsize{10pt}{12pt}\selectfont },
			columns={halign=r},
			row{1}   = { rowsep=2pt, halign=c, valign=m,font = \fontsize{10pt}{12pt}\selectfont  }
		}
		$L$    &  $\left \langle \abs{\Aut } \right \rangle_\text{dec}  $ &    $\left \langle \ln \abs{\Aut } \right \rangle_\text{dec}  $  & $\left \langle \frac{1}{\abs{\Aut}} \right \rangle_\text{dec}  $ &$\left \langle D^\text{rel}_\text{pla} \right \rangle $ & $\left \langle \abs{\Aut } \right \rangle_\text{pla} $ &    $\left \langle \ln \abs{\Aut } \right \rangle_\text{pla}   $  & $\left \langle \frac{1}{\abs{\Aut}} \right \rangle_\text{pla}   $ \\
		5 &   10.000000  & 1.98354752 & 0.2152778 & 0.14285714 & 9.0000000 & 1.73286795 & 0.2812500 \\
		6 &  17.300000  & 1.43861918 & 0.3673611 & 0.12500000 & 3.0000000 & 0.91286964 & 0.4833333 \\
		7 & 3.6818181 & 0.86643398 & 0.5390625 & 0.15079365 & 4.4210526 & 0.91203576 & 0.5476974 \\
		8 &   2.4112903 & 0.55107975 & 0.6838458 & 0.11836735 & 1.9310345 & 0.39437684 & 0.7640086 \\	
		9 &  1.8329383 & 0.35800954 & 0.7856715 & 0.09411294 & 1.9148936 & 0.34142627 & 0.7981161 \\
		10 &   1.4272949 & 0.22310430 & 0.8581177 & 0.05416051 & 1.3034091 & 0.16068412 & 0.8964489 \\
		11 &  1.2842496 & 0.15583254 & 0.8982601 & 0.02866615 & 1.3066520 & 0.13870596 & 0.9116431 \\
		12 &  1.1953716 & 0.11434124 & 0.9234890 & 0.01282297 & 1.1508633 & 0.08456242 & 0.9435972 \\
		13 &  1.1473675 & 0.08968771 & 0.9390163 & 0.00530868 & 1.1300771 & 0.07254937 & 0.9512887 \\
		14 &  1.1172909 & 0.07338441 & 0.9495301 & 0.00201826 & 1.0919890 & 0.05557029 & 0.9620677 \\ 
		15 &  1.0972817 & 0.06198054 & 0.9570355 & 0.00072221 & 1.0823823 & 0.04984923 & 0.9658697 \\
		16 &  1.0827565 & 0.05342383 & 0.9627510 & 0.00024319 & 1.0692237 & 0.04316497 & 0.9702493 \\
	\end{tblr}
	\caption{Averages of automorphisms of decompletions according to \cref{def:mean_dec}. Loop order \|  3 columns referring to all decompletions \| \|  4 columns   referring to planar decompletions.}
	\label{tab:aut_planar}
\end{table}
 
\begin{figure}[htb]
	\centering
	\begin{subfigure}[b]{.47 \textwidth}
		\includegraphics[width=\textwidth]{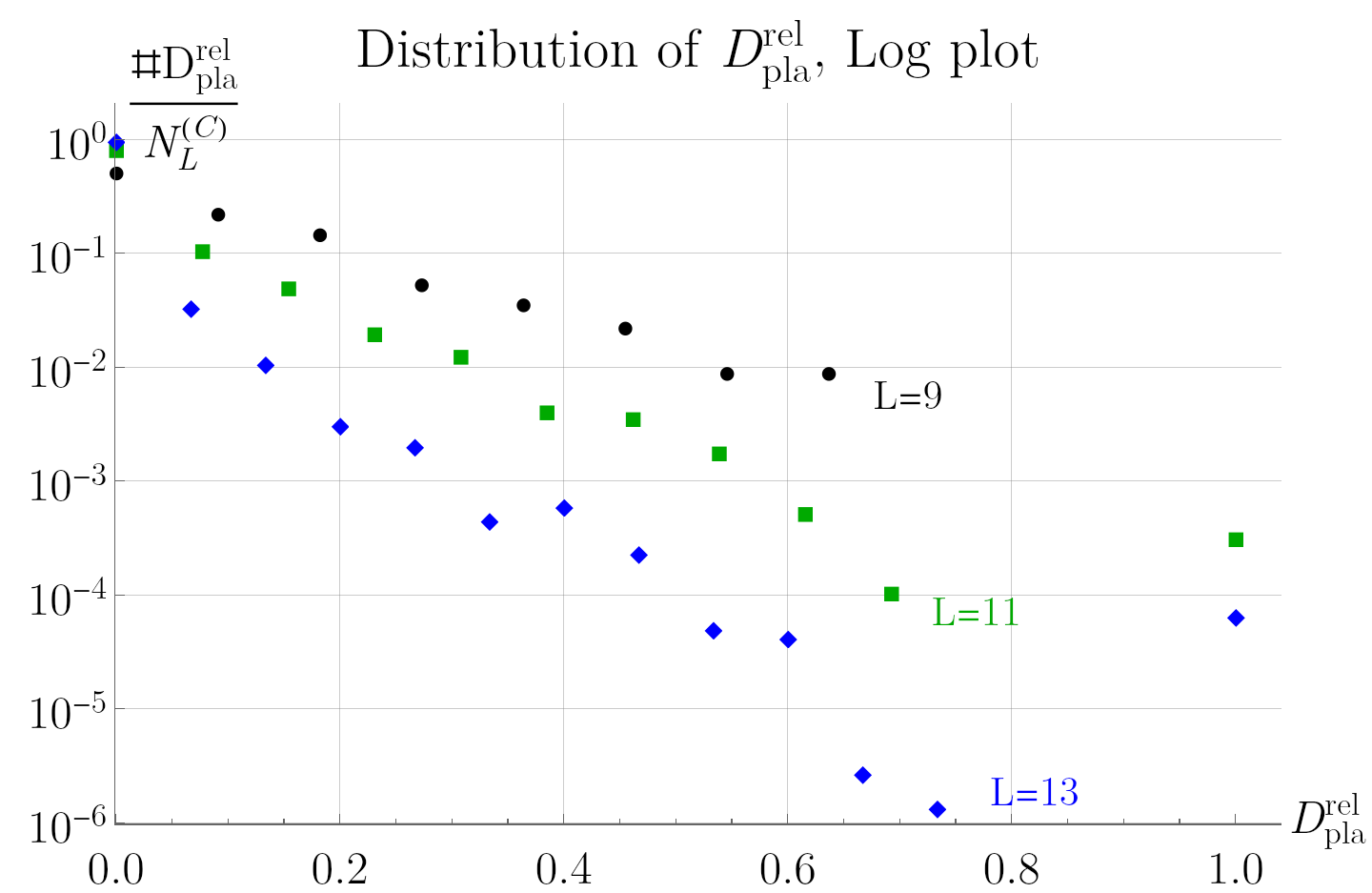}
		\subcaption{}
		\label{fig:Drelpla_distribution}
	\end{subfigure}
	\begin{subfigure}[b]{.47 \textwidth}
		\includegraphics[width=\textwidth]{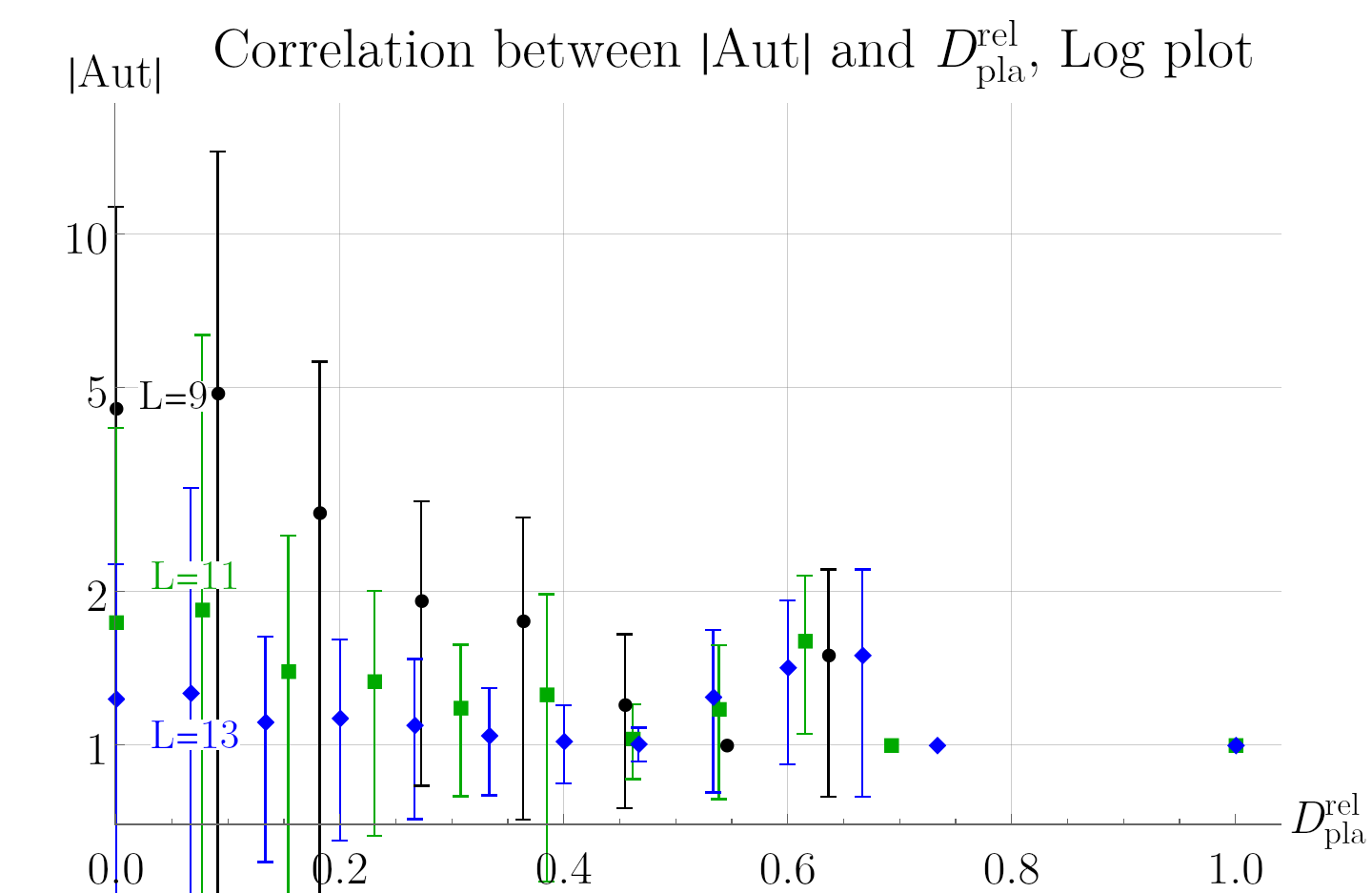}
		\subcaption{}
		\label{fig:Drelpla_correlation}
	\end{subfigure}
	\caption{Plot analogous to \cref{fig:Drel}, but for \emph{planar} decompletions (\cref{def:Drel_planar}). Note that the slope of (a) is reversed compared to \cref{fig:Drel_distribution}, and that the correlation with $\abs{\Aut}$ shown in (b) is much weaker than in \cref{fig:Drel_correlation}.}
	\label{fig:Drelpla}
\end{figure}

In analogy to \cref{def:Drel}, we define the relative count of planar decompletions,
\begin{align}\label{def:Drel_planar}
	 D^\text{rel}_\text{pla} (G)   := \frac{\#( \text{planar decompletions of }G)}{L_G + 2}   \leq D^\text{rel}(G).
\end{align}
This quantity converges to zero as $L\rightarrow \infty$. As seen in \cref{fig:Drelpla_distribution}, the distribution indeed is concentrated near $ D^\text{rel}_\text{pla}=0$.  Note that $ D^\text{rel}_\text{pla}$ does not show a local maximum at $\frac 12$ as does $D^\text{rel}$ (\cref{fig:Drel_distribution}). Moreover, the correlation between $D^\text{rel}_\text{pla}$ and $\abs{\Aut}$, plotted in \cref{fig:Drelpla_correlation}, is almost negligible, again in contrast to the behavior of $D^\text{rel}$ in \cref{fig:Drel_correlation}.

\FloatBarrier

\section{Period symmetries and numerical integration}\label{sec:symmetries}

\subsection{Period symmetries}\label{sec:period_symmetries}

By a \emph{symmetry of the period}, we mean an operation $ G_1 \mapsto G_2$ such that $\period(G_1)=\period(G_2)$ and both $G_1$ and $G_2$ are primitive decompletions of $\phi^4$-theory and not isomorphic to each other.
Several symmetries of the period are known, we briefly list them below. A more precise description and illustrations can be found in the literature cited or in \cite{panzer_hepp_2022}.
\begin{description}
	\item[Completion] \cite{schnetz_quantum_2010} If two decompletions $G_1$ and $G_2$ have the same completion, then    $\period(G_1)=\period(G_2)$. In view of this identity, we define $\period(G)$, for a completion $G$, to mean the period of any, and hence all, of its decompletions.
	\item[Product] \cite{schnetz_quantum_2010} Consider two completions $G_1$ and $G_2$, each of which has at least one triangle $t_1\in G_1,t_2\in G_2$. Removing the edges of the triangles produces three 2-valent vertices in each graph. Merge those vertices pairwise to obtain a graph $G_1 \times G_2$, called the product, which has $\period(G_1\times G_2)=\period(G_1) \cdot \period(G_2)$. Conversely, every completed graph which has a 3-vertex cut is a product of two smaller graphs. 
	\item[Fourier] \cite{broadhurst_exploiting_1986} If the decompletion $G_2$ is isomorphic to the planar dual $\tilde G_1$ of a decompletion $G_1$, then $\period(G_2)=\period(G_1)$. This symmetry expresses a Fourier transform of the Feynman integral.
	\item[Extended Fourier]\cite{schnetz_quantum_2010} In full generality,   planar duality operates on integrands non-unit edge weights, where a negative weight corresponds to a factor in the numerator, see e.g. \cite{golz_graphical_2017,borinsky_graphical_2022}.   Fourier symmetry holds even in this more general setting, where the planar dual $\tilde G_1$ of a decompletion might not be  a $\phi^4$ decompletion itself. One may still construct the completion  $\tilde G_1\cup \left \lbrace w \right \rbrace $, and pick a vertex $v\neq w$ to construct the decompletion $G_2$ of this graph. Even if $G_2$ still is not a decompletion of $\phi^4$ theory, one might now be able to reach a valid decompletion $\tilde G_2$ by computing the planar dual of $G_2$, and iterating this procedure. 
	\item[Schnetz Twist]\cite{schnetz_quantum_2010} Consider a 4-vertex cut  of a completion $G_1=G_A \cup \left \lbrace v_1,v_2,v_3,v_4 \right \rbrace \cup G_B$.  Pick two of the vertices $\left \lbrace v_1,v_2 \right \rbrace $, and, in $G_B$, change those edges going to $v_1$ to go to $v_2$, and those originally going to $v_2$ to go to $v_1$. Repeat with the remaining two vertices $\left \lbrace v_3,v_4 \right \rbrace $. One might need to introduce integral numerators (=edges with negative weight) to make the resulting graph 4-regular, see e.g. \cite{borinsky_graphical_2022}. If the resulting graph is 4-regular without such negative weights, it is a $\phi^4$ period which coincides with $\period(G_1)$.
	\item[Fourier split] \cite{hu_further_2022} Consider a 4-vertex cut  of a completion $G_1=G_A \cup \left \lbrace v_1,v_2,v_3,v_4 \right \rbrace \cup G_B$.  Remove $v_4$ and split the remaining graph at the three vertices.   Introduce three new edges $\left \lbrace e_1,e_2,e_3 \right \rbrace   $ to $G_B$, forming a triangle at the three cut vertices such that $e_i$ is not incident to $v_i$. If the resulting graph $G'_B$ is planar, take its dual. The triangle becomes a star $\left \lbrace e_1', e_2', e_3' \right \rbrace   $. Remove the central vertex of the star, and let $v_i'$ be the unique remaining vertex incident to $e_i'$.  Join these three vertices pairwise to the three cut vertices of $G_A$, where $v_i'$ is joined to $v_i$.  If the resulting graph is a decompletion of a 4-regular graph $G_2$ then $\period(G_1)=\period(G_2)$. The Fourier split can be promoted to an \enquote{extended} version similarly to the ordinary Fourier symmetry.
\end{description}

\noindent
All the above symmetries are known, and all except the \enquote{extended} versions had been implemented in previous work. Our own contribution is threefold:
\begin{enumerate}
	\item 	We wrote a new \texttt{C++} implementation of all known symmetries, using \texttt{nauty} \cite{mckay_practical_2014} to check for graph isomorphisms and the \texttt{boost graph library} \cite{siek_boost_2001} for planar duals. Being independent of all previous implementations, this serves as a check for the previously obtained counts of symmetries.
	\item We include a version of the \enquote{extended} Fourier symmetry. Concretely, we compute the planar dual $\tilde G_1$ of a decompletion, and if $\tilde G_1$ has exactly one vertex $v$ of valence greater than 4, we complete $\tilde G_1$ and remove $v$, reaching a new graph $\tilde G_2$. We then compute its dual $G_2$. If $G_2$ is a decompletion of $\phi^4$ theory, we are done. If $G_2$ again has one vertex of too high degree, we iterate this procedure until either we either reach a decompletion of $\phi^4$ theory, or  we are back to a graph that we already know, or we have visited more than 1000 intermediate graphs without finding a valid $\phi^4$ decompletion (the latter is rare and happened only for $L\geq 14$ loops). We use a similar iteration   for the Fourier split symmetry, with Fourier splits instead of planar duals for the intermediate non-$\phi^4$ decompletions.
	
	Our implementation is not the most general conceivable extension of the period symmetries, for three reasons: (1) We do not allow for intermediate graphs to have more than one vertex of valence $>4$, nor to have non-trivial edge weights, (2) we do not consider Twist symmetries for these intermediate non-$\phi^4$ graphs, and (3) we do not consider the possibility of having a Fourier-like symmetry for non-planar graphs, where intermediate objects are matroids. 
	\item We compute the symmetries for higher loop numbers than previous authors.
\end{enumerate}

Note that the computational effort for the individual symmetries is vastly different: Completions  and Fourier transforms are very fast. Extended Fourier, Product, and Twist take considerably longer because they require trying out many possible operations. The extended Fourier split is slower yet because it typically requires to construct thousands of transformation candidates for a single input graph, each of which involves non-trivial operations such as planarity tests or finding and merging particular vertices. 
Our algorithm heavily relies on caching of the graphs that have already been visited, this entails that every intermediate step needs to transform the resulting graph to \texttt{nauty}'s standard form in order to compare with the cached graphs, and it limits the performance that can be gained by parallelization as the caches need to be synchronized.

\subsection{Invariants}\label{sec:invariants}
Over the last years, there has been effort to find quantities which share the symmetries of the period, but are easier to compute or more directly related to properties of the underlying graphs. Notable examples are the $c_2$  invariant \cite{schnetz_quantum_2011,hu_further_2022} and the permanent \cite{crump_period_2016,crump_properties_2017}.

The \emph{Hepp bound} \cite{panzer_hepp_2022} is defined as the tropical approximation of the period integral \cref{def:period}. In that approximation, the nested integrals can be solved analytically and the computation of the Hepp bound reduces to a purely combinatoric operation on the subgraphs of the graph in question. The Hepp bound $\mathcal H(G)$ is a rational number. We wrote another \texttt{C++} program to compute the Hepp bound approximately (to 64-bit floating point precision) using the flag formula \cite{panzer_hepp_2022}:
\begin{align}\label{hepp_recursive}
	\mathcal H(G) &= \sum_{\substack{\text{1 PI }\gamma \subset G \\ L_\gamma= L_G-1}} \frac{\mathcal H(\gamma)}{\omega(\gamma)}= \sum_{ \text{hyperplane }\gamma \subset G } \frac{\mathcal H (\gamma)}{\omega(\gamma)}.
\end{align}
Here, $\omega(\gamma)$ is the superficial degree of divergence, a subgraph is 1 PI (bridgeless in graph theory language) if it is 2-edge connected, and a \emph{hyperplane} is an induced subgraph $\gamma \subset G$ such that both $\gamma$ and the subgraph $\gamma'$ induced from the vertices of $(G \setminus \gamma)$ are connected. Note that even if \cref{hepp_recursive} is a sum over graphs of lower loop number, these subgraphs are not superficially log divergent, therefore the subgraphs do not coincide with the graphs in our period samples at lower loop number.

The linear coefficient of the Martin polynomial (\cref{circuit_martin}) of the graph has been identified as an invariant only recently  \cite{panzer_feynman_2023}. This \emph{Martin invariant} is a sequence of invariants which are obtained from the graph by duplicating edges. This sequence is a generalization of earlier period invariants. It is conjectured \cite{panzer_feynman_2023} that the full sequences of Martin invariants  of two graphs concide if and only if the periods coincide. In the present work, we only compute  the first Martin invariant $M^{[1]}$ of the sequence (\enquote{first Martin} in \cref{tab:symmetries}). We verified for all graphs in our samples that it coincides whenever two graphs are symmetric. Conversely, the first Martin invariant alone is not sufficient to indicate the opposite direction, there are many pairs of graphs where the first Martin invariant coincides but the periods do not.

\subsection{Counts of symmetries}\label{sec:symmetries_count}

We computed the symmetries as described for all completions with $L \leq 14$ loops. For $L=15$, we computed Schnetz Twists and extended Fourier symmetries, but not the Fourier splits. Due to the product- and completion symmetries, the most interesting counts are the numbers of symmetries for 3-vertex irreducible completions, reported in \cref{tab:symmetries}. Quite often, one relation between two graphs can be obtained by more than one symmetry, this is why the counts of the individual symmetries in \cref{tab:symmetries} do not add up to the total number of independent periods.

\begin{table}[htbp]
	\centering 
		\begin{tblr}{  vlines, 
				hline{1}={solid},
				hline{2,Z}={solid},
				cells = {  font= \fontsize{11pt}{12pt}\selectfont  },
				rowsep=0pt,
				columns={halign=r},
				row{1}   = {  halign=c,valign=m,font = \fontsize{10pt}{12pt}\selectfont  }
			}
			$L$  &    Twists &F &{ Twists \& \\ Fourier}  & { Fourier \\ Split}  &  {only \\ Split}  &  {independ. \\ periods}  &   Hepp & {dupl. \\Hepp} & {first\\ Martin} \\
			5  &         0 &      0 &         0 &        0 & 0  &        1 &      1 & 0 & 1\\
			6  &         0 &      0 &         0 &        0 &  0 &        4 &      4 & 0 &4\\
			7  &         1 &      2 &         2 &        2 & 0  &        9 &      9 & 0  & 9 \\
			8  &         9 &      3 &        10 &        9 & 0  &       31 &     29 & 2 & 25 \\
			9  &        48 &     14 &        55 &       53 & 1  &      134 &    129 & 5  & 100  \\
			10 &       336 &     21 &       350 &      334 &  13 &      819 &    776 & 42 & 409 \\
			11 &      2387 &     43 &      2420 &     2276 & 70&       6197 &   6030 & 158 & 1622 \\
			12 &     18680 &     60 &     18728 &    17040 & 280  &    55196 &  54552 & 618 & 5799  \\
			13 &    155547 &     90 &    155630 &   138164 &  1077 &   543535 & 541196 & 2246 & 19278  \\
			14 &   1386809 &    117 &   1386919 &  1209177 & 4581  &  5769143 &        &  & 61757 \\
			15 &  13153096 &    184 &  13153272 &     &  & 65117118 &      &   &  192113 \\
		\end{tblr}
		\caption{Prevalence of symmetries for  3-vertex irreducible completions. \enquote{One identity} means a reduction of the  number of independent graphs by one. Loop order of decompletion \| Total number of\ldots Schnetz twist \| extended Fourier identities \| Twist and extended Fourier combined \| extended Fourier splits \| Identities which can only be explained as (extended) Split, not as combinations of non-Split symmetries \| upper bound of independent 3-vertex irreducible periods, exploiting known symmetries as described in the text \| numerically distinct Hepp bounds of irreducible graphs at machine precision \|  Unique Hepp bounds coinciding for  sets of graphs not related by known symmetries \| distinct Martin invariants $M^{[1]}$.
	 }
	\label{tab:symmetries}
\end{table}

Within a symmetry orbit, not every pair of graphs is directly related via a known symmetry, but some only via intermediate graphs.

For comparison, we list the number of distinct Hepp bounds (\cref{hepp_recursive}), and the number of Hepp bounds which coincide in cases where the period is not known to be symmetric.
The numbers of Hepp bounds in \cref{tab:symmetries} are not entirely reliable for high loop orders because we computed the Hepp bounds numericlaly to floating-point precision and counted them as equal if their relative difference was smaller than $10^{-14}$. Our result for $L\leq 11$ coincides with \cite{panzer_hepp_2022} and we expect the results for $L\in \left \lbrace 12,13 \right \rbrace $ to be at least very close to the true numbers.

We have verified in every single case that those graphs which are expected to be symmetric have the same Hepp bound (within rounding errors). Further, in many cases we numerically computed the periods of graphs which are known to be symmetric (see \cref{tab:samples}). In these cases, we compared the estimated accuracy of numerical results to the observed differences for symmetric graphs. We found that the difference are approximately normal distributed with a standard deviation close to the expected uncertainty. This indicates that the expected uncertainty is a good approximation of the deviation between the numerical result and the true period. Moreover, we take these findings as evidence that our implementation of the symmetries does not label any graphs as symmetric which really are not. 

The number of independent periods has been reported in the literature for $L\leq 11$ loops. We find our results to be  consistent  with  (i.e. smaller or equal than)   \cite{schnetz_quantum_2010} when excluding the Fourier split, and  with \cite{hu_further_2022} when including Fourier split. In fact, we find agreement if we do not use the \emph{extended} version of Fourier symmetry and Fourier split.

Since we take into account an \enquote{extended} version of the Fourier symmetry, we find more graphs related by Fourier identities than previously authors. The number of Fourier splits grows even more by including the extended Fourier symmetry, so that the number of Fourier splits which can not be explained by other identities  (\enquote{only Split} in \cref{tab:symmetries}) comes out larger than in \cite{hu_further_2022}.
Moreover, the number of coinciding Hepp bound without underlying known symmetry, reported as \enquote{dupl. Hepp} in \cref{tab:symmetries}, is slightly lower than the numbers given in \cite[Table 3]{panzer_hepp_2022} for $L\leq 10$. We note in passing that the circulant graphs (of which the zigzag \cref{zigzag_amplitude} is the most prominent one) for $6 \leq L \leq 18$ do not have any non-trivial period symmetries.

Our numbers of \cref{tab:symmetries} extend the previously known results both in terms of higher loop orders, and in terms of systematically  including extended Fourier transforms. The overall picture is that Fourier symmetries are rare (in accordance with planar graphs being rare, see \cref{tab:periods_count}), that the symmetries obtained by twist show a large overlap with the ones obtained by Fourier split, and that there are graphs where the Hepp bound coincides even if they are not known to be symmetric.

\subsection{Numerical integration}\label{sec:numerical_integration}

For the  numerical integration in this work, we have used the reference implementation of the tropical Monte Carlo quadrature for Feynman integrals developed recently by Borinsky \cite{borinsky_tropical_2023a}.\footnote{The implementation used in the present work is  available from \url{https://github.com/michibo/tropical-feynman-quadrature/}.} We slightly optimized this program for the present case of massless log-divergent graphs in $D=4$ spacetime dimensions, by removing computation steps which are only required for more general Feynman integrals. Further,  we added an automatic adjustment of the number of sample points to meet the target precision. Note that the original algorithm has recently been generalized to a much larger class of amplitudes beyond periods \cite{borinsky_tropical_2023}.\footnote{Available from \url{https://github.com/michibo/feyntrop}.}

The scale of the the project requires that  many intermediate steps be automated. In order to handle large tables of graphs or results, most of our programs directly operate on bzip2-compressed text files, where graphs are described by their Nickel index (\cref{def:nickel}). We found this format to be comparable in size to binary formats such as the native graph files of \texttt{nauty} \cite{mckay_practical_2014}, but having the advantage that files can be inspected or edited in an ordinary text editor after decompression. 

Conceptually, the workflow looks as follows:
\begin{enumerate}
	\item We use \texttt{nauty}  to generate a list of completed graphs to be integrated, and filter it for 1PI primitive ones as described in \cref{sec:samples,sec:asymptotics}.  
	\item We split this list across many computers, generate a decompletion of each graph, and integrate it with the tropical Feynman quadrature algorithm where  the number of sampling points is adjusted automatically to reach a relative target precision.
	\item We collect the results, re-do missing graphs, sort, and check if symmetries are fulfilled  as described in \cref{sec:symmetries_count}.
	\item We exploit the known symmetries to improve the numerical accuracy.
	\item We compute symmetry factors (\cref{sec:automorphism}), $O(N)$-factors (\cref{sec:beta}) and various other   properties (\cref{sec:relations}), and write the results to one file per loop order.
\end{enumerate}
All steps are implemented as \texttt{C++} programs, using \texttt{nauty} version 2.7 \cite{mckay_practical_2014} and the \texttt{boost graph library} version 1.82 \cite{siek_boost_2001}. The first item of the workflow has been discussed in \cref{sec:number_of_graphs}. Points 3 and 4 use pre-computed tables of symmetries (\cref{sec:period_symmetries}). In step 4, the program combines all graphs of the same loop order which are known to be symmetric (and whose numerical results pass step 3) by computing an average weighted by the individual uncertainties, see \cref{sec:statistics}. This program is capable of utilizing duplicate computations of the same graph  without wasting previous results. For example, we can gradually increase the accuracy of a selected set of graphs by integrating them multiple times, and the resulting improvement will consistently propagate through the whole data set of that loop order.

Apart from checking the symmetries, we compared our numerical results to the (analytical) data in \cite{panzer_galois_2017} and found agreement, where the differences have a standard deviation that matches our estimate. 

For the non-complete samples at $L \geq 14$, we furthermore used the symmetries to infer the periods of a large number of additional graphs which were not computed in the original samples. The numbers are reported in \cref{tab:samples}, but we do not use those graphs for the statistical evaluation in the subsequent chapters since they would systematically distort the random samples in favor of graphs with larger symmetry orbits. 

The numerical integrations ran on multiple machines with significantly different performance, most notably  on servers of the institute of Mathematics and a cluster of the institute of Physics at Humboldt-Universität zu Berlin. A smaller batch was done on the servers of the Faculty of Mathematics at University of Waterloo.  The total time spent on integration was well above 2 million CPU-core hours. The other tasks such as generating graphs, constructing period symmetries, or merging the results, were considerably faster, but still had run times of several days to a few weeks for the largest data sets.

\FloatBarrier

\section{Distribution of periods}\label{sec:distribution}

\subsection{Mean and Moments}\label{sec:moments}

Unless specified otherwise, we define a mean to extend over completions of a given loop order, normalized to the number of  completions  from \cref{def:NL}, 
\begin{align}\label{def:period_mean}
 \left \langle \period  \right \rangle  :=  \frac{1}{N^{(C)}_L}\sum_{\substack{ \text{completions }G \\ L \text{ loops} }} \period (G) . 
\end{align}
If our sample does not include all completions, the mean is computed by summing the graphs in the sample, and using $N$,  the number of graphs in the sample, instead of $N^{(C)}_L$. 

The samples generated by \texttt{genrang} are not distributed uniformly, but proportionally to their symmetry factor. This has been corrected for, as described in  \cref{sec:uniform_sampling}.
The influence of the non-uniform distribution on the sample mean $\left \langle \period \right \rangle $ (\cref{def:period_mean}) is smaller than  one might expect. Firstly,  at $L \geq 14$  most graphs have trivial symmetry factor (\cref{sec:symmetries_count}), and secondly, there is no strong correlation between the symmetry factor and the period (\cref{sec:relations_symmetry_factor}). For example, ignoring the effect and plainly summing over all graphs, the sample $14s$ gives rise to $\left \langle \period \right \rangle = (553.2 \pm 1.0) \cdot 10^2$, which is $4.5\%$ smaller than the true value (\cref{tab:means}).  Adding the duplicated periods, for $14s$, the sample size increases from $N=216189$ (the true sample size reported in \cref{tab:samples}) to $N_\text{corr}=230884$, but the relevant factor $\frac{1}{\sqrt{N}}$ for estimating the uncertainty decreases  by only 3\%.

The mean of the period grows quickly with the loop order, see \cref{tab:means}. In order to compare the distributions at different loop orders, we scale the results with respect to the mean,
\begin{align}\label{mean_scaling}
p(G)   &= \frac{1}{\left \langle \period\right \rangle }\period(G).
\end{align}
By construction, $\left \langle p \right \rangle =1$, and deviations from the mean can be characterized by the central moments
\begin{align}\label{def:central_moments}
	C_j  &:= \left \langle (p-1)^j \right \rangle , \qquad c_k :=  C_k\cdot C_2^{-\frac k 2}, \quad k\geq 2.
\end{align}
The standardized moments $c_3$ and $c_4$ are known as skewness and kurtosis, the latter is a measure for how \enquote{heavy-tailed} the distribution is. All moments can be converted back to the distribution prior to the scaling \cref{mean_scaling} using the value $\left \langle \period \right \rangle $. In particular, the  standard deviation of the un-scaled sample is  
\begin{align}\label{C2_std}
	\sigma &= \left \langle   \period  \right \rangle   \cdot\sqrt{ C_2}.
\end{align}

An alternative measure for deviations from the mean are quantile boundaries. In \cref{tab:means}, we give the boundaries [1\%,99\%], that is, the minimum and maximum of a subset that includes all periods except for the smallest and the largest 1\%.

\begin{table}[htb]
	\centering
	\begin{tblr}{vlines,
			hline{1}={solid},
			hline{2,Z}={solid},
			rowsep=0pt,
			cells={font=\fontsize{10pt}{12pt}\selectfont },
			columns={halign=r},
			column{1}={halign=c,mode=math},
			row{1}={halign=c,rowsep=2pt, font=\fontsize{12pt}{14pt}\selectfont, mode=math }, 
			cell{1}{5}={font=\fontsize{11pt}{13pt}\selectfont }
		}
			
		L &   \left \langle \period  \right \rangle     &  \Delta_\text{num}  &  \Delta_\text{samp} &  [1\% ~; ~~99\%]   & C_2 & c_3 & c_4  \\
		5        & $53.8013 \pm 0.0003$             & $ 3 \cdot 10^{-4}$ &                  0 &               52.02 ; 55.58  & 0.001 & 0.000 & 1.000 \\
		6        & $125.8743 \pm 0.0003$            & $ 3 \cdot 10^{-4}$ &                  0 &                71.51 ; 168.3 & 0.072 & -0.399& 1.917 \\
		7 		 & $ 315.875 \pm  0.0007 $          & $ 7 \cdot 10^{-4}$ &                  0 &                183.0 ; 527.7 & 0.089 & 0.572 & 2.613 \\
		8   	 & $ 727.559 \pm 0.001 $            & $ 1 \cdot 10^{-3}$ &                  0 &                 311.0 ; 1716 & 0.147 & 0.982 & 4.374 \\
		9   	 & $1654.858 \pm 0.003$             & $ 3 \cdot 10^{-3}$ &                  0 &                 671.0 ; 3620 & 0.185 & 1.402 & 6.940 \\
		10  	 & $3581.796 \pm 0.003$             & $ 3 \cdot 10^{-3}$ &                  0 &                  1346 ; 8757 & 0.241 & 1.629 & 8.917 \\
		11  	 & $7516.003 \pm 0.003$             & $ 3 \cdot 10^{-3}$ &                  0 & $(2.617 ; 21.17) \cdot 10^3$ & 0.311 & 1.838 & 10.46 \\
		12  	 & $15235.118  \pm 0.003$           & $ 3 \cdot 10^{-3}$ &                  0 & $(4.916 ; 50.27) \cdot 10^3$ & 0.402 & 2.124 & 12.25 \\
		13  	 & $30023.445  \pm 0.007$           & $ 7 \cdot 10^{-3}$ &                  0 & $(8.992 ; 111.8) \cdot 10^3$ & 0.511 & 2.512 & 15.12 \\
		13\text{s}   	 & $29963  \pm 298$  		        & $ 7 \cdot 10^{-2}$ &  $3  \cdot 10^{2}$ & $(8.947 ; 104.1) \cdot 10^3$ & 0.508 & 2.345 & 11.14 \\
		13 \star & $30091   \pm 217$                & $ 6 \cdot 10^{-2}$ &   $3 \cdot 10^{2}$ & $(9.056 ; 112.6) \cdot 10^3$ & 0.520 & 2.390 & 11.04 \\
		14\text{s}  	 & $(578.1  \pm 1.0) \cdot 10^{2}$  & $ 4 \cdot 10^{-2}$ &   $1 \cdot 10^{2}$ & $(1.605 ; 24.44) \cdot 10^4$ & 0.637 & 2.908 & 16.30 \\
		14 \star & $(577.9  \pm 1.4) \cdot 10^{2}$  & $ 5 \cdot 10^{-2}$ &   $2 \cdot 10^{2}$ & $(1.613 ; 24.44) \cdot 10^4$ & 0.635 & 2.880 & 15.58 \\
		15\text{s}  	 & $(1080.3 \pm 2.9) \cdot 10^{2}$  & $ 9  \cdot 10^{-2}$ &  $3 \cdot 10^{2}$ & $(2.847 ; 50.08) \cdot 10^4$ & 0.774 &3.660 & 25.73 \\
		15 \star & $(1085.8 \pm 3.8)\cdot 10^{2}$   & $ 2 \cdot 10^{-1}$ &   $4 \cdot 10^{2}$ & $(2.854 ; 50.22) \cdot 10^4$ & 0.790 & 3.560 & 24.23 \\
		16\text{s}  	 & $(196.8 \pm 1.8) \cdot 10^{3}$   & $ 6 \cdot 10^{-1}$ &   $2 \cdot 10^{3}$ & $(4.997 ; 94.29) \cdot 10^4$ & 0.853 & 3.742 & 25.02 \\
		17\text{s}  	 & $(35.8  \pm 4.9) \cdot 10^{4}$   & $ 2 \cdot 10^{0~}$ &   $5 \cdot 10^{3}$ &  $(8.327 ;181.0) \cdot 10^4$ & 1.002 & 4.238 & 30.54 \\
		18\text{s}  	 & $(65.4  \pm 2.3) \cdot 10^{4}$   & $ 4 \cdot 10^{0~}$ &   $3 \cdot 10^{4}$ & $(1.588 ; 35.37) \cdot 10^5$ & 1.122  & 4.373 & 30.15 \\
	\end{tblr}
	\caption{Mean of the period (\cref{def:period_mean}) with combined numerical and sampling uncertainty. The exact values \cite{schnetz_numbers_2018} for $L\in \left \lbrace 5,6,7 \right \rbrace $ are, to 10 digits, $\left \lbrace  53.80156133, ~125.8743179,~ 315.8747617 \right \rbrace   $   \| numerical uncertainty \| sampling uncertainty  \| 1\% and 99 \% quantiles \| second central moment of the normalized distribution \| skewness of the normalized distribution \| kurtosis of normalized distribution.  }
	\label{tab:means}
\end{table}

There are two different sources of uncertainty for our results. Firstly, the  uncertainty of the numerical integration, stated in \cref{tab:samples}, give rise to an uncertainty  $\Delta_\text{num}$.
Secondly, for the non-complete samples, there is a sampling uncertainty $\Delta_\text{samp}$ which expresses that the empirical mean of a sample can deviate from the mean of the underlying population. Details are discussed in \cref{sec:statistics}. We find that for all non-complete samples, even if they contain 100000s of graphs, the statistical uncertainty by far dominates the numerical one, as evidenced   in \cref{tab:means}.

For the   perturbation series in physics (\cref{perturbation_series}), we need the sum over decompletions, weighted by their symmetry factors. Owing to \cref{aut_decompletion}, this sum can equivalently be written as a sum over completed graphs, where we omit the trivial factor  $4! (L_G+2)$ in order to facilitate comparisons with \cref{def:period_mean}. 
\begin{align}\label{def:period_mean_aut}
	\left \langle \frac{\period}{\abs{\Aut}}  \right \rangle  :=  \frac{1}{N^{(C)}_L}\sum_{\substack{ \text{completions }G \\ L \text{ loops} }} \frac{\period (G)}{\abs{\Aut (G)}} = \frac{1}{(L+2) N^{(C)}_L  } \sum_{\substack{ \text{decompletions }g \\ L \text{ loops} }} \frac{\period (g)}{\abs{\Aut (g)}}  . 
\end{align}
For the \enquote{s}-samples, the correction \cref{period_sampling_rescaling} cancels the symmetry factors from \cref{def:period_mean_aut} as should be expected: A sum of $\frac{\period}{\Aut}$ is equivalent to a sum of $\period$, but distributed proportionally to $\frac{1}{\Aut}$.

Analogously to \cref{mean_scaling}, we normalize the distribution of symmetry-factor-weighted periods with respect to the mean \cref{def:period_mean_aut} and examine the quantiles and moments of the resulting distribution. The results are reported in \cref{tab:means_Aut}. The numerical- and sampling accuracy behave completely analogously to \cref{tab:means}, we have omitted them in \cref{tab:means_Aut} and instead included the median, i.e. the 50\% quantile.  Note the very small 1\% quantile at $L=6$ and $L=8$ in \cref{tab:means_Aut,} caused by graphs with very large symmetry factors. For $L \geq 10$, those graphs lie outside the 1\% quantile boundary. The minimum symmetry-factor weighted period at 10 loops is $1.982$ and at 12 loops it is $4.483$.

\begin{table}[htb]
	\centering
	\begin{tblr}{vlines,
			hline{1}={solid},
			hline{2,Z}={solid},
			rowsep=0pt,
			cells={font=\fontsize{11pt}{12pt}\selectfont },
			columns={halign=r},
			column{1}={halign=c,mode=math},
			row{1}={halign=c,rowsep=2pt, font=\fontsize{12pt}{14pt}\selectfont ,mode=math}, 
			cell{1}{3}={font=\fontsize{11pt}{13pt}\selectfont },
			cell{1}{4}={font=\fontsize{11pt}{13pt}\selectfont }
		}
		L &  \left \langle \frac{\period }{\abs{\Aut }} \right \rangle     & 50\%  & [1\% ~; ~~99\%]  & C_2 & c_3 & c_4 \\
		5        & $ 2.52702  \pm 0.00002 $ & 1.0837 & 1.084 ; 3.970 & 0.3262 & 0.000 & 1.000 \\
		6        & $ 12.56811 \pm 0.00005 $ & 10.521 & 0.062 ; 33.06 & 0.7782 & 0.945 & 2.674 \\
		7 		 & $  61.3302 \pm 0.00015 $ & 31.845 & 3.538 ; 190.4 & 0.9729 & 0.966 & 2.413 \\
		8   	 & $ 257.7329 \pm 0.0005 $  & 211.48 & 0.972 ; 966.8 & 0.7060 & 1.180 & 4.164 \\
		9   	 & $ 871.8986 \pm 0.0017 $  & 690.62 & 27.52 ; 2631  & 0.5265 & 0.857 & 3.045 \\
		10  	 & $ 2422.493 \pm 0.003 $   & 2181.7 & 76.30 ; 7351  & 0.4158 & 0.936 & 3.951  \\
		11  	 & $ 5880.015 \pm 0.003 $   & 5119.3 & 609.8 ; 17770 & 0.3699 & 1.362 & 5.750 \\
		12  	 & $ 12834.033 \pm 0.003 $  & 10701  &  $(1.865 ; 41.94) \cdot 10^3$  & 0.4019 & 1.905 & 8.713 \\
		13  	 & $ 26320.414 \pm 0.007 $  & 21098    & $(5.250 ; 95.65) \cdot 10^3$   & 0.4718 & 2.439 & 12.69 \\
		13\text{s}   	 & $ (264.9  \pm 1.9) \cdot 10^2 $  &20618    & $(8.241 ; 95.93) \cdot 10^3$   & 0.4860 & 2.501 & 12.92 \\
		13\star  	 & $ (263.2  \pm 1.9) \cdot 10^2 $  & 20952   & $(5.228 ; 97.27) \cdot 10^3$   & 0.4838 & 2.513 & 13.03\\
		14\text{s}  	 & $(518.0  \pm 0.9)\cdot 10^{2}$  & $3.911 \cdot 10^{4}$ &  $(15.03 ; 212.7) \cdot 10^3$ & 0.5975 & 2.959 & 17.13 \\
		14\star& $(517.3 \pm 1.2) \cdot 10^{2}$  & $4.015 \cdot 10^{4}$ &   $(1.191 ; 20.69) \cdot 10^4$ & 0.5622 & 2.904 & 16.71 \\
		15\text{s}  	 & $(981.7 \pm 2.7) \cdot 10^{2}$  & $7.252 \cdot 10^{4}$ & $(2.688 ; 44.13) \cdot 10^4$ & 0.7186 & 3.642 & 26.07 \\
		15\star& $(985.7 \pm 3.2) \cdot 10^{2}$ & $7.412 \cdot 10^{4}$ & $(24.00 ; 441.9) \cdot 10^3$ & 0.6943 &3.677 & 28.30 \\
		16\text{s}  	 & $(181.8  \pm 1.7) \cdot 10^{3}$& $1.307 \cdot 10^{5}$   & $(4.756 ; 86.66) \cdot 10^4$ & 0.8212 & 3.877 & 27.55 \\
		17\text{s}  	 & $(334.9  \pm 4.6) \cdot 10^{3}$    & $2.365 \cdot 10^{5}$   & $(8.060 ; 171.0 ) \cdot 10^4$ & 0.9593 & 4.257 & 31.08 \\
		18\text{s}  	 & $(60.3 \pm 2.1) \cdot 10^{4}$    & $4.156 \cdot 10^{5}$  & $(1.531 ; 29.59 ) \cdot 10^5$ & 1.047 & 4.524 & 33.48 \\
	\end{tblr}
	\caption{Analogous data as \cref{tab:means}, but for the periods weighted with symmetry factors.  \enquote{50\%} denotes the median. The results do not include the overall factor $4!(L+2)$. The exact values \cite{schnetz_numbers_2018} for $L\in \left \lbrace 5,6,7 \right \rbrace $ are, to 10 digits, $\left \lbrace  2.527040439, ~12.56808867, ~61.33012120 \right \rbrace   $. }
	\label{tab:means_Aut}
\end{table}

The data in \cref{tab:means,tab:means_Aut} shows that the average period   grows significantly with loop order. Moreover,  the normalized variance $C_2$ of $\left \langle \period \right \rangle $ grows by more than one order of magnitude between $L=6$ and $L=18s$. For both  $\left \langle \period \right \rangle  $ and $\left \langle \frac{\period}{\abs{\Aut}} \right \rangle  $,  skewness and kurtosis grow at a similar rate. Recall that these central moments refer to the normalized distribution \cref{mean_scaling}, the observed growth is not caused by the growth of $\left \langle \period \right \rangle $. Similarly, due to the normalization \cref{def:central_moments}, the growth of $c_j$ is not explained by a growing standard deviation, but conversely, we find that the higher moments grow \emph{faster} than what should be expected from the growth of $C_2$. This leads us to the conclusion that in the limit $L\rightarrow \infty$, the higher moments $C_{j \geq 2}$ are very likely infinite. 

\begin{figure}[htb]
	\begin{subfigure}[b]{.48 \textwidth}
		\includegraphics[width=\linewidth]{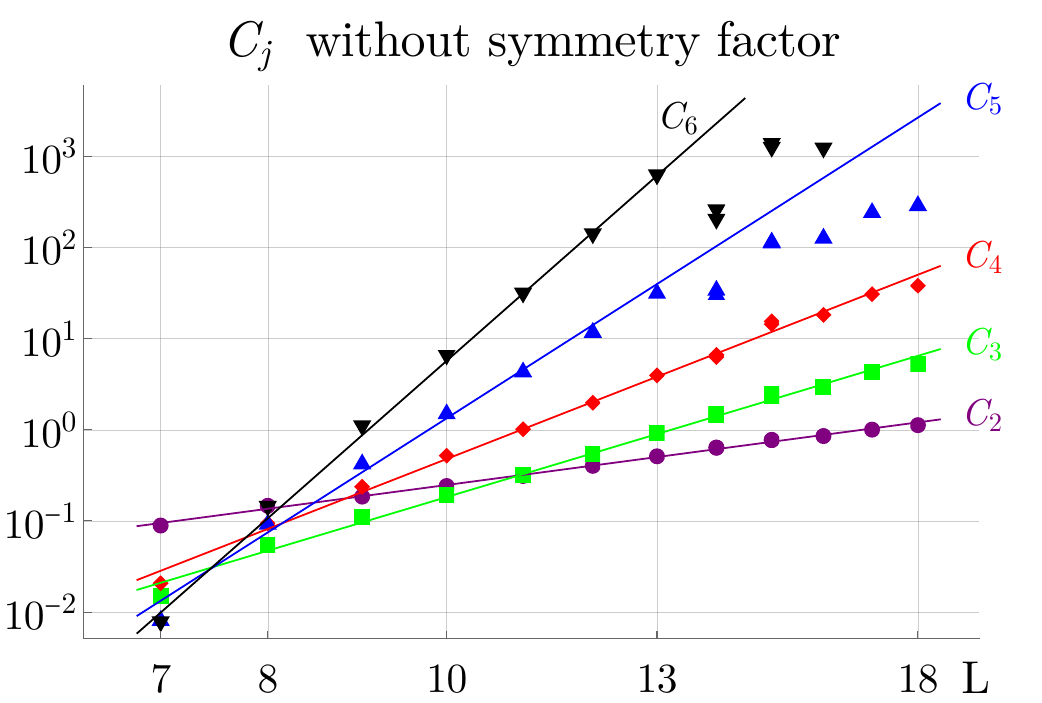}
		\subcaption{}
	\end{subfigure}
	\begin{subfigure}[b]{.48 \textwidth}
		\includegraphics[width=\linewidth]{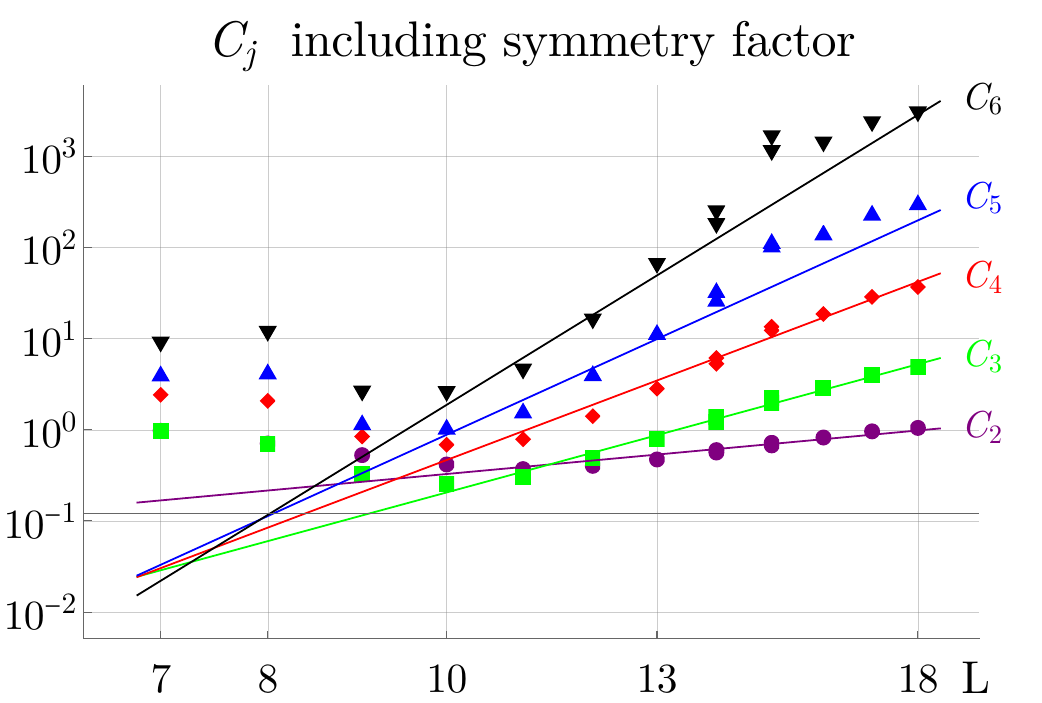}
		\subcaption{}
	\end{subfigure}
	
	\caption{Dependence of the central moments $C_j$ on the loop order $L$. The lines are  \cref{cumulant_growth} with coefficients in \cref{tab:centralMoments}. For $L\geq 14$, our samples are incomplete and their observed higher moments $C_j$ are increasingly unreliable for growing $j$. (a) for periods without symmetry factor, the growth is well described by the fit functions. (b) Including the symmetry factor results in a significant deviation for $L\leq L_\text{crit}\approx 9$.   }
	\label{fig:Cj_growth}
\end{figure}

A similar growth, but not shown in the tables, is  observed for the non-central moments $\left \langle p^j \right \rangle $ and the cumulants $\left \langle e^{jp} \right \rangle $. To examine the growth of the central moments $C_j$ more systematically, we plot them as a function of $L$ in  \cref{fig:Cj_growth}. Apparently, they can be approximated by a power law
\begin{align}\label{cumulant_growth}
C_j \approx e^{a_j} \cdot \left(\frac L 8\right)^{b_j} .
\end{align}
The offset 8 is motivated by the approximate intersection observed   in \cref{fig:Cj_growth}. Note that this value is close to $L_\text{crit}\approx 9$ of \cref{fig:ratio_correction}. The empirical values of $a_j,b_j$ are shown in \cref{tab:centralMoments}. The parameter $a_j$ is approximately constant for growing $j$, while $b_j$ increases approximately linearly.

\begin{table}[htb]
	\centering
	\begin{tblr}{vline{1}={solid},
			vline{3-Z}={solid},
			hlines,
			rowsep=0pt,
			cells={halign=c,valign=m  },
			columns={halign=r},
			column{1}={halign=c},
			row{1}={halign=c,rowsep=2pt, font=\fontsize{12pt}{14pt}\selectfont,mode=math },
			column{1}={halign=c,font=\fontsize{12pt}{14pt}\selectfont, mode=math}
			}
		& & C_2 & C_3& C_4 & C_5& C_6 \\
		\period &  {$a_j$\\ $b_j$}  &   {$-2.00 $\\ $2.69$}      &   {$-3.05 $\\$ 6.07$  }   &{ $-2.50 $\\$ 7.91$}    &  {$-2.59 $\\$ 12.9$ }   &   {$-2.24$ \\  $17.8 $}   \\
		 \frac{\period}{\abs{\Aut}} & {$a_j$ \\  $b_j$ }   &  { $-1.54$ \\  $1.87$}    &  { $-2.81 $\\ $5.51$}     &  {$ -2.47 $\\ $ 7.65 $}  &  {$-2.18 $\\ $ 9.21$ } & {$ -2.15$ \\  $12.5$ }  \\
	\end{tblr}

	\caption{Growth parameters of the central moments $C_j$ according to \cref{cumulant_growth}, as shown in \cref{fig:Cj_growth}}
	\label{tab:centralMoments}
\end{table}

The presence of symmetry factors has a notable influence on the growth parameters, see \cref{tab:centralMoments}, but they do not change the overall picture that all higher moments seem to diverge as $L\rightarrow \infty$. 
The crucial property is \emph{not} that the $C_j$ grow with growing $j$, for a fixed $L$, but conversely, that every $C_{j>1}$, for fixed $j$, tends to infinity as $L$ grows.

The overall very large standard deviations (and higher moments) imply that the sampling uncertainty (see \cref{sec:statistics}) is high even for large samples. We discuss a concrete example for our $L=13$ data set in \cref{sec:sampling_example}.

\subsection{Mean of decompletions and planar graphs}\label{sec:mean_decompletions}

So far, we have normalized all means with respect to the number of completions. Alternatively, we can consider the mean of decompletions. Using \cref{aut_decompletion,def:period_mean_aut,def:Drel}, we define 
\begin{align}\label{def:period_mean_decompletion_aut}
\left \langle \frac{\period  }{\abs{\Aut }} \right \rangle _\text{dec} := \frac{1}{  N^{(D)}_L  } \sum_{\substack{ \text{decompletions }g \\ L \text{ loops} }} \frac{\period (g)}{\abs{\Aut (g)}} = 	\frac 1 {\left \langle D^\text{rel}\right \rangle} \cdot \left \langle \frac{\period}{\abs{\Aut}}  \right \rangle \geq \left \langle \frac{\period}{\abs{\Aut}}  \right \rangle .
\end{align}
The correction factor $\left \langle D^\text{rel} \right \rangle $ converges to unity as $L \rightarrow \infty$. Note that the numbers in  \cref{tab:aut} for $L \geq 14$ refer to all completions, not just to our samples. 
\begin{align}\label{def:period_mean_decompletion}
	\left \langle \period    \right \rangle _\text{dec} &:= \frac{1}{  N^{(D)}_L  } \sum_{\substack{ \text{decompletions }g \\ L \text{ loops} }}  \period (g) .
\end{align}
In this case, we can not leverage \cref{aut_decompletion} and, unlike \cref{def:period_mean_decompletion_aut}, the sum in \cref{def:period_mean_decompletion} can not be rewritten in terms of completions. Nevertheless, we expect $\left \langle \period \right \rangle_\text{dec} \rightarrow \left \langle \period \right \rangle $ as automorphisms become increasingly unimportant in the limit $N \rightarrow \infty$, see \cref{sec:planar_graphs}.

Analogous to \cref{sec:planar_graphs}, dividing by the number of \emph{planar} decompletions, we define
\begin{align}\label{def:period_mean_planar_aut}
	\left \langle \frac{\period  }{\abs{\Aut }} \right \rangle _\text{pla} := \frac{1}{  \# \text{planar decompletions} } \sum_{\substack{ \text{planar dec. }g \\ L \text{ loops} }} \frac{\period (g)}{\abs{\Aut (g)}} =: F^{(\Aut)}_\text{pla} \cdot \left \langle \frac{\period}{\abs{\Aut}}  \right \rangle .
\end{align}
We do not have any a priori information about how the factor $F_\text{pla}$ introduced in  \cref{def:period_mean_planar_aut}  behaves in the limit $L\rightarrow \infty$. In particular, $F_\text{pla}$ does not (only) encode information about automorphisms of the graphs, but about whether or not the numerical value of the period of a planar graph is representative for the period of all graphs. Finally, we define
\begin{align}\label{def:period_mean_planar}
		\left \langle \period  \right \rangle _\text{pla} &:= \frac{1}{  \# \text{planar decompletions} } \sum_{\substack{ \text{planar dec. }g \\ L \text{ loops} }} \period (g) =: F_\text{pla} \cdot \left \langle \period  \right \rangle.  
\end{align}
In all cases, the non-uniform samples are corrected according to \cref{sec:uniform_sampling}.

\begin{table}[htb]
	\centering
	\begin{tblr}{vlines,
			vline{4}={1}{-}{solid},
			vline{4}={2}{-}{solid},
			hline{1}={solid},
			hline{2,Z}={solid},
			rowsep=0pt,
			cells={font=\fontsize{11pt}{12pt}\selectfont },
			columns={halign=r},
			row{1}={halign=c,rowsep=2pt, font=\fontsize{12pt}{14pt}\selectfont },column{1}={r}
			}
		$L$ & $\left \langle \period \right \rangle  _\text{dec} $& $\left \langle \frac{\period }{\abs{\Aut }} \right \rangle _\text{dec}   $& $\left \langle \period \right \rangle  _\text{pla} $& $\left \langle \frac{\period }{\abs{\Aut }} \right \rangle _\text{pla}   $ & $F_\text{pla} $ &$F^{(\Aut)}_\text{pla} $\\
		5        & 53.206837  & 11.792786 & 53.801340 & 15.521770 & 21.29 & 6.142 \\
		6        & 130.07186  & 50.272453 & 146.39450 & 68.971371 &11.65 & 5.488 \\
		7 		 & 318.74446  & 175.62729 & 355.88168 & 196.84586 & 5.803 & 3.210 \\
		8   	 & 744.30225  & 509.23037 & 901.14361 & 684.50231 & 3.496 & 2.656 \\
		9   	 & 1664.3771  & 1289.7695 & 2225.0521 & 1763.3185 & 2.552 & 2.022 \\
		10  	 & 3591.9116  & 3006.0073 & 5490.1060 & 4853.5088 & 2.266 & 2.004  \\
		11  	 & 7486.4986 & 6517.9587 & 13373.880 & 12028.176  & 2.274 & 2.046 \\
		12  	 & 15143.497  & 13507.354 & 32494.348 & 30213.802 & 2.532& 2.354 \\
		13  	 & 29822.743  & 27001.114 & 78580.447 & 73666.453 & 2.986 & 2.799 \\
		13s  	 & $2.9801 \cdot 10^4$  & $2.7094 \cdot 10^4$ & $8.1908 \cdot 10^4$ & $7.8462 \cdot 10^4$ & 3.022 & 2.825 \\
		13$\star$& $2.9904 \cdot 10^4$  & $2.7022 \cdot 10^4$ & $7.9377 \cdot 10^4$ & $7.4925 \cdot 10^4$ & 3.016 & 2.847 \\
		14s 	 & $5.7436 \cdot 10^4$  & $5.2546 \cdot 10^4$ & $1.9039 \cdot 10^5$ & $1.8045 \cdot 10^5$ & 3.627 & 3.437  \\
		14$\star$& $5.7419 \cdot 10^4$  & $5.2471 \cdot 10^4$ & $1.8696 \cdot 10^5$ & $1.7716 \cdot 10^5$ & 3.614 & 3.425  \\
		15s 	 & $1.0740 \cdot 10^5$  & $0.9900 \cdot 10^5$ & $4.6530 \cdot 10^5$ & $4.3709 \cdot 10^5$ & 4.709 & 4.423 \\
		15$\star$& $1.0773 \cdot 10^5$  & $0.9933 \cdot 10^5$ & $4.5417 \cdot 10^5$ & $4.3269 \cdot 10^5$ & 4.612 & 4.394 \\
		16s 	 & $1.959 \cdot 10^5$ & $1.823 \cdot 10^5$ & $8.896 \cdot 10^5$ & $8.725 \cdot 10^5$ & 4.865 &  4.772 \\
		17s 	 & $3.567 \cdot 10^5$ & $3.363 \cdot 10^5$ & $2.434 \cdot 10^6$ & $2.434 \cdot 10^6$ & 7.264 & 7.264 \\
	\end{tblr}
	\caption{ Average period of decompletions \cref{def:period_mean_decompletion} \| \cref{def:period_mean_decompletion_aut} \| \| Average period of planar  decompletions \cref{def:period_mean_planar} \| \Cref{def:period_mean_planar_aut}\| Ratios between average of planar decompletions and average of completions as defined in \cref{def:period_mean_planar,def:period_mean_planar_aut}.  All uncertainties  are similar to the ones in \cref{tab:means,tab:means_Aut}. Note that the factors $F_\text{pla}$ are significantly larger than unity. }
	\label{tab:means_dec}
\end{table}

The data in \cref{tab:means_dec} shows that indeed, the factors $ F_\text{pla}$ and $F^{(\Aut)}_\text{pla}$ are significantly larger than unity, and grow with the loop order. This means that the period of planar graphs is much larger than the average, and the effect gets even stronger for higher loop orders.

\FloatBarrier

\subsection{Largest and smallest periods}\label{sec:largest}

\begin{figure}[htb]
	\begin{subfigure}[b]{.24 \textwidth}
		\includegraphics[width=\linewidth]{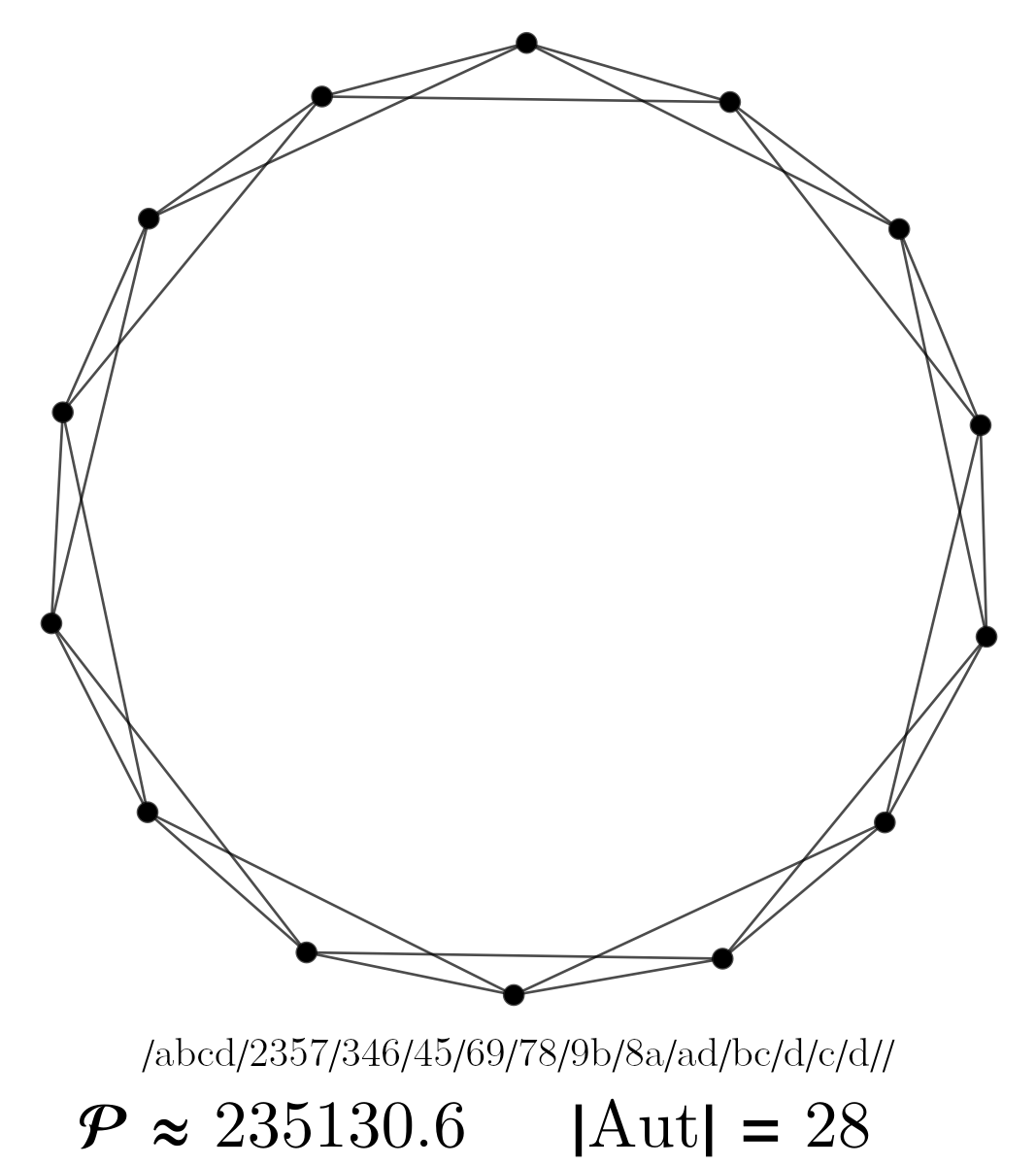}
		\subcaption{}
	\end{subfigure}
	\begin{subfigure}[b]{.24 \textwidth}
		\includegraphics[width=\linewidth]{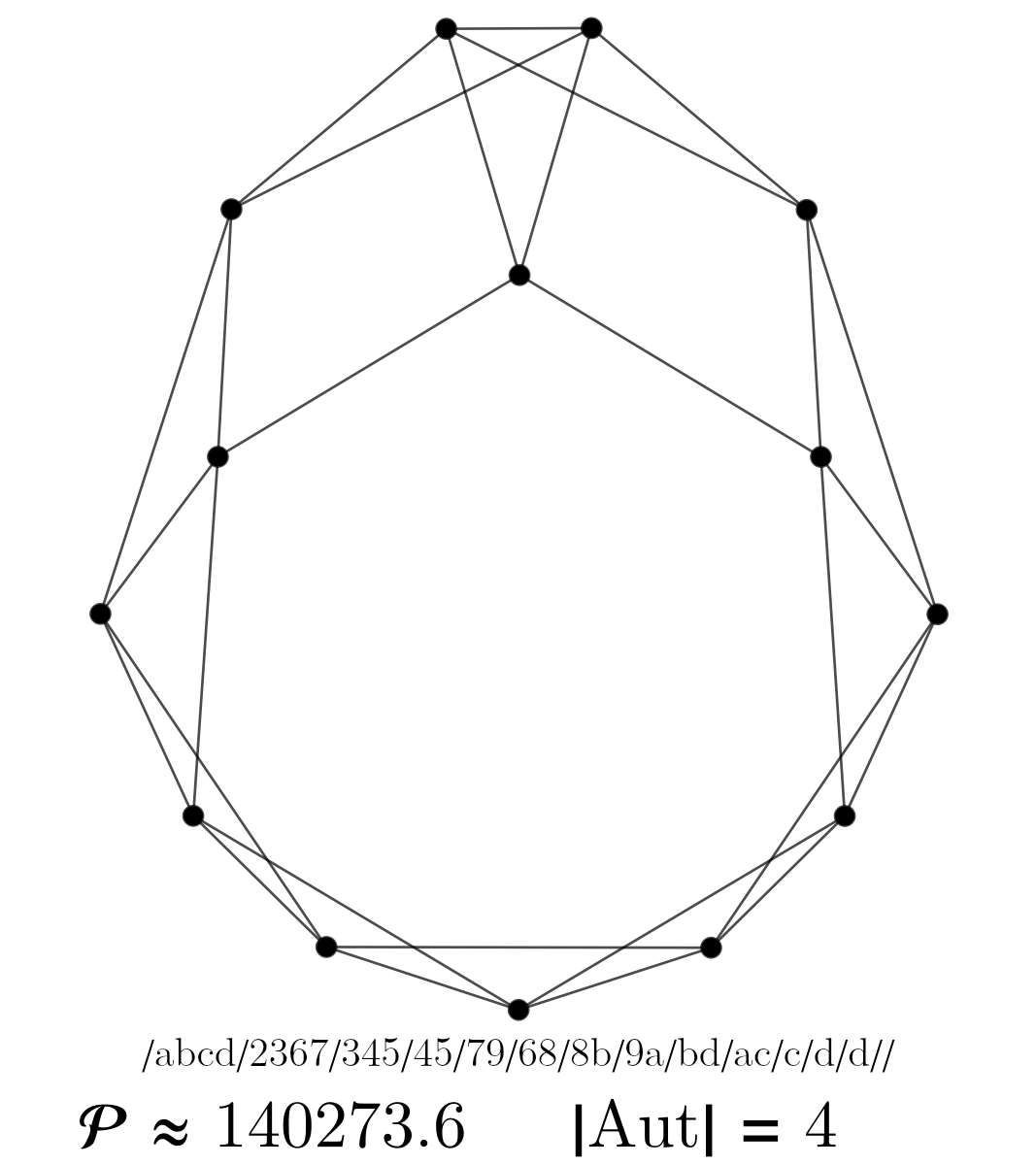}
		\subcaption{}
	\end{subfigure}
	\begin{subfigure}[b]{.24 \textwidth}
		\includegraphics[width=\linewidth]{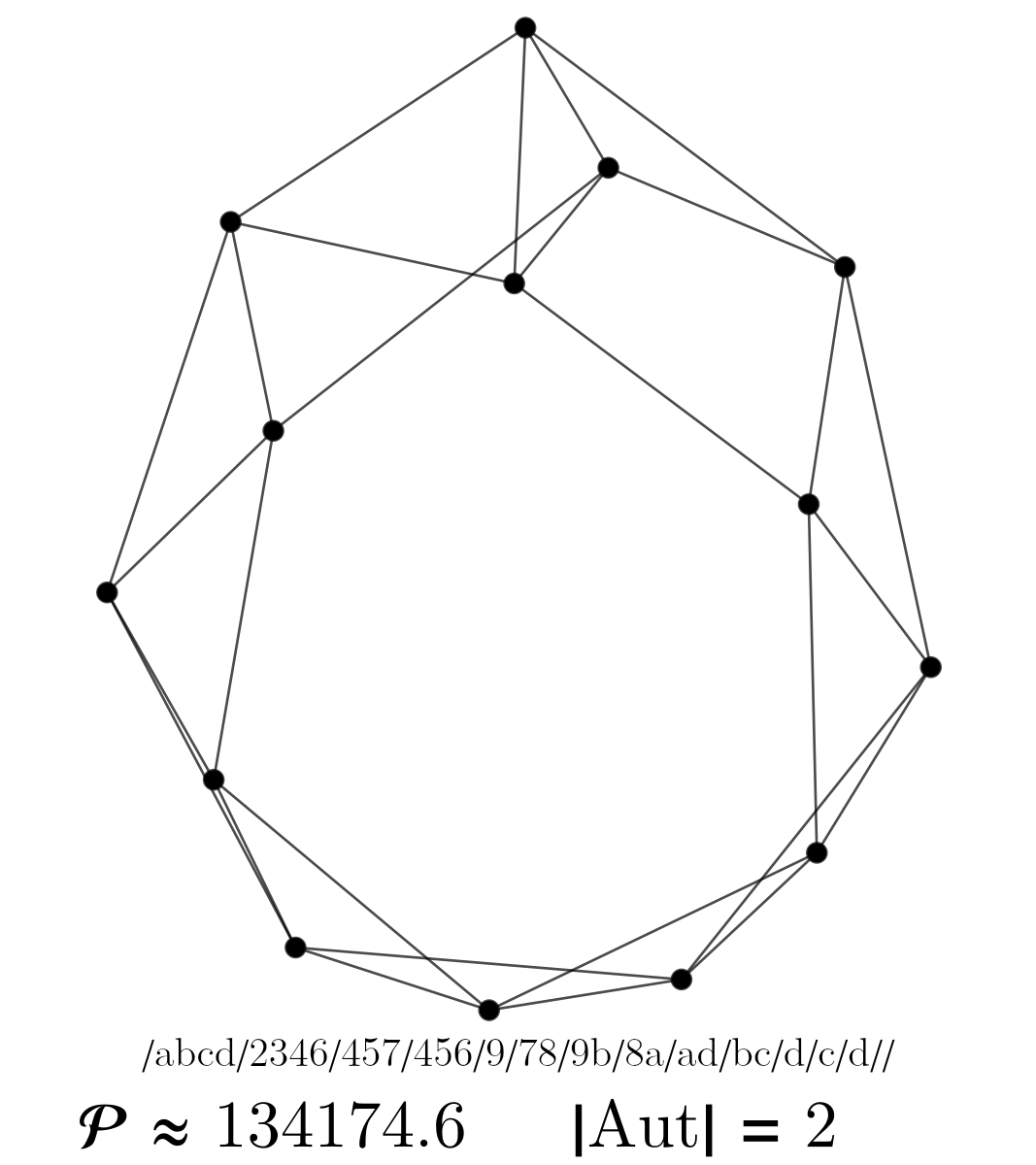}
		\subcaption{}
	\end{subfigure}
	\begin{subfigure}[b]{.24 \textwidth}
		\includegraphics[width=\linewidth]{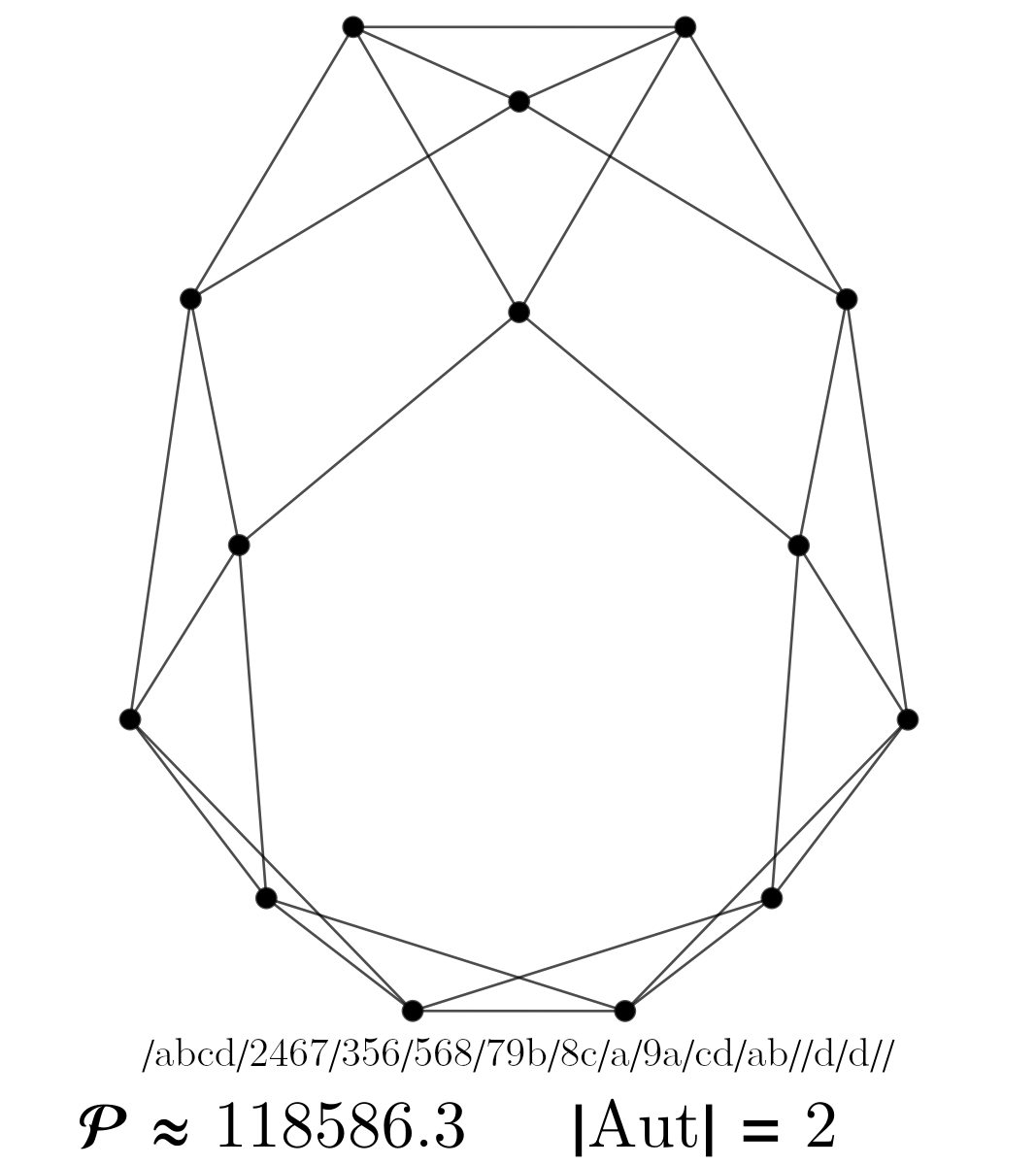}
		\subcaption{}
	\end{subfigure}
	
	\caption{The four 12-loop graphs with the largest period.}
	\label{largest_12}
\end{figure}

\begin{figure}[htb]
	\begin{subfigure}[b]{.24 \textwidth}
		\includegraphics[width=\linewidth]{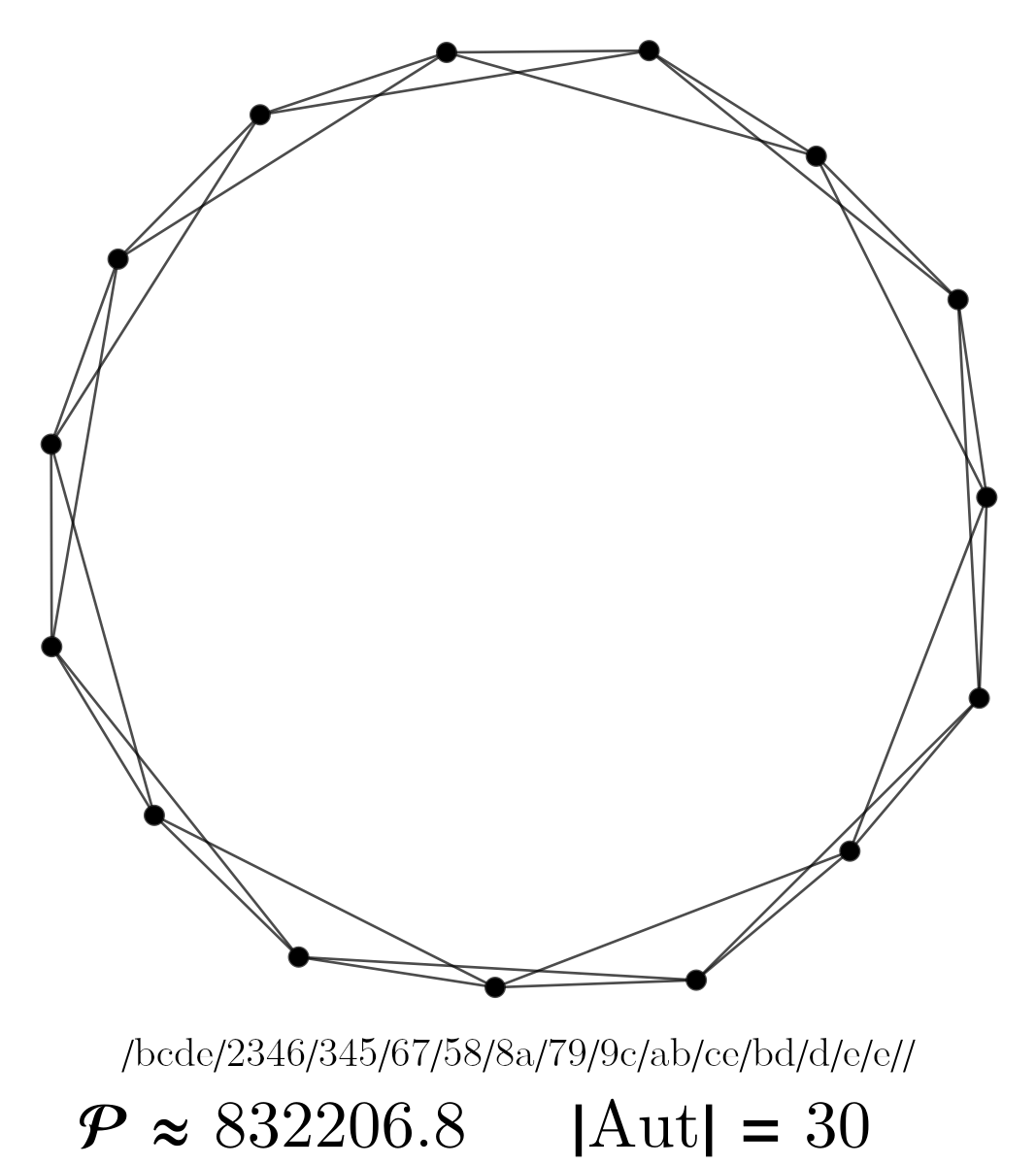}
		\subcaption{}
	\end{subfigure}
	\begin{subfigure}[b]{.24 \textwidth}
		\includegraphics[width=\linewidth]{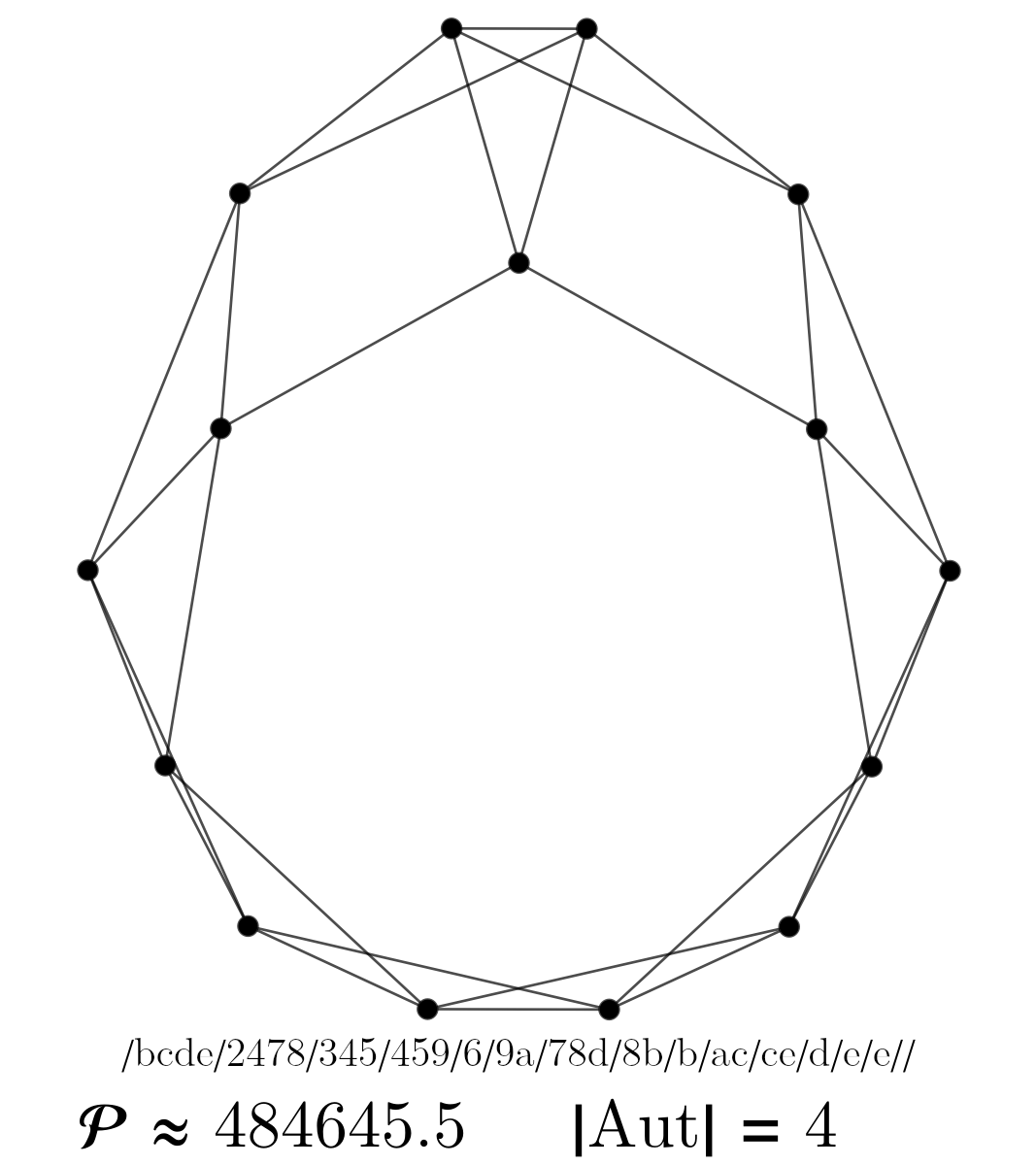}
		\subcaption{}
	\end{subfigure}
	\begin{subfigure}[b]{.24 \textwidth}
		\includegraphics[width=\linewidth]{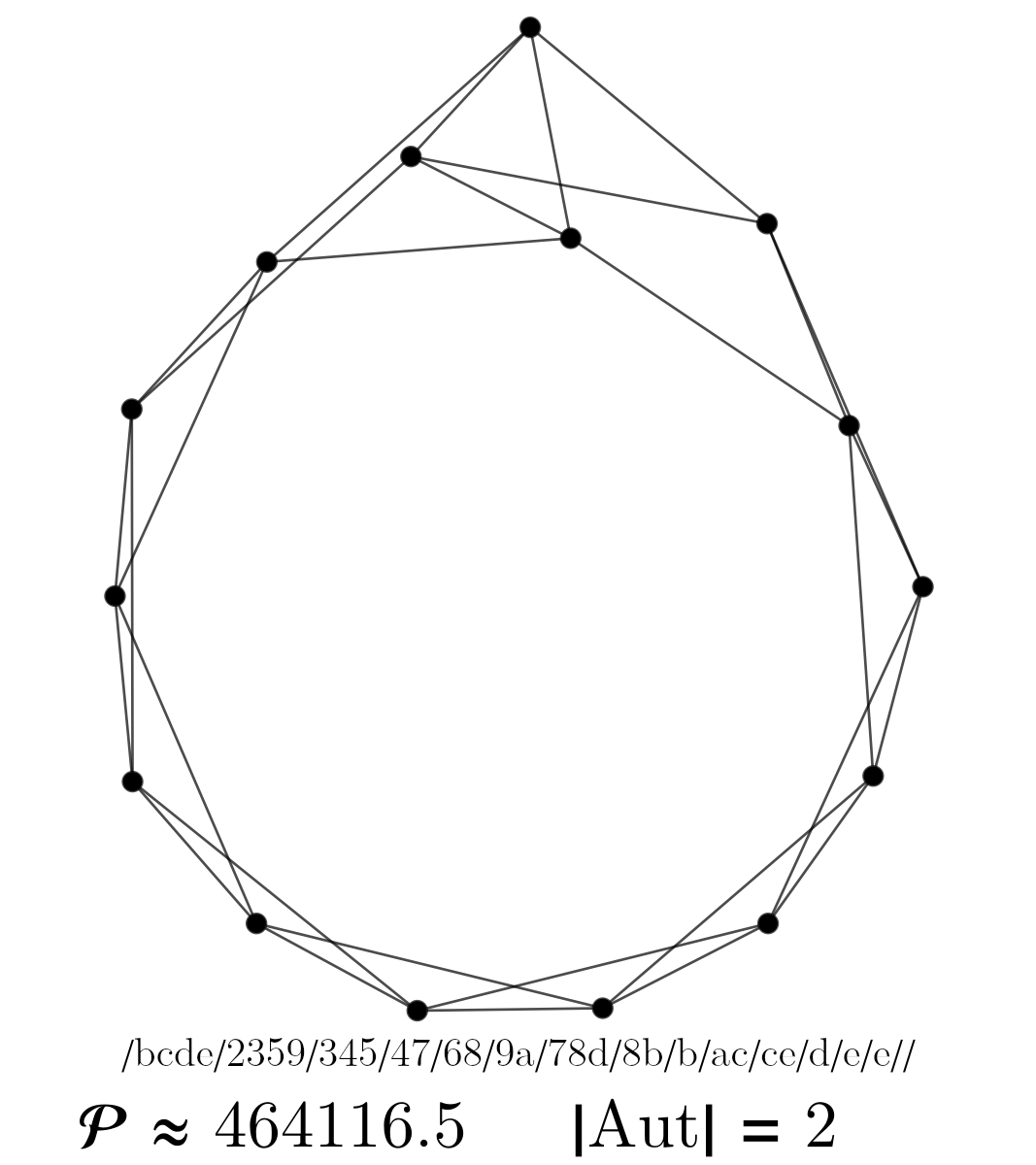}
		\subcaption{}
	\end{subfigure}
	\begin{subfigure}[b]{.24 \textwidth}
		\includegraphics[width=\linewidth]{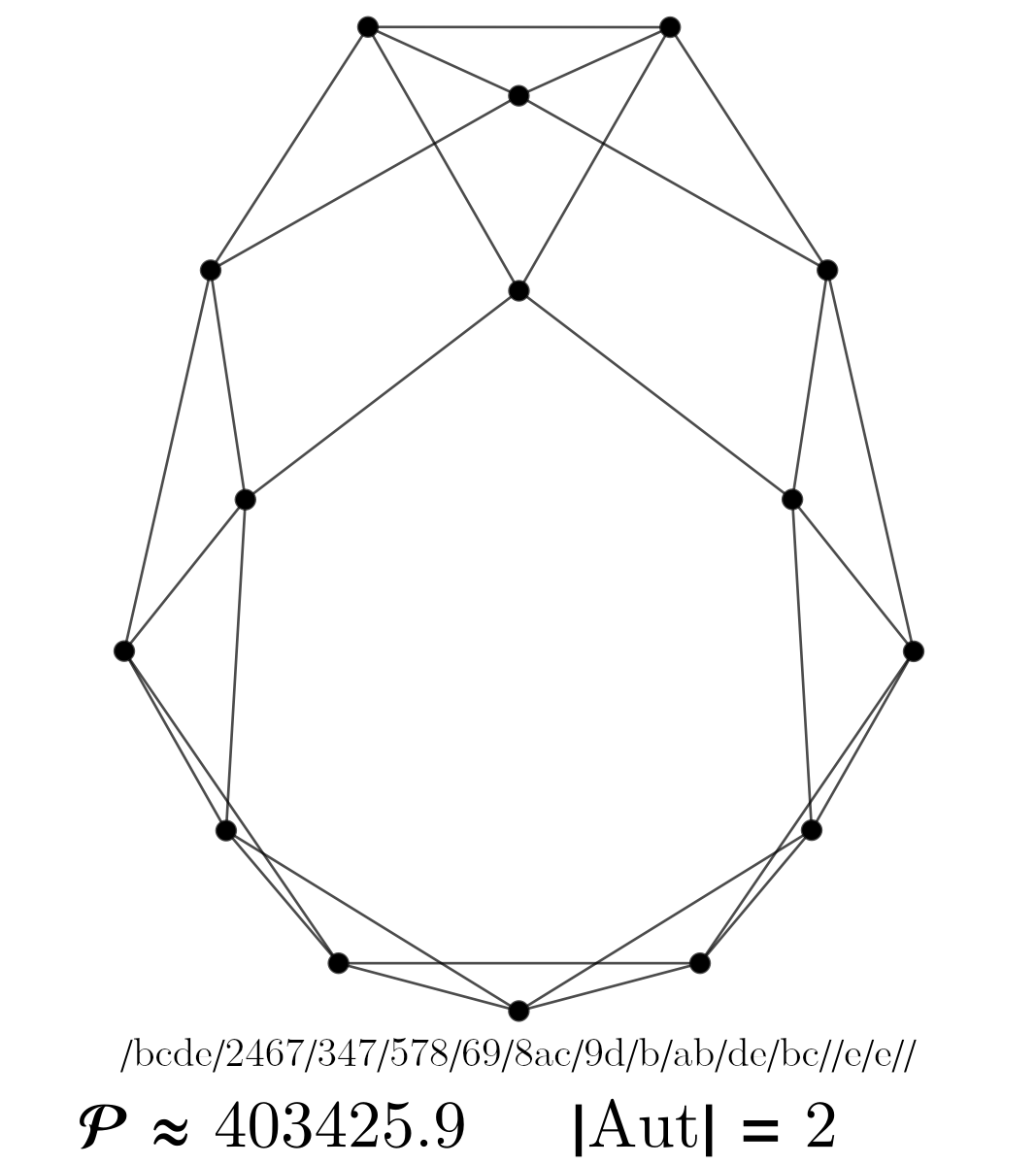}
		\subcaption{}
	\end{subfigure}
	
	\caption{The four 13-loop graphs with the largest period. Note that each of them corresponds   to one of the largest 12-loop graphs, shown in \cref{largest_12}.}
	\label{largest_13}
\end{figure}

Having computed all graphs up to 13 loops, we can identify the largest and smallest periods. 
In all our samples, the graphs with the largest period per loop order are the $(1,2)$-circulants, known as zigzags $Z_L$, see graphs (a) in \cref{largest_12,largest_13}. Their period is known to be \cite{broadhurst_knots_1995,brown_singlevalued_2015}
\begin{align}\label{zigzag_amplitude}
	\period \big( ~(a)~\big)=\period(Z_L) &= 4 \frac{(2L-2)!}{L! (L-1)!} \left( 1- \frac{1-(-1)^L}{2^{2L-3}} \right) \zeta(2L-3).
\end{align}
The next-largest graphs are almost of the zigzag shape, concretely, they are either 3-vertex products of zigzags or they are zigzags with \enquote{local} distortions that only affect a few vertices. For $L\geq 8$, the second-largest graphs, (b) in \cref{largest_12,largest_13}, is a 3-vertex product of   zigzags $Z_{L-2}\times Z_3$, 
\begin{align*}
	\period \big(  ~(b)~\big) &= \period(Z_{L-2}) \period(Z_3)=24 \frac{(2L-6)!}{(L-2)! (L-3)!} \left( 1- \frac{1-(-1)^L}{2^{2L-7}} \right) \zeta(2L-7) \zeta(3).
\end{align*}
Graphs of type (c) are not 3-vertex reducible, but we find empirically that $(c)_L \approx (a)_{L-1}$ when they are scaled to the mean of the respective loop order, concretely
\begin{align*}
	 \frac{\frac{\period \big( (c)_{11} \big)}{\left \langle \period \right \rangle _{11}}}{\frac{\period \big( (a)_{10} \big)}{\left \langle \period \right \rangle _{10}}}= 0.9655, \qquad , \qquad \frac{\frac{\period \big( (c)_{12} \big)}{\left \langle \period \right \rangle _{12}}}{\frac{\period \big( (a)_{11} \big) }{\left \langle \period \right \rangle _{11}}} = 0.9852, \qquad\frac{\frac{\period \big( (c)_{13} \big)}{\left \langle \period \right \rangle _{13}}}{\frac{\period \big( (a)_{12} \big)}{\left \langle \period \right \rangle _{12}}} &= 1.0016.
\end{align*} 
The 4\textsuperscript{th} largest graph, type (d), for $L\geq 9$, is again a product
\begin{align*}
	\period \big(  ~(d)~\big) &= \period(Z_{L-3}) \period(Z_4)=80 \frac{(2L-8)!}{(L-3)! (L-4)!} \left( 1- \frac{1+(-1)^L}{2^{2L-9}} \right) \zeta(2L-9) \zeta(5).
\end{align*}

This pattern can be continued, among the next-larges periods is $\period (Z_{L-4}) \period (Z_5)$ and so on. Also, more graphs of \enquote{almost-zigzag} shape, such as (c), and their products, start to appear. The period of zigzags grows much faster than the mean period, for example
\begin{align}\label{zigzag_growth}
	\period \big( Z_L\big) &\sim \frac{4^L}{4\sqrt{L\pi}}\left( \frac 1 L +  \mathcal{O}\left( \frac 1 {L^2} \right)   \right) , \qquad \qquad 
	\frac{\period (Z_L)}{\period \big(  ~(b)~\big)} \sim \frac{8}{3\zeta(3)}+ \mathcal{O}\left( \frac 1 L \right)  .
\end{align}
Concretely, if we normalize to unit mean and consider the relative period $x \cdot \left \langle \period \right \rangle $, then at $L=11$ loops, the zigzag is at $\period( (a)_{11} )=8.94 \cdot \left \langle \period \right \rangle _{11}$, while at 12 loops it is at $x=15.4$ and the 13-loop zigzag is at $x=27.7$, compare \cref{fig:distribution} in the introduction.

At every fixed loop number, there is only one zigzag, one graph of type (b), and so on. That means that for $L\rightarrow \infty$, there is a comparatively small set of periods which are arbitrarily much larger than the average, and also arbitrarily far apart (i.e. they differ by asymptotically constant factors such as $\frac 8{3\zeta(3)}$ in \cref{zigzag_growth}).
These observations make it plausible that, even if the distribution of periods is normalized to unit mean as in \cref{mean_scaling}, still all higher moments diverge as $L\rightarrow \infty$. Conversely, we can not expect to find a continuous, monotonically decaying probability density function for the largest periods.

\begin{figure}[htb]
	\begin{subfigure}[b]{.24 \textwidth}
		\includegraphics[width=\linewidth]{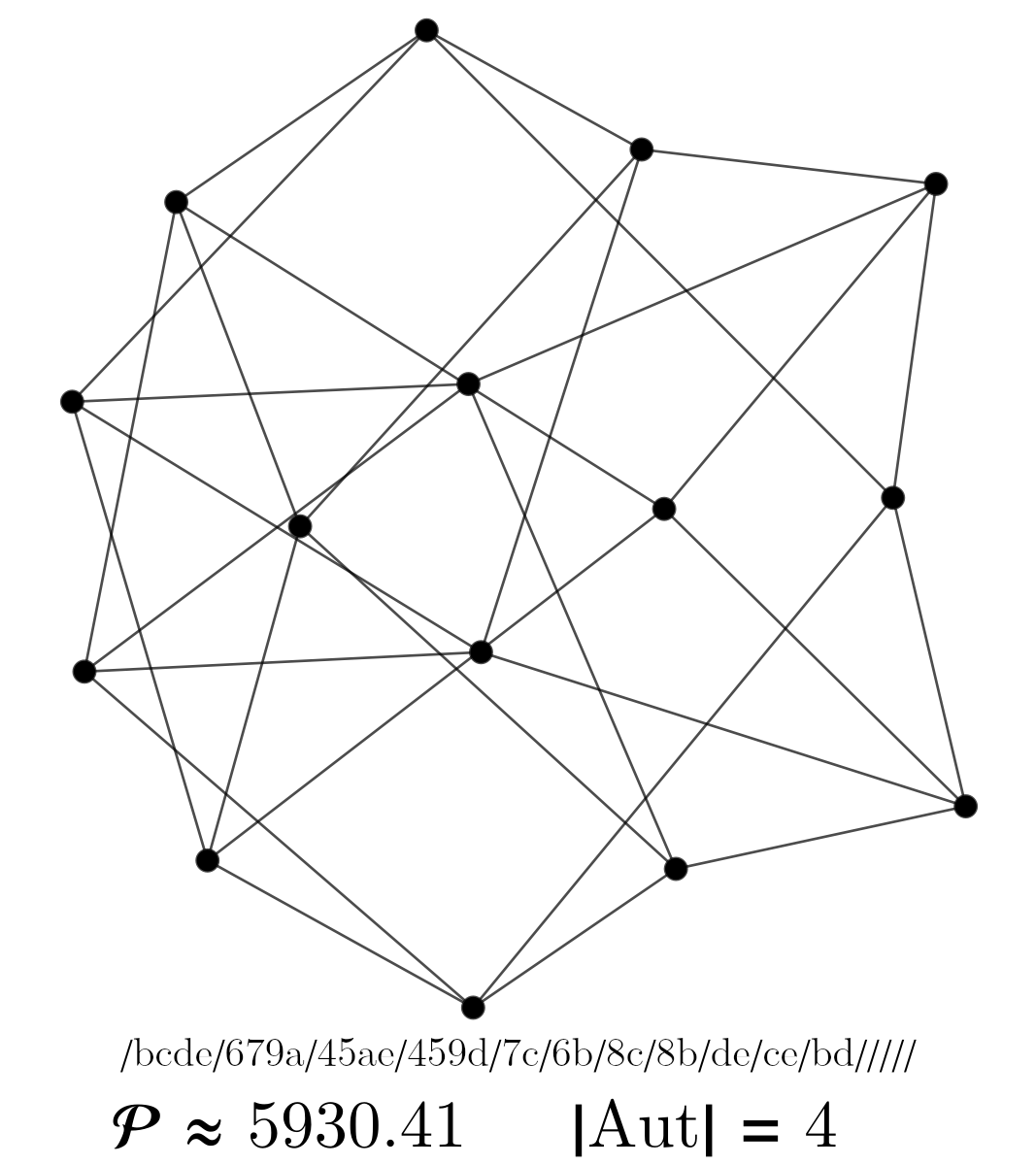}
		\subcaption{}
	\end{subfigure}
	\begin{subfigure}[b]{.24 \textwidth}
		\includegraphics[width=\linewidth]{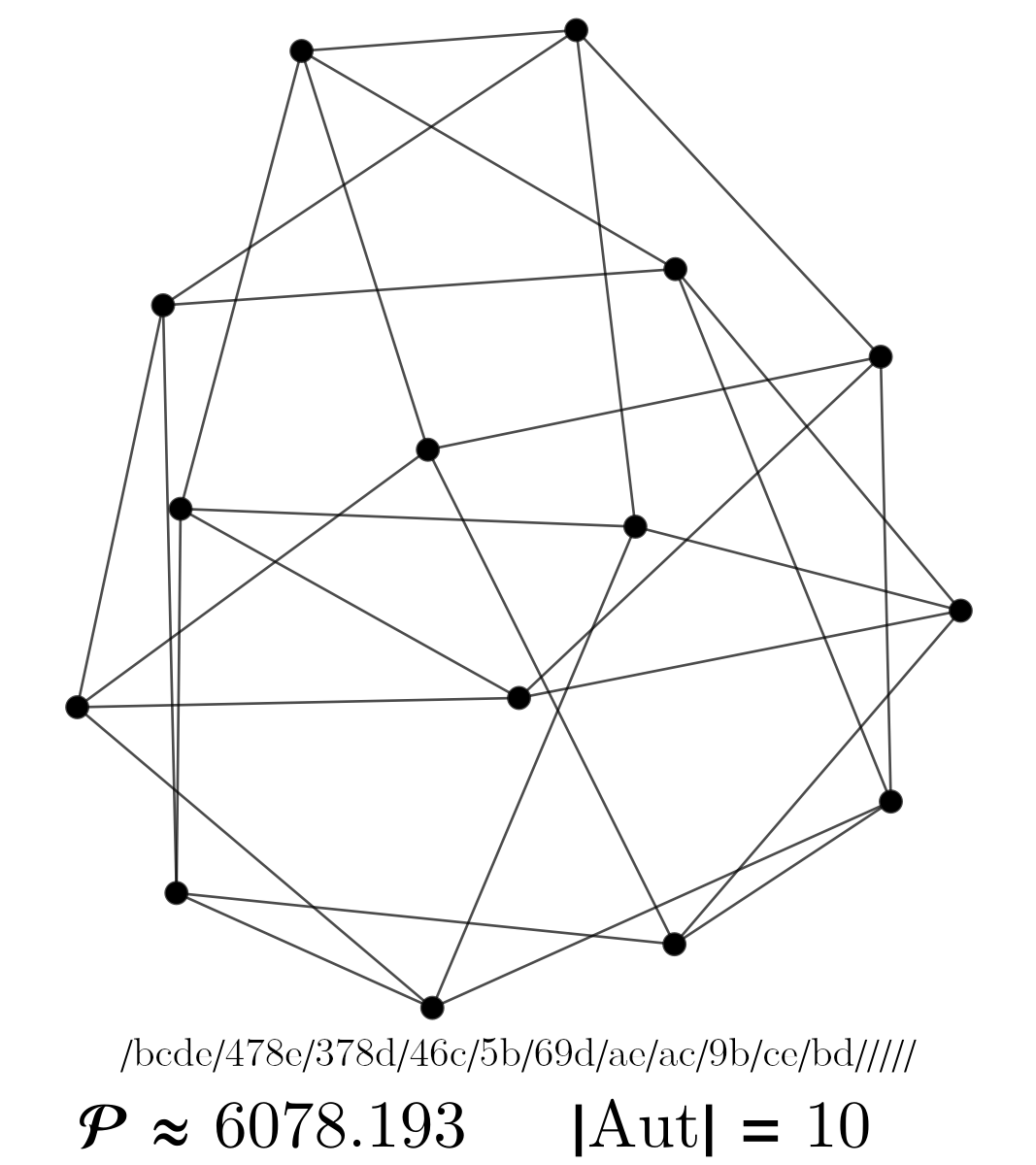}
		\subcaption{}
	\end{subfigure}
	\begin{subfigure}[b]{.24 \textwidth}
		\includegraphics[width=\linewidth]{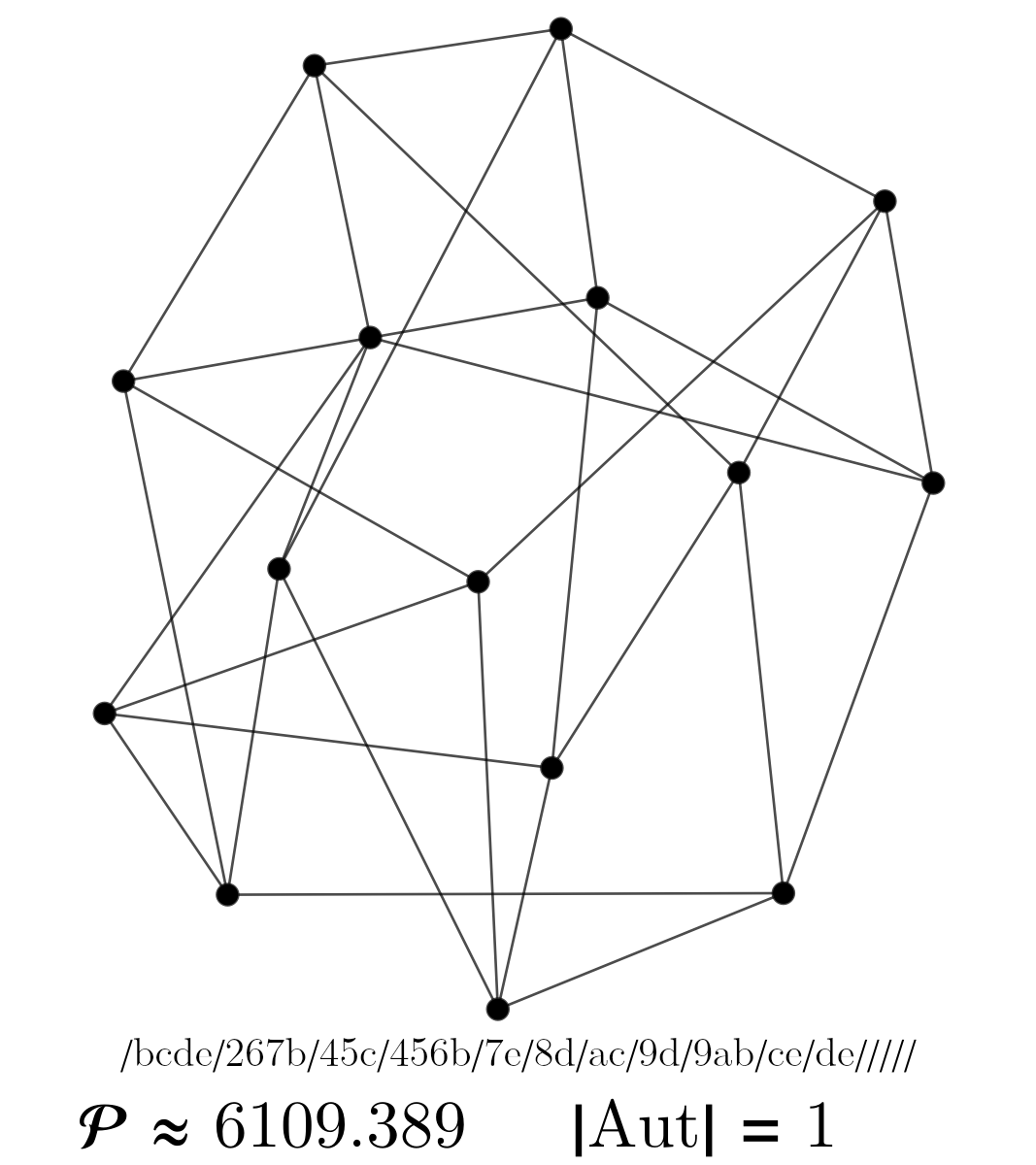}
		\subcaption{}
	\end{subfigure}
	\begin{subfigure}[b]{.24 \textwidth}
		\includegraphics[width=\linewidth]{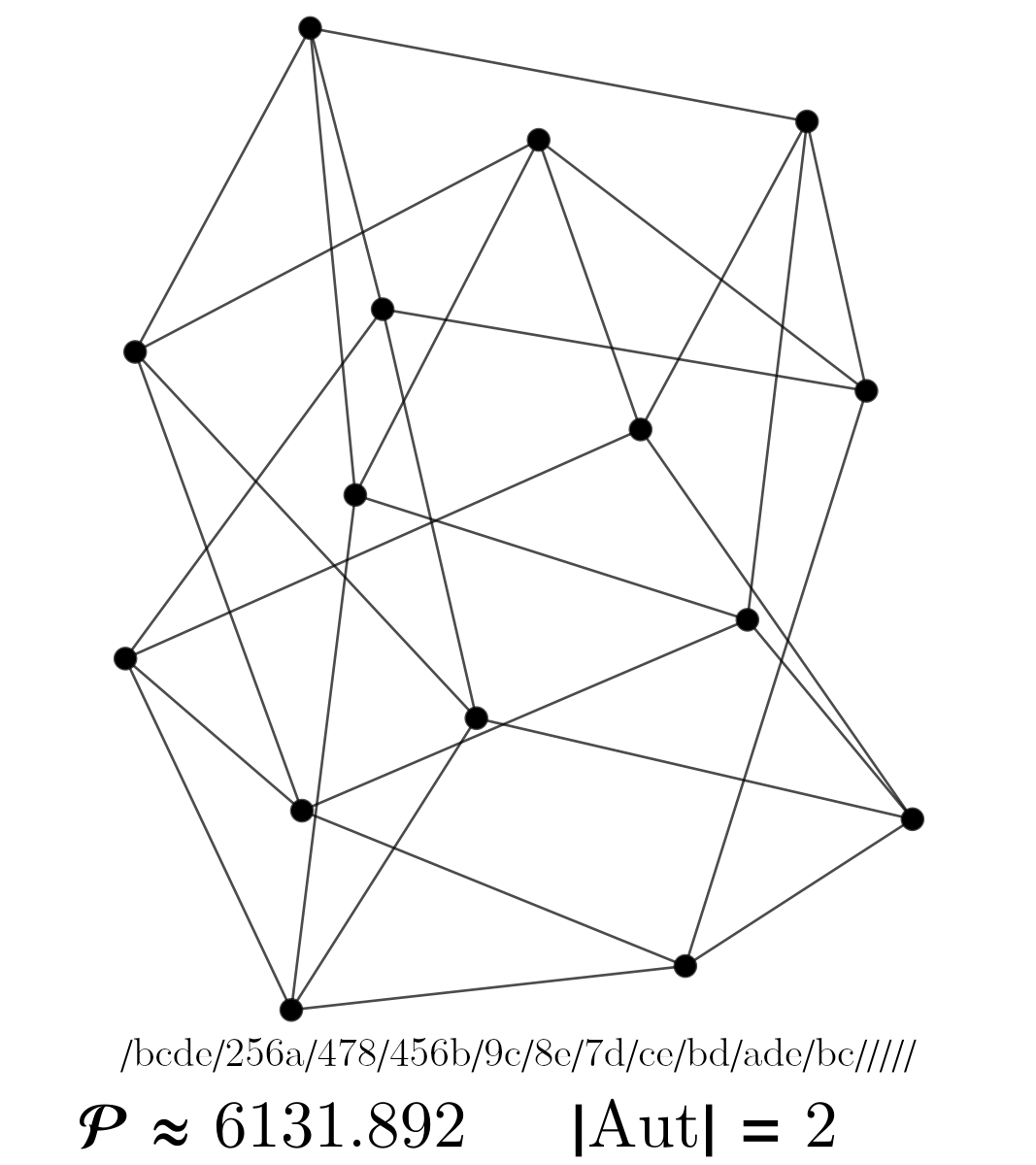}
		\subcaption{}
	\end{subfigure}
	
	\caption{The four 13-loop graphs with the smallest period.}
	\label{smallest_13}
\end{figure}

\begin{figure}[htb]
	\begin{subfigure}[b]{.24 \textwidth}
		\includegraphics[width=\linewidth]{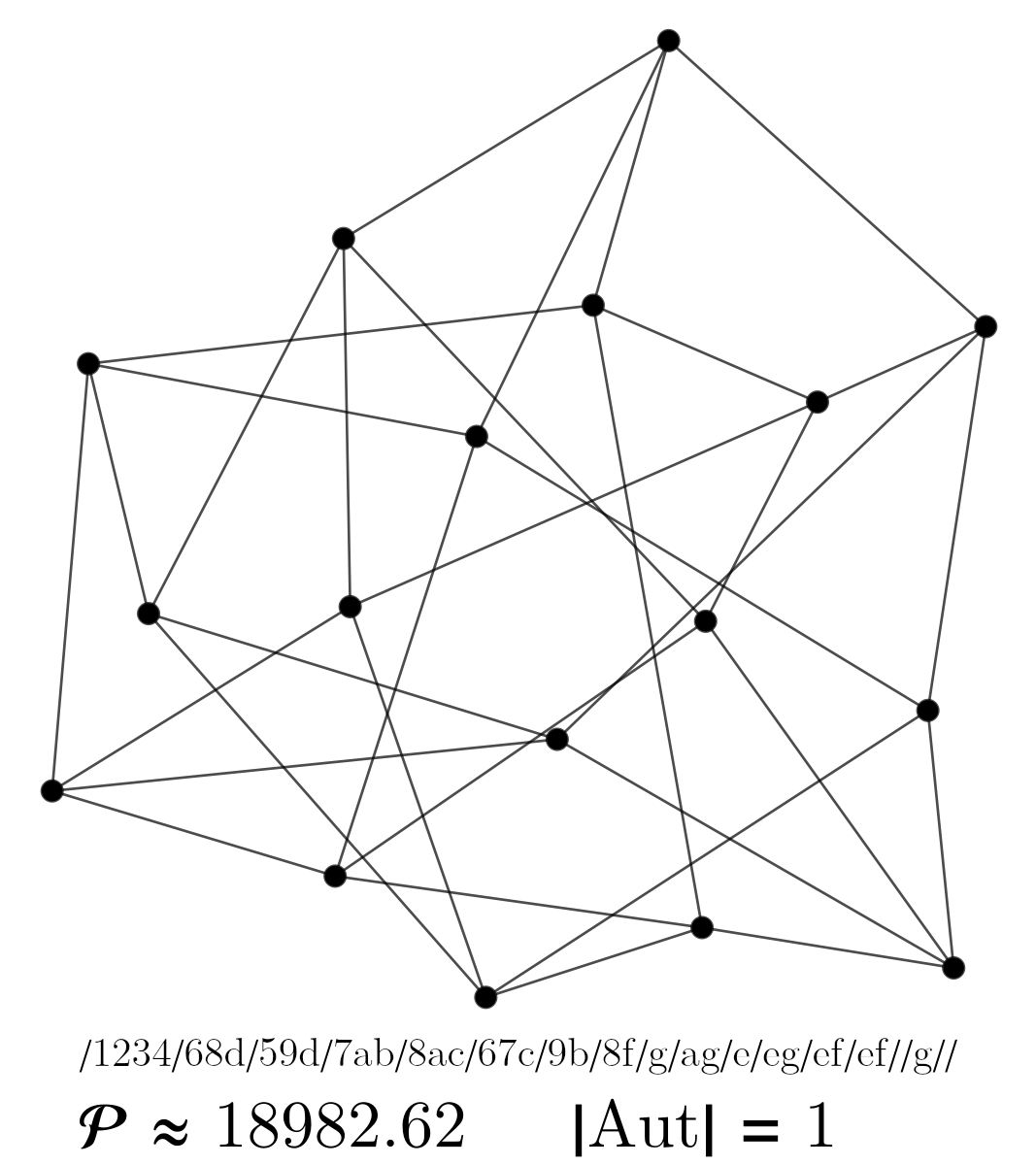}
		\subcaption{}
	\end{subfigure}
	\begin{subfigure}[b]{.24 \textwidth}
		\includegraphics[width=\linewidth]{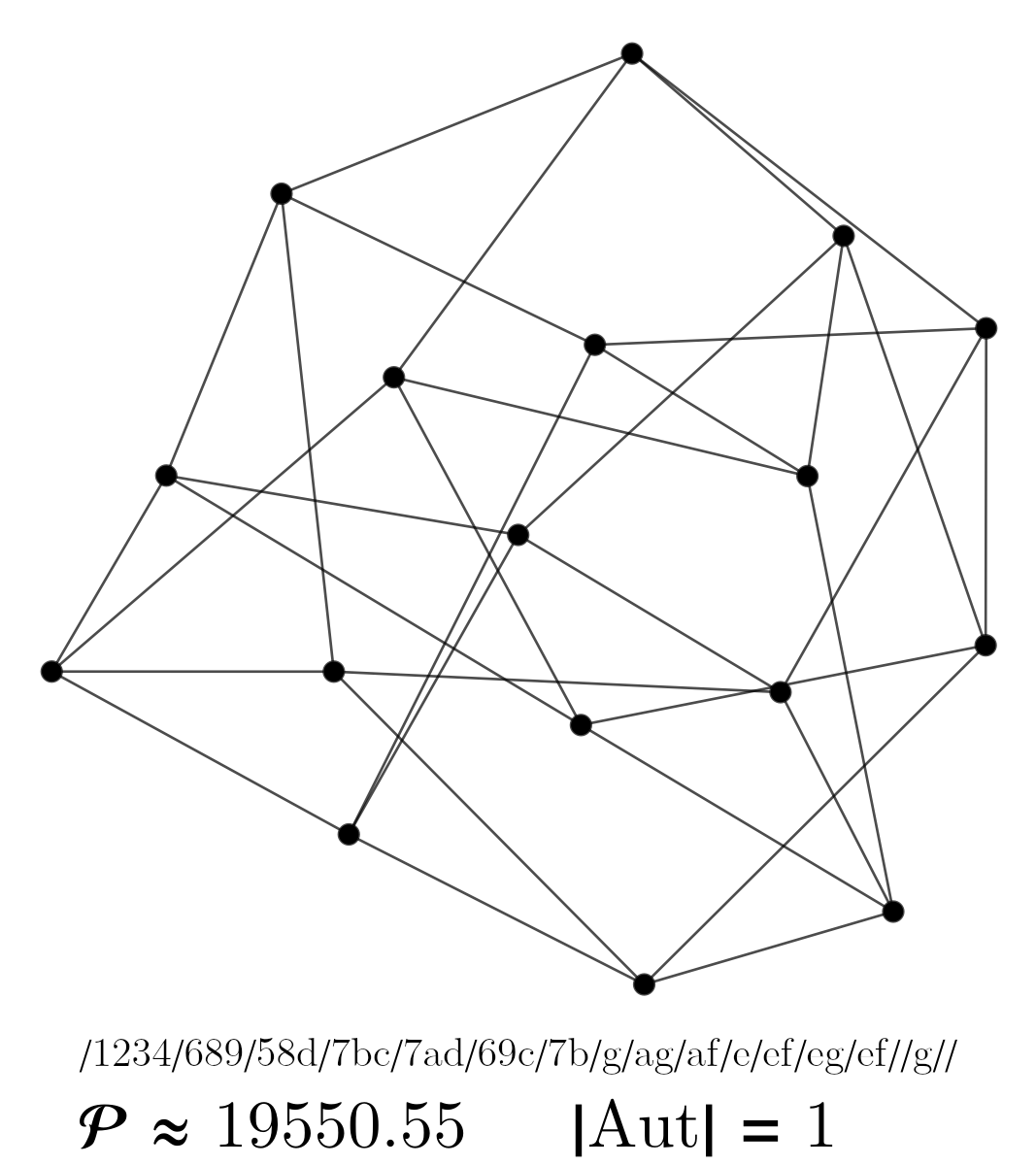}
		\subcaption{}
	\end{subfigure}
	\begin{subfigure}[b]{.24 \textwidth}
		\includegraphics[width=\linewidth]{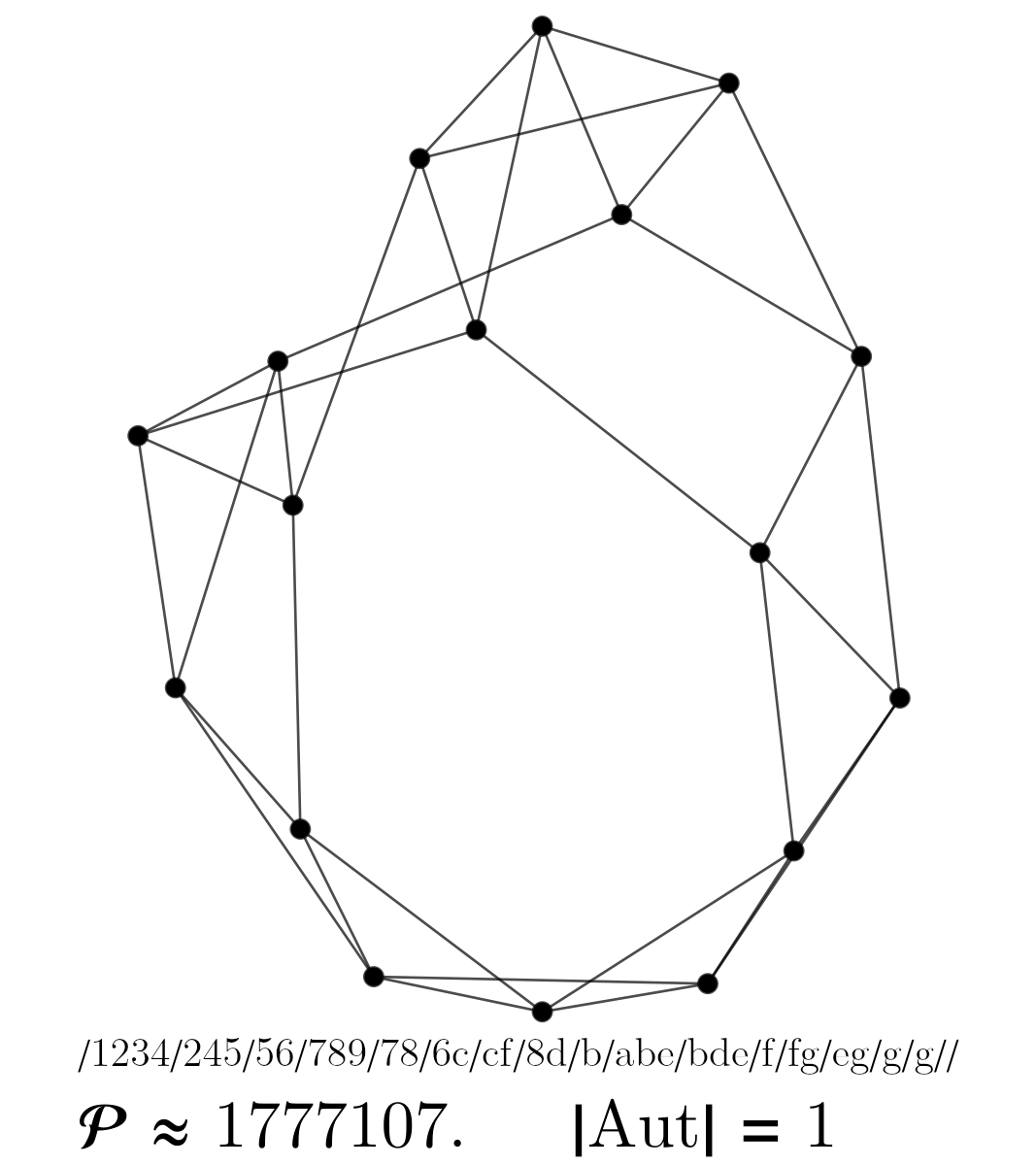}
		\subcaption{}
	\end{subfigure}
	\begin{subfigure}[b]{.24 \textwidth}
		\includegraphics[width=\linewidth]{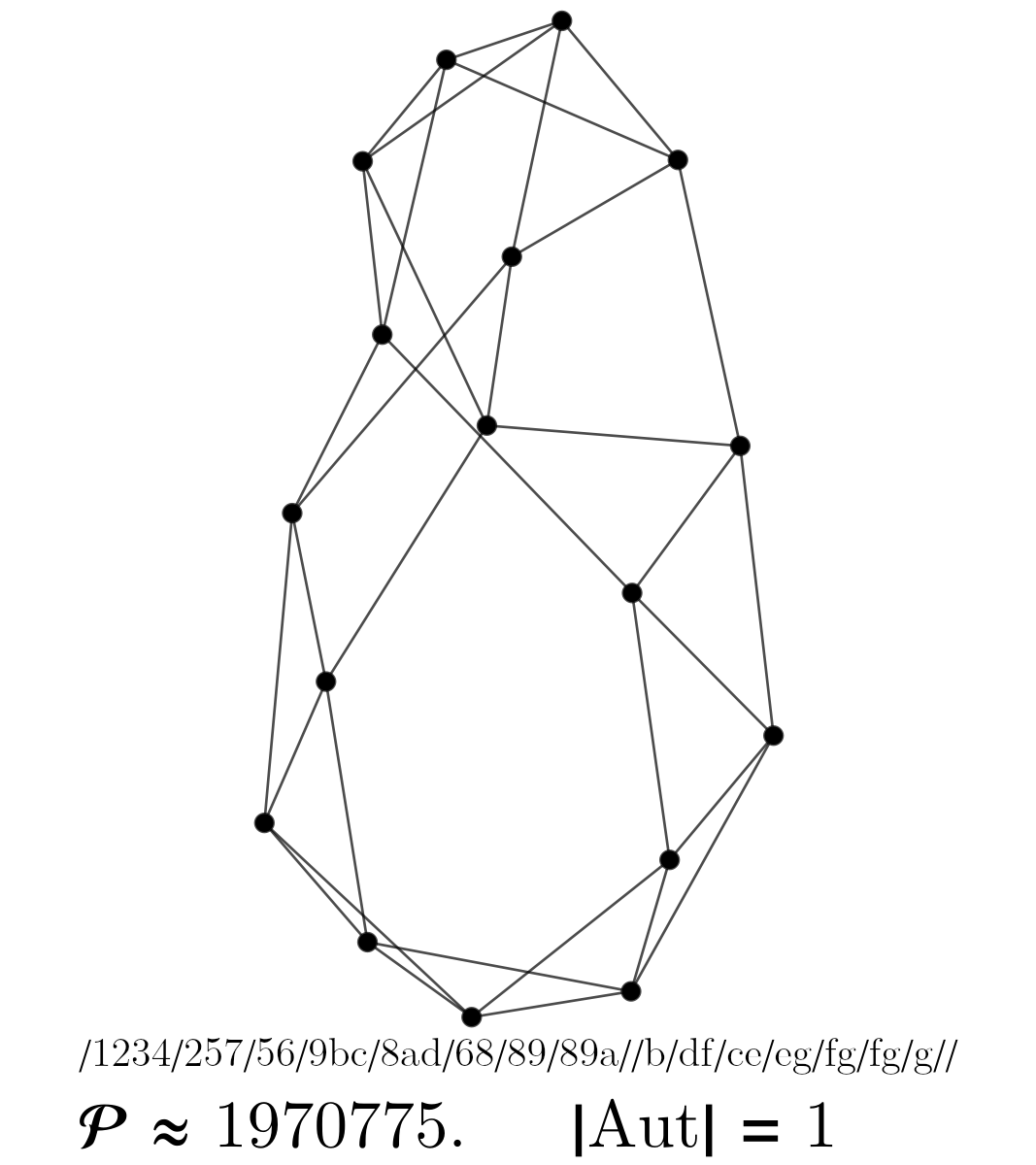}
		\subcaption{}
	\end{subfigure}
	
	\caption{The two smallest and the two largest graphs of the 15-loop samples.}
	\label{smallest_15}
\end{figure}

\begin{figure}[htb]
	\begin{subfigure}[b]{.24 \textwidth}
		\includegraphics[width=\linewidth]{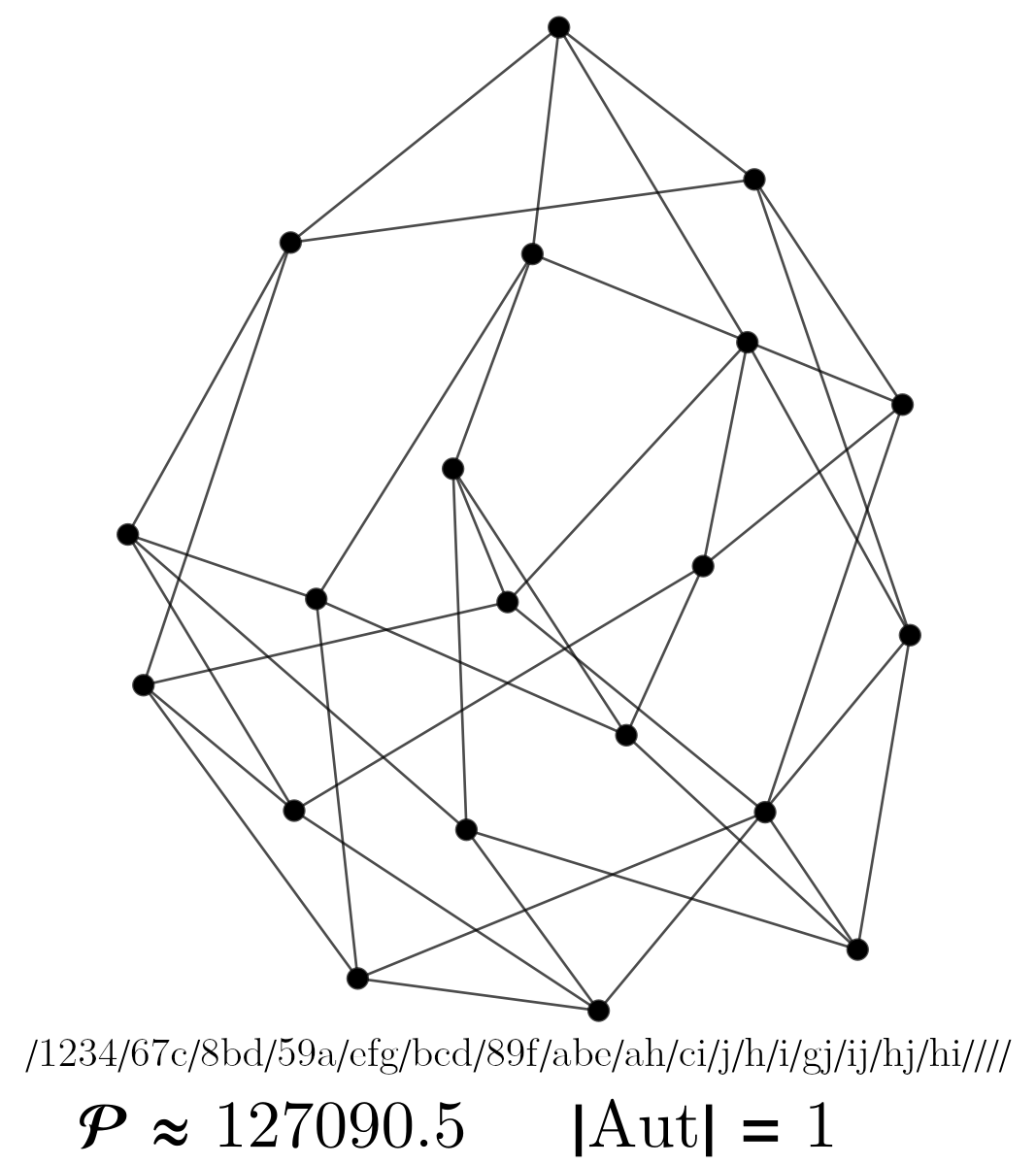}
		\subcaption{}
	\end{subfigure}
	\begin{subfigure}[b]{.24 \textwidth}
		\includegraphics[width=\linewidth]{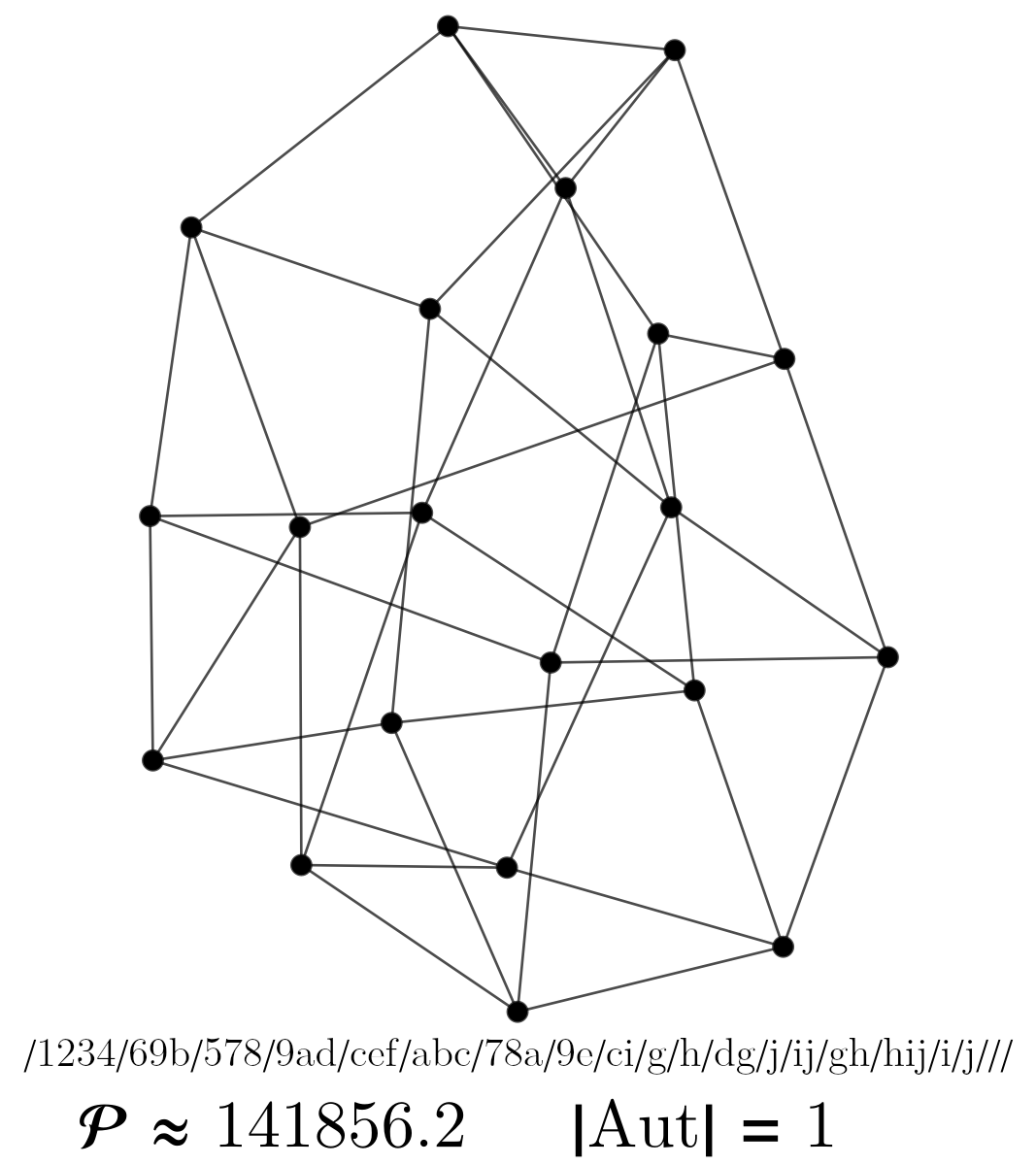}
		\subcaption{}
	\end{subfigure}
	\begin{subfigure}[b]{.24 \textwidth}
		\includegraphics[width=\linewidth]{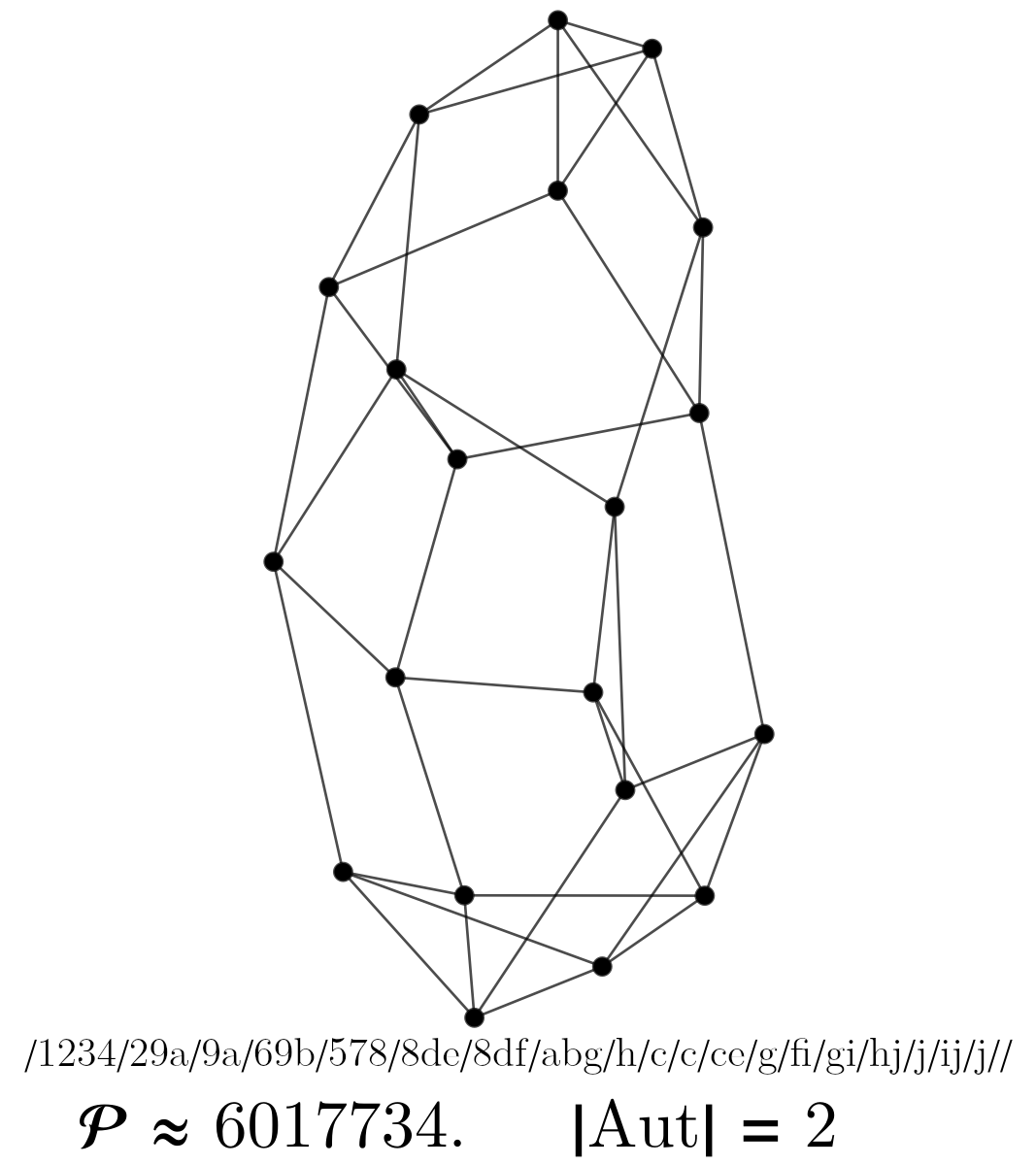}
		\subcaption{}
	\end{subfigure}
	\begin{subfigure}[b]{.24 \textwidth}
		\includegraphics[width=\linewidth]{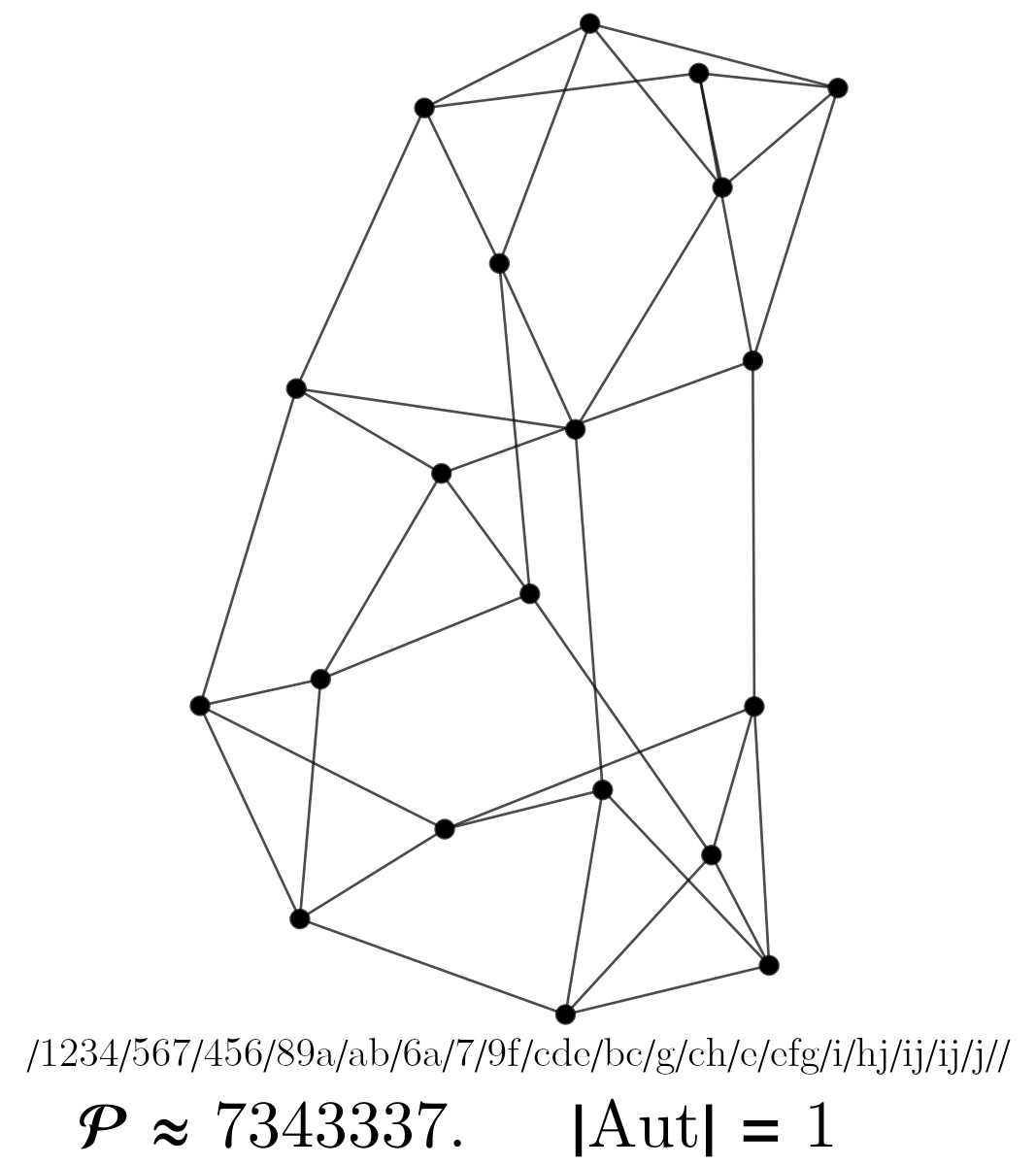}
		\subcaption{}
	\end{subfigure}
	
	\caption{The two smallest and the two largest graphs of the 18-loop samples. Note that the difference in shape is less pronounced than in the previous examples because this sample is very small, compared to the total number of 18-loop graphs, see \cref{tab:samples}.}
	\label{smallest_18}
\end{figure}

The smallest periods, shown in \cref{,smallest_13}, do not show any obvious regular pattern for $L \geq 12$. Conversely, they appear to be particularly \enquote{dense} or \enquote{irregular}. 
For $L\geq 14$ loops,  we can not expect to find the overall smallest and largest periods of the populations in our non-complete samples.
Even in the non-complete samples, the smallest graphs  appear more dense than the largest ones, see \cref{smallest_15,smallest_18}.

The fact that the $(1,2)$-circulants constitute the largest periods of each loop order motivates to examine the other $(i,j)$-circulants as well. To this end, we express their period relative to the average of the corresponding loop order like in \cref{mean_scaling}, $\period = x \cdot \left \langle \period \right \rangle $. The results are given in \cref{tab:circulants} in the form $(i,j,x)$. Note that many circulants are isomorphic, the choice of $(i,j)$ is generally not unique.

\begin{table}[htb]
	\centering
	\begin{tblr}{ width=\linewidth,hlines,rowsep=0pt,row{1}={rowsep=3pt}, colspec={|Q[r,m]|Q[l]|}, columns={m}
		}
		L & Circulants in the form $\big(i, ~j, ~\period/\left \langle \period \right \rangle  \big)$ \\
		5   & $  (1,~2,~1.033)$ \\
		6   & $ (1,~3,~0.568), \quad (1,~2,~1.337) $ \\
		7 	& $ (1,~3,~0.634), \quad (1,~2,~1.671)$  \\
		8   & $  (1,~4,~0.427), \quad (1,~3,~0.446),\quad (1,~2,~2.359) $ \\
		9   & $ (1,~3,~0.366),\quad (1,~2,~3.456) $ \\
		10  & {$ (2,~3,~ 0.284), \quad (1,~3,~ 0.365), \quad (1,~5,~0.425),$\\ $  (1,~4,~0.661), \quad (3,~4,~0.676), \quad (1,~2,~5.430)$ }  \\
		11  & $ (1,~ 5,~ 0.254), \quad (1,~ 3,~ 0.360),\quad (1,~ 2,~ 8.939) $\\
		12  & $(1,~ 4,~ 0.238),\quad (1,~ 3,~ 0.374), \quad (1,~ 6,~ 0.527), \quad (1,~ 2,~ 15.43)  $ \\
		13  &  {$ (1,~ 4,~ 0.219), \quad (1,~ 6,~ 0.228), \quad (1,~3,~ 0.413) $\\ $  (1,~5,~1.128), \quad (3,~5,~1.130),\quad (1,~2,~27.71) $} \\
		14 	& {$ (1,~ 6,~ 0.230), \quad (1,~4,~ 0.241), \quad (1,~3,~ 0.469) $\\ $ (1,~7,~0.772), \quad (1,~ 2,~ 51.41)    $}  \\
		15 	& $ (1,~ 4,~ 0.227),\quad (1,~ 5,~ 0.241), \quad (1,~3,~0.555), \quad (1,~ 2,~ 99.03)     $ \\
		16 	& {$ (1,~ 5,~ 0.195), \quad (1,~ 4,~ 0.216), \quad (2,~ 3,~ 0.261), \quad (1,~ 3,~ 0.681)  $ \\$ (1,~ 8,~ 1.306), \quad (1,~ 6,~ 2.778), \quad (1,~ 2,~ 197.1) $} \\
		17 	& $ (1,~4,~0.221), \quad (1,~ 7,~ 0.287), \quad (1,~ 3,~ 0.844), \quad (1,~ 2,~ 395.1)   $ \\
		18 	& {$(1, ~ 8,~ 0.195), \quad (1,~ 4,~ 0.234), \quad (2,~ 5,~ 0.274), \quad (1,~ 5,~ 0.277) , \quad (4, ~5,~ 0.282) $ \\$  (1,~ 6,~ 0.314), \quad (1,~ 3,~ 1.047), \quad (1,~ 9,~ 2.327), \quad (1,~ 2,~ 792.2) $}\\ 
	\end{tblr}
	\caption{Periods of non-isomorphic circulants. The zigzag $(1,2,x)$ is by far the largest period of each loop order, see also \cref{fig:distribution}. For $L \leq 11$, the smallest circulant is also the smallest period of that loop order, but this is false for $L \geq 12$.}
	\label{tab:circulants}
\end{table}

The data in \cref{tab:circulants} shows that even if the zigzag is by far the largest period, many other circulants are smaller than the average. The smallest circulant, for $L \geq 12$, does not coincide with the smallest period of the same loop order. The difference grows with the loop order, such that even in our $L=18s$ sample, which is tiny compared to the number of 18-loop graphs, we find a graph with smaller period than the smallest 18-loop circulant.

\FloatBarrier

\subsection{Probability distribution function near the mean}\label{sec:distribution_central}

\begin{figure}[htbp]
	\begin{subfigure}[b]{.48 \textwidth}
		\includegraphics[width=\linewidth]{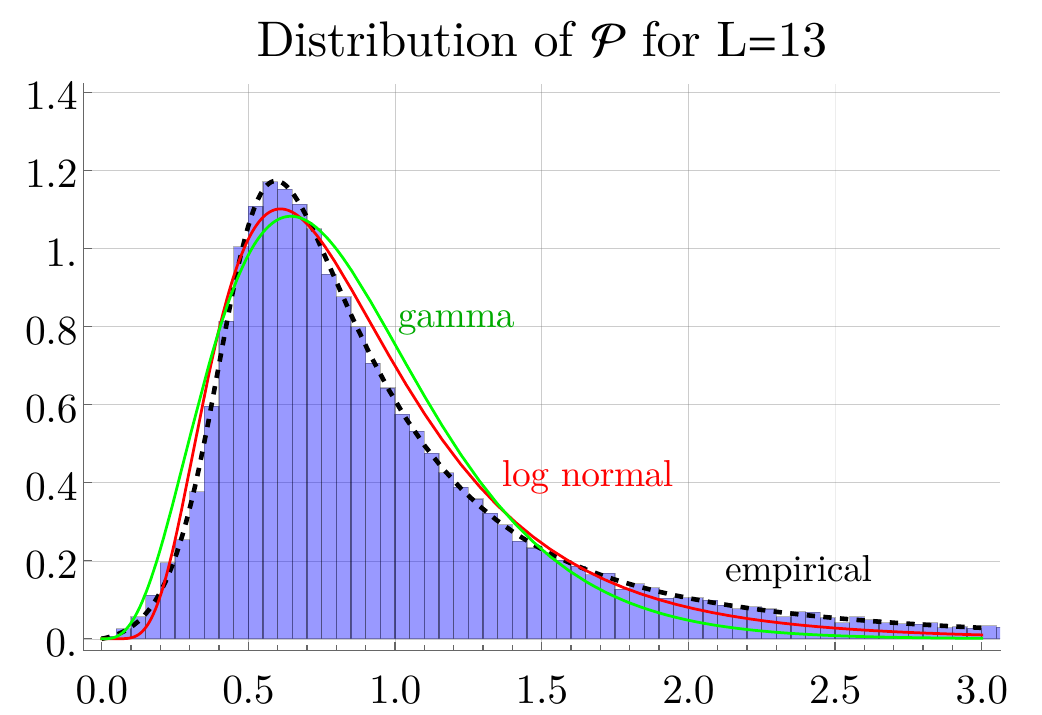}
		\subcaption{}
		\label{histograms_comparisonA}
	\end{subfigure}
	\begin{subfigure}[b]{.48 \textwidth}
		\includegraphics[width=\linewidth]{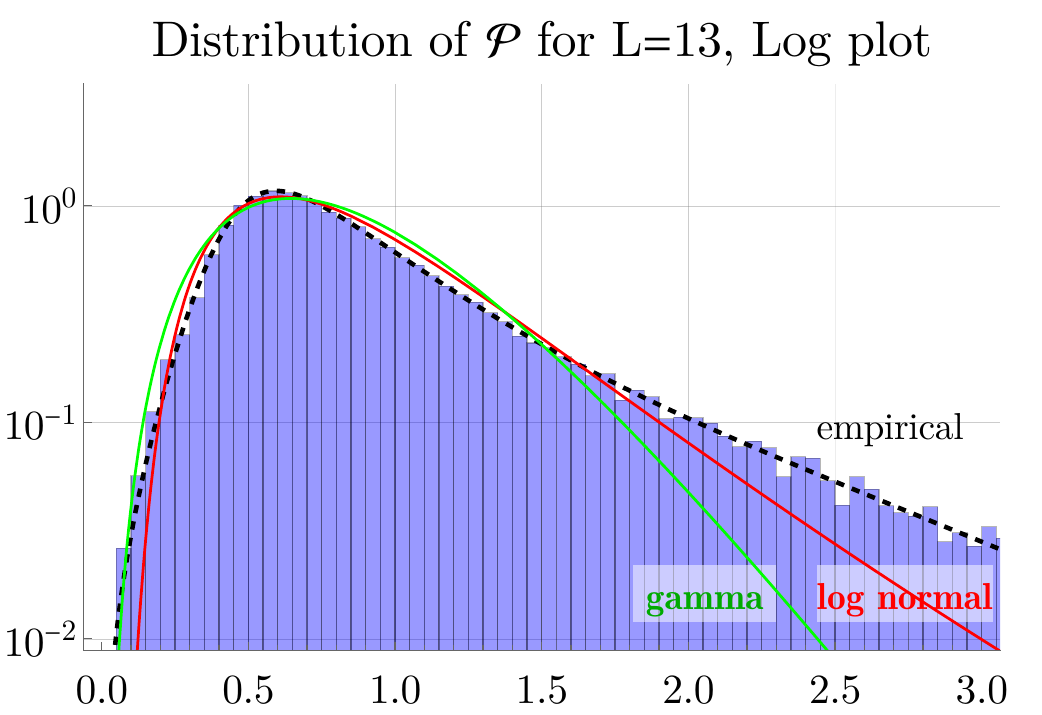}
		\subcaption{}
		\label{histograms_comparisonB}
	\end{subfigure}
	
	\caption{Histograms of the symmetry-factor-weighted period, normalized to unit mean, for $L=13$, with $y$-axis scaled linearly (\ref{histograms_comparisonA}) or logarithmically (\ref{histograms_comparisonB}). The red resp. green lines indicate the best fits for log-normal resp. gamma distributions. Both clearly fail to describe the empirical distribution. The dashed black line is the observed distribution function \cref{P_distribution}. }
	\label{fig:histograms_comparison}
\end{figure}

In the present subsection, we consider the distribution of periods normalized to unit mean as in \cref{mean_scaling}.
We have established in \cref{sec:moments,sec:largest} that the higher moments of this  distribution  are most likely infinite in the limit $L\rightarrow \infty$. Nevertheless, our data suggests that in the vicinity of the mean, the distribution  converges to a non-degenerate limiting distribution, shown in \cref{fig:histograms_comparison}. Our goal is to describe this distribution near the mean, ignoring the extreme values which cause the moments to diverge. 

We have tried fitting several common distributions to the empirical data, but none of them was satisfying. For illustration, the plot \cref{fig:histograms_comparison} shows the best fits for a gamma distribution and a log-normal distribution, both of them obviously miss qualitative features of the distribution.

We construct an empirical probability distribution function $\rho(x)$ by inspecting the characteristic features of the histogram in a logarithmic plot such as \cref{histograms_comparisonB}. We find that 
\begin{enumerate}
	\item The log-density grows approximately linearly for small $x$,
	\item The log-density then falls of approximately linearly, but with different slope, at $x \sim 1$, and
	\item For $x \gg 1$, the log-density falls of linearly with yet another slope.
	\item The density vanishes at $x=0$ since periods are positive.
\end{enumerate}
One possible function with the sought-after properties is
\begin{align}\label{P_distribution}
\rho(x) &= s \cdot x \frac{e^{l\cdot (a-b)  } + e^{m x}}{e^{-l \cdot (x-b)} + e^{(m+r)x}}.
\end{align}
Here, the parameters $a$ and $b$ are defined such that they take values near unity for our data, and the slopes $l,m,r$ are positive.
Out of the six parameters in \cref{P_distribution}, only four are independent; the overall scale $s$ is determined by total  probability being unity,
and one of the remaining parameters $\{a,b,m,l,r\}$  can be eliminated by enforcing unit expectation.   See \cref{tab:fit} for fit parameters.

\begin{table}
	\centering
	\begin{tblr}{vlines,
			vline{7}={1}{-}{solid},
			vline{7}={2}{-}{solid},
			hline{1}={solid},
			hline{2,Z}={solid},
			rowsep=0pt,
			cells={font=\fontsize{11pt}{12pt}\selectfont,mode=math },
			columns={halign=r},
			column{1}={halign=c},
			row{1}={halign=c,rowsep=0pt, font=\fontsize{12pt}{14pt}\selectfont } 
		}
		L &  a  &b & m & l & r & a & b & m& l& r  \\
		11  	 & 3.521 & 1.818 & 1.636 & 1.636 & 1.738 & 0.741 & 0.610 & 2.907 & 13.21 & 2.066 \\
		12  	 & 2.463 & 1.496 & 3.166 & 3.167 & 1.872 & 0.677 & 0.508 & 2.325 & 15.16 & 1.622 \\
		13  	 & 1.114 & 0.696 & 2.362 & 7.142 & 1.572 & 0.656 & 0.458 & 2.373 & 16.19 & 1.361 \\
		13\text{s}   	 & 0.711 & 0.493 & 2.392 & 15.37 & 1.357 & 0.701 & 0.485 & 2.576 & 15.06 & 1.386 \\
		13\star & 1.001 & 0.613 & 2.222 & 8.508 & 1.407 & 0.604 & 0.436 & 2.374 & 19.01 & 1.355 \\
		14\text{s}  	 & 0.684 & 0.445 & 2.378 & 15.826 & 1.176 & 0.642 & 0.426 & 2.489 & 16.44 & 1.235 \\
		14\star & 0.883 & 0.531 & 2.263 & 10.11 & 1.285 & 0.645 & 0.428 & 2.461 & 16.25 & 1.254 \\
		15\text{s}  	 & 0.740 & 0.449 & 2.256 & 12.94 & 1.177 & 0.589 & 0.393 & 2.459 & 17.89 & 1.260 \\
		15\star& 0.739 & 0.444 & 2.313 & 13.39 & 1.118 & 0.590 & 0.389 & 2.546 & 18.63 & 1.176 \\
		16\text{s}  	 & 0.708 & 0.414 & 2.389 & 14.14 & 1.041 & 0.626 & 0.402 & 2.648 & 15.59 & 1.295 \\
		17\text{s}  	 & 0.749 & 0.424 & 2.500 & 12.91 & 1.065 & 0.631 & 0.387 & 2.652 & 15.86 & 1.169 \\
	\end{tblr}
	\caption{Best fit parameters for the distribution \cref{P_distribution}. Loop order \|   5 columns for periods weighted by symmetry factors \| \|  5 columns for periods without symmetry factors. Observe that the parameters are similar for all loop orders, the distributions have similar shapes. For $L=\left \lbrace 11,12 \right \rbrace $ with symmetry factor, the parameters $m$ and $l$ are degenerate, this corresponds to the strong distortion visible in \cref{fig:histograms_log_small}. }
	\label{tab:fit}
\end{table}

\begin{figure}[htbp]
	\centering
	\begin{subfigure}[b]{.48 \textwidth}
		\includegraphics[width=\linewidth]{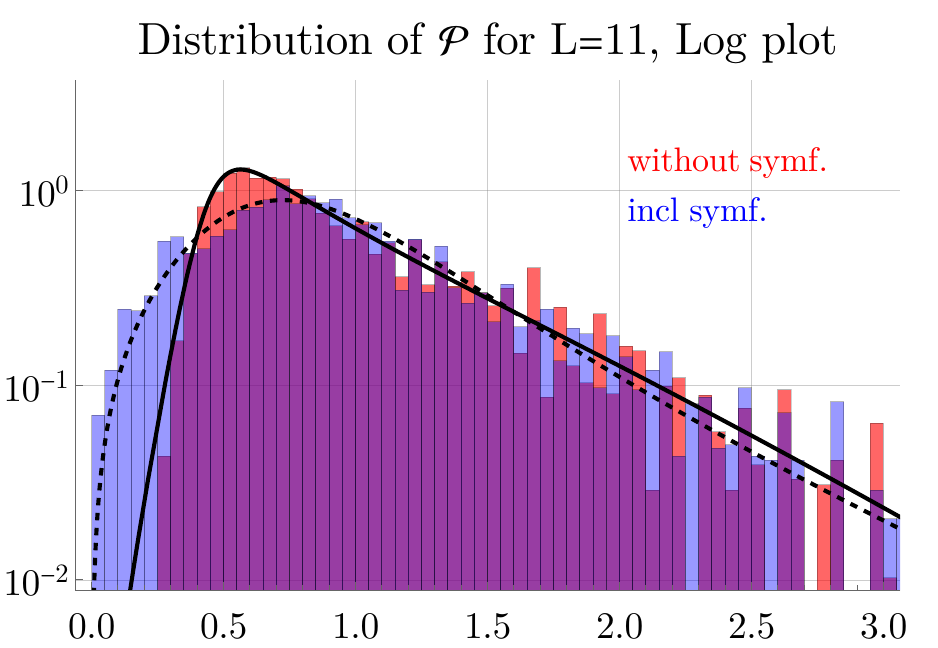}
		\subcaption{}
	\end{subfigure}
	\begin{subfigure}[b]{.48 \textwidth}
		\includegraphics[width=\linewidth]{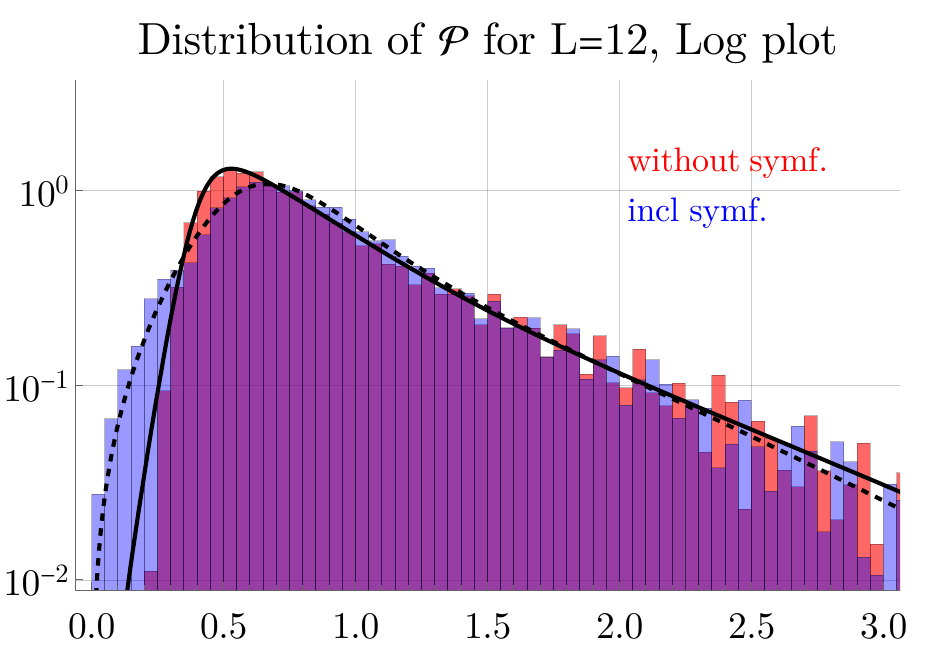}
		\subcaption{}
	\end{subfigure}
	
	\caption{Histograms of the   period for smaller loop numbers. The empirical distribution function \cref{P_distribution} still captures the data, but it is visibly distorted from its shape at higher loop orders. }
	\label{fig:histograms_log_small}
\end{figure}

\begin{figure}[htbp]
	\centering
	\begin{subfigure}[b]{.48 \textwidth}
		\includegraphics[width=\linewidth]{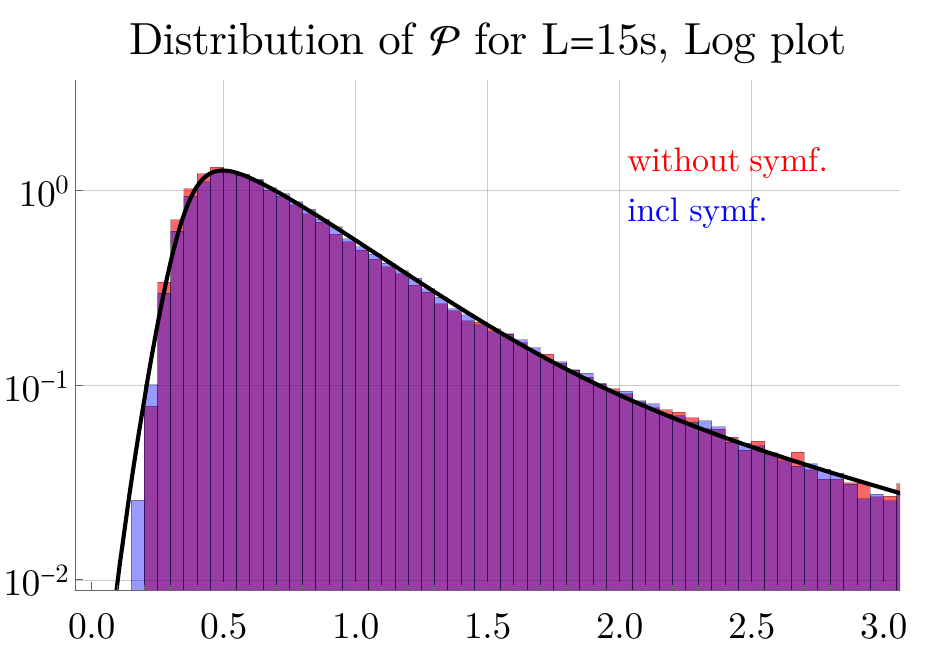}
		\subcaption{}
	\end{subfigure}
	\begin{subfigure}[b]{.48 \textwidth}
		\includegraphics[width=\linewidth]{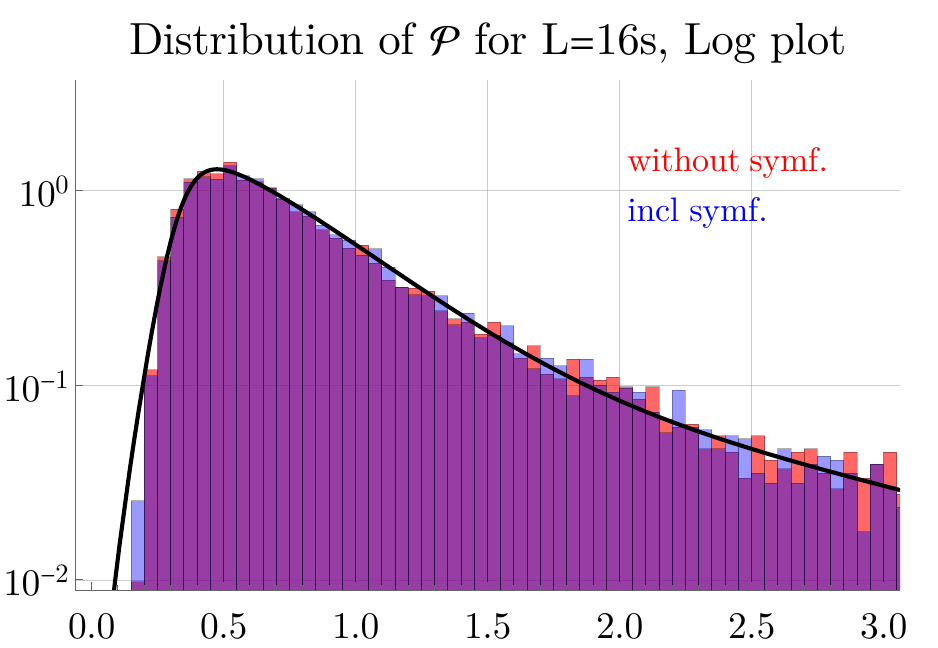}
		\subcaption{}
	\end{subfigure}
	
	\caption{Histograms of the period for higher loop numbers, together with best fit empirical distribution. The shape  is very similar to the one for $L=13$ loops (\cref{fig:histograms_comparison}), apart from larger fluctuations in the histograms. At 16 loops, the influence of symmetry factors is barely visible. }
	\label{fig:histograms_log_big}
\end{figure}

As shown in \cref{fig:histograms_comparison} (dashed black line), this density function describes the data well for $L=13$ loops. 
Even for small loop orders, the distribution \cref{P_distribution} is a fairly good approximation to the empirical histograms, see \cref{fig:histograms_log_small}, but the shape is visibly different from higher loop orders and automorphisms have a large influcence. 
Conversely, for high loop orders $L\geq 15$, our data sets are smaller and the fit parameters are less reliable. Nonetheless, the fitted empirical distribution resembles the histograms acceptably well, see \cref{fig:histograms_log_big}.

The fit parameters of \cref{tab:fit} are similar in magnitude across all loop orders, but not quite identical. In particular, the left slope $l$ varies significantly, this might indicate that an exponential (linear in the log plot) growth is not an accurate model for the limiting distribution. Conversely, the middle slope $m \approx 2.3$ and the right slope $r \approx 1.2$ are fairly consistent. This implies that the probability for a period to be $x$-times the mean of the corresponding loop order falls of approximately $\propto \exp(-1.2 x)$. Nevertheless, this description is accurate only for moderately large $x$, not for the outliers (\cref{sec:largest}). In particular, the distribution function \cref{P_distribution} has finite moments for all values of its parameters, it therefore arbitrarily much underestimates the probability of large outliers  in the limit $L\rightarrow \infty$.

\FloatBarrier

\subsection{Almost-Gaussian model for the period distribution} \label{sec:nonlinear}

Consider the distribution of the (normalized) $t$\textsuperscript{th} power of the period, 
\begin{align}\label{period_transformed}
	\frac{1}{\left \langle  \period  ^t \right \rangle  }  \period   ^t, \qquad t\in \mathbb R.
\end{align}
Here, $t=1$ reproduces the normalized period examined in \cref{sec:distribution_central}. Another obvious choice is  $t=-1$. The distribution of  the inverse period $\period^{-1}$  has an overall wider, more round peak, compared to $\period$ in \cref{fig:histograms_comparison}, but we skip the details at this point.

\begin{figure}[htbp]
	\begin{subfigure}[b]{.48 \textwidth}
		\includegraphics[width=\linewidth]{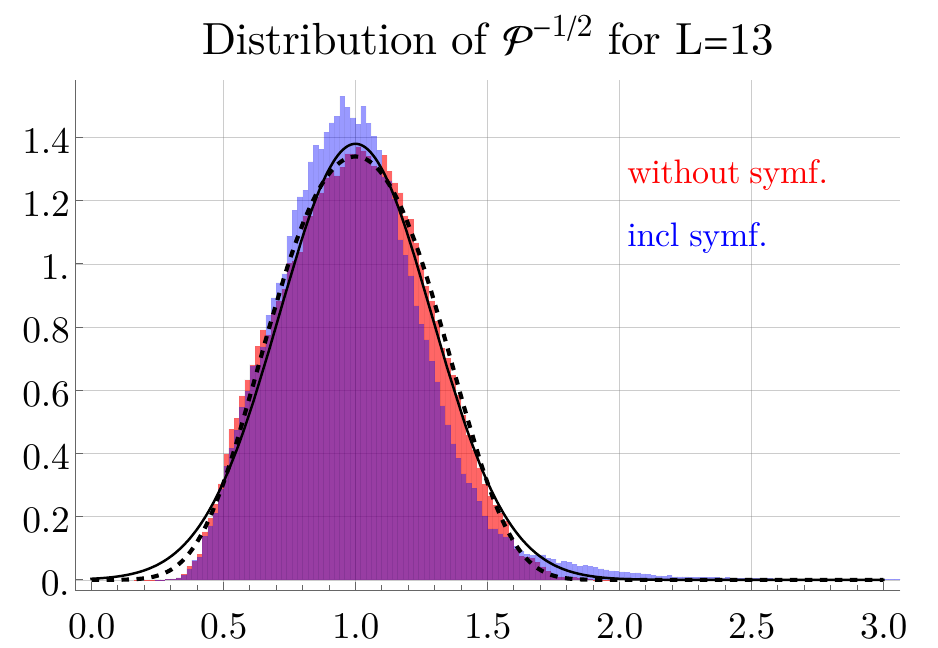}
		\subcaption{}
		
	\end{subfigure}
	\begin{subfigure}[b]{.48 \textwidth}
		\includegraphics[width=\linewidth]{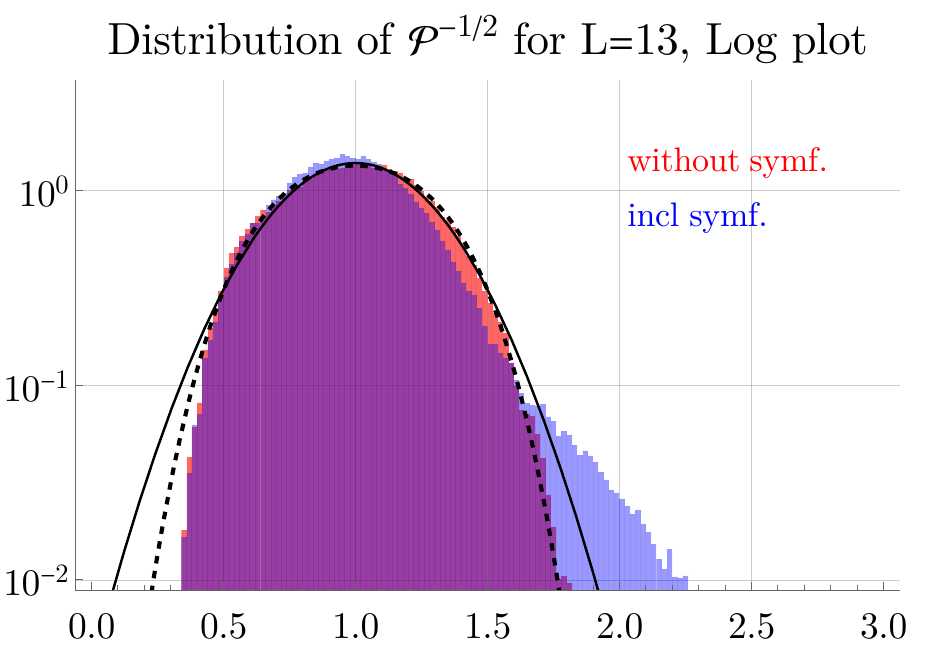}
		\subcaption{}
		\label{fig:histogram_sqrt_log}
	\end{subfigure}
	
	\caption{Histogram of $\period^{-\frac 12}$ for 13 loops. The solid line is the best fit normal distribution to the periods without symmetry factor.  A quartic correction \cref{quartic_distribution} (dashed line) improves the fit, but still overestimates the presence of large periods (small values of $\period^{-\frac 12}$). }
	\label{fig:histogram_sqrt}
\end{figure}

A remarkable, but unexpected, phenomenon occurs at $t=-\frac 12$.  If no symmetry factors are taken into account, the quantity   $\frac{1}{\sqrt{\period}}$ almost follows a normal distribution, see \cref{fig:histogram_sqrt} for  $L=13$. We scanned the parameter range   $t\in (-3,+3)$ to confirm that $t=-\frac 12$ is in fact the only choice where the distribution is close to a Gaussian.

As seen  in \cref{fig:histogram_sqrt}, the normal distribution (i.e. a parabola in the log plot) is a good description near the center $x=1$. An even better fit is obtained with a quartic distribution 
\begin{align}\label{quartic_distribution}
	\rho(x) &= \frac{b e^{-\frac{c^4}{b^4}}}{2c^2 K_{\frac 14} \left( \frac{c^4}{b^4} \right)  } \cdot \exp \left( - \frac{(x-1)^2}{2b^2}  - \frac{(x-1)^4}{32 c^4} \right)  , \qquad c>0.
\end{align}
Here, the first factor is a normalization to unit total probability, $K_\frac 14$ is the modified Bessel function of the second kind, and the parameters $b,c$ are generalizations of the parameter $\sigma$ of the normal distribution. We remark that the empirical model \cref{quartic_distribution}   can not be exactly correct since it assigns a (tiny) non-zero probability to negative periods.

Best fit parameters, both for a normal distribution and for \cref{quartic_distribution}, are shown in \cref{tab:sqrt}. 
Even at high loop orders, the transformed distribution continues to be well approximated by a normal distribution. Our data suggests a diminishing influence of the quartic correction with higher loop order.

\begin{table}
	\centering 
	\begin{tblr}{vlines,
			hline{1}={solid},
			hline{2,Z}={solid},
			rowsep=0pt,
			cells={font=\fontsize{11pt}{12pt}\selectfont },
			columns={halign=l},
			column{1}={halign=l},
			row{1}={halign=c,rowsep=2pt, font=\fontsize{12pt}{14pt}\selectfont,mode=math } 
		}
		L &   1 /  \left \langle \frac{1}{\sqrt{\period }} \right \rangle   &  \sigma   &  b  & c  \\
		11  	 & 78.860132 & 0.2621 & 0.4383 & 0.1902 \\
		12  	 & 110.22109 & 0.2776 & 0.3788 & 0.2215 \\
		13  	 & 151.94024 & 0.2892 & 0.3510 & 0.2554 \\
		13s  	 & 151.774 & 0.289  & 0.351  & 0.254 \\
		13$\star$& 151.874 & 0.293  & 0.372  & 0.245 \\
		14s 	 & 207.131 & 0.299  & 0.342  & 0.287 \\
		14$\star$& 207.258 & 0.298  & 0.340  & 0.288 \\
		15s 	 & 279.405 & 0.306  & 0.348  & 0.297 \\
		15$\star$& 279.543 & 0.307  & 0.346  & 0.303 \\
		16s 	 & 373.295 & 0.313  & 0.346  & 0.320 \\
		17s 	 & 497.744 & 0.311  & 0.325  & 0.386 \\
		18s 	 & 666.063 & 0.318  & 0.334  & 0.385 \\
	\end{tblr}
	\caption{Best fit parameters for the distribution of $  \period  ^{-\frac 12}$ without symmetry factors. Loop order  \| normalization factor \| standard deviation of normal distribution \| parameters $b,c$ of \cref{quartic_distribution}.  }
	\label{tab:sqrt}
\end{table}

The central moments (\cref{def:central_moments}) of the distribution of $ \period  ^{-\frac 12}$   behave much better than those of $\period$ in \cref{tab:means,tab:centralMoments,fig:Cj_growth}. Firstly, the odd moments $C_3, C_5$ are significantly smaller than the even ones $C_2, C_4, C_6$, which confirms that the distribution is almost symmetric around its mean. Secondly, the moments $C_j$ overall become smaller as $j$ increases, in contrast to the behavior for the non-inverted distribution. Thirdly, the odd moments stay constant with growing loop number $L$, while the even ones grow only moderately. Our data data makes it plausible, but not certain, that all moments of $  \period ^{-\frac 12}$ have a finite limit as $L\rightarrow \infty$. 

The models of the distribution of $\period^{-\frac 12}$ imply   corresponding models for the distribution of $\period$. For a normal distribution, where we introduce a lower cutoff $\delta >0$ to exclude negative periods,  a transformation $t \mapsto t^{-2}$ results in
\begin{align}\label{fitN}
	\rho_N(x) &=\left(\left \langle \frac{1}{\sqrt {\period (G)}} \right \rangle \right)^2\cdot  \frac{1}{\sqrt{2\pi} \sigma \left( 2-\operatorname{erfc}\left( \frac{1-\delta }{\sqrt 2 s} \right)   \right)  } \frac{e^{- \frac{\left( \sqrt x-1 \right) ^2}{2 x \sigma^2} }}{ x^{\frac 32}} .
\end{align}
For the quartic distribution, we find (if no lower limit  $\delta$ is imposed) 
\begin{align}\label{fitQ}
	\rho_Q(x) &= \frac{b e^{-\frac{c^4}{b^4}}}{4c^2 K_{\frac 14} \left( \frac{c^4}{b^4} \right)  } \cdot\frac{\exp \left( - \frac{(\sqrt x-1)^2}{2xb^2}  - \frac{(\sqrt x-1)^4}{32 x^2c^4} \right)}{x^{\frac 32}}.
\end{align}
A corresponding distribution centered around the the empirical distribution mean can be obtained as usual by scaling,
\begin{align}\label{fitQ_scaling}
	\rho(u)\cdot \left(\left \langle \frac{1}{\sqrt {\period (G)}} \right \rangle \right)^{-2} , \qquad \text{where} \qquad u= x\cdot  \left(\left \langle \frac{1}{\sqrt {\period (G)}} \right \rangle \right)^{-2} .
\end{align}

The distribution functions in  \cref{fitN,fitQ} constitute   alternative empirical models, called \enquote{inverse models} hereafter, for the distribution of periods. These models complement the \enquote{exponential model} \cref{P_distribution}. 
As they are continuous, both types of models fail to describe the outliers $\period \gg \left \langle \period \right \rangle $ and neither of them is a candidate for an exact limiting distribution as $L \rightarrow \infty$. In fact, the presence of a set of discrete outliers (which is a small set, but growing nonetheless) makes it very unlikely that a limiting distribution can be given in closed form at all.

The two different empirical models have quite different features:
\begin{itemize}
	\item For $x\rightarrow\infty$, both   \cref{fitN} and \cref{fitQ} decay   $\sim x^{-\frac 3 2}$, consequently, their expectation $\int \d x \, x\cdot \rho(x)$ and all higher moments are infinite. The inverse models correctly reproduce a divergence of higher moments in the limit $L \rightarrow \infty$.
	\item As the higher moments of $\period^{-\frac 12}$ are  finite, the inverse models have a well-defined limit  for $L \rightarrow \infty$. Moreover, the inverse models can be improved systematically by measuring these moments in samples and constructing a corresponding distribution. Conversely, the exponential model has no obvious way to increase the accuracy, apart from introducing ad-hoc terms and fitting them to a histogram.
	\item The inverse models are very accurate for the distribution of small values $\period<\left \langle \period \right \rangle  $, whereas the exponential model is superior for values of $\period>\left \langle \period \right \rangle  $ Compare the performance for the left resp. right side of the histogram \cref{fig:fit_comparison} in the introduction section. 
\end{itemize}

\FloatBarrier

\subsection{Logarithm of the period}\label{sec:logarithm}

Although the distribution of periods is well-behaved near the mean (\cref{sec:distribution_central}), a systematic examination is considerably hindered by the fact that the empirical moments diverge. The presence of very large outliers suggests considering the logarithm $\ln \period(G)$ instead of the period itself. Analogously to \cref{sec:distribution_central}, we normalize our distribution to unit mean by dividing by $\left \langle \ln \period \right \rangle $. The normalization factor is given explicitly in \cref{tab:log_means} as it can not be inferred   from $\left \langle \period \right \rangle $ given in \cref{tab:means,tab:means_Aut}. The weighting with symmetry factors can be reconstructed from  
\begin{align*}
	\left \langle \ln \frac{\period }{\abs{\Aut }} \right \rangle = \left \langle  \ln \period  \right \rangle -\left \langle \ln \abs{\Aut } \right \rangle ,
\end{align*}
where the second summand is listed in \cref{tab:aut}.  Furthermore, the standard deviation can be obtained from the mean and $C_2$ according to \cref{C2_std}. 
We computed the central moments (\cref{def:central_moments}) for the distribution of $\ln \period$, both with and without symmetry factors, results are again reported in \cref{tab:log_means}. The median and quantile boundaries of $\ln \period$ can be reconstructed from \cref{tab:means,tab:means_Aut}.

\begin{figure}[htbp]
	\centering
	\begin{subfigure}[b]{.48 \textwidth}
		\includegraphics[width=\linewidth]{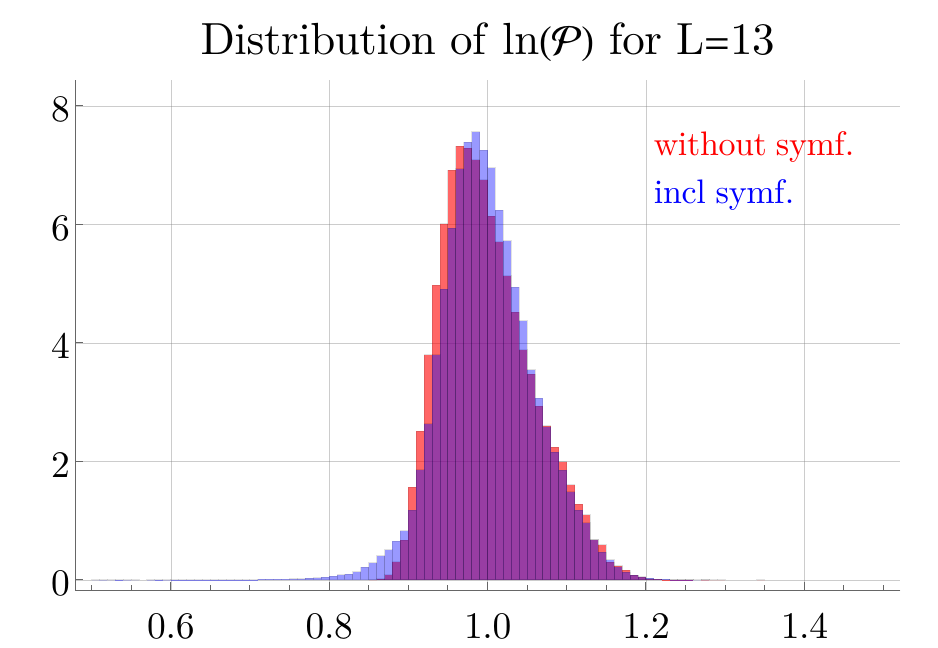}
		\subcaption{}
	\end{subfigure}
	\begin{subfigure}[b]{.48 \textwidth}
		\includegraphics[width=\linewidth]{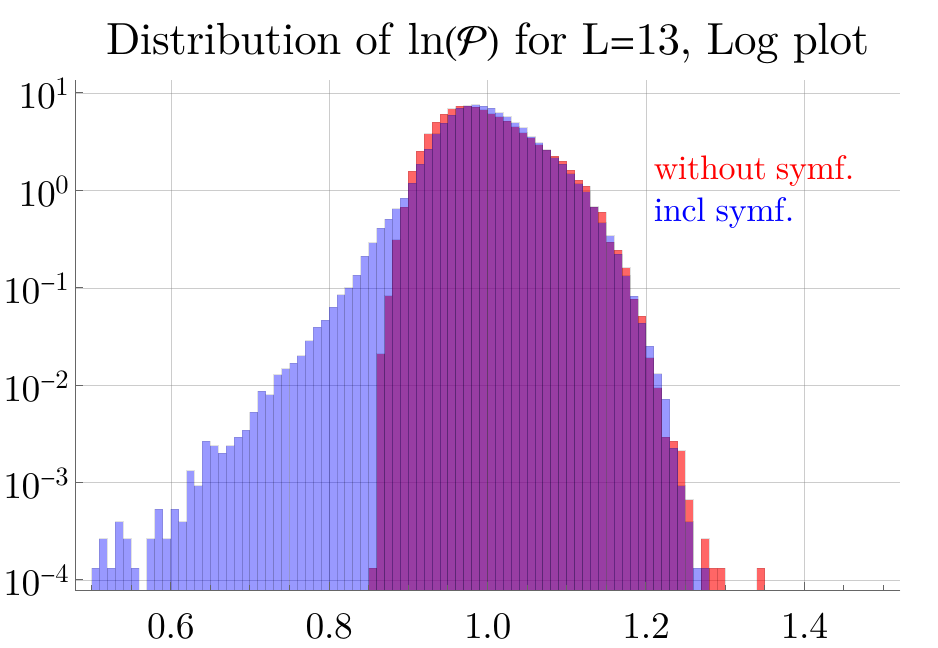}
		\subcaption{}
	\end{subfigure}
	
	\caption{Histograms of the logarithm of the period for 13 loops. Note the different scale of the $x$-axis compared to previous plots, and the large influence of symmetry factors in (b). }
	\label{fig:histograms_logP}
\end{figure}

\Cref{fig:histograms_logP} shows the normalized distribution of the logarithm of the period for $L=13$ loops. Comparing to   \cref{fig:histograms_comparison}, we see that the influence of symmetry factors is much more pronounced for $\ln \period$ than for $\period$.  Although, at 13 loops, already 84\% of all symmetry factors are unity (see \cref{tab:aut}), their presence significantly alters the distribution for small values of $\ln \period $.

\begin{table}[htb]
	\centering
	\begin{tblr}{vlines,
			vline{6}={1}{-}{solid},
				vline{6}={2}{-}{solid},
			hline{1}={solid},
			hline{2,Z}={solid},
			rowsep=0pt,
			cells={font=\fontsize{11pt}{12pt}\selectfont,mode=math },
			columns={halign=r},
			column{1}={halign=c},
			row{1}={halign=c,rowsep=2pt, font=\fontsize{12pt}{14pt}\selectfont } 
		}
		L &  \left \langle \ln  \period   \right \rangle   &  C_2\cdot 10^4  &  c_3\cdot 10^2  & c_4    &  C_2\cdot 10^4  &  c_3 \cdot 10^2  &  c_4  \\
		5        &  3.9847486 & 0.693 & 0 & 1 & 7918& 0 & 1 \\
		6        &  4.7935274 & 39.35 & -71.97 & 2.197 & 21540 & -128.6 & 3.003  \\
		7 		 & 5.7115799       & 26.86 &   5.173 & 2.132 & 1040 & -16.03 & 2.152  \\
		8   	 & 6.5197021       & 33.01 &   2.769  & 2.476 & 773.2 & -174.2 & 6.715 \\
		9   	 & 7.3279257      & 30.63 &  15.80 & 2.574  & 216.9 & -101.0& 4.355 \\
		10  	 & 8.0781075       & 31.50 &  24.09 & 2.529 & 137.9 & -173.4 & 9.166 \\
		11  	 & 8.7952253        & 31.96 &  35.49 & 2.564 & 64.21 & -99.85 & 5.747  \\
		12  	 & 9.4747560        & 32.50 &  46.66 & 2.718 & 44.25 & -56.40 & 5.731 \\
		13  	 & 10.125860      & 32.55 & 56.65   & 2.928  & 35.06 &  7.075 & 3.894 \\
		13\text{s}   	 &  10.124   & 32.57 & 57.95   & 2.914 & 31.20 & 60.29 & 3.026 \\
		13\star&   10.126   & 32.88 & 58.76  & 2.927 & 35.16 & 9.231 & 3.945 \\
		14\text{s}   	 &   10.754   & 32.27 & 64.77   & 3.137 & 30.73 & 65.89 & 3.217 \\
		14\star&   10.755  & 32.09 & 65.03   & 3.154 & 31.43 & 38.57 & 3.521  \\
		15\text{s}   	 &   11.358   & 31.07 & 70.66   & 3.377 & 29.66 & 70.56 &  3.419 \\
		15\star&   11.359  & 31.56 & 72.90   & 3.380 & 29.66 & 54.84 &  3.516 \\
		16\text{s}   	 &   11.918 & 28.48 & 74.15   & 3.503& 28.69 & 74.15 & 3.503  \\
		17\text{s}   	 &12.522 & 28.35 & 79.30  & 3.695 & 27.40 & 80.16 & 3.763  \\
		18\text{s}   	 & 13.109 & 27.07 & 85.06   & 3.782 & 25.73 & 84.40 &3.830  \\
	\end{tblr}
	\caption{Mean and central moments (\cref{def:central_moments}) of the logarithm of the period. Four columns refering to $\ln \period$ \| \| three columns refering to $\frac{\ln \period.}{\abs{\Aut}}$. Note that $C_2$ stays bounded as $L$ increases, unlike for the non-logarithmic period in \cref{tab:means}. }
	\label{tab:log_means}
\end{table}

The parameters in \cref{tab:log_means} indicate that  the distribution of $\ln \period$ differs considerably whether the symmetry factor is taken into account or not. In particular, for $L\leq 12$, the skewness for the symmetry-weighted periods is negative, reflecting a tail to the left of the distribution, caused by symmetry factors. For larger loop orders, the difference between the two distributions becomes smaller and the normalized variance seems to converge to a value $C_2 \approx 25 \cdot 20^{-4}$. The skewness and the kurtois, on the other hand, seem to grow with loop order, albeit less quickly as for the non-logarithmic period in \cref{tab:means,tab:means_Aut}. 

From the present data, we can not decide whether the higher moments stay finite as $L\rightarrow\infty$, but it seems plausible that the limiting distribution at least has finite variance. This has important consequences for sampling since it allows to estimate the uncertainty of the sample mean as outlined in \cref{sec:statistics}. Unfortunately, a reliable mean and uncertainty of the distribution of $\ln \period$ does not directly translate to a mean of the distribution of $\period$.

\FloatBarrier

\section[Beta function and O(N)-dependence]{Beta function and $O(N)$-dependence}\label{sec:beta}

\subsection{Background}\label{sec:beta_background}

In the present section, we consider the implications of our data for the beta function of $O(N)$-symmetric massless $\phi^4$-theory in 4 spacetime dimensions. The scalar field $\phi$ is promoted to a $N$-component vector $(\phi_1, \ldots, \phi_N)$ and the Lagrangian density reads
\begin{align}\label{lagrangian_phi4}
\mathcal L &= -\frac 12 \left( \partial_\mu \phi_1 \partial^\mu \phi_1 + \partial_\mu \phi_2 \partial^\mu \phi_2 + \ldots + \partial_\mu \phi_N \partial^\mu \phi_N \right)- \frac{\lambda}{4!} \left( \phi_1^2 + \phi_2^2 + \ldots + \phi_N^2 \right) ^2.
\end{align}
See \cite{pelissetto_critical_2002} for a review of the background and physical significance of $O(N)$-symmetry.

The beta function encodes the overall scale dependence of a theory. For $\phi^4$-theory, it is given by the derivative of the 1PI 4-point vertex Green function $G^{(4)}$with respect to logarithmic energy scale $\ln s$ at the renormalization point, 
\begin{align*}
\beta(g) := 2\cdot \frac{\partial}{\partial \ln s}G^{(4)} \Big|_{s=s_0}.
\end{align*}
Equivalently, the beta function is the coefficient of the simple pole of the 4-point counterterm in dimensional regularization. The beta function depends on the chosen renormalization scheme, see \cite{balduf_dyson_2023} for a recent discussion. Analytical results for the beta function up to 7 loops in the  minimal subtraction scheme can be found in \cite{schnetz_numbers_2018,schnetz_phi_2023}.

In the present work, we restrict ourselves to  primitive graphs. 
Their amplitude has the form \cref{amplitude_period}, and different renormalization schemes merely amount to adding finite constants. According to \cref{period_derivative}, the dependence on the logarithmic momentum is given, in all renormalization schemes, by the period. We redefine the coupling constant $\lambda\rightarrow g=\frac{\lambda}{(4\pi)^2}$ of the Lagrangian  in order to absorb the factor $\Lambda$ in \cref{period_derivative}. 
We define a series expansion of the contribution of primitive graphs to the beta function according to
\begin{align}\label{beta_expansion}
	\beta^{\text{prim}}(N,g) :=   \sum_{L \geq 1} (-g)^{L+1} \beta^{\text{prim}}_{L}(N) := 2\sum_{ \substack{G \text{ primitive }\\ \text{compl.}}  } (-g)^{L_G+1}\cdot \frac{ 4!(L+2) \cdot 3T(G,N)\cdot \period (G)}{N(N+2)\abs{\Aut(G)}}.
\end{align}
Here, $\period(G)$ is the period \cref{def:period} and $T(G,N)$ is a polynomial in $N$ which encodes the the $O(N)$-symmetry of the completion, see \cref{TGN_decompletion}. The automorphism group $\Aut(G)$ has no fixed vertices and the sum is over all non-isomorphic completions, we include the factor $4!$ to reproduce the usual definition in physics as discussed in \cref{sec:automorphism}. As explained in \cref{sec:uniform_sampling}, we do not need to know the total number of graphs $N^{(C)}_L$, but only $N^{(\Aut)}_L$, to estimate the beta function from the \enquote{s}-samples. For the \enquote{$\star$}-samples, this is irrelevant as we know $N^{(C)}_L$ for $L\leq 16$ (\cref{tab:periods_count}).

The $O(N)$-symmetric 4-valent interaction in \cref{lagrangian_phi4} describes a coupling between two  pairs of fields with identical indices. Graphically, every 4-valent vertex is a sum of 3 terms representing the 3 ways to connect its 4 adjacent edges into pairs. The Euclidean vertex Feynman rule is
\begin{align*}
	- \lambda \frac{\delta_{ij}\delta_{kl} + \delta_{ik}\delta_{jl} + \delta_{il}\delta_{jk}}{3}.
\end{align*}
The polynomial $T(G,N)$ is the product of vertex Feynman rules, without the factor $(-i\lambda)$, and summed over all indices $  i,j,\ldots  \in [1,N]$, where the indices of the   external edges are set to 1. The resulting polynomial is closely related to the Martin polynomial $M(G,x)$ and the circuit partition polynomial $J(G,x)=r_G(x)$  \cite{martin_enumerations_1977,ellis-monaghan_new_1998,bollobas_evaluations_2002} by 
\begin{align}\label{circuit_martin}
3^{\abs{V_G}}  T(G,x)  = J(G,x)= xM(G,x+2).
\end{align}
The linear coefficient of $M(G,x)$ is the \emph{Martin invariant} \cite{bouchet_connectivity_1996,panzer_feynman_2023}. 
Computation, and examples, of $T(G,N)$ can be found in \cite{kleinert_critical_2001}.
We compute $T(G,N)$ recursively and store the intermediate results in a cache, similarly to our computation of the Hepp bound (\cref{sec:symmetries_count}).
For example, for a single vertex $T(v,N)=1$ and if $G$ is the 3-loop multiedge arising from joining all 4 edges of a vertex to a new vertex, we find $T(G,N)=\frac{N(N+2)}{3}$. A 1-loop multiedge $G$ with 4 external edges has $T(G,N)=\frac{N+8}{9}$ and if the 4 external edges are joined to a new vertex, the resulting triangular graph $G$ has $T(G,N)=\frac{16N+10N^2+N^3}{27}$ and the graph $G=(a)$ in \cref{fig:completion_decompletion} has 
\begin{align*}
	T\left( G,N \right) &=\frac{  3004 N + 2740 N^2 + 749 N^3 + 67 N^4 + N^5}{3^8}.
\end{align*}
In fact, the recursive structure of $T(G,N)$  implies that the result for a completion $G$ is related to that of its decompletion $G\setminus \left \lbrace v \right \rbrace $ by 
\begin{align}\label{TGN_decompletion}
T(G,N) &= \frac{N(N+2)}{3} T(G\setminus \left \lbrace v \right \rbrace , N ).
\end{align}
The order of $N$ appearing in the polynomial $T(G,N)$ is bounded. Concretely, a factor $N$ arises from each closed path (circuit) in the graph, subject to each edge of the graph appearing in exactly one circuit. The maximum number of circuits is reached if they are as short as possible,  that is, 3 edges long. A $L$-loop completion has $2L+4$ edges, therefore, for a completion $G$ with $L\geq 3$ loops
\begin{align}\label{TGN_bound}
\text{Order of } N \text{ of the polynomial } T(G,N)  \leq  \left \lfloor \frac{\abs{E_G}}{3}\right \rfloor  =\left \lfloor \frac{2}{3}L_G + \frac 4 3\right \rfloor.
\end{align}
All graphs in our data set respect this inequality. For   $L \leq 15$ the data set contains graphs where equality holds and these graphs do not always have planar decompletions. 

Taking into account \cref{TGN_decompletion}, the inequality \cref{TGN_bound} implies that terms of higher order in $N$ start to appear only later in the perturbation series, limiting the reliability of asymptotic predictions from our data. 
For example, terms of order $N^8$ start to appear only at $L\geq 13$ loops, so we effectively only have the first 6 loop orders, not the first 18, which contribute to the order-$N^8$-term in $\beta^\text{prim}$. Moreover, only a single graph in our $L=15$ data sets contributes to the order $N^9$. This implies that in order to reach high accuracy for large $N$, one not only needs high loop orders $L$, but also large samples.

The $L$-loop coefficient $\beta_L$ of the beta function in the minimal subtraction (MS) scheme of an $O(N)$-symmetric $\phi^4$-theory grows asymptotically as $\beta_L \sim \bar \beta_L \left( 1+ \mathcal{O}\left( L^{-1} \right)   \right)  $, where \cite{mckane_nonperturbative_1984,mckane_perturbation_2019}
\begin{align}\label{beta_asymptotics}
	 \bar \beta_L &:=(L+1)^{3+\frac{N}{2}}\Gamma(L+2) \frac{36 \sqrt 3\cdot 3^{ \frac{N}{2}}  }{\pi   \Gamma \left( 2+\frac{N}{2} \right) A^{2N+4} } e^{-\frac 32 - \frac{N+8}{3}\left( \frac 3 4 + \gamma_\text{E} \right)   }    .
\end{align}
Here, $A\approx 1.28242713$ is the Glaisher-Kinkelin constant \cite{kinkelin_ueber_1860} and $\gamma_\text E\approx 0.57721566$ is the Euler-Mascheroni constant. 
The  coefficient $\beta_L$ in \cref{beta_asymptotics} is not the same as the primitive coefficient $\beta^\text{prim}_L$ in \cref{beta_expansion}, but it is conjectured \cite{mckane_perturbation_2019} that their leading asymptotic growth coincides for $L\rightarrow\infty$, or equivalently, that primitive graphs constitute the leading contribution to the beta function in minimal subtraction. We will use our data to assess the validity of this conjecture.

\subsection{Numerical results}\label{sec:beta_numerical}

\Cref{tab:beta} contains our results for the primitive contribution to the beta function \cref{beta_expansion}. Note that even if the individual periods  for $L=13$ have $\approx 250$ppm numerical uncertainty (\cref{tab:samples} ,  we reach an expected uncertainty of $0.25$ppm, i.e. more than 6 significant digits, for the beta function, reflecting the large number of graphs in that sample.

\begin{table}[h]
	\begin{tblr}{vlines,
			hline{1}={solid},
			hline{2,Z}={solid},
			rowsep=.5pt,
			cells={font=\fontsize{11pt}{12pt}\selectfont,mode=math },
			columns={halign=r},
			column{1}={halign=c},
			column{X}={font=\fontsize{10pt}{12pt}},
			column{Y}={font=\fontsize{10pt}{12pt}},
			column{Z}={font=\fontsize{10pt}{12pt}},
			row{1}={halign=c,valign=m,rowsep=2pt, font=\fontsize{11pt}{12pt}\selectfont ,mode=text},
			cell{1}{1}={font=\fontsize{12pt}{14pt}\selectfont },
			cell{1}{2}={font=\fontsize{12pt}{14pt}\selectfont },
			cell{1}{X}={font=\fontsize{9pt}{11pt}\selectfont }, 
			cell{1}{Y}={font=\fontsize{9pt}{11pt}\selectfont }, 
			cell{1}{Z}={font=\fontsize{9pt}{11pt}\selectfont }, 
		}
		$L$ & $\beta^\text{prim}_{L} (N=1)$   & {relative \\ accuracy}  &  {$\beta^\text{prim}_{L} (N=1)$  \\ exact  }    &  {estimate \\   from Hepp}   & {only planar \\  decompl.}  & { $\beta_{L} (N=1)$  \\ exact}   \\
		1        & 3.000002 \cdot 10^{0~}     & 4.5 \cdot 10^{-6} & 3\cdot 10^0         &                     & 3.000 \cdot 10^{0~} & 3\cdot 10^0 \\
		3        & 1.4424675 \cdot 10^{1~}   & 2.1 \cdot 10^{-6} & 1.4424683\cdot 10^1 &                     & 1.442 \cdot 10^{1~} & 3.25\cdot 10^1 \\
		4        & 1.2443163\cdot 10^{2~}    & 2.6 \cdot 10^{-6} & 1.2443133\cdot 10^2 &                     & 1.244 \cdot 10^{2~} & 2.72\cdot 10^2 \\
		5        & 1.6981612 \cdot 10^{3~}   & 5.7 \cdot 10^{-6} & 1.6981712\cdot 10^3 &                     & 1.490 \cdot 10^{3~} & 2.85\cdot 10^3 \\
		6        & 2.4130778 \cdot 10^{4~}   & 3.3 \cdot 10^{-6} & 2.4130730\cdot 10^4 & 2.41 \cdot 10^{4~}  & 1.655 \cdot 10^{4~} & 3.48\cdot 10^4 \\
		7        & 3.7092484\cdot 10^{5~}     & 2.4 \cdot 10^{-6} & 3.7092457\cdot 10^5 & 3.71 \cdot 10^{5~}  & 1.795 \cdot 10^{5~} & 4.75\cdot 10^5 \\
		8        & 6.0618784 \cdot 10^{6~}    & 1.8 \cdot 10^{-6} &                     & 6.06 \cdot 10^{6~}  & 1.906 \cdot 10^{6~} &  \\
		9        & 1.0450228 \cdot 10^{8~}    & 1.9 \cdot 10^{-6} &                     & 1.05 \cdot 10^{8~}  & 1.989 \cdot 10^{7~} &  \\
		10       & 1.8893120 \cdot 10^{9~}   & 8.9 \cdot 10^{-7} &                     & 1.89 \cdot 10^{9~}  & 2.050 \cdot 10^{8~} &  \\
		11       & 3.5671276 \cdot 10^{10}    & 4.0 \cdot 10^{-7} &                     & 3.57 \cdot 10^{10}  & 2.092 \cdot 10^{9~} &  \\
		12       & 7.0121257 \cdot 10^{11}   & 2.1 \cdot 10^{-7} &                     &                     & 2.117 \cdot 10^{10} &  \\
		13       & 1.4319963 \cdot 10^{13}   & 2.5 \cdot 10^{-7} &                     &                     & 2.128 \cdot 10^{11} &  \\
		13\text{s}       & 1.43237\cdot 10^{13}       & 7.1 \cdot 10^{-3} &                     &                     & 2.308 \cdot 10^{11} &  \\
		13\star& 1.43445 \cdot 10^{13}        & 7.0 \cdot 10^{-3} &                     &                     & 2.295 \cdot 10^{11} &  \\
		14\text{s}       & 3.03718 \cdot 10^{14}      & 2.3 \cdot 10^{-3} &                     &                     & 2.205 \cdot 10^{12} &  \\
		14\star& 3.03473 \cdot 10^{14}       & 1.2 \cdot 10^{-3} &                     &                     & 2.155 \cdot 10^{12} &  \\
		15\text{s}       & 6.62502  \cdot 10^{15}    & 2.7 \cdot 10^{-3} &                     &                     & 2.189 \cdot 10^{13} &  \\
		15\star& 6.65288 \cdot 10^{15}      & 3.3 \cdot 10^{-3} &                     &                     & 2.107 \cdot 10^{13} &  \\
		16\text{s}       & 1.49588 \cdot 10^{17}    & 9.0 \cdot 10^{-3} &                     &                     & 1.498 \cdot 10^{14} &  \\
		17\text{s}       & 3.535 \cdot 10^{18}      & 1.4 \cdot 10^{-2} &                     &                     & 1.766 \cdot 10^{15} &  \\
		18\text{s}       & 8.671\cdot 10^{19}       & 3.4 \cdot 10^{-2}  &                     &                     &                     &  \\
	\end{tblr}
	\caption{Primitive contribution to the beta function for $N=1$ \| Estimated relative uncertainty \| Analytic value, to 8 significant digits, from \cite{panzer_galois_2017} \| Estimate of the primitive $\beta$-function from Hepp bounds in \cite{kompaniets_minimally_2017} \| Contribution of  planar primitive decompletions, for $N=1$  \| Analytic beta function, including non-primitive graphs,  to 3 digits, from \cite{schnetz_numbers_2018}.}
	\label{tab:beta}
\end{table}

Our numerical results for $L \leq 7$ are consistent\footnote{Using the data provided in \cite{panzer_galois_2017}, the coefficient of  $\beta^\text{prim}$ at $L=6$ loops and $N=1$ is  $24130.7302\ldots$, consistent with our estimate $(24130.778 \pm 0.080)$, despite being rounded to $24130$ in \cite{kompaniets_minimally_2017}.} with the exact results of  \cite{kompaniets_minimally_2017}.
Moreover, \cite[Appendix B]{kompaniets_minimally_2017} contains an estimate for the primitive beta function up to 11 loops, based on the Hepp bounds of the corresponding graphs, which also is confirmed by our results. 
For the choice $N=1$, the $O(N)$-symmetric theory reduces to the ordinary $\phi^4$-theory. We have verified for every single graph in our samples that the factor \cref{circuit_martin} satisfies $T(G,1)=1$.

\begin{figure}[htb]
	\centering 
	\begin{subfigure}[b]{.49 \textwidth}
		\includegraphics[width=\linewidth]{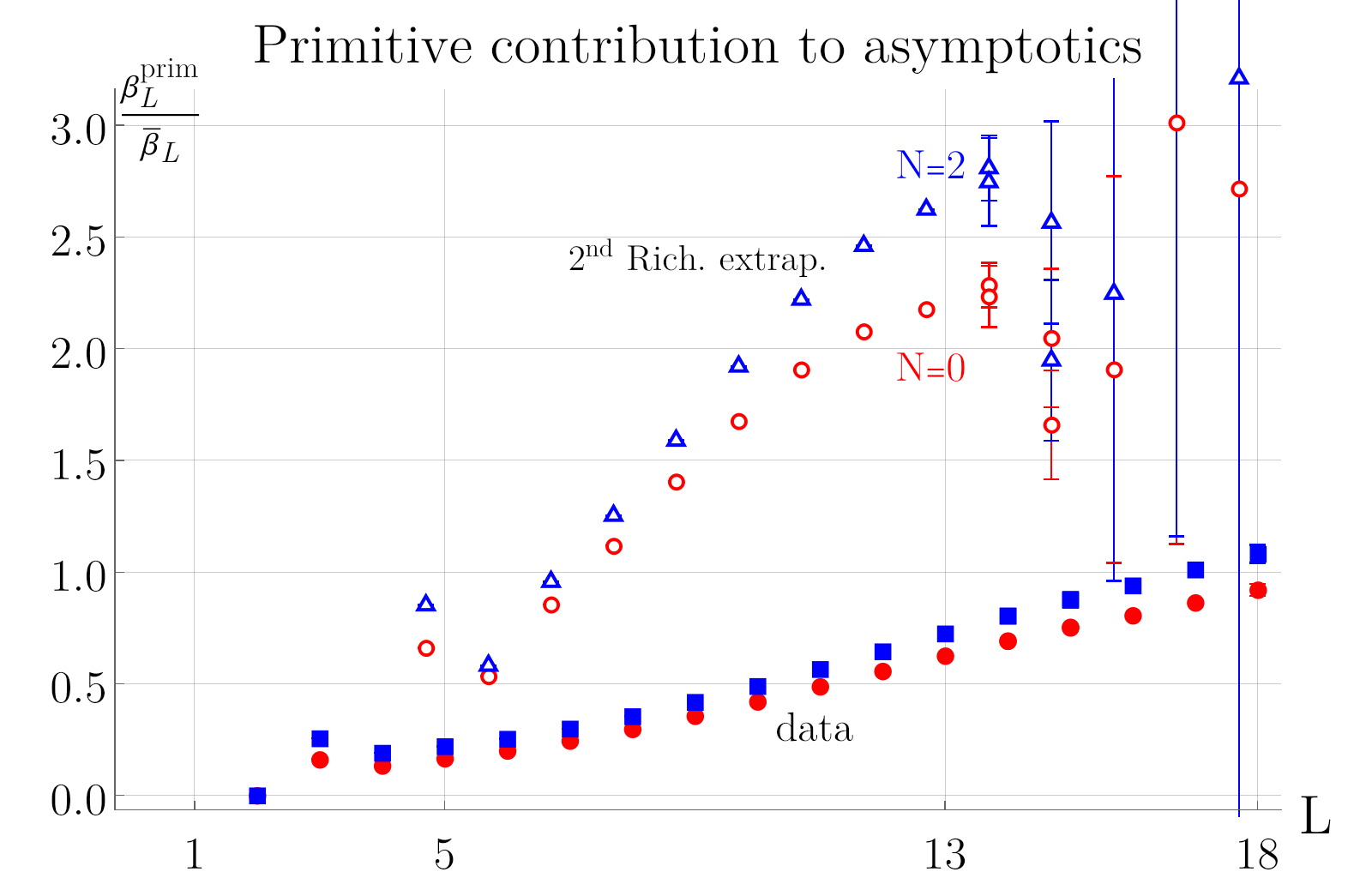}
		\subcaption{}
		\label{fig:beta_relative}
	\end{subfigure}
	\begin{subfigure}[b]{.49 \textwidth}
		\includegraphics[width=\linewidth]{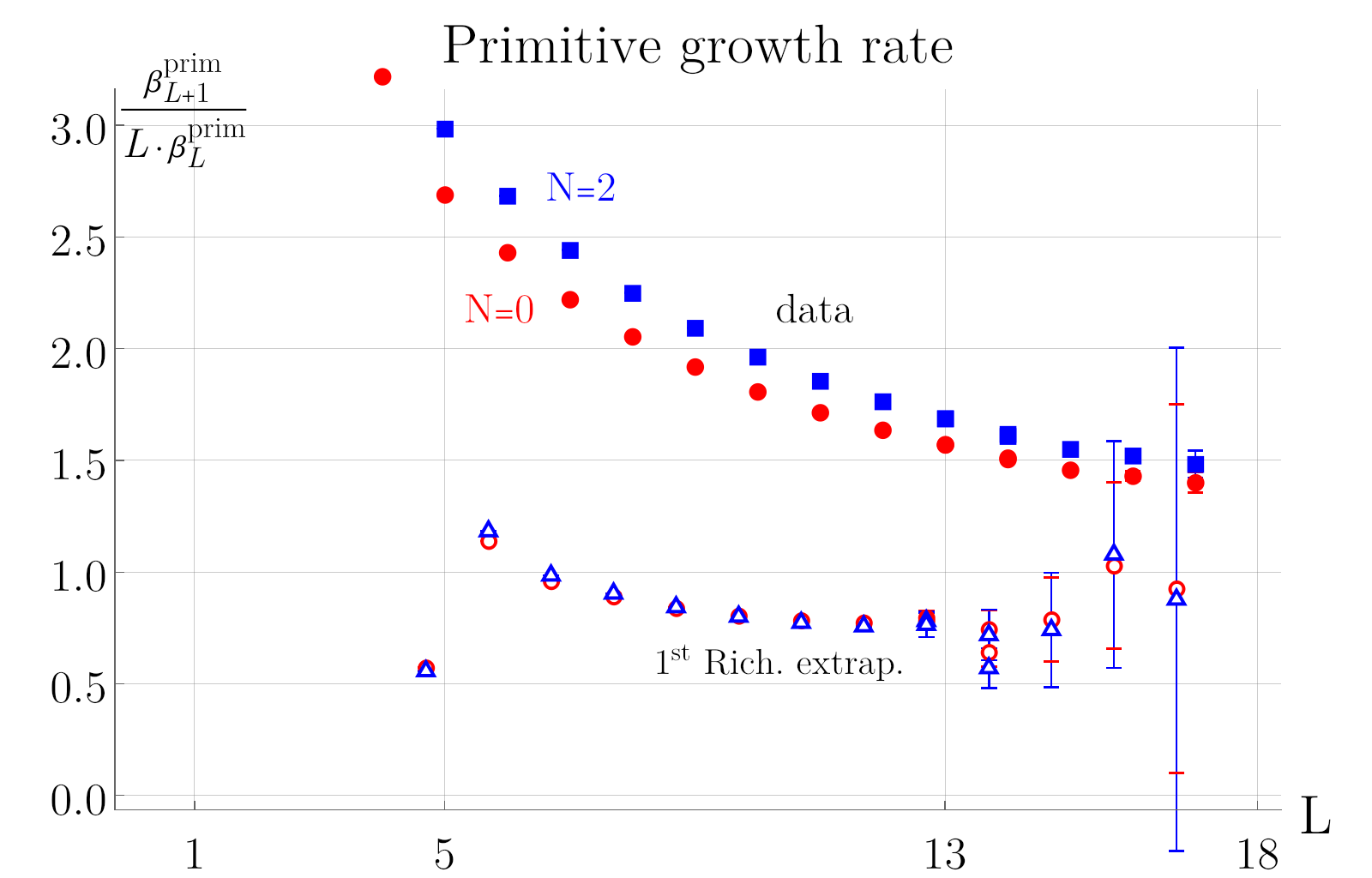}
		\subcaption{}
		\label{fig:beta_growth_rate}
	\end{subfigure}
	
	\caption{(a) Ratio \cref{beta_asymptotic_ratio} between the observed primitive beta function and the supposed asymptotics, for $N=0$ (red),  and $N=2$ (blue). Richardson extrapolation  suggests a limit $\approx 2\neq 1$ of this ratio, with significant uncertainty. (b) Ratio $\frac{\beta^{\text{prim}}_{L+1}}{L\cdot \beta^{\text{prim}}_{L}}$ and its Richardson extrapolation. Multiple data points at $L=13$ and $L=14$ loops represent the difference between samples \enquote{$\star$} and \enquote{s}. The data is compatible with the expected limit unity (\cref{beta_limit}). }
	\label{fig:beta_ratios}
\end{figure}

If the conjecture is true that \cref{beta_asymptotics} coincides with the asymptotics of the primitive beta function, then their ratio should converge to unity,
\begin{align}\label{beta_asymptotic_ratio}
\frac{\beta^{\text{prim}}_L}{\bar \beta_L} &\overset ?\sim 1 + \mathcal O\left( L^{-1} \right) .
\end{align}
A plot of this ratio for $L \leq 11$ loops in  \cite[Figure 1]{kompaniets_minimally_2017} indicated that the ratio is below unity at 11 loops, but growing. We show our data for the choices $N=0$ and $N=2$ in \cref{fig:beta_relative}, the data for $N=1$ lies between those points and has been omitted for clarity. We confirm the observation that the ratio \cref{beta_asymptotic_ratio} grows with the loop order. It appears to reach unity around $L\approx 18$, but with a non-vanishing slope. To confirm the limit unity, we computed a second-order Richardson extrapolation and found that it approaches values far above unity, see the hollow points in \cref{fig:beta_relative}. Unfortunately, the large fluctuations of  the non-complete samples   severely distort the extrapolation and we   can not  determine with certainty how the extrapolation behaves for $14 \leq L \leq 18$. The fact that our numerical data coincides with the known coefficients of the beta function for $L \leq 7$ makes it improbable that the values of $\beta^\text{prim}_L$ are severely wrong, at least for $L \leq 13$. Therefore, our findings suggest one of three different interpretations: 
\begin{enumerate}
	\item Either,   the asymptotics \cref{beta_asymptotics} is missing a factor $\approx 2$, or
	\item the conjecture is false that primitive graphs dominate the beta function in MS,  or
	\item  the convergence to the asymptotic regime is very slow, such that \cref{beta_asymptotic_ratio} reaches the value unity only for $L\gg 20$ loops.
\end{enumerate}
The first possibility can not be excluded, but we were unable to spot a concrete error, or un-adjusted difference in conventions, in neither of \cite{mckane_nonperturbative_1984,mckane_perturbation_2019,kompaniets_minimally_2017}.  However, findings in \cref{sec:asymptotics_mean} support this hypothesis. 

For the second possibility,   note that   our numerical values tend to be \emph{larger} than the supposed asymptotics. Therefore, in this scenario, the non-primitive graphs would have to produce a relative contribution with opposite sign, which does not vanish at $L \rightarrow\infty$. However,    the exact (non just primitive) beta function for $L\leq 7$ in \cref{tab:beta} suggests that non-primitive graphs contribute with the same sign, but smaller magnitude, compared to the primitive ones. From this data it is plausible that primitive graphs dominate in the limit, and the above second possibility  seems unlikely.

To assess the third possibility, we consider the growth rate of coefficients.
If \cref{beta_asymptotics} describes the correct asymptotics of primitive graphs,   the ratio of successive coefficients should be
\begin{align}\label{beta_limit}
	\frac{\beta^\text{prim}_{L+1}}{L\cdot \beta^\text{prim}_L} \overset{?}{\sim }  \frac{\bar \beta_{L+1}}{L \cdot \bar \beta_L}&=1 + \mathcal{O} \left( L^{-1} \right) .
\end{align}
\Cref{fig:beta_growth_rate} shows that for all our data points, the growth rate \cref{beta_limit} is well above unity, but it becomes smaller with growing loop order. We can use Richardson extrapolation \cite{richardson_approximate_1911} as explained in e.g. \cite{aniceto_primer_2019}, to obtain a more accurate estimate of the asymptotic value of the ratio. The first order Richardson extrapolation (hollow points in \cref{fig:beta_growth_rate}) lies below unity.  This behavior is compatible with the third hypothesis above. Concretely, our data suggests that at first, $\beta^\text{prim}_L$ grows too quickly, overshoots its supposed asymptotics, but then grows slower than the asymptotics. As we are merely considering rations, this observation does neither prove nor disprove the existence of a missing absolute factor of 2 in the asymptotics. Note that a slow convergence of $\beta^\text{prim}_L$ to $\bar \beta_L$ has been observed already in \cite{kompaniets_minimally_2017}. 
In a slightly different analysis, the authors of \cite{komarova_asymptotic_2001} conclude that the asymptotic  approximation at least is not useful below poles of order $\epsilon^{-10}$ in dimensional regularization, i.e. 10-loop graphs. It thus seems possible that our numerical data $L \leq 18$ is not yet representative for the truly asymptotic regime.

All in all, we do observe that primitive graphs contribute significantly to the conjectured asymptotic beta function in MS. But our data does not allow to clearly confirm, nor exclude, that the leading asymptotics of $\beta^{\text{prim}}_L$ follows the concrete formula \cref{beta_asymptotics}. Note that the form \cref{beta_asymptotics} is chosen to match \cite{mckane_perturbation_2019,kompaniets_minimally_2017}, but the same leading growth can be expressed by replacing the first two factors by, for example, $L^{4+\frac N 2}\Gamma(L+1)$. The difference between the two is a subleading term $\propto L^{-1}$, but it amounts to a deviation of $\approx 40\%$ for $L=13$. Consequently, even if \cref{beta_asymptotics} is the correct leading growth, it is perfectly reasonable that subleading corrections dominate for our data. This situation is slightly tautological because for \emph{any} finite sequence of coefficients, one can introduce subleading terms to make the coefficients match almost any supposed asymptotics. The question, therefore, should not be \enquote{Does our data follow the supposed leading asymptotics?}, but rather \enquote{Given the data and the asymptotics, can we determine subleading coefficients and are they small?} We will see in \cref{sec:asymptotics_mean} that the answer to the second question is \enquote{yes} as soon as we introduce a factor of 2 into the asymptotics.

\subsection[O(N)-dependence and planar graphs]{$O(N)$-dependence and planar graphs}\label{sec:beta_N_dependence}

\begin{figure}[htb]
	\centering
	\begin{subfigure}[b]{.49 \textwidth}
		\includegraphics[width=\linewidth]{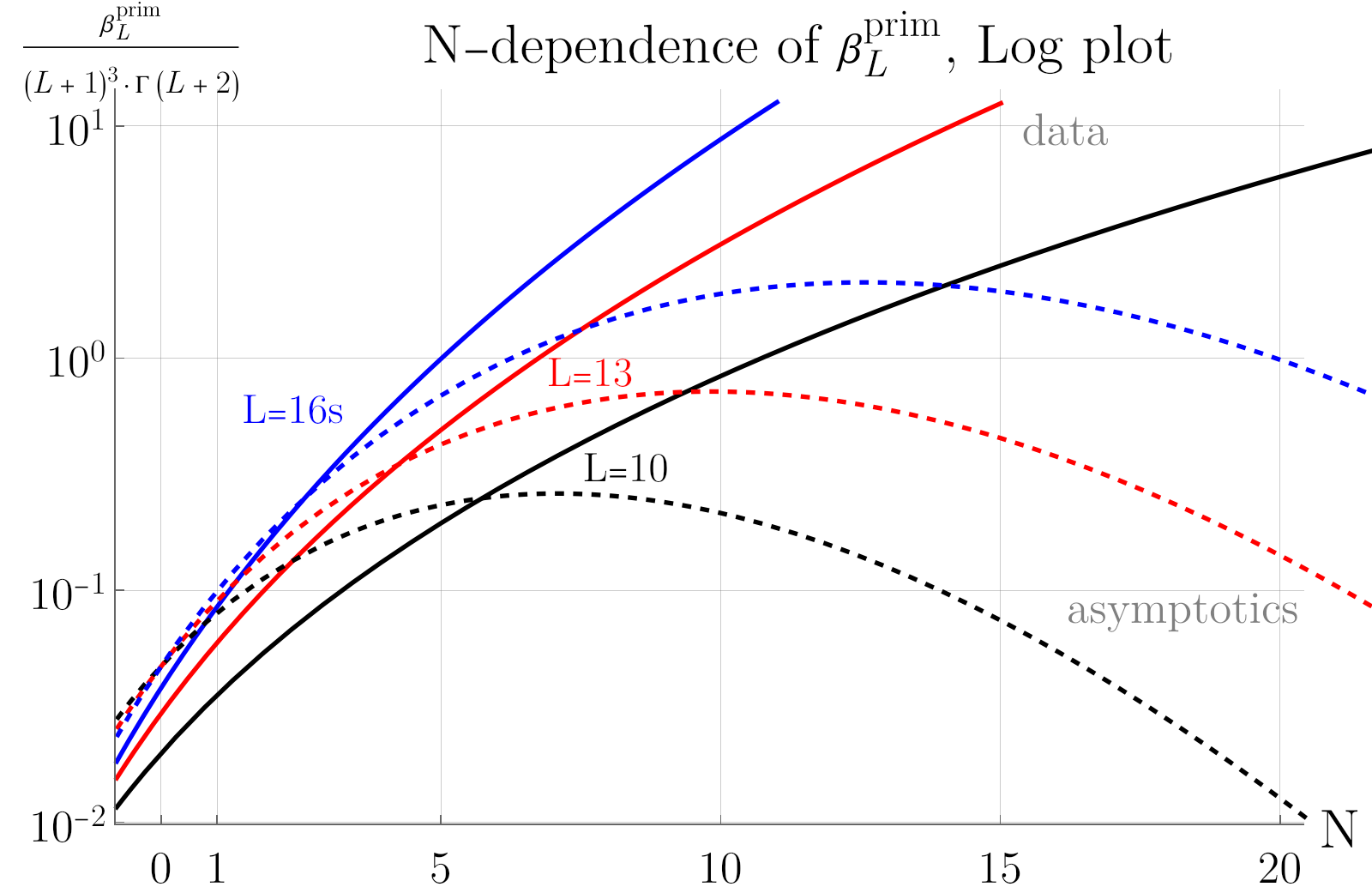}
		\subcaption{}
		\label{fig:beta_N_dependence}
	\end{subfigure}
	\begin{subfigure}[b]{.49 \textwidth}
		\includegraphics[width=\linewidth]{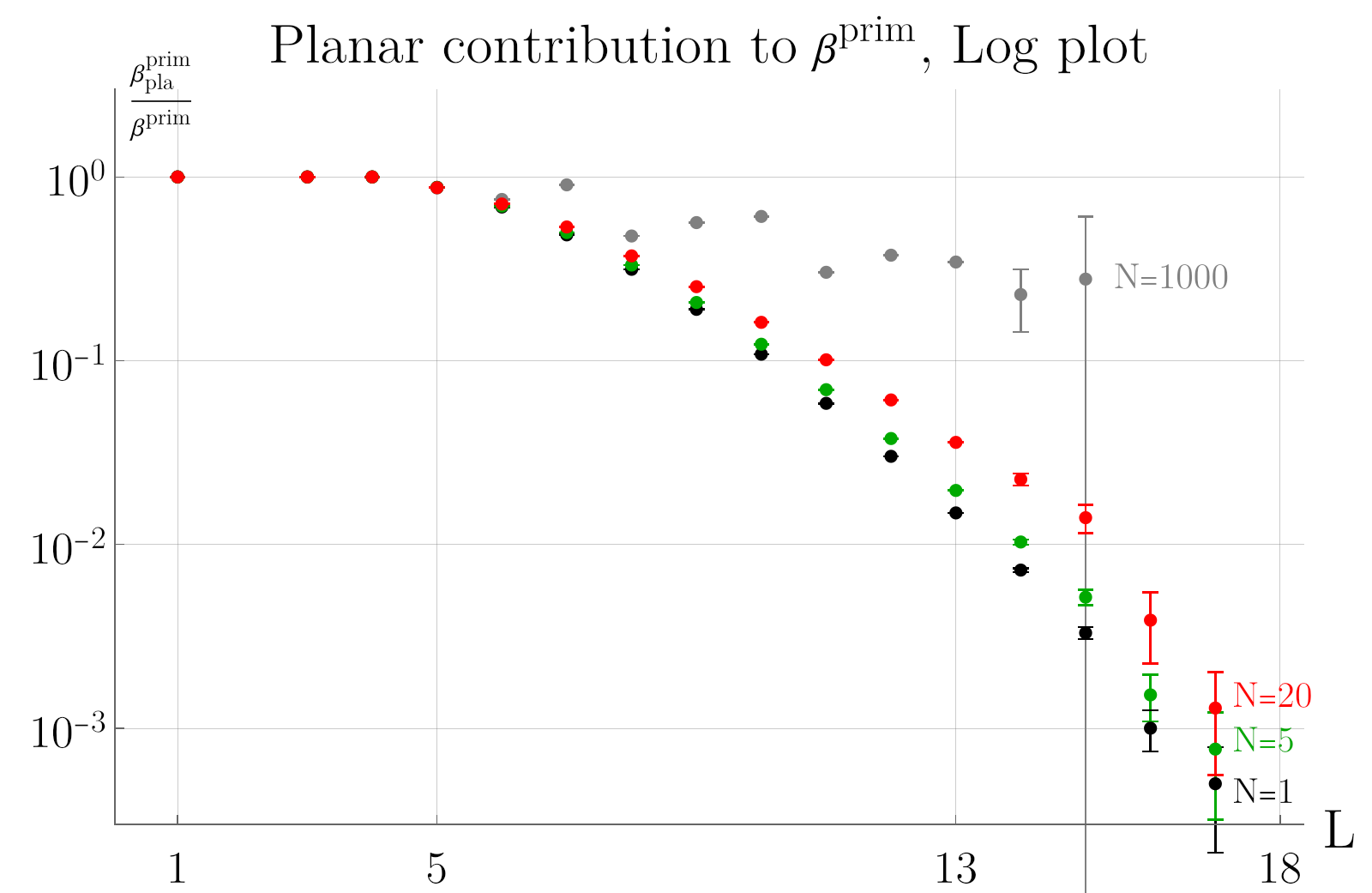}
		\subcaption{}
		\label{fig:beta_planar_relative}
	\end{subfigure}
	
	\caption{(a) $N$-dependence of the ratio \cref{beta_N_ratio} between   $\beta^{\text{prim}}_L(N)$ and the conjectured asymptotics $\bar \beta_L$ for fixed loop order, log plot. Solid lines: data, dashed: asymptotic prediction. With growing loop order, the empirical data begins to resemble the asymptotic curve, but  only near $N=1$.
		(b) Relative contribution of planar decompletions to the primitive beta function. While this ratio is small for large $L$, the influence of planar graphs grows with $N$.  
	}
\end{figure}

In \cref{fig:beta_ratios}, the growth of coefficients differs visibly between the cases $N=0$ and $N=2$ of the $O(N)$-symmetry.  To examine this effect, we normalize the coefficients with respect to the (supposed) leading growth in $L$ \cref{beta_asymptotics}, according to 
\begin{align}\label{beta_N_ratio}
\frac{\bar \beta_L}{(L+1)^3 \Gamma(L+2)} &=  \frac{36 \sqrt 3\cdot (L+1)^{ \frac{N}{2}}3^{ \frac{N}{2}}  }{\pi   \Gamma \left( 2+\frac{N}{2} \right) A^{2N+4} } e^{-\frac 32 - \frac{N+8}{3}\left( \frac 3 4 + \gamma_\text{E} \right)   } .
\end{align}
Firstly, we note   that the right hand side is independent of $L$ for $N=0$, so all plots of this ratio for different $L$ will intersect at $N=0$.
Secondly, for fixed $L$, \cref{beta_N_ratio}, has a finite maximum and approaches zero as $N\rightarrow\infty$. Conversely, for finite loop order $L$, the $O(N)$-factor $T(G,N)$ is a polynomial (\cref{TGN_bound}) and hence unbounded for $N\rightarrow \infty$. Therefore, for large enough $N$, the finite-order coefficient $\beta_L(N)$  deviates from the asymptotics $\bar \beta_L$ arbitrarily much, 
\begin{align}\label{beta_asymptotic_ratio_N}
\lim_{N \rightarrow \infty} ~\frac{\beta_L(N)}{\bar \beta_L(N)} &= \infty \quad \text{for fixed }L<\infty.
\end{align}
The asymptotc ratio \cref{beta_N_ratio}, together with our empirical data, is shown in \cref{fig:beta_N_dependence}. We see that the $N$-dependence of our data, even at $L=13$, differs  substantially from the predicted asymptotic behavior. The deviation grows with larger $N$ according to \cref{beta_asymptotic_ratio_N}. Only at high loop order, $L=16s$, our data begins to resemble the asymptotic shape at least for small $N$. We conclude that even knowledge of a sample of 18-loop graphs is insufficient to reach the large-$L$ regime for values of $N$ substantially different from unity.

In some theories which have an internal $SU(N)$, $O(N)$, or similar gauge symmetry,  the leading contribution to   scattering amplitudes,  in the limit $N\rightarrow\infty$, is given by  planar (decompleted) graphs, see e.g. \cite{hooft_planar_1974,gurau_colored_2012}. It is interesting to examine whether we see such an effect for the $O(N)$-symmetry in our data, even if we have graphs at leading order in $N$ which are not planar. Note that unlike \cref{beta_asymptotic_ratio_N}, this statement does not concern the $L\rightarrow\infty$ asymptotics, but the relation between graphs at fixed finite loop order as $N\rightarrow \infty$.
Firstly, at $N=1$, the contribution of planar decompletions is relatively small, and quickly decreases with growing loop order (\cref{tab:beta}).  This is not surprising since the number of planar graphs grows much slower than the total number of graphs (\cref{tab:periods_count}). 

\Cref{fig:beta_planar_relative} shows the relative contribution of planar decompletions to the primitive beta function, as a function of the loop order for different values of $N$. Clearly, the relative importance of planar graphs increases with increasing $N$, but even as high as $N=20$, planar graphs account for less than 1\% of the total value at $L=15$ loops.   For illustration, \cref{fig:beta_planar_relative} shows the data points for $N=1000$, which contribute $\sim 10\%$ at 15 loops.   Overall, it seems plausible that planar graphs will eventually dominate in the limit $N\rightarrow \infty$ for fixed $L$, but in the light of \cref{TGN_bound} and \cref{fig:beta_N_dependence}, this regime is far outside the validity of our perturbative data, so we should not expect our concrete numerical values to be accurate in the limit.

The dominance of planar graphs in the limit $N\rightarrow\infty$ is caused entirely by their $O(N)$-factors $T(G,N)$. Regardless of these factors, we have seen in \cref{tab:means_dec} that the actual amplitude, i.e. the period, of planar graphs is systematically larger than the average by a factor $F^{(\Aut)}\sim 3$. This has the effect that even for $N=1$, the contribution of planar graphs to the beta function, albeit small, is larger than what should be expected from merely the number of planar graphs. Conversely, if one were to extrapolate the amplitude of planar graphs to all graphs, the resulting beta function would come out significantly too large, even for $N=1$.

\subsection{Asymptotics of the period mean}\label{sec:asymptotics_mean}

The coefficients \cref{beta_expansion} of the $N$-dependent primitive beta function is not directly related to the symmetry-factor weighted mean  \cref{def:period_mean_aut} because the factor $T(G,N)$ is  different for each graph,
\begin{align}\label{beta_PLAut}
	\beta^\text{prim}_{L} &= 2 N^{(C)}_L \cdot 4! (L+2)  \left \langle \frac{T(G,N)\cdot\period (G)	}{\abs{\Aut(G)}} \right \rangle.
\end{align}
We have seen  below \cref{beta_asymptotic_ratio_N} that for $N>1$, our results are far from their large-$L$-asymptotic values. We therefore restrict our attention to $N=1$, where $T(G,1)=1$. Then, combining the (conjectured) asymptotic expression for $\beta^\text{prim}_L$ (\cref{beta_asymptotics}) with the known asymptotic growth of $N^{(\Aut)}_L$ (\cref{thm:asymptotic_symmetry_factor})) and the fact that the leading growth of  $N^{(C)}_L$ coincides with that of $N^{(\Aut)}_L$ (\cref{def:average_symfactor}), we obtain the following expression for the leading growth $\bar \period^{(\Aut)}_L$ of the mean period, up to terms $\mathcal{O}(L^{-1})$: 
\begin{align}\label{P_mean_asymptotics}
\left \langle \frac{ \period  	}{\abs{\Aut}} \right \rangle &\sim \bar \period^{(\Aut)}_L + \mathcal O \left( \frac 1 L \right) , \qquad  \bar \period^{(\Aut)}_L  =	  \frac{ \bar \beta_{L}}{ 2\bar N^{(\Aut)}_L}
=  \left( \frac 3 2 \right) ^{L+3}  L^{\frac 52}  \cdot \frac{2 \sqrt{\frac 2 \pi} e^{-3 \gamma_\text{E}}}{A^6} .
\end{align}
Owing to the vanishing influence of automorphisms as $L \rightarrow \infty$, the asymptotic growth $\bar \period_L$ of $\left \langle \period \right \rangle $ coincides with \cref{P_mean_asymptotics} to leading order. In particular, the  growth rate per loop order is
\begin{align}\label{P_mean_asymptotic_ratio}
\frac{\bar \period_{L+1}}{\bar \period_L}= \frac 3 2 + \delta\frac{1}{L}, \qquad \qquad \frac{\bar \period^{(\Aut)}_{L+1}}{\bar \period^{(\Aut)}_L}= \frac 3 2 + \delta^{(\Aut)}\frac{1}{L},
\end{align}
where we have introduced correction coefficients $\delta,\delta^{(\Aut)}$.

We can currently not determine subleading corrections to $\bar \period_L$ and $\bar \period^{(\Aut)}_L$, or the coefficients  $\delta,\delta^{(\Aut)}$, theoretically. Firstly, we are missing the corrections to various ratios, and secondly, $\beta_L$ and $\beta^{\text{prim}}_L$ are not expected to coincide beyond the leading term \cref{beta_asymptotics}. The naive $\frac 1 L$ correction to the right hand side of \cref{P_mean_asymptotics}, computed using $\bar \beta_L$ to order $L^0$ and $\bar N^{(\Aut)}_L$ to order $L^{-1}$, is $\frac{105}{16 }\frac{1}{L}$, this corresponds to $\delta = \frac{15}{4}\frac 1 L$ in \cref{P_mean_asymptotic_ratio}. At $L=18$, these  terms still amount to a relative correction as large as 36\%, or 21\%, respectively. From this estimate, and our observations in \cref{sec:beta_numerical}, we expect that our data considerably deviates from the leading asymptotic growth.

\begin{figure}[htbp]
	\centering
		\begin{subfigure}[b]{.49 \textwidth}
		\includegraphics[width=\linewidth]{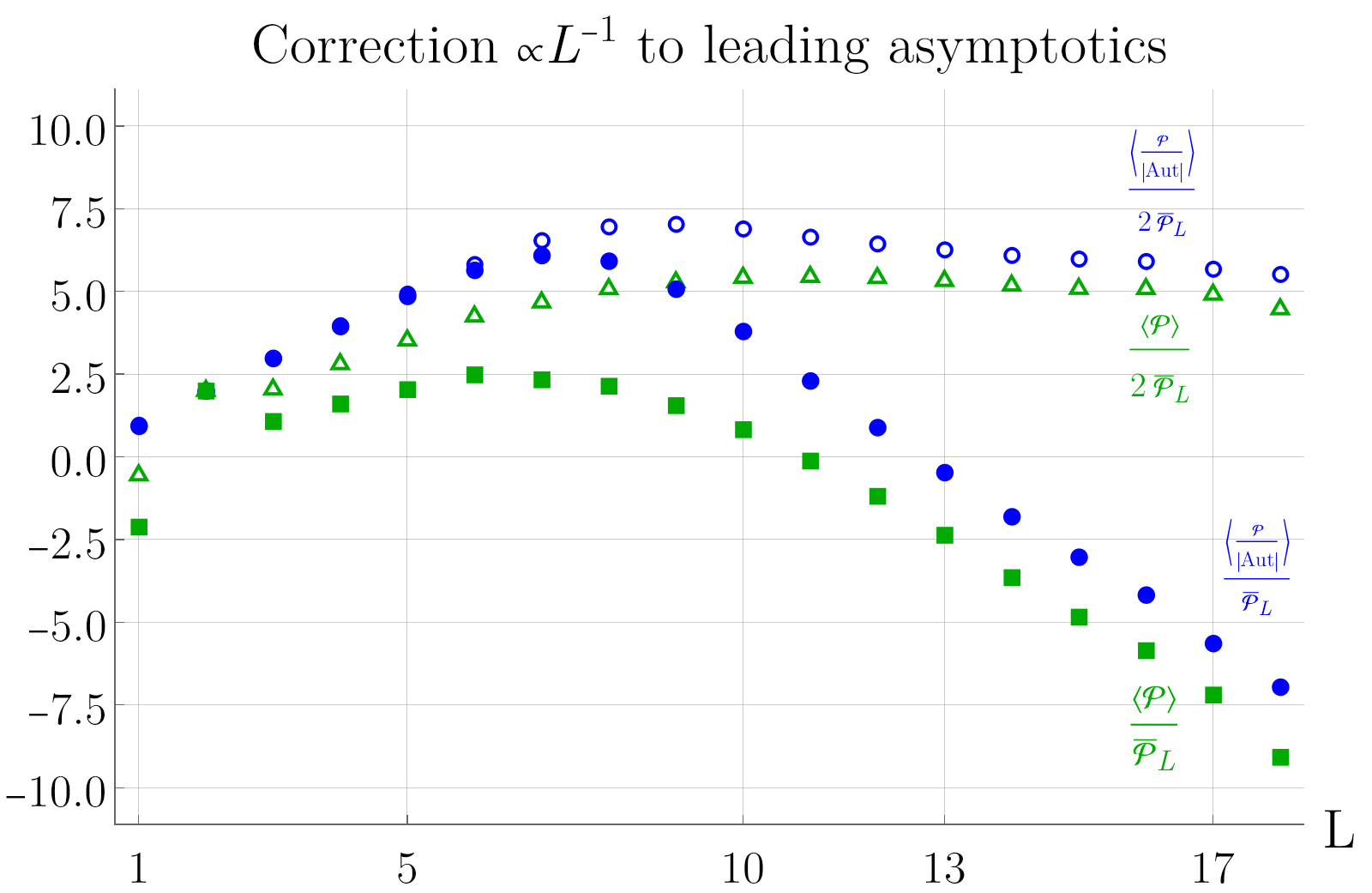}
		\subcaption{}
		\label{fig:period_mean_correction}
	\end{subfigure}
	\begin{subfigure}[b]{.49 \textwidth}
		\includegraphics[width=\linewidth]{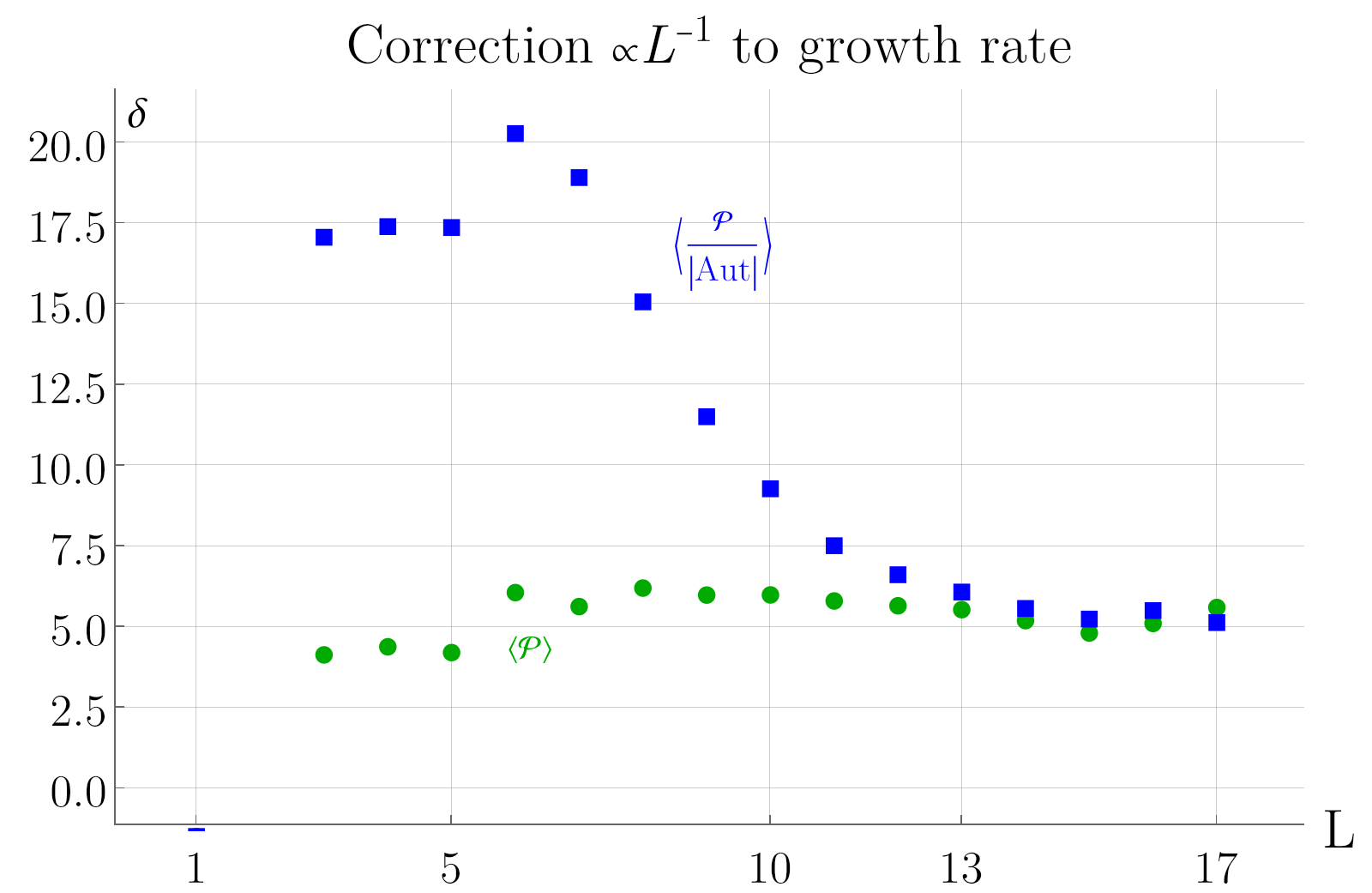}
		\subcaption{}
		\label{fig:period_growth_correction}
	\end{subfigure}
	\caption{
		(a) Correction \cref{mean_growth_correction}. (filled symbols). The coefficient seems to fall off linearly, which suggests that the ratio $\left \langle \period \right \rangle / \bar \period_L$ is not unity. If we, however, introduce a factor of 2 into the asymptotics, the subleading correction converges to a value $\Delta \approx 5$ (hollow symbols). 
		(b) Correction $\delta$ to the growth ratio of the period as defined in \cref{P_mean_asymptotic_ratio}. This ratio is not affected by overall factors in the asymptotics, the limiting value is $\delta \approx 5$.
	}
\end{figure}

The period averages are reported in \cref{tab:means,tab:means_Aut} and shown in the plot \cref{fig:period_mean} in the introduction section. Much like the beta function (\cref{fig:beta_relative}), we see that the leading asymptotics \cref{P_mean_asymptotics} matches the data only roughly, but the observed  asymptotics appears to be larger than the expected one  by a factor 2. 
In \cref{fig:period_mean_correction}, we plot the subleading  coefficient $\Delta$, defined by
\begin{align}\label{mean_growth_correction}
\frac{\left \langle \period \right \rangle }{\bar \period_L}= 1 + \Delta \frac{1}{L}.
\end{align}
The plot shows $L \cdot \left(  \left \langle \period \right \rangle / \bar \period_L -1  \right) = \Delta + \mathcal O(L^{-1}) $. This quantity is expected to converge to a constant value $\Delta$ if \cref{P_mean_asymptotics} is the correct leading term, but instead it seems to fall off linearly.  However, if we normalize with respect to $2 \bar \period_L$ (i.e. assume that \cref{P_mean_asymptotics} is missing a factor of 2), the subleading correction comes out finite and we find $\Delta \approx 5$. Of course, it is perfectly possible that the true asymptotic regime is only reached at $L\gg 20$, and in that case there is no contradiction. But if we require that the asymptotics should be a good approximation for $L \approx 15$, then our  finding suggests that a factor of two is missing in \cref{P_mean_asymptotics}. Note that, once we introduce that factor, the correction in \cref{fig:period_mean_correction} assumes its asymptotic value starting from $L_\text{crit}\approx 9$, which is approximately the same regime where the graph-theoretical quantities in \cref{sec:asymptotics_symmetry} start to follow their asymptotics.

The growth ratio \cref{P_mean_asymptotic_ratio} is insensitive to overall factors, and it seems to be accurate, the leading correction is shown in \cref{fig:period_growth_correction}. Empirically, we find that both coefficients approximately coincide, $\delta \approx \delta^{(\Aut)}\approx 5$.

\FloatBarrier

\section{Relations of the period to other graph properties}\label{sec:relations}

We have found above, in \cref{sec:moments}, that a planar decompletion on average has larger period than a non-planar one. In the present section, we examine the correlation of the period with several other properties of the graph.

\subsection{Symmetry factor}\label{sec:relations_symmetry_factor}

One the one hand, for $L\rightarrow \infty$, most graphs have trivial symmetry factor, $\abs{\Aut(G)}\rightarrow 1$ (\cref{sec:asymptotics}). On the other hand, in \cref{sec:largest}, we found that the largest periods appear with graphs which \enquote{look symmetric}. This motivates the study of the relation between the period and the symmetry factor of a graph.

\begin{figure}[htbp]
	\begin{subfigure}[b]{.49 \textwidth}
		\includegraphics[width=\linewidth]{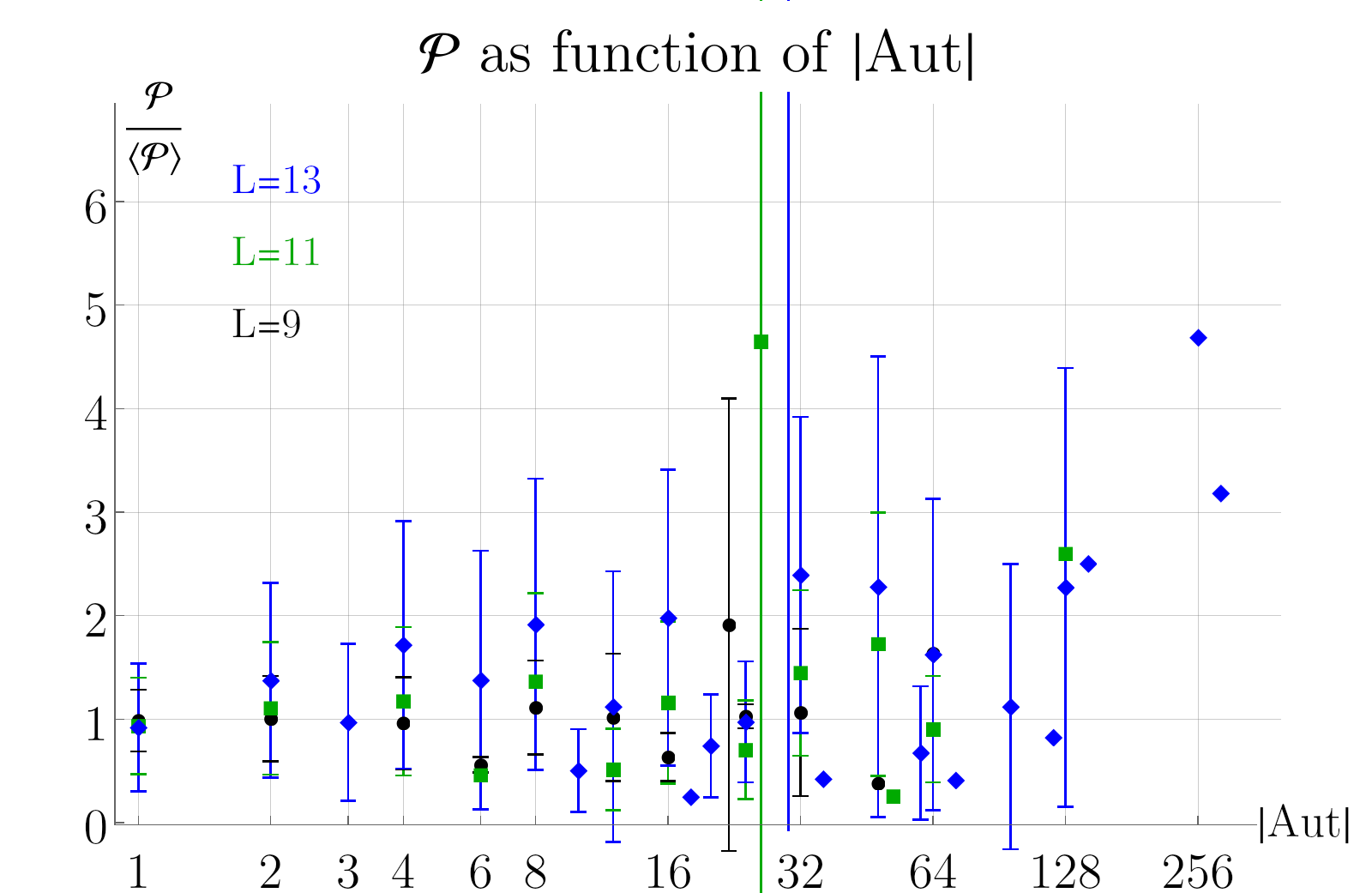}
		\subcaption{}
		\label{fig:P_Aut}
	\end{subfigure}
	\begin{subfigure}[b]{.48 \textwidth}
	\includegraphics[width=\linewidth]{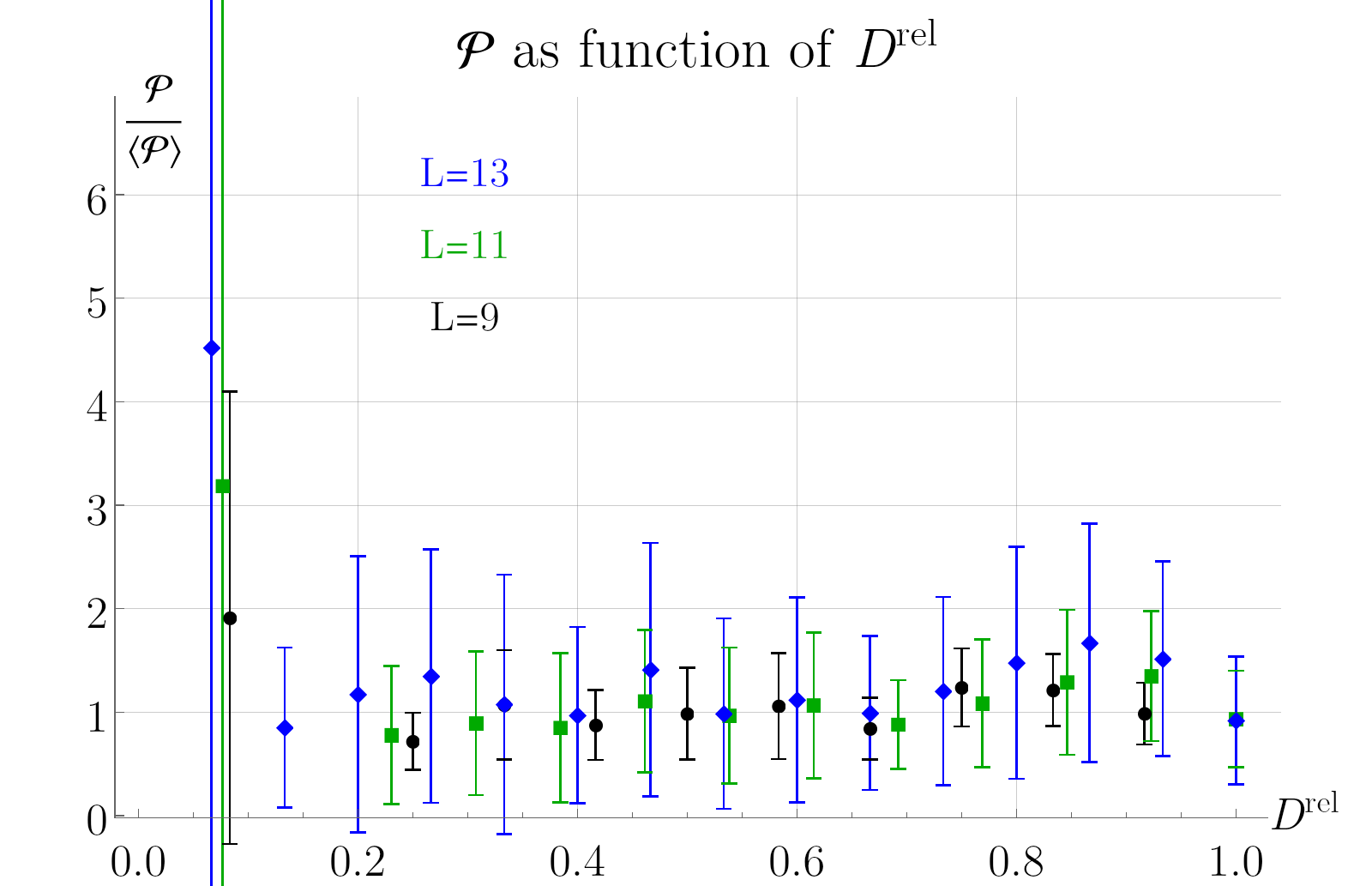}
	\subcaption{}
	\label{fig:P_Drel}
\end{subfigure}
	
	\caption{(a) Period, normalized to unit mean, as a function of the size of the automorphism group.  Shown is the mean of the corresponding sets of graphs, error bars indicate the standard deviation within the set, which in almost all cases  includes the value unity. (b) The same quantity, but as a function of non-isomorphic completions (\cref{def:Drel}). There is almost no correlation. }
	
\end{figure}

In \cref{fig:P_Aut}, we have plotted the period of completed graphs, normalized to its mean $\left \langle \period \right \rangle $ from \cref{tab:means}, as a function of the size of the automorphism group. There are some patterns, for example the graphs where $\abs{\Aut}$ is $2^n$ tend to have larger periods than others, but no significant overall trend. As a consequence, the contribution of a graph to physical observables, i.e. its period weighted by the symmetry factor, is small for graphs with large automorphism groups.

In \cref{def:Drel}, we introduced the number of non-isomorphic decompletions
as an alternative measure for the automorphisms of a graph, it is inversely correlated with $\log \abs{\Aut}$ (\cref{fig:Drel_correlation}). Consequently, we find that the period is almost uncorrelated with $D^\text{rel}$, see \cref{fig:P_Drel}. The large standard deviation at the smallest value of $D^\text{rel}$ is caused by zigzag graphs. 
Note that the \enquote{bump} in \cref{fig:Drel_correlation} at $D^\text{rel}=\frac 12$ translates to the fact that  graphs with $D^\text{rel}\approx\frac 12$ constitute a slightly larger contribution to the symmetry-factor weighted sum than what would be expected (i.e. those graphs have a smaller symmetry factor than expected from their value of $D^\text{rel}$, and hence contribute more).

\begin{figure}[htbp]
	\begin{subfigure}[b]{.48 \textwidth}
		\includegraphics[width=\linewidth]{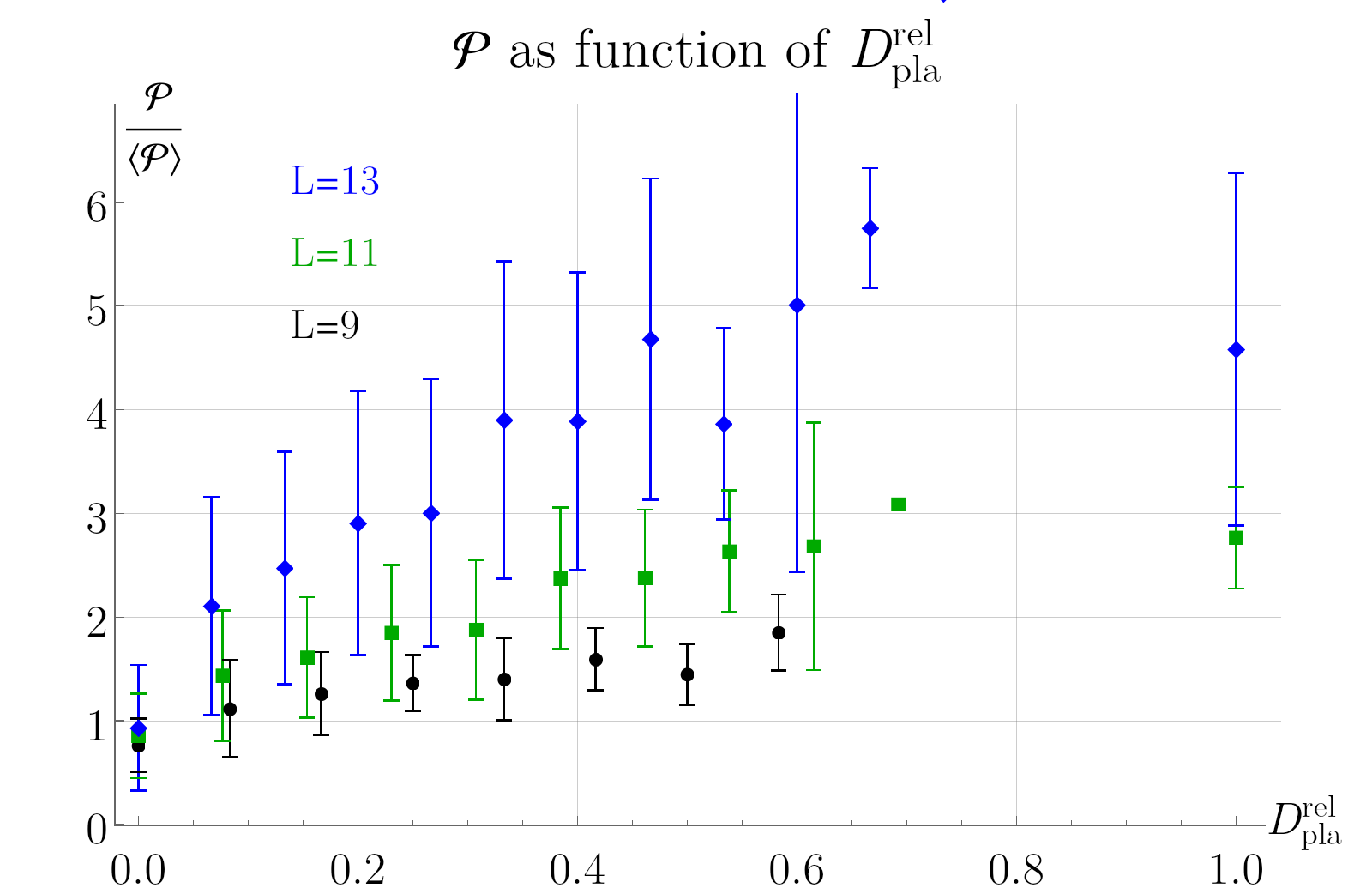}
		\subcaption{}
	\end{subfigure}
	\begin{subfigure}[b]{.48 \textwidth}
		\includegraphics[width=\linewidth]{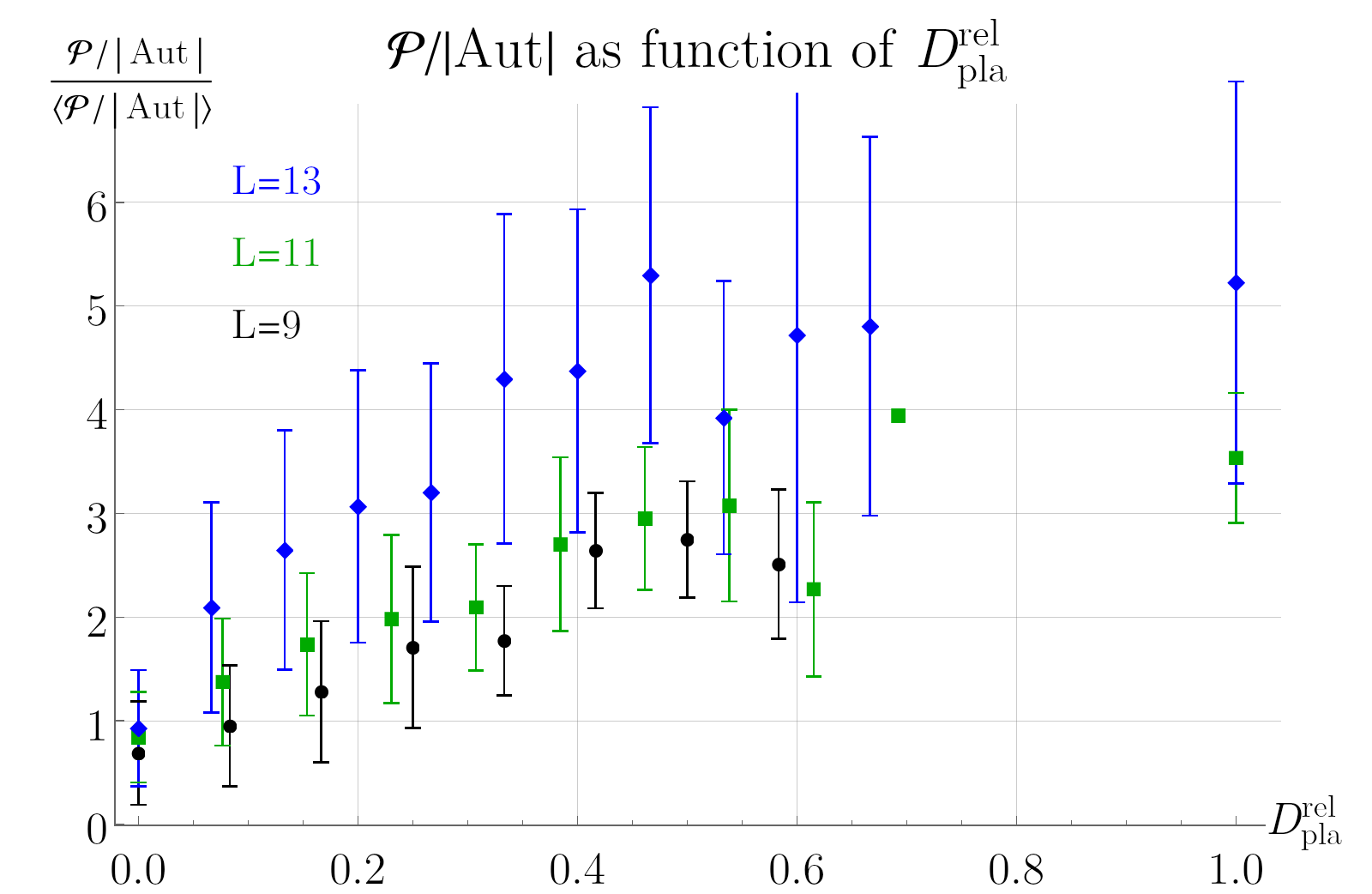}
		\subcaption{}
	\end{subfigure}
	
	\caption{
		Period, rescaled to its average, as a function of the number of non-isomorphic planar decompletions (\cref{def:Drel_planar}). There is an almost linear relation, which prevails even if the graphs are weighted by symmetry factors and gets stronger at higher loop orders.}
	\label{fig:P_Drelpla}
\end{figure}

The picture changes if we consider the ratio of \emph{planar} decompletions  $D^\text{rel}_\text{pla}$ as defined in \cref{def:Drel_planar}. We know from \cref{fig:Drelpla} that this quantity is essentially uncorrelated with the symmetry factor. Conversely, there is a significant, almost linear correlation of $D^\text{rel}_\text{pla}$ with the period, see \cref{fig:P_Drelpla}. This correlation prevails even if the period is weighted with its symmetry factor. Those completed graphs which have a large number of non-isomorphic \emph{planar} decompletions   contribute significantly more to the perturbation series than the graphs with few planar decompletions.

\subsection{Diameter and mean distance}\label{sec:diameter}

The findings in \cref{sec:largest} can also be interpreted along the lines that, within the same loop number, the period of a graph tends to be larger if the graph is visually  \enquote{larger}. 
The \emph{diameter} of a graph is defined as the maximum distance between any two vertices, where edges are assumed to have unit length:
\begin{align*}
\diam(G) &:= \max_{v_1,v_2\in V_G} \left( \text{length of the shortest path from $v_1$ to $v_2$} \right) .
\end{align*}
Assuming, as  in \cref{sec:asymptotics}, that the space of random 4-regular simple graph $G\in \mathcal G_{n,4}$ captures the leading asymptotic behavior of completed primitives, the diameter of an $L-$loop completion $G$ is expected to grow like $\diam(G)\sim \frac{1}{\ln 3} \ln \left( (L+2)\ln (L+2) \right) $  \cite{bollobas_diameter_1982}.

We have examined the relation between the diameter and the period, but it turns out to be of little use for our case. The reason is that  the diameter of a graph is an integer and, for $L \rightarrow \infty$, it is known to lie almost surely within $[-2,+3]$ of its asymptotics, i.e. to only assume very few different values. For example, all 755643 completions with $L=13$ have diameter 2, 3, or 4. 

Instead of the maximum distance between two vertices, we rather consider the mean distance
\begin{align*}
d (G) := \underset{v_1,v_2 \in V_G}{\operatorname{mean}}\left( \text{length of the shortest path from $v_1$ to $v_2$} \right) .
\end{align*}
This quantity is still granular, but much less so, compared to the diameter. For the 13-loop graphs, it takes 51 different values between $1.714$ and $2.286$. In particular, we find that the graph with the largest mean distance is the zigzag and the next smaller distances belong to the \enquote{almost-zigzag} shapes discussed in \cref{sec:largest}. 

\begin{figure}[htb]
	\centering
	\includegraphics[width=.9\linewidth]{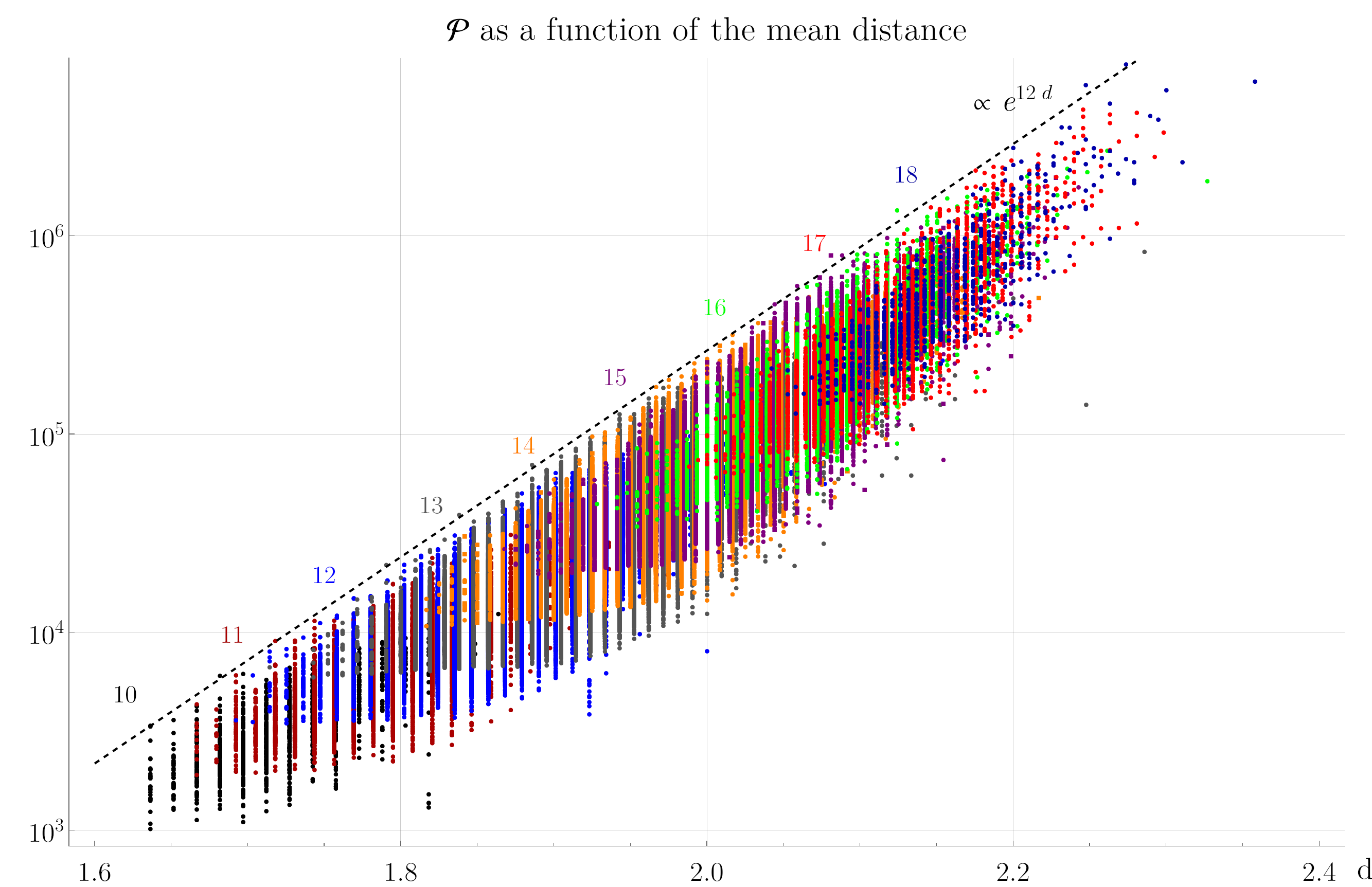}
	\caption{The period as a function of the mean distance $d$, log plot. Numbers indicate the loop order, graphs of the same order are plotted in the same color. Function $10^{-5} e^{12  d}$ (black dashed) to guide the eye. Note that   periods  tend to be similar when their mean distance is similar, even if the loop order is different.  }
	\label{fig:period_distance}
\end{figure}

\Cref{fig:period_distance} shows the period as a function of the mean distance $d$ of a graph. Clearly, the period is correlated with $d$, and both $\period$ and $d$ grow with growing loop order $L$. What is more remarkable is that, for a fixed value of  $d$, the graphs of different loop orders tend to have similar periods. This shows that the mean distance is not merely a proxy for the loop order, but it carries additional non-trivial information.
To guide the eye, we have included in \cref{fig:period_distance} the function $10^{-5} e^{12d}$. A few periods lie above this line, but it is remarkable that the overall distribution has a rather sharp upper boundary.

\subsection{Girth and small circuits}\label{sec:small_circuits}

Another straightforward property of a graph is the girth, that is, the length of the shortest circuit in the graph. Unfortunately, the girth is either 3 (the graph contains a triangle) or 4 (no triangles, but at least one square) for almost all graphs in our samples. This is expected from \cref{thm:random_graphs} (3). Instead, it proves more interesting to consider $n_3$, the number of triangles in the graph. 

\begin{figure}[htb]
	\centering
	\includegraphics[width=.9\linewidth]{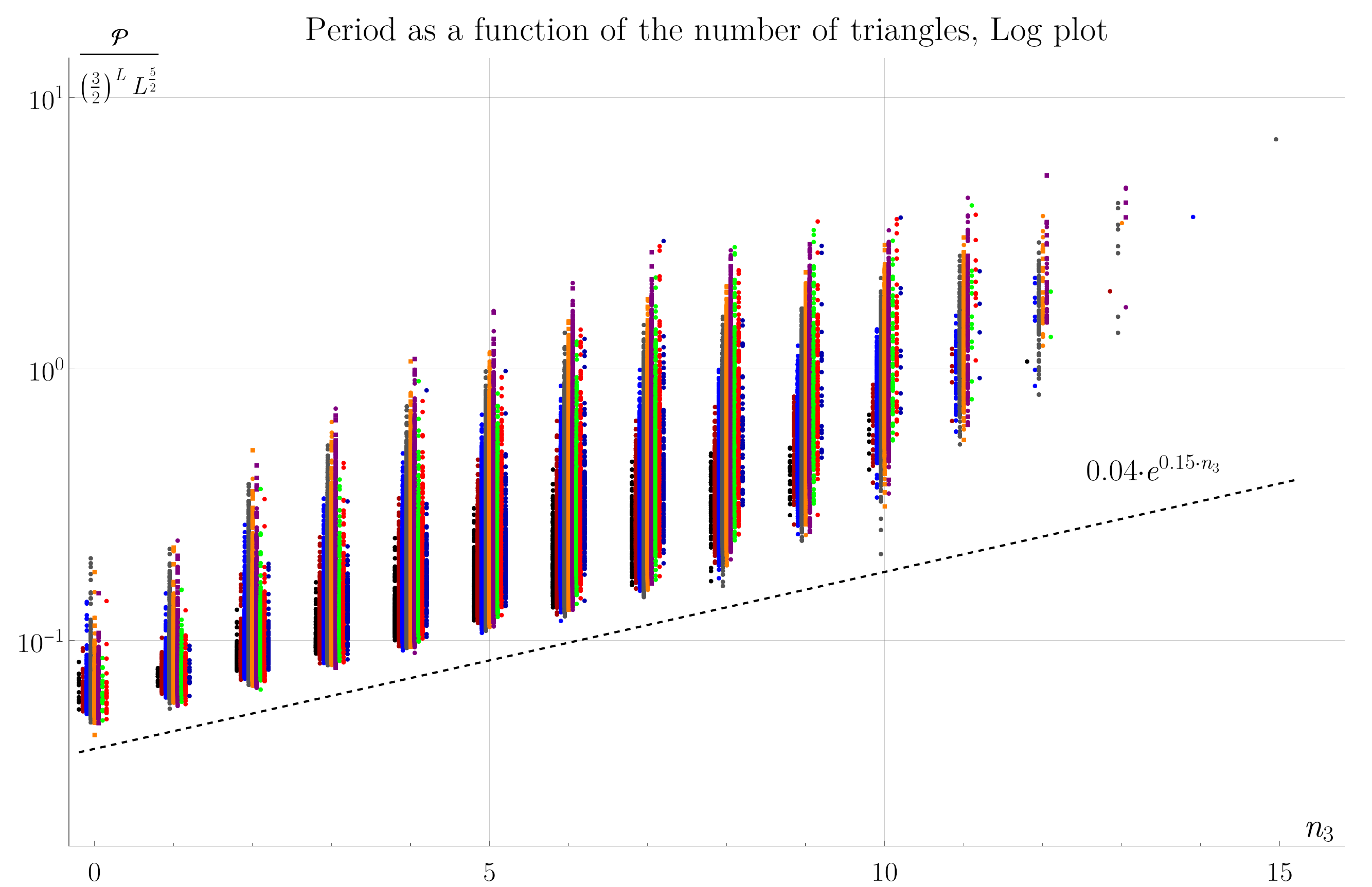}
	\caption{Period, scaled proportional to its leading growth, as a function of the number of triangles $n_3$. The plot contains all data points for $L \in \left \lbrace 10, \ldots, 18 \right \rbrace $, colors are the same as in \cref{fig:period_distance}. $n_3$ only takes integer values, the $x$-coordinate in the plot is slightly shifted to reduce overlapping.  }
	\label{fig:period_triangles}
\end{figure}

Unlike with the distance in \cref{fig:period_distance}, the plots depending on $n_3$ do not align for different loop orders. To eliminate the trivial dependence on loop order, we divide by $\left( \frac 32 \right) ^L L^{\frac 52}$, which is the leading growth rate \cref{P_mean_asymptotics} (up to constants). The resulting distribution of periods is shown in \cref{fig:period_triangles}. The plot suggests an exponential function as a lower bound, empirically we find that all our results for $L \geq 5$ satisfy
\begin{align}\label{period_triangle_bound}
\period > 0.04 \cdot  \left( \frac 32 \right) ^L L^{\frac 52} \exp \left(  0.15 \cdot n_3 \right)  .
\end{align}

One could suspect a similar correlation for the number of 4-circuits $n_4$ and the number of 5-circuits $n_5$. But this is not the case. The correlation of $n_4$ with the period is much weaker than that of $n_3$, and $n_5$ is even anticorrelated; graphs with many 5-circuits tend to have smaller periods.

\subsection{6-edge cuts}\label{sec:relations_6cuts}

Every 3-circuit passes through 3 vertices of the graph, each of which has 2 edges not in the circuit. Hence, every triangle is connected to the remainder of the graph by exactly 6 edges and it can be disconnected by a 6-edge cut. The converse is not true. In particular,  removing from a completion any one of its edges, together with the two adjacent vertices, gives rise to a 6-edge cut. Denote the number of 6-edge cuts by $c_6$, then $c_6 \geq \abs{E_G}= 2 L_G + 4$.  
 
From a physics perspective, the number of 6-edge cuts is more interesting than the number of triangles. This is because the Feynman graphs in quantum field theory correspond to the same physical observable if they have the same number of external edges. Hence, $c_6$ can be understood as the number of subgraphs in $G$ which contribute to a 6-point amplitude. In the spirit of Dyson-Schwinger equations, a primitive 4-point amplitude (i.e. a decompletion) can be written as an integral over the 6-point function,   where 3 of the legs are joined by a new vertex. This   consideration  motivates us to examine the relation between the number of 6-edge cuts and the period.

\begin{figure}[htb]
	\centering
	\includegraphics[width=.9\linewidth]{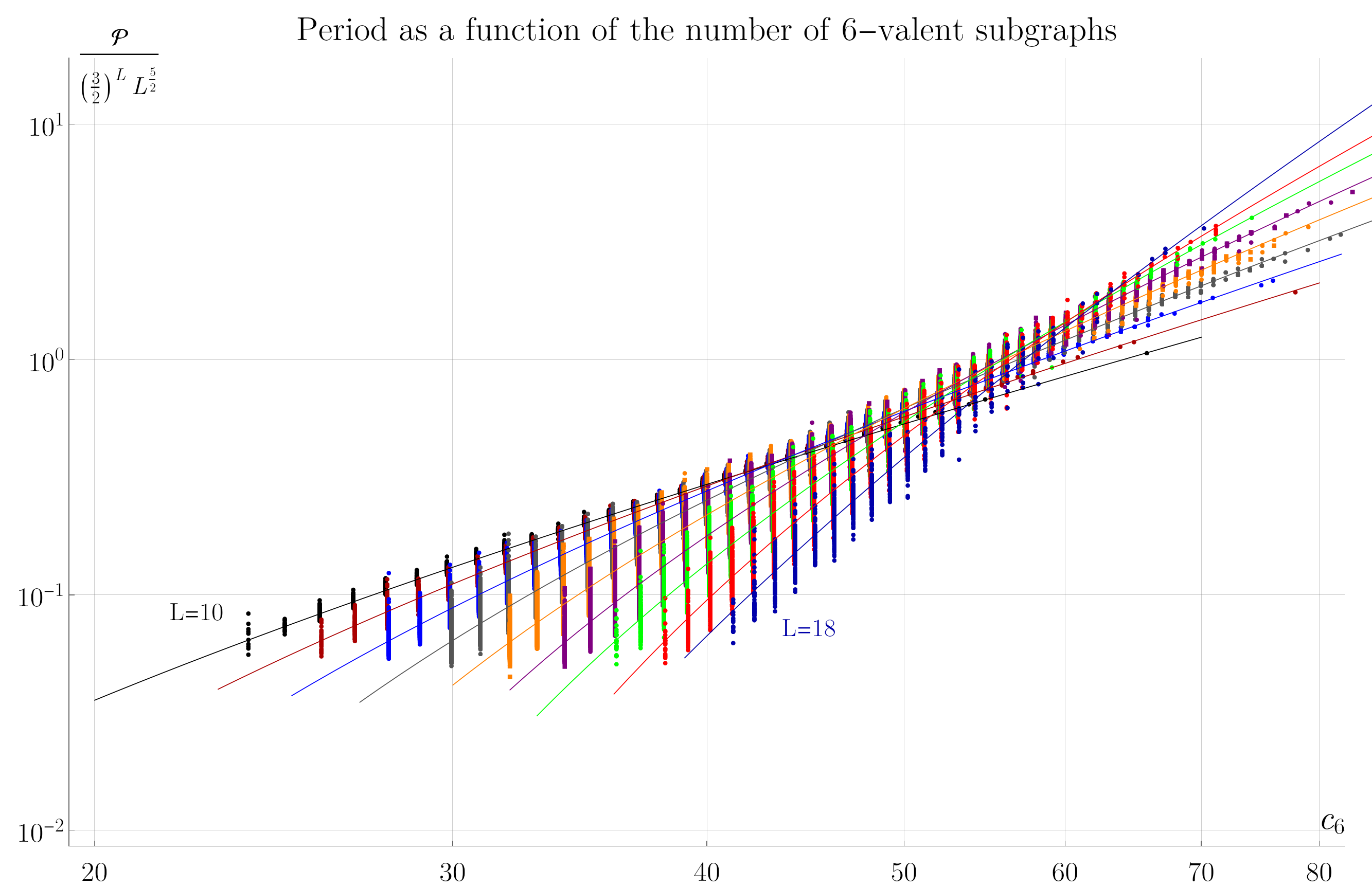}
	\caption{Period  as a function of the number of 6-edge cuts $c_6$, double logarithmic plot. Colors and scaling as in \cref{fig:period_triangles}. Solid lines show best fit functions \cref{6cuts_prediction}.}
	\label{fig:period_6cuts}
\end{figure}

As can be seen from \cref{fig:period_6cuts}, $c_6$ strongly correlates with the period. The precise relationship depends on the loop number, for a fixed loop number, it can be approximated by a function
\begin{align}\label{6cuts_prediction}
	\frac{\period}{\left( \frac 32 \right) ^L L^{\frac 52}}\approx f(c_6) &:= \left( \frac{c_6-d}{s}  \right) ^p,
\end{align}
where the constants $s,d,p$ are listed in  \cref{tab:6cuts}. With growing loop number, the relation between $c_6$ and $\period$ gets steeper. For our samples, the approximate dependence of the parameters of \cref{6cuts_prediction} on the loop number is $s \approx 550/L$, $d \approx 30 \ln (L/8)$, and $p \approx 1.2 e^{L/16}$, but no systematic approach has been made to justify these functional forms. 

Even if the number of 6-edge cuts has a physical interpretation, its predictive power for the period is remarkable, keeping in mind that we are merely \emph{counting} the number of 6-valent subgraphs, not actually modeling their contribution to recursive integrals. As indicated in \cref{tab:6cuts}, simply counting $n_6$ allows to predict an individual period with an average uncertainty of less than 15\%.

\begin{table}
	\centering
	\begin{tblr}{vlines,
			hline{1}={solid},
			hline{2,Z}={solid},
			rowsep=0pt,
			cells={font=\fontsize{11pt}{12pt}\selectfont,mode=math },
			columns={halign=r},
			column{1}={halign=c},
			row{1}={halign=c,rowsep=2pt, font=\fontsize{12pt}{14pt}\selectfont } 
		}
		L & s & d & p & \langle \abs{\frac{\period-f(c_6)}{\period}} \rangle & \max \abs{\frac{\period-f(c_6)}{\period}}  \\ 
		8  & 71.810  &  1.2410 & 2.0045 & 0.007243  & 0.02946 \\
		9  &  63.566 &  4.6942 & 2.0804  & 0.013341& 0.05797  \\
		10 & 56.368  &  7.6495 & 2.1934 & 0.020592 & 0.20688  \\
		11 & 50.268  &  10.475 & 2.3188  & 0.028544 & 0.25617 \\
		12 &  45.354 & 13.035  & 2.4653  & 0.039924 & 0.47640 \\
		13 &  41.446 & 15.410  & 2.6290   &0.054229  &0.50639\\
		13\text{s}  & 41.570 & 15.244  & 2.6478  & 0.053728  & 0.29752 \\
		13\star & 41.700 &15.156 &2.6520  & 0.054431&0.31633 \\
		14\text{s}  & 38.356 & 17.613 & 2.8151   & 0.070503 & 0.41371\\
		14\star &38.662 & 17.342&2.8360 & 0.070642  & 0.43935\\
		15\text{s}  & 35.923  & 19.770 &2.9976  & 0.087821 & 0.48885 \\
		15\star & 36.047 & 19.671 & 3.0043  & 0.088337 & 0.45205 \\
		16\text{s}  & 35.412  & 20.452 & 3.3562  &  0.10536 & 0.49715 \\
		17\text{s}  & 34.041  & 22.312& 3.5891  & 0.12145 & 0.55486  \\
		18\text{s}  & 39.530  & 17.736 & 4.7019 & 0.13513  & 0.55301 \\
	\end{tblr}
	\caption{Best fit parameters for the function \cref{6cuts_prediction} to approximate the period  \| average relative error of the approximation \| maximum relative error. }
	\label{tab:6cuts}
\end{table}

\FloatBarrier

\subsection{Hepp bound}\label{sec:hepp}

The Hepp bound $\mathcal H(G)$ enters the tropical integration algorithm \cite{borinsky_tropical_2023a} as a weighting factor for the contribution of sector integrals, consequently, we obtain this number automatically for every graph which is computed numerically. As explained in \cref{sec:symmetries_count}, we also wrote our own program to compute Hepp bounds, based on the flag formula \cref{hepp_recursive}.

\begin{figure}[htbp]
	\begin{subfigure}[b]{.48 \textwidth}
		\includegraphics[width=\linewidth]{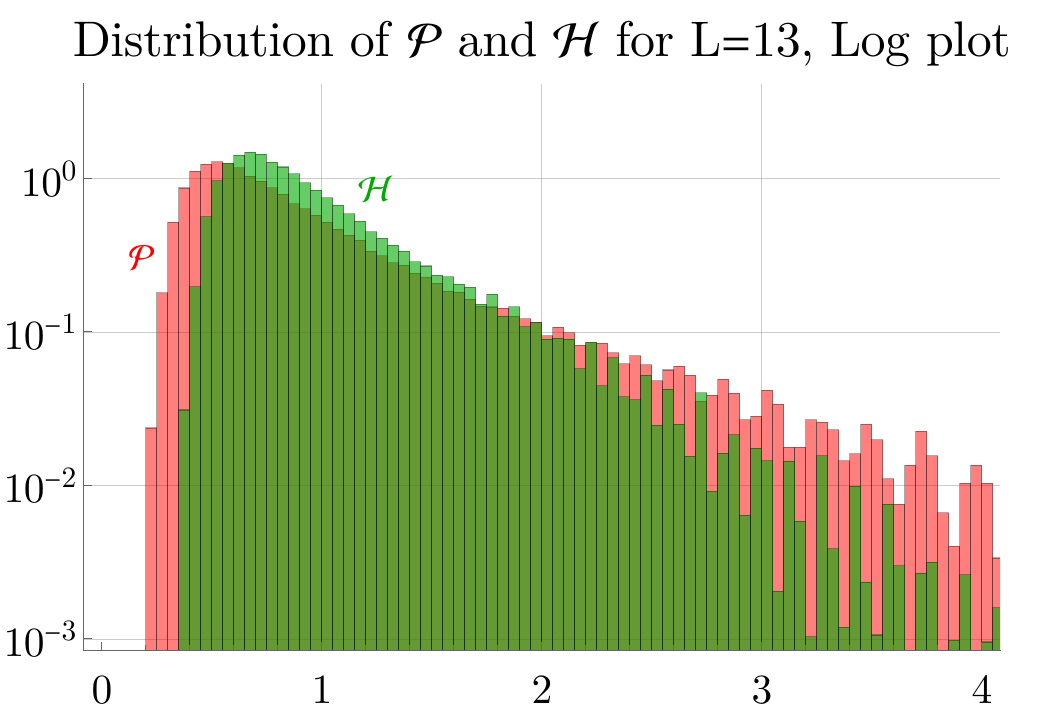}
	\end{subfigure}
	\begin{subfigure}[b]{.48 \textwidth}
		\includegraphics[width=\linewidth]{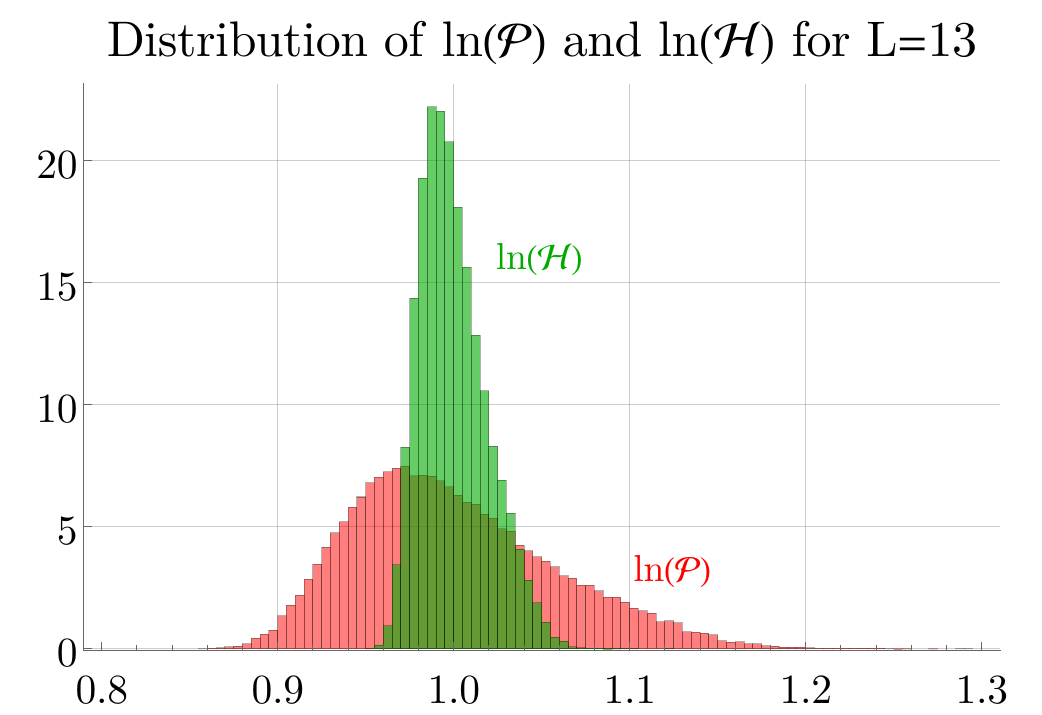}
	\end{subfigure}
	
	\caption{Histograms of the Hepp bound (green) and the period (red), both without symmetry factors. }
	\label{histograms_H}
\end{figure}

Before we turn to the correlation with the period, it is interesting to inspect the distribution of the Hepp bound itself. In \cref{histograms_H}, we see that the Hepp bound, or its logarithm, are more concentrated around the mean, and have lower influence of tails, compared to the distribution of periods.

Much like the period in \cref{sec:distribution}, we can find mean and moments  for the Hepp bound, they are reported in in \cref{tab:hepp}, both for $\mathcal H$ and for $\ln \mathcal H$. The overall trend is similar to that of the period in \cref{tab:means,tab:log_means}, but the value of the Hepp bound is considerably larger, and grows faster, than that of the period. Conversely, the Hepp bound has smaller moments, indicating that its distribution is overall more narrow, less skew, and more concentrated around the mean. Nevertheless, moments grow monotonically with the loop order, indicating that the limiting distribution has infinite moments even for the Hepp bound. 
We  note in passing that, similar to the period in \cref{sec:nonlinear}, the distribution of the transformed Hepp bound $\mathcal H^t$ comes close to a normal distribution, but for a power $t \approx 0.72$.

\begin{table}
	\centering
	\begin{tblr}	{vlines,
		vline{6}= {1}{-}{solid},vline{6}= {2}{-}{solid},
		hline{1}={solid},
		hline{2,Z}={solid},
		rowsep=0pt,
		cells={font=\fontsize{11pt}{12pt}\selectfont,mode=math },
		columns={halign=r},
		column{1}={halign=c},
		row{1}={halign=c,rowsep=2pt, font=\fontsize{12pt}{14pt}\selectfont } 
	}
		L & \left \langle \mathcal H \right \rangle   & C_2 & c_3 & c_4 & \left \langle \ln \mathcal H \right \rangle   & C_2\cdot 10^6 & c_3 & c_4 \\
		7  & 1.29107 \cdot 10^{5~} & 0.0490 &0.452 &2.443& 11.7442  & 349.6& 0.06959 & 2.132 \\
		8  & 7.42608 \cdot 10^{5~} & 0.0792 &0.739 & 3.633 & 13.4796 & 421.4& 0.05641 &2.484 \\
		9  & 4.24179 \cdot 10^{6~}  & 0.0970 &1.045 & 5.052 & 15.2152 & 384.8& 0.1924 & 2.599 \\
		10 & 2.33357 \cdot 10^{7~} & 0.1238 &1.196 & 5.742 & 16.9091 &384.2 & 0.2781 & 2.556 \\
		11 & 1.25530 \cdot 10^{8~} & 0.1549 & 1.362 & 6.211& 18.5800 & 378.4 & 0.3949 & 2.602 \\
		12 & 6.58540 \cdot 10^{8~} & 0.1925 & 1.592 &7.000 & 20.2248 & 371.5 & 0.5114 & 2.772 \\
		13 & 3.38768 \cdot 10^{9~} & 0.2334 & 1.878 & 8.405 & 21.8507 & 358.7 & 0.6168 & 3.004 \\
		13\text{s}  & 3.38236 \cdot 10^{9~} & 0.2332 & 1.812 & 7.411 & 21.8490 & 358.9 & 0.6284 & 2.986 \\
		13\star & 3.39141 \cdot 10^{9~} & 0.2374 &1.853 & 7.575 & 21.8505 & 362.4 & 0.6373 & 3.005  \\
		14\text{s}      & 1.71560 \cdot 10^{10} & 0.2762 & 2.170 & 9.944 & 23.4616 & 342.7 & 0.7031 & 3.237 \\
		14\star &1.71555 \cdot 10^{10} &0.2750 & 2.166 &9.774 & 23.4621 & 340.8 & 0.7060 & 3.255 \\
		15\text{s}      & 8.52439 \cdot 10^{10} & 0.3139 & 2.594 & 13.86 &25.0570 & 317.5&0.7676 &3.508\\
		15\star & 8.54667 \cdot 10^{10} & 0.3213 & 2.554 & 13.23 &25.0574 & 322.7 & 0.7896 & 3.509 \\
		16\text{s}      & 4.17260 \cdot 10^{11} & 0.3345 &2.661 &13.94 & 26.6400 & 292.2 & 0.7910 & 3.564 \\
		17\text{s}      & 2.03998 \cdot 10^{12} & 0.3700 & 2.974 & 16.65 & 28.2210 & 269.3 & 0.8587 & 3.852 \\
		18\text{s}      & 1.00353 \cdot 10^{13} & 0.3984 & 3.084 & 17.05 & 29.8084 & 249.9 & 0.9120 & 3.934 \\
	\end{tblr}
	\caption{Distribution parameters of the Hepp bound in our samples, without symmetry factors \| \| Parameters of the logarithm of the Hepp bound. In both cases, moments $C_j,c_j$ (\cref{def:central_moments}) refer to a normalized distribution with unit mean and the samples have been adjusted to be uniformly distributed (\cref{sec:uniform_sampling}). }
	\label{tab:hepp}
\end{table}

\begin{figure}[h]
	\centering
	\includegraphics[width=.85\linewidth]{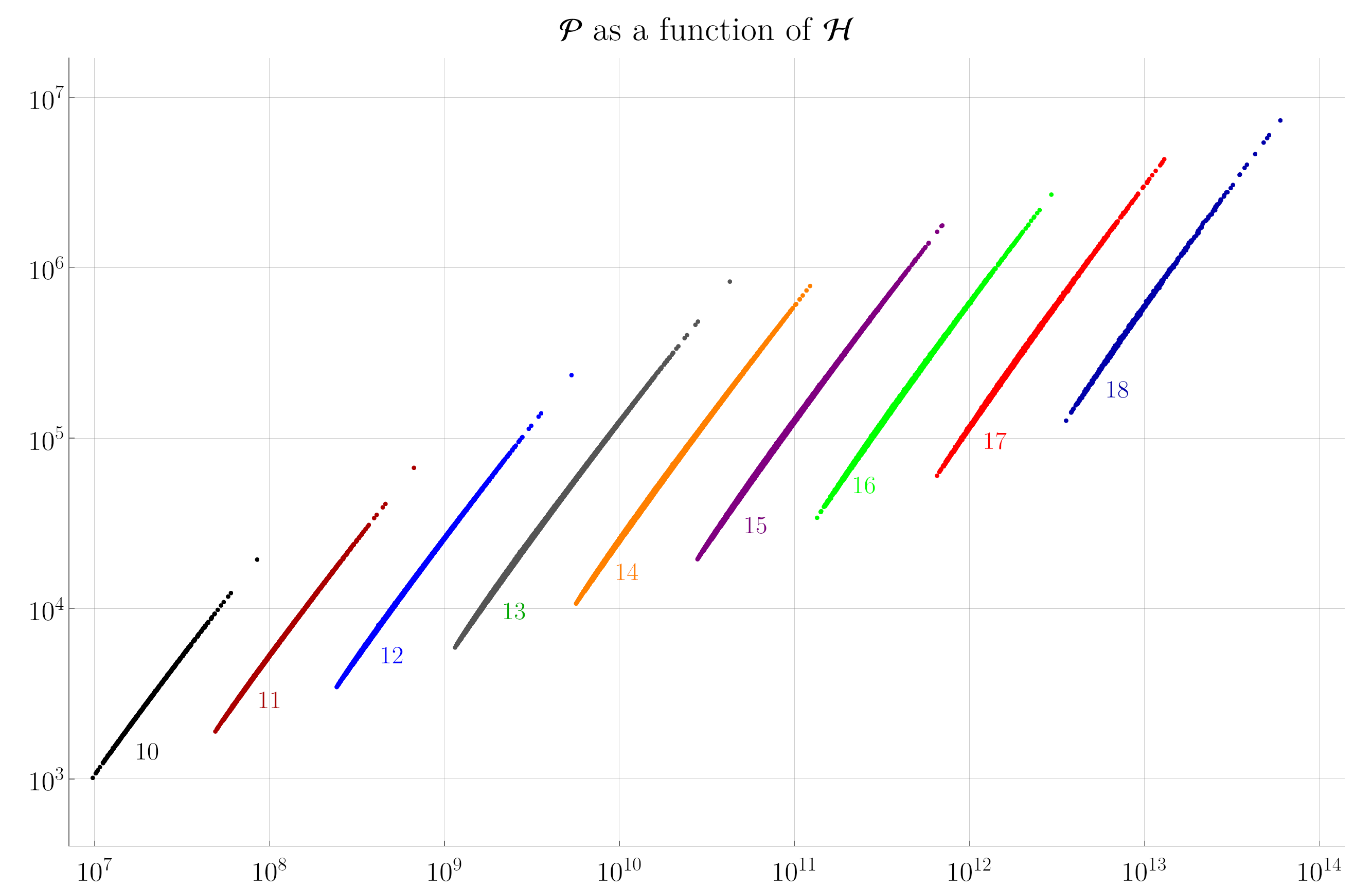}
	\caption{The period as a function of the Hepp bound, for various loop orders, Double-logarithmic plot.   The single large  outlier point for $10 \leq L \leq 13$ is the zigzag graph  (\cref{zigzag_amplitude}). }
	\label{P_H}
\end{figure}

It is known that the Hepp bound constitutes a strong prediction for the period, and that their relationship takes almost the same form across different loop orders \cite{panzer_hepp_2022}. Concretely, the points
\begin{align}\label{P_H_scaling}
\left(  \frac{\ln \mathcal H - \ln 2}{L-1}, \quad \frac{\ln (\period)}{L-1} \right)  
\end{align}
almost lie on the same curve for different loop orders $L$.  
In \cref{P_H}, we plot $\period$ vs $\mathcal H$  in logarithmic coordinates, but without the rescaling \cref{P_H_scaling} in order to avoid overlap.

The almost-linear relation \cref{P_H_scaling} of the logarithms is equivalent to an almost-power-law relation $\period \sim a \cdot \mathcal H^b$. In  \cite{kompaniets_minimally_2017}, the function $\period \approx a \mathcal H^b (1-c\mathcal H )$, with appropriate parameters $a,b,c$, was introduced to describe the relationship. This choice works well empirically, but it has the drawback of parameters which are far from unity and strongly vary with loop order, making the fit procedure unstable and obscuring the regularity  between different loop orders. Instead, we use the function
\begin{align}\label{hepp_fit_function}
\period &\approx g(\mathcal H) := \left( \mathcal H e^{-a(L+1)} - e^{c(L+1)} \right) ^b.
\end{align}
This function is not equivalent to $a \mathcal H^b (1-c\mathcal H )$, and we have found that our choice gives rise to lower average fit residues in all  samples.  As anticipated, the fit parameters of \cref{hepp_fit_function} are of order unity, see \cref{tab:hepp_fit}. 
We remark that although the relationship \cref{P_H} of the logarithms is almost linear, we have determined the parameters of \cref{tab:hepp_fit} with respect to $\mathcal H$ and $\period$, not with respect to their logarithms. This is because fitting with respect to logarithms would put disproportional weight on the smaller periods, which have only small importance in the perturbation series \cref{perturbation_series}.

\begin{table}
	\centering
	\begin{tblr}	{vlines,
			hline{1}={solid},
			hline{2,Z}={solid},
			rowsep=0pt,
			cells={font=\fontsize{11pt}{12pt}\selectfont,mode=math },
			columns={halign=r},
			column{1}={halign=c},
			row{1}={halign=c,rowsep=2pt, font=\fontsize{12pt}{14pt}\selectfont } 
		}
		L & a & b & c &  \langle \abs{\frac{\period-g(\mathcal H)}{\period}} \rangle   &   {\max \abs{\frac{\period-g(\mathcal H)}{\period}}}   \\
		7        & 0.9098 & 1.2915 & 0.16039 & 0.0008 & 0.0031 \\
		8        & 0.9312 & 1.2909 & 0.22236 & 0.0012 & 0.0038 \\
		9        & 0.9480 & 1.2899 & 0.27444 & 0.0017 & 0.0071 \\
		10       & 0.9638 & 1.2930 & 0.30877 & 0.0022 & 0.0163 \\
		11       & 0.9786 & 1.2986 & 0.33197 & 0.0028 & 0.0197 \\
		12       & 0.9927 & 1.3063 & 0.34686 & 0.0035 & 0.0374 \\
		13       & 1.0056 & 1.3146 & 0.35726 & 0.0044 & 0.0405 \\
		13\text{s}       & 1.0070 & 1.3175 & 0.35294 & 0.0043 & 0.0236 \\
		13\star& 1.0067 & 1.3169 & 0.35348 & 0.0044 & 0.0272 \\
		14\text{s}       & 1.0183 & 1.3252 & 0.36197 & 0.0055 & 0.0350 \\
		14\star& 1.0179 & 1.3243 & 0.36276 & 0.0055 & 0.0336 \\
		15\text{s}       & 1.0284 & 1.3322 & 0.36929 & 0.0069 & 0.0385 \\
		15\star& 1.0278 & 1.3310 & 0.37085 & 0.0070 & 0.0390 \\
		16\text{s}       & 1.0399 & 1.3442 & 0.36836 & 0.0081 & 0.0405 \\
		17\text{s}       & 1.0489 & 1.3523 & 0.37138 & 0.0096 & 0.0477 \\
		18\text{s}       & 1.0611 & 1.3689 & 0.36250 & 0.0102 & 0.0398 \\
	\end{tblr}
	\caption{Loop order \| 3 columns best fit parameters for the approximation of the period by the Hepp bound according to \cref{hepp_fit_function} \| Average and maximum relative error.}
	\label{tab:hepp_fit}
\end{table}

\begin{figure}[htbp]
	\begin{subfigure}[b]{.48 \textwidth}
		\includegraphics[width=\linewidth]{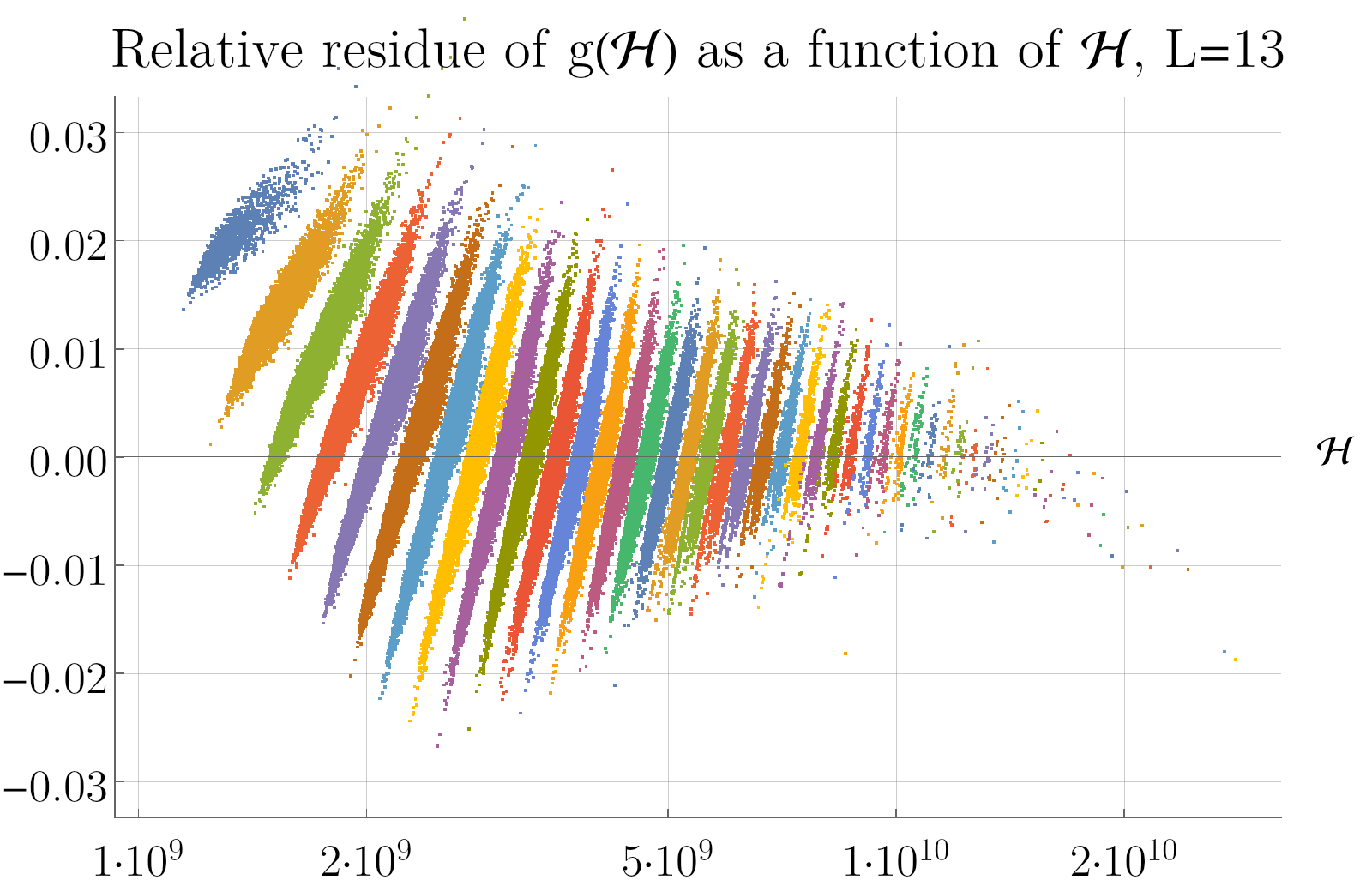}
		\subcaption{}
		\label{fig:residue_g_13}
	\end{subfigure}
	\begin{subfigure}[b]{.48 \textwidth}
		\includegraphics[width=\linewidth]{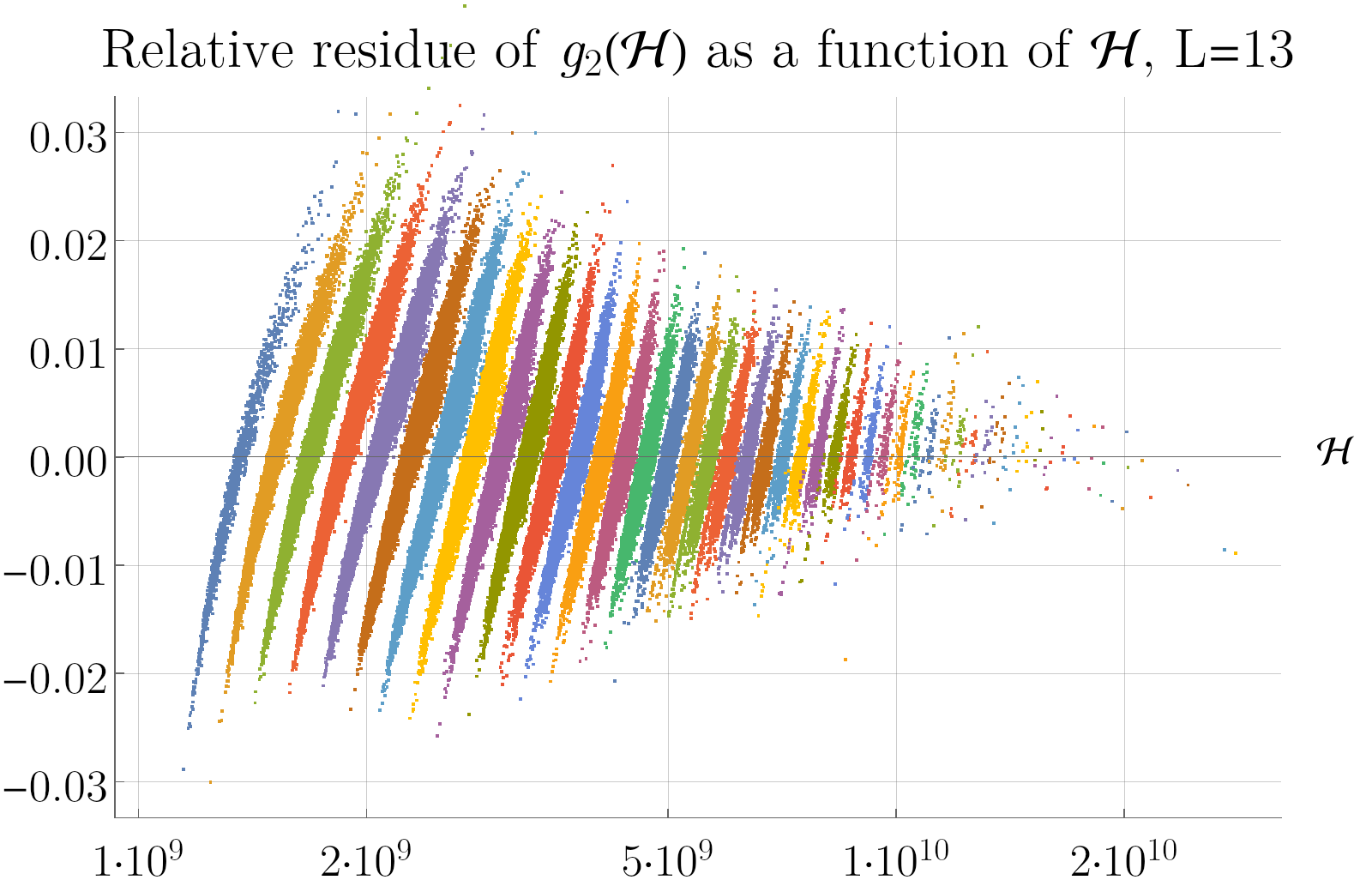}
		\subcaption{}
		\label{fig:residue_g2_13}
	\end{subfigure}
	\caption{(a) Residue of the best fit function $g$  (\cref{hepp_fit_function}) approximating the period in  \cref{P_H}. The plots show the quantity $\frac{\period -g(\mathcal H)}{\period}$. Colors indicate the number $c_6$ of 6-edge cuts, this number determines which \enquote{band} a residue lies in. Overall, the residues show an S-shaped pattern. (b) The S-shape is absent in the residues of $g_2$ (\cref{hepp_fit_function2}).}
	\label{fig:residue_hepp}
\end{figure}

The residues of the approximation \cref{hepp_fit_function} are shown in \cref{fig:residue_g_13}. The most remarkable feature of that plot is that the residues fall into discrete bands. As was pointed out by Erik Panzer,\footnote{Talk at research seminar \emph{structure of local quantum field theories},  Humboldt-Universität zu Berlin, December 20\textsuperscript{th}, 2021} all periods within one band share the same value of $c_6$, the number of 6-edge cuts. This is  indicated by colors in \cref{fig:residue_g_13}.
Besides the band structure, the residues in \cref{fig:residue_g_13} display an overall S-shape which suggests to add a correction term cubic in $\ln(\mathcal H)$ to the fit function \cref{hepp_fit_function}. We define
\begin{align}\label{hepp_fit_function2}
	\period &\approx g_2(\mathcal H) := \left( \mathcal H e^{-a(L+1)} - e^{c(L+1)} \right) ^b - \left( d \ln (\mathcal H) -e (L+1) \right) ^3 .
\end{align}
Best fit parameters for this function are shown in \cref{tab:hepp_fit2}. They refer to the samples \enquote{as is}, without adjustment for non-uniform sampling.

\begin{table}
	\centering
	\begin{tblr}	{vlines,
			hline{1}={solid},
			hline{2,Z}={solid},
			rowsep=0pt,
			cells={font=\fontsize{11pt}{12pt}\selectfont,mode=math },
			columns={halign=r},
			column{1}={halign=c},
			row{1}={halign=c,rowsep=2pt, font=\fontsize{12pt}{14pt}\selectfont } 
		}
		L & a & b & c &  d & e & \langle \abs{\frac{\period-g_2(\mathcal H)}{\period}} \rangle  & \max \abs{\frac{\period-g_2(\mathcal H)}{\period}}     \\
		7        & 0.6246 & 1.0019 & 0.62034 &  2.7385 & 3.0879 & 0.0007 & 0.0028 \\
		8        & 0.8052 & 1.1318 & 0.45202 &  2.5920 & 3.1158 & 0.0012 & 0.0042  \\
		9        & 0.9429 & 1.2813 & 0.33590 & -0.0391 & 0.3138 & 0.0017 & 0.0070 \\
		10       & 0.9540 & 1.2765 & 0.39125 &  0.0670 & 0.6185 & 0.0022 & 0.0147 \\
		11       & 0.9669 & 1.2784 & 0.41731 &  0.1305 & 0.8447 & 0.0027 & 0.0193 \\
		12       & 0.9801 & 1.2841 & 0.43058 &  0.1644 & 1.0321 & 0.0034 & 0.0369 \\
		13       & 0.9931 & 1.2920 & 0.43684 &  0.1807 & 1.2030 & 0.0042 & 0.0417 \\
		13\text{s}       & 0.9962 & 1.2978 & 0.42483 &  0.1496 & 1.0959 & 0.0041 & 0.0233 \\
		13\star& 0.9963 & 1.2979 & 0.42323 &  0.1720 & 1.1180 & 0.0042 & 0.0256 \\
		14\text{s}       & 1.0070 & 1.3040 & 0.43235 &  0.2177 & 1.3740 & 0.0052 & 0.0356 \\
		14\star& 1.0070 & 1.3038 & 0.43251 &  0.2230 & 1.3837 & 0.0052 & 0.0389 \\
		15\text{s}       & 1.0168 & 1.3100 & 0.43816 &  0.3269 & 1.7507 & 0.0063 & 0.0388 \\
		15\star& 1.0161 & 1.3087 & 0.44043 &  0.3337 & 1.7798 & 0.0064 & 0.0384 \\
		16\text{s}       & 1.0291 & 1.3227 & 0.43285 &  0.3012 & 1.8896 & 0.0074 & 0.0386 \\
		17\text{s}       & 1.0374 & 1.3288 & 0.43523 &  0.4865 & 2.4404 & 0.0086 & 0.0375 \\
		18\text{s}       & 1.0510 & 1.3476 & 0.42446 & -0.0372 & 1.8750 & 0.0098 & 0.0412 \\
	\end{tblr}
	\caption{Loop order \| 5 columns best fit parameters for the approximation of the period by the Hepp bound according to \cref{hepp_fit_function2} \| Relative errors. Despite having two additional degrees of freedom, and looking visually more appropriate, the fit function $g_2$ produces only marginally smaller approximation errors than  $g$ in \cref{tab:hepp_fit}. }
	\label{tab:hepp_fit2}
\end{table}

The resulting residues are shown in \cref{fig:residue_g2_13}. As expected, the S-shape has been removed, but in terms of mean error, the fit function \cref{hepp_fit_function2} is only marginally superior to \cref{hepp_fit_function}, whereas the fit is numerically much more unstable. The reason for this seeming contradiction is that, even if the extremely large or small values of $\mathcal H$ in the plots \cref{fig:residue_g2_13} have been moved towards zero, the large majority of data points lies in the central region and is unaffected by the introduction of a cubic term in \cref{hepp_fit_function2}.

\subsection{Martin invariant}\label{sec:martin}

As observed in \cite{panzer_feynman_2023}, the  Martin invariant (\cref{sec:invariants}) is correlated with the period.

\begin{figure}[h]
	\centering
	\includegraphics[width=.85\linewidth]{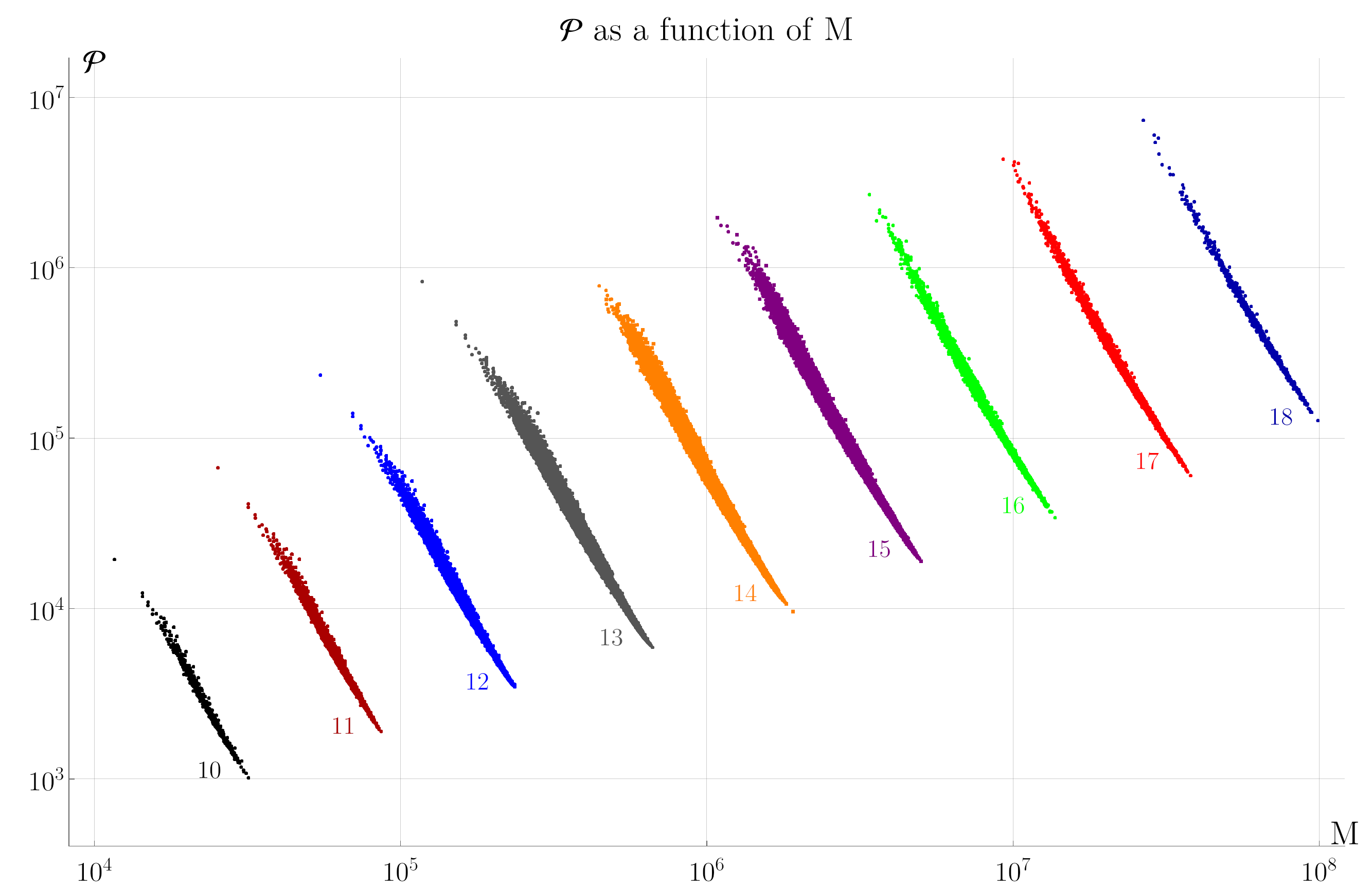}
	\caption{The period as a function of the (first) Martin invariant, for various loop orders, Double-logarithmic plot.  Compared to the Hepp bound  (\cref{P_H}), the relation is not as strict, and the slopes are reversed.  }
	\label{P_M}
\end{figure}

For our data, the relation between the period and the Martin invariant is shown in \cref{P_M}. 
The correlation is close, but not quite as close as for the period and the Hepp bound (\cref{P_H}). Moreover, the slope is inverted, i.e. a large Martin invariant indicates a small period. Analogously to \cref{sec:hepp}, it is possible to define functions that depend on only few parameters and approximate the relationship $M \mapsto \period$ to good accuracy. Again, the residues fall into bands indexed by $c_6$, but the structure is not as pronounced as in \cref{fig:residue_hepp}. More details are subject of an upcoming article of the author.

\bigskip

This concludes our survey of statistical properties of Feynman periods in $\phi^4$-theory. A summary as been given already in \cref{sec:results} in the introduction, and a discussion in \cref{sec:discussion}.

\FloatBarrier

\appendix

\section{Non-uniform sampling}\label{sec:uniform_sampling}
The samples generated by \texttt{genrang} are not distributed uniformly, but proportionally to their symmetry factor. The weight $\rho$ of a graph $G$ in the sample $S$ is 
\begin{align*}
	\rho(G) &= \frac{\frac{N}{\abs{\Aut(G)}}}{\sum_{g\in S} \frac{1}{\abs{\Aut(g)}}}.
\end{align*}
Note that this weight is normalized to the total number of graphs $N$, and not to unity. Naively averaging over the sample $S$ leads to the weighted mean
\begin{align*}
	\left \langle \period \right \rangle _\rho := \frac 1 N\sum_{G\in S} \rho(G)\period(G).
\end{align*}
To produce the un-weighted mean (\cref{def:period_mean}), we can replace each graph $G$ in the sample by $\abs{\Aut(G)}$ copies of $G$. 
\begin{align}\label{def:Scorr}
S_\text{corr}:= \left \lbrace \abs{\Aut(G)}\text{ copies of }G \, \Big| G\in S \right \rbrace  .
\end{align}
$S_\text{corr}$ has uniform distribution and contains $N_\text{corr}=\sum_{g\in S}\abs{\Aut(g)}$ graphs, the mean is
\begin{align*}
	\left \langle \period \right \rangle &=  \frac{1}{ N_\text{corr} } \sum_{G\in S_\text{corr}}    \period(G) = \frac{1}{ N_\text{corr}  } \sum_{G\in S} \rho(G) \abs{\Aut(G)}  \period(G) = \left( \frac{N}{\sum_{g\in S}\abs{\Aut(g)}} \right) \left \langle \abs{\Aut } \cdot \period  \right \rangle _\rho.
\end{align*}
Using
\begin{align*}
	\sum_{G \in S_\text{corr}} \frac{1}{\abs{\Aut(G)}} = N , \qquad  \sum_{G\in S} \abs{\Aut(G)}=N_\text{corr}=\abs{S_\text{corr}},
\end{align*}
the correction factor can be identified as the average symmetry factor,
\begin{align}\label{average_symfactor}
\frac{N}{\sum_{g\in S}\abs{\Aut(g)}} =	\frac{N}{N_\text{corr}} = \frac{1}{\abs{S_\text{corr}}}\sum_{G\in S_\text{corr}} \frac{1}{\abs{\Aut(G)}} =\left \langle \frac{1}{\abs{\Aut}} \right \rangle.
\end{align}

Adding copies of a graph to the sample as in \cref{def:Scorr} has the disadvantage that we slightly underestimate the sampling uncertainty, because the copies are not independent additional samples. 
A different approach to obtain the correct mean is   by scaling each individual period according to
\begin{align}\label{period_sampling_rescaling}
	\period(G)&\mapsto\left \langle \frac{1}{\abs{\Aut}} \right \rangle \cdot  \abs{\Aut(G)} \cdot \period (G) . 
\end{align}
While this scaling does not influence the sample size, it significantly distorts the higher moments  because it artificially introduces very large \enquote{effective periods}. Hence, we use the first approach.

For the mean weighted by symmetry factors (\cref{def:period_mean_aut}), the graphs generated by \texttt{genrang} have the correct relative weighting, but still the normalization factor is wrong. To see this, note that if we were to weight the graphs in the set  $S_\text{corr}$ (\cref{def:Scorr}) with respect to their symmetry factor, we reproduce  the relative weight within $S$, but the number of elements is $N_\text{corr}=\abs{S_\text{corr}}>\abs{S}=N$. By \cref{average_symfactor}, the ratio of the two sizes is precisely the average symmetry factor, so   \cref{def:period_mean_aut} can be computed as
\begin{align*}
\left \langle \frac{\period}{\abs{\Aut}} \right \rangle &= \frac{1}{N_\text{corr}}\sum_{G\in S_\text{corr}} \frac{\period(G)}{\abs{\Aut(G)}} = \left \langle \frac{1}{\abs{\Aut}} \right \rangle  \left \langle \period \right \rangle _\rho.
\end{align*}

The beta function in \cref{sec:beta} requires not the average, but the sum of periods. Let $U$ be the sought-after sum over \emph{all} graphs $G$ (not just the sample), a priori, it is 
\begin{align*}
U :=\sum_G \frac{\period(G)}{\abs{\Aut(G)}} = N^{(C)}_L \left \langle \frac{\period}{\abs{\Aut}} \right \rangle  =  N^{(C)}_L\left \langle \frac{1}{\abs{\Aut}} \right \rangle \left \langle \period \right \rangle _\rho.
\end{align*}
However, the total number of completions $N^{(C)}_L$ is unknown for large $L$, therefore, it is useful to express $U$ in terms of $N^{(\Aut)}_L$ (\cref{def:NLAut}) according to 
\begin{align}\label{sum_P_rho}
U &=  \frac{N^{(\Aut)}_L}{4!(L+2)} \cdot \left \langle \period \right \rangle _\rho. 
\end{align}
\cref{sum_P_rho} is exact, this means that it does not involve any additional uncertainty from e.g. approximating $N^{(C)}_L$ by $N^{(\Aut)}_L$ or $\left \langle \abs{\Aut} \right \rangle \approx 1$. These effects precisely cancel in the end result.

\section{Sampling uncertainties and higher moments}\label{sec:sampling_uncertainty}

This section is a review of basic facts of sampling statistics, for more details see e.g.  \cite{shenton_development_1975,davison_bootstrap_1997,tille_sampling_2020,evans_probability_2023}. To simplify notation, we will call the random variable $X$. 

It is important to distinguish between true properties of the underlying distribution, and the observed properties of finite samples. By $\mathbb E(X)$, we denote the  expectation,  by $\bar M_j(X)$ the   $j$\textsuperscript{th} moment and by $\bar C_j(X)$ the   $j$\textsuperscript{th}  central moment,
\begin{align*}
	\bar M_j=\bar M_j(X) = \mathbb{E}\left( X^j \right)  , \qquad \bar C_j =\bar C_j(X)=  M_j \left( X-M_1(X) \right)  = \mathbb E \left( X -\mathbb E(X)\right) ^j.
\end{align*}
These quantities are constant numbers (not random variables), and if the symbols appear without argument they refer to  $X$. 
By construction $\bar M_0 = \bar C_0=1$, $\bar C_1=0$, and
\begin{align}\label{moments_central_moments}
	\bar C_n &= \sum_{j=0}^n \binom n j (- \bar M_1)^{n-j} \bar M_j, \qquad \bar C_2  = \bar M_2-\bar M_1^2\\
	  \bar C_3 &= \bar M_3 - 3 \bar M_1 \bar M_2 + 2 \bar M_1^3, \quad \bar C_4 = \bar M_4 - 4 \bar M_1 \bar M_3 + 6 \bar M_1^2 \bar M_2 - 3 \bar M_1^4.\nonumber 
\end{align}
We further define the \emph{skewness} and the \emph{kurtosis} like in \cref{def:central_moments}
\begin{align}\label{def:skewness_kurtosis}
	\bar c_3 &= \bar C_3 \bar C_2^{-\frac 3 2}, \qquad \bar c_4 = \bar C_4 \bar C_2^{-2}.
\end{align}
We refer to $X$ as the \emph{population}. 
Now consider a sample $X_n=\left \lbrace x_1, \ldots, x_n \right \rbrace $ of $n$ elements drawn from that population \emph{with replacement}. This means that each element follows the distribution of $X$ and all elements are independent. In particular, for a finite population, an element is allowed to be present in the sample multiple times.  By $\left \langle X_n \right \rangle $ we denote the sample average according to \cref{def:period_mean}, and by $M_j(X_n)$ and $C_j(X_n)$ the sample moments (note that in the main text \cref{def:central_moments}, we report sample moments scaled to the sample mean).
\begin{align}\label{def:sample_mean}
	\left \langle X_n \right \rangle = \frac 1 n \sum_{k=1}^n x_k, \qquad M_j(X_n) = \left \langle X_n^j \right \rangle , \qquad C_j(X_n) =  \left \langle \big(X_n- \left \langle X_n \right \rangle  \big)^j \right \rangle  .
\end{align}

We can probe the population only through samples, therefore, we need to know how the moments of the sample are related to the ones of the population.  
The \emph{characteristic function} of a random variable is the Fourier transform of its density $\rho_Y(y)$, or the expectation of its complex exponential function,
\begin{align}\label{def:characteristic_function}
	\phi_X(t) = \int \d x \, \rho_X(x) e^{itx} =\mathbb E e^{itX}   = \sum_{j=0}^\infty  \mathbb E\left(  \frac{(it)^j X^j}{j!} \right)   =\sum_{j=0}^\infty   \frac{(it)^j}{j!} \bar M_j.
\end{align}
We have used the linearity of the expectation to recognize the   moments $\bar M_j=\mathbb E(X^j)$.
For the sum of $n$ copies of $X$, using the fact that these copies are independent and thus the expectation of the product is the product of expectations, one finds the characteristic function
\begin{align*}
	\phi_{\left \langle X_n \right \rangle  }(t)=  \mathbb E e^{it \left \langle X_n \right \rangle  }=\mathbb E e^{it \frac 1 n \sum_{j=1}^n x_j}  = \mathbb E \prod_{j=1}^n e^{it\frac 1 n x_j}   = \prod_{j=1}^n \mathbb E e^{i \frac t n X}   = \left( \phi_X\left(\frac t n\right) \right) ^n.
\end{align*}
By \cref{def:characteristic_function}, the moments of a distribution are given by derivatives of the corresponding characteristic function. Evaluating the derivatives of $\phi_{\left \langle X_n \right \rangle  }(t)$, we find the moments $\bar M_j\big(\left \langle X_n \right \rangle  \big)$ of the $n$-element sample:  
\begin{align}\label{sample_mean_moments}
	\bar M_j(  \left \langle X_n \right \rangle  ) &= (-i)^j\frac{\d^j \; \phi_{\left \langle X_n \right \rangle  }(t)}{\d t^j} \Big|_{t=0} =  \frac{n!}{n^j}\sum_{k=0}^j \frac{1}{(n-k)!}B_{j,k}\left( \bar M_1 ,   \bar M_2 ,  \bar M_3  , \ldots  \right)  .
\end{align}
For the last equality, we have used  Fa\`{a} di Bruno's formula \cite{johnson_curious_2002} and introduced the incomplete Bell polynomials  $B_{j,k}$ \cite{bell_exponential_1934}. The first moments are
\begin{align}\label{sample_moments_relations}
	\bar M_1 \big( \left \langle X_n \right \rangle   \big) &= \bar M_1, \qquad \qquad \bar M_2 \big( \left \langle X_n \right \rangle  \big)  = \frac{n-1}{n} \bar M_1^2 + \frac 1 n \bar M_2 \\
	\bar M_3 \big(\left \langle X_n \right \rangle  \big) &= \frac{(n-1)(n-2)}{n^2} \bar M_1^3 +   \frac{3(n-1)}{n^2} \bar M_1 \bar M_2 + \frac{1}{n^2} \bar M_3. \nonumber
\end{align}

We define the \emph{sampling uncertainty} of the sample mean $\left \langle X_n \right \rangle $  as the square root of the variance of the sample mean, it is related to moments according to \cref{moments_central_moments}:
\begin{align}\label{def:sampling_uncertainty}
	\left(\Delta_{\text{samp}} \left \langle X_n \right \rangle \right)^2&= \bar C_2 \big(\left \langle X_n \right \rangle  \big)  = \bar M_2 \big(\left \langle X_n \right \rangle  \big) -\left( \bar M_1 \big(\left \langle X_n \right \rangle  \big) \right)  ^2.
\end{align}
By \cref{sample_moments_relations}, it is proportional to the population standard deviation  $\sqrt{\bar C_2}$:
\begin{align}\label{uncertainty_sampling_theo} 
	\Delta_{\text{samp}} \left \langle X_n \right \rangle  &= \sqrt{\frac{n-1}{n} \bar M_1^2 + \frac 1 n \bar M_2 - \bar M_1^2 } = \frac{1}{\sqrt n} \sqrt{\bar M_2-\bar M_1^2} = \frac{1}{\sqrt n } \sqrt{\bar C_2}.
\end{align}

We need to estimate $\sqrt{\bar C_2}$ from a finite sample. The naive sample variance according to \cref{def:sample_mean}, $C_2(X_n   )=\left \langle X_n^2 \right \rangle  $, is systematically biased because it relies on the sample mean, which  is not independent. The sample moments itself have the correct expectation,
\begin{align*}
	\mathbb E \left( M_j(X_n) \right)   &=\mathbb E \left \langle X_n^j \right \rangle = \frac 1 n \sum_{k=1}^n \mathbb E(x_k^j) = \frac 1 n \sum_{k=1}^n \bar M_j = \bar M_j,
\end{align*}
but products of sample moments with individual sample elements are biased,
\begin{align*}
	\mathbb E \left( x_i^l \left \langle X_n^j \right \rangle   \right) &= \mathbb E \left( x_i^l \frac 1 n \sum_{k=1}^n x_k^j \right) = \mathbb E \left(  \frac 1 n \sum_{k \neq i} x_i^l x_k^j +\frac 1 n x_i^{j+l}\right) = \frac{n-1}{n} \bar M_l \bar M_j + \frac 1 n \bar M_{j+l}.
\end{align*}
This implies that the expectation of the sample variance contains  a non-trivial factor,
\begin{align*}
	\mathbb E \left( C_2(X_n) \right) &= \frac 1 n \sum_{i=1}^n \left( \mathbb E x_i^2 - 2 \mathbb E x_i \left \langle X_n \right \rangle + \left \langle X_n \right \rangle ^2 \right) = \frac{n-1}{n} \bar C_2.
\end{align*}
Including this correction factor, the unbiased estimator of the population variance is
\begin{align}\label{def:sample_variance}
	\tilde C_2(X_n) &= \frac{n}{n-1}C_2 (X_n)=  \frac{1}{n-1}\sum_{j=1}^n \left( x_j  -  \sum_{k=1}^n x_j  \right) ^2 .
\end{align}
Similar corrections exist for higher moments.
The sample moments reported in the paper contain these corrections, but the influence is minuscule since the sample sizes  are $n \geq 900$. 

The sample variance \cref{def:sample_variance} is itself a random variable, its value depends on the particular sample, while the population variance $\bar C_2$ is a constant. This implies that powers of the variance, such as the standard deviation or the denominators in \cref{def:skewness_kurtosis}, are generally not equal to (the expectation of) the same powers of the sample variance. The necessary correction factors are known analytically if $X$ is a normal distribution \cite{holtzman_unbiased_1950}. To find them for arbitrary distributions, rewrite the sample variance as
\begin{align*}
	C_2(X_n)^r &= \bar C_2 ^r \left( \frac{  C_2(X_n)}{\bar C_2 } \right) ^r =  \bar C_2 ^r  \left(  1 + \frac{  C_2(X_n) -\bar C_2 }{\bar C_2 } \right) =\bar C_2^r \sum_{k=0}^\infty  \binom r k \frac{1}{\bar C_2^k} \left( C_2(X_n)-\bar C_2 \right) ^k.
\end{align*}
Consequently, the expectation of $C_2(X_n)^r$ is a power series  of its $k$\textsuperscript{th} central moments,
\begin{align*}
	\mathbb E \left( C_2(X_n)^r \right)   &=  \bar C_2^r \sum_{k=0}^\infty  \binom r k \frac{1}{\bar C_2^k} \mathbb E \left( C_2(X_n)-\bar C_2 \right) ^k= \bar C_2^r \sum_{k=0}^\infty  \binom r k \frac{1}{\bar C_2^k} \bar C_k \Big(   C_2(X_n)\Big). 
\end{align*}
Formulas for these moments can be found in \cite{angelova_moments_2012}. The leading term    $\bar C_2\left( C_2(X_n) \right) $ is determined by the population kurtosis $\bar c_4= \frac{\bar C_4}{\bar C_2^2}$ (\cref{def:skewness_kurtosis}),
\begin{align}\label{kurtosis_correction}
	\mathbb E \left( C_2^r \right) &=  \bar C_2^r \left( 1 +   \frac{r(r-1)}{2}  \frac{\bar c_4 -1}{n}  +\mathcal O \left( \frac{1}{n^2} \right)     \right) . 
\end{align} 
\Cref{kurtosis_correction} is one example for how the knowledge of distribution parameters can be used to improve the error estimates of small samples. Corrections of this type are not used in the main text since they are vanishing small for our sample sizes, compare \cref{sec:sampling_example}.

Everything above is valid for sampling \emph{with replacement}. Conversely, sampling from a finite population of size $N$ \emph{without replacement} means that every element can be selected at most once. At $n=N$, the sample contains the entire population. In that setting, the individual sample elements are no longer independent because their possible value depends on which elements are contained in the sample already.  The variance of the sample mean in this case requires a \emph{finite population correction} factor  \cite{davison_bootstrap_1997} and \cref{uncertainty_sampling_theo} becomes
\begin{align}\label{uncertainty_sampling2} 
	\Delta_{\text{samp}} \left \langle X_n \right \rangle  &= \sqrt{\frac{N -n}{N -1}} \frac{1}{\sqrt n } \sqrt{\bar C_2}.
\end{align}
This factor is irrelevant for the main text since $n\ll N$ for all our samples, but we demonstrate its effect in \cref{sec:sampling_example}.

\section{Practical example: Sampling from $L=13$ periods} \label{sec:sampling_example}
In this section, we use the $L=13$ data set, weighted with symmetry factors, to illustrate sampling accuracy.  From \cref{tab:periods_count,tab:means_Aut,C2_std}, the relevant numbers are
\begin{align*}
	\bar M_1=  \left \langle 	\frac{\period}{\abs{\Aut}} \right \rangle  \approx26320 , \qquad \sigma =\sqrt{\bar C_2}\approx 18079, \qquad N=N^{(C)}_L=755643.
\end{align*}

Concretely, we fix a sample size $n$ and draw 1000 random samples of $n$ periods without replacement. For each of the 1000 samples, we compute the sample mean $ \left \langle  X_n  \right \rangle  $ (\cref{def:sample_mean}) and the sample variance $\tilde C_2(X_n)$ (\cref{def:sample_variance}). We then compute the empirical standard deviation of the 1000 sample means, which is the sampling uncertainty $\Delta_{\text{samp}} \left \langle X_n \right \rangle $ (\cref{def:sampling_uncertainty}). This procedure is repeated for various sample sizes $n$.

\begin{figure}[htb]
	\centering
	\includegraphics[width=.75\linewidth]{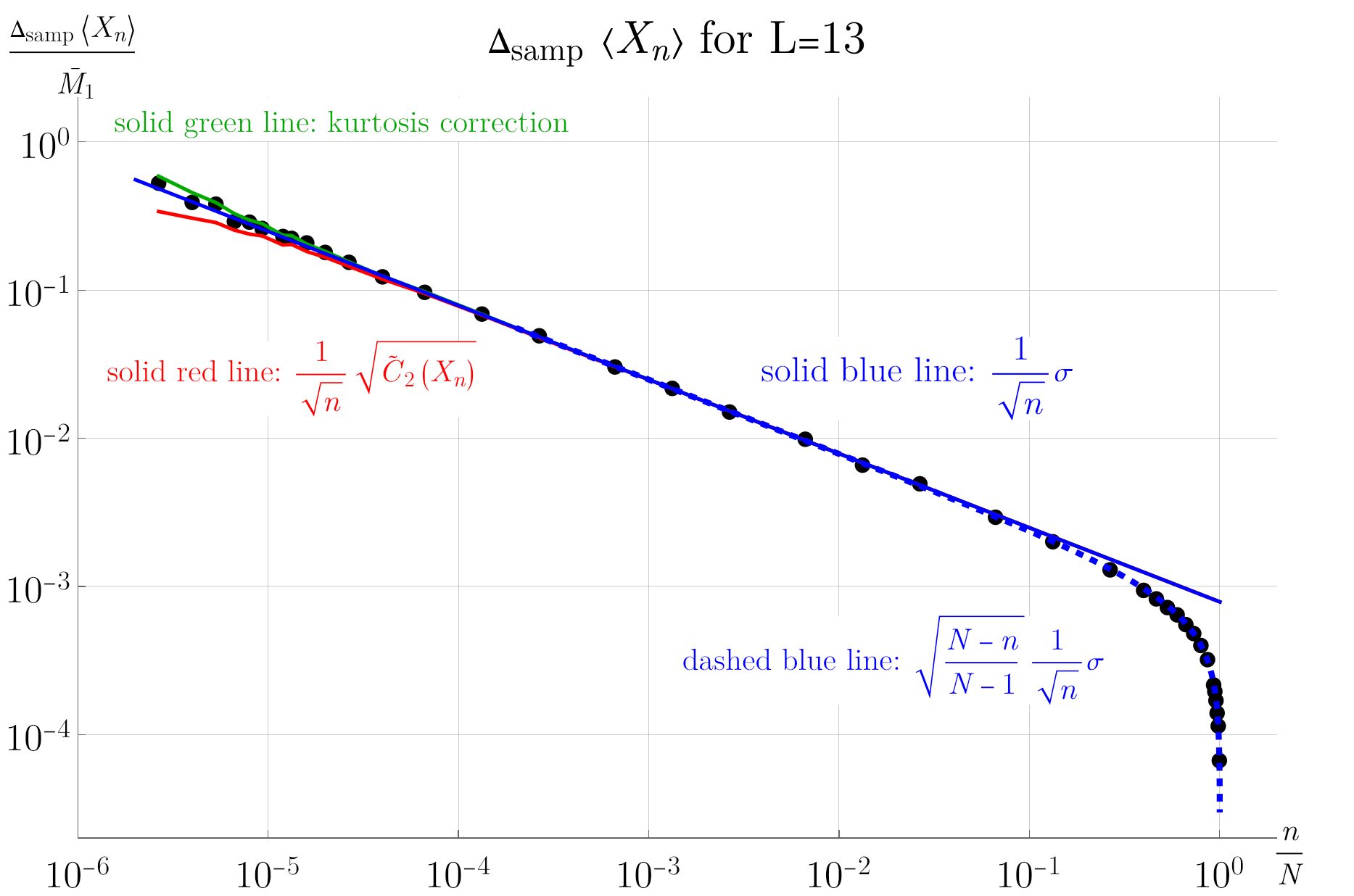}
	\caption{Black dots: Empirical sampling uncertainty for different sample sizes of the data set $\frac{\period}{\abs{\Aut}}$ for $L=13$ loops. As long as $n \ll N^{(C)}_L$, the observed accuracy is accurately described by \cref{uncertainty_sampling_theo} (solid blue line). Only when $n$ is similar to the population size $N=N^{(C)}_L$, the finite population correction \cref{uncertainty_sampling2} (dashed blue line) is necessary. For very small samples $n \leq 10$, the sample standard deviation $\sqrt{\tilde C_2(X_n)}$ (\cref{def:sample_variance}) underestimates the true standard deviation and thus the sample uncertainty (red line). This can be corrected with the help of the population kurtosis according to \cref{kurtosis_correction} (green line). }
	\label{fig:sample_uncertainty}
\end{figure}

The outcomes are shown in \cref{fig:sample_uncertainty} as black data points, where the axes are scaled relative to the population size $N^{(C)}_L$ respective the population mean $\bar M_1$. For all $n \ll N^{(C)}_L$,   these dots lie on the solid blue line, that is, the observed sampling uncertainty of the mean is described accurately by \cref{uncertainty_sampling_theo}. In particular, this formula is still true for very small samples as long as one uses the population standard deviation $\sigma$.

If the sample is very small, $n \approx 10$, then the naive sample standard deviation $\sqrt{\tilde C_2(X_n)}$  (\cref{def:sample_variance}) systematically underestimates the true standard deviation and consequently the sampling uncertainty is estimated too small. This is shown as a red line in \cref{fig:sample_uncertainty}. In this regime, we can use the kurtosis correction \cref{kurtosis_correction} to obtain a much better estimate of the sampling uncertainty (green line in  \cref{fig:sample_uncertainty}). Note that we used the true population kurtosis $\bar c_4 = 12.7$, not the kurtosis measured in the small sample itself. This highlights the importance of the population higher moments for estimating the reliability of small samples. For $n > 100$, this effect is barely visible.

The kurtosis correction might appear somewhat tautological because from a small sample it is very hard to obtain an accurate kurtosis. However, we do not necessarily need an accurate value. For example, suppose we want to sample periods at $L=20$ loops. By \cref{cumulant_growth,tab:centralMoments}, the kurtosis at higher loop order roughly grows like
\begin{align*}
c_4 &= \frac{C_4}{C_2^2} \approx \frac{e^{a_4} \left( \frac L 8 \right) ^{b_4}}{e^{2 a_2} \left( \frac L 8 \right) ^{2 b_2}} \approx e^{1.5} \left( \frac L 8 \right) ^{2.53}, \quad c_4(L=20) \approx 46.
\end{align*}
This allows us to correct our sample standard deviation according to \cref{kurtosis_correction}.

We stress that the intuitive notion of standard deviation as \enquote{uncertainty} like in \cref{def:sampling_uncertainty} is often done with a normal distribution in mind. By the central limit theorem, the distribution of the mean of a large sample is almost normal, but the smaller the sample, the more the distribution of $\left \langle X_n \right \rangle $ resembles that of $X$ itself, including outliers (indeed, we know this relation explicitly from \cref{sample_mean_moments}). While our numerical data confirms that  \cref{uncertainty_sampling_theo} gives the correct expected standard deviation, one might not want to interpret this quantity as a reliable sampling accuracy for very small samples. 

Conversely, for very large samples $n \approx N$, the observed uncertainty is increasingly smaller than estimated from \cref{uncertainty_sampling_theo}. In that case, the finite population correction \cref{uncertainty_sampling2} gives an accurate description.  We stress again that this effect only appears if one draws samples without replacement. As soon as replacement is allowed, \cref{uncertainty_sampling_theo} stays correct even for $n>N^{(C)}_L$. 

We conclude that for all samples in the main part of this paper, \cref{uncertainty_sampling_theo} is an accurate estimate of the sampling standard deviation. Note also the quantitative scale of the uncertainties: To reach an accuracy of  just 1\%, we need  $n \geq \left( \frac{\sigma}{1\%\, \bar M_1} \right) ^2=4120$ periods. In an upcoming paper, we examine how the empirical data from the present work allows to reduce the required sample sizes.

\section{Weighted averages and estimation of combined uncertainties} \label{sec:statistics}

In this section, we discuss the influence of numerical uncertainty and the combination of uncertainties.
The numerical integration procedure returns an estimate for the uncertainty $\Delta_\text{num} \period(G)$ of each period, measured as the   standard deviation. These uncertainties are statistically independent  since each period  results from a different sum of   Monte-Carlo summands. Comparing to the analytically known periods, we confirmed that the deviations are approximately normal distributed with standard deviation $  \Delta_\text{num} \period(G)$ as expected.  Let $S_n$ be the sum of $n$ periods, weighted by symmetry factor. The uncertainty $ \Delta_\text{num} S_n$ is the square root of the variance, and the latter is the sum of the variances of the individual summands. Consequently, the uncertainty of the mean \cref{def:period_mean_aut} is given by
\begin{align}\label{uncertainty_numerical}
	\left( \Delta_\text{num} S_n \right) ^2 &=\sum_{ n \text{ completions }G } \left(\frac{\Delta \period(G)} {  \abs{\Aut(G)}}  \right)^2,  \qquad  \Delta_\text{num} \left \langle \frac{\period}{\abs{\Aut }} \right \rangle  = \frac{\sqrt{\left( \Delta_\text{num} S_n \right) ^2 }}{n}    .
\end{align} 
The sum   has $n$ summands which all contribute roughly equally, hence the numerical uncertainty falls off  like $n ^{-\frac 12}$ with growing $n$. This explains why the means in \cref{tab:means,tab:means_Aut} have much smaller uncertainty than the individual periods in \cref{tab:samples}. 

If we use a non-complete sample of $n$ periods of a given loop order, the sample average is subject to a sampling uncertainty. In \cref{sec:sampling_uncertainty}, we derived \cref{uncertainty_sampling_theo},
\begin{align}\label{uncertainty_sampling}
	\Delta_{\text{samp}} \left \langle X_n \right \rangle  &= \frac{1}{\sqrt n } \sigma,
\end{align}
where $\sigma$ is the population standard deviation. In \cref{sec:sampling_example}, we confirmed that this formula is accurate for all samples in the main text. 
The two uncertainties \cref{uncertainty_sampling,uncertainty_numerical} are  independent, therefore the total uncertainty is
\begin{align}\label{uncertainty_combined}
\Delta \period_n &= \sqrt{\left(\Delta_\text{num} \period_n\right) ^ 2  +   \left( \Delta_\text{samp}  \period_n \right)^2 }  .
\end{align}

A third type of uncertainty can arise if the sampling itself is not uniform, for example due to a low-quality random number generator. We have included the $\star$-samples in order to verify the accuracy of \texttt{nauty}'s random graph generator, and found no significant systematic deviations as soon as the \enquote{s} samples are corrected with regard to their symmetry factor as explained in \cref{sec:uniform_sampling}.

The procedure for combining multiple numerical values of \emph{the same} period (that is, graphs which are either isomorphic or related by a known period symmetry) is similar to a sample mean, but the individual contributions are weighted with factors $w_j$:
\begin{align*} 
	\bar \period  &=  \left( \sum \nolimits_{j=1}^n w_j \right) ^{-1}\cdot \sum \nolimits_{j=1}^n w_j\period(G_j), \qquad \Delta \bar \period = \left( \sum \nolimits_{j=1}^n w_j \right) ^{-1}\cdot \sqrt{\sum \nolimits_{j=1}^n ( w_j\Delta \period(G_j))^2}.
\end{align*}
In principle, we can freely chose the $w_j$. The uncertainty of the average is minimized by choosing  $w_i:= (\Delta \period(G_j))^{-2}$. With this choice, the  estimate and uncertainty are 
\begin{align}\label{weighted_average} 
	\bar \period &:= \left(  \sum\nolimits_{i=1}^n w_i\right)^{-1}\cdot  \sum\nolimits_{i=1}^n w_i \period(G_i) ,\qquad  \Delta \bar \period =  \left(\sum\nolimits_{i=1}^n w_i \right)^{-\frac 12}.
\end{align}

Note that this  weighted mean is not equivalent to introducing non-trivial factors $w_i$ for distinct periods, such as the $O(N)$ symmetry factors in \cref{sec:beta_N_dependence}. In the latter case, different choices of weights $w_i$ do change the value of the sum because we are summing \emph{different} periods and the sum is normalized with respect to their number $n$,
\begin{align}\label{app:Pn_weighted}
	\period_n &=  \frac{1}{n } \sum_{j=1}^n \frac{  w_j \period(G_j)} {\abs{\Aut(G_j)}} \neq \frac{1}{ \sum_{j=1}^n w_j } \sum_{j=1}^n \frac{  w_j \period(G_j)} {\abs{\Aut(G_j)}} .
\end{align}
Conversely, in \cref{weighted_average}, the sum is over periods which should be equal and the sum is normalized with respect to the sum of weights.

\FloatBarrier

\bibliography{Periods}

\providecommand{\href}[2]{#2}\begingroup\raggedright\begin{thebibliography}{10}

\bibitem{dunne_resurgence_2012}
G.V.~Dunne and M.~Unsal, \emph{Resurgence and {{Trans-series}} in {{Quantum
  Field Theory}}: {{The CP}}({{N-1}}) {{Model}}},
  \href{https://doi.org/10.1007/JHEP11(2012)170}{\emph{Journal of High Energy
  Physics} {\bfseries 2012} (2012) 170}
  [\href{https://arxiv.org/abs/1210.2423}{{\ttfamily 1210.2423}}].

\bibitem{borinsky_nonperturbative_2020}
M.~Borinsky and G.V.~Dunne, \emph{Non-perturbative completion of
  {{Hopf-algebraic Dyson-Schwinger}} equations},
  \href{https://doi.org/10.1016/j.nuclphysb.2020.115096}{\emph{Nuclear Physics
  B} {\bfseries 957} (2020) 115096}
  [\href{https://arxiv.org/abs/2005.04265}{{\ttfamily 2005.04265}}].

\bibitem{bellon_resurgent_2021}
M.P.~Bellon and E.I.~Russo, \emph{Resurgent analysis of
  {{Ward-Schwinger-Dyson}} equations},
  \href{https://doi.org/10.3842/SIGMA.2021.075}{\emph{SIGMA. Symmetry,
  Integrability and Geometry: Methods and Applications} {\bfseries 17} (2021)
  075} [\href{https://arxiv.org/abs/2011.13822}{{\ttfamily 2011.13822}}].

\bibitem{borinsky_semiclassical_2021}
M.~Borinsky, G.~Dunne and M.~Meynig, \emph{Semiclassical {{Trans-Series}} from
  the {{Perturbative Hopf-Algebraic Dyson-Schwinger Equations}}: $\phi^3$
  {{QFT}} in 6 {{Dimensions}}},
  \href{https://doi.org/10.3842/SIGMA.2021.087}{\emph{Symmetry, Integrability
  and Geometry: Methods and Applications} {\bfseries 17} (2021) 087}
  [\href{https://arxiv.org/abs/2104.00593}{{\ttfamily 2104.00593}}].

\bibitem{clavier_borelecalle_2021}
P.J.~Clavier, \emph{Borel-{{\'Ecalle Resummation}} of a {{Two-Point
  Function}}}, \href{https://doi.org/10.1007/s00023-021-01057-w}{\emph{Annales
  Henri Poincar\'e} {\bfseries 22} (2021) 2103}
  [\href{https://arxiv.org/abs/1912.03237}{{\ttfamily 1912.03237}}].

\bibitem{borinsky_resonant_2022}
M.~Borinsky and D.~Broadhurst, \emph{Resonant resurgent asymptotics from
  quantum field theory},
  \href{https://doi.org/10.1016/j.nuclphysb.2022.115861}{\emph{arXiv:2202.01513
  [hep-th, physics:math-ph]} {\bfseries 981} (2022) 115861}
  [\href{https://arxiv.org/abs/2202.01513}{{\ttfamily 2202.01513}}].

\bibitem{nakanishi_graph_1971}
N.~Nakanishi, \emph{Graph {{Theory}} and {{Feynman Integrals}}}, vol.~11 of
  \emph{Mathematics and Its Applications}, {Gordon and Breach, Science
  Publishers, Inc}, {New York London Paris} (1971).

\bibitem{panzer_feynman_2015}
E.~Panzer, \emph{Feynman Integrals and Hyperlogarithms}, doctoral {{Thesis}},
  Humboldt-Universit\"at zu Berlin, {Berlin}, Mar., 2015.
\newblock \href{https://arxiv.org/abs/1506.07243}{{\ttfamily 1506.07243}}.

\bibitem{bogner_feynman_2010}
C.~Bogner and S.~Weinzierl, \emph{Feynman graph polynomials},
  \href{https://doi.org/10.1142/S0217751X10049438}{\emph{International Journal
  of Modern Physics A} {\bfseries 25} (2010) 2585}
  [\href{https://arxiv.org/abs/1002.3458}{{\ttfamily 1002.3458}}].

\bibitem{bloch_motives_2006}
S.~Bloch, H.~Esnault and D.~Kreimer, \emph{On {{Motives Associated}} to {{Graph
  Polynomials}}},
  \href{https://doi.org/10.1007/s00220-006-0040-2}{\emph{Communications in
  Mathematical Physics} {\bfseries 267} (2006) 181}
  [\href{https://arxiv.org/abs/math/0510011}{{\ttfamily math/0510011}}].

\bibitem{schnetz_quantum_2010}
O.~Schnetz, \emph{Quantum periods: {{A}} census of $\phi^4$-transcendentals},
  {\emph{Commun.Num.Theor.Phys.} {\bfseries 4} (2010) 1}
  [\href{https://arxiv.org/abs/0801.2856}{{\ttfamily 0801.2856}}].

\bibitem{brown_angles_2013}
F.~Brown and D.~Kreimer, \emph{Angles, {{Scales}} and {{Parametric
  Renormalization}}},
  \href{https://doi.org/10.1007/s11005-013-0625-6}{\emph{Letters in
  Mathematical Physics} {\bfseries 103} (2013) 933}.

\bibitem{wilson_renormalization_1974}
K.G.~Wilson and J.~Kogut, \emph{The renormalization group and the {$\epsilon$}
  expansion}, \href{https://doi.org/10.1016/0370-1573(74)90023-4}{\emph{Physics
  Reports} {\bfseries 12} (1974) 75}.

\bibitem{broadhurst_knots_1995}
D.J.~Broadhurst and D.~Kreimer, \emph{Knots and {{Numbers}} in $\phi^4$
  {{Theory}} to 7 {{Loops}} and {{Beyond}}},
  \href{https://doi.org/10.1142/S012918319500037X}{\emph{International Journal
  of Modern Physics C} {\bfseries 06} (1995) 519}
  [\href{https://arxiv.org/abs/hep-ph/9504352}{{\ttfamily hep-ph/9504352}}].

\bibitem{brown_singlevalued_2015}
F.~Brown and O.~Schnetz, \emph{Single-valued multiple polylogarithms and a
  proof of the zig-zag conjecture},
  \href{https://doi.org/10.1016/j.jnt.2014.09.007}{\emph{Journal of Number
  Theory} {\bfseries 148} (2015) 478}
  [\href{https://arxiv.org/abs/1208.1890}{{\ttfamily 1208.1890}}].

\bibitem{panzer_analytic_2013}
E.~Panzer, \emph{On the analytic computation of massless propagators in
  dimensional regularization},
  \href{https://doi.org/10.1016/j.nuclphysb.2013.05.025}{\emph{Nuclear Physics
  B} {\bfseries 874} (2013) 567}
  [\href{https://arxiv.org/abs/1305.2161}{{\ttfamily 1305.2161}}].

\bibitem{panzer_galois_2017}
E.~Panzer and O.~Schnetz, \emph{The {{Galois}} coaction on $\phi^4$ periods},
  \href{https://doi.org/10.4310/CNTP.2017.v11.n3.a3}{\emph{Communications in
  Number Theory and Physics} {\bfseries 11} (2017) 657}
  [\href{https://arxiv.org/abs/1603.04289}{{\ttfamily 1603.04289}}].

\bibitem{schnetz_numbers_2018}
O.~Schnetz, \emph{Numbers and {{Functions}} in {{Quantum Field Theory}}},
  \href{https://doi.org/10.1103/PhysRevD.97.085018}{\emph{Physical Review D}
  {\bfseries 97} (2018) 085018}
  [\href{https://arxiv.org/abs/1606.08598}{{\ttfamily 1606.08598}}].

\bibitem{schnetz_hyperlogprocedures_2023}
O.~Schnetz, \emph{{{HyperlogProcedures}}},  June, 2023.

\bibitem{kontsevich_periods_2001}
M.~Kontsevich and D.~Zagier, \emph{Periods},  in \emph{Mathematics
  {{Unlimited}} - 2001 and {{Beyond}}}, pp.~771--808, {Springer} (2001).

\bibitem{belkale_periods_2003}
P.~Belkale and P.~Brosnan, \emph{Periods and {{Igusa Zeta}} functions},
  {\emph{arXiv:math/0302090} (2003) }
  [\href{https://arxiv.org/abs/math/0302090}{{\ttfamily math/0302090}}].

\bibitem{brown_multiple_2006}
F.C.S.~Brown, \emph{Multiple zeta values and periods of moduli spaces
  $\mathfrak{M}_{0,n}$}, {\emph{arXiv:math/0606419} (2006) }
  [\href{https://arxiv.org/abs/math/0606419}{{\ttfamily math/0606419}}].

\bibitem{brown_periods_2010}
F.C.S.~Brown, \emph{On the periods of some {{Feynman}} integrals},
  {\emph{arXiv:0910.0114 [math-ph]} (2010) }
  [\href{https://arxiv.org/abs/0910.0114}{{\ttfamily 0910.0114}}].

\bibitem{kreimer_quantum_2015}
D.~Kreimer, \emph{Quantum fields, periods and algebraic geometry},
  \href{https://arxiv.org/abs/1405.4964}{{\ttfamily 1405.4964}}.

\bibitem{todorov_polylogarithms_2014}
I.~Todorov, \emph{Polylogarithms and {{Multizeta Values}} in {{Massless Feynman
  Amplitudes}}},  in \emph{Lie {{Theory}} and {{Its Applications}} in
  {{Physics}}}, V.~Dobrev, ed., vol.~111, ({Tokyo}), pp.~155--176, {Springer
  Japan} (2014), \href{https://doi.org/10.1007/978-4-431-55285-7_10}{DOI}.

\bibitem{crump_period_2016}
I.~Crump, M.~DeVos and K.~Yeats, \emph{Period preserving properties of an
  invariant from the permanent of signed incidence matrices},
  \href{https://doi.org/10.4171/aihpd/35}{\emph{Annales de l'Institut Henri
  Poincar\'e D} {\bfseries 3} (2016) 429}.

\bibitem{nasrollahpoursamami_periods_2016}
E.~Nasrollahpoursamami, \emph{Periods of {{Feynman Diagrams}} and {{GKZ
  D-Modules}}},  June, 2016.
\newblock 10.48550/arXiv.1605.04970.

\bibitem{crump_properties_2017}
I.~Crump, \emph{Properties of the {{Extended Graph Permanent}}},
  {\emph{arXiv:1608.01414 [math-ph]} (2017) }
  [\href{https://arxiv.org/abs/1608.01414}{{\ttfamily 1608.01414}}].

\bibitem{hu_further_2022}
S.~Hu, O.~Schnetz, J.~Shaw and K.~Yeats, \emph{Further investigations into the
  graph theory of $\phi^4$-periods and the $c_2$ invariant},
  \href{https://doi.org/10.4171/AIHPD/123}{\emph{Annales de l'Institut Henri
  Poincar\'e D} {\bfseries 9} (2022) 473}.

\bibitem{laradji_results_2021}
M.~Laradji, M.~Mishna and K.~Yeats, \emph{Some results on double triangle
  descendants of \${{K}}\_5\$},
  \href{https://doi.org/10.4171/aihpd/110}{\emph{Annales de l'Institut Henri
  Poincar\'e D} {\bfseries 8} (2021) 537}.

\bibitem{borinsky_graphical_2022}
M.~Borinsky and O.~Schnetz, \emph{Graphical functions in even dimensions},
  \href{https://doi.org/10.4310/CNTP.2022.v16.n3.a3}{\emph{arXiv:2105.05015
  [hep-th, physics:math-ph]} {\bfseries 16} (2022) 515}
  [\href{https://arxiv.org/abs/2105.05015}{{\ttfamily 2105.05015}}].

\bibitem{borinsky_recursive_2022}
M.~Borinsky and O.~Schnetz, \emph{Recursive computation of {{Feynman}}
  periods},  June, 2022.
\newblock 10.48550/arXiv.2206.10460.

\bibitem{borinsky_tropical_2023a}
M.~Borinsky, \emph{Tropical {{Monte Carlo}} quadrature for {{Feynman}}
  integrals}, \href{https://doi.org/10.4171/AIHPD/158}{\emph{Annales de
  l'Institut Henri Poincar\'e D} {\bfseries 10} (2023) 635}
  [\href{https://arxiv.org/abs/2008.12310}{{\ttfamily 2008.12310}}].

\bibitem{mckay_practical_2014}
B.D.~McKay and A.~Piperno, \emph{Practical graph isomorphism, {{II}}},
  \href{https://doi.org/10.1016/j.jsc.2013.09.003}{\emph{Journal of Symbolic
  Computation} {\bfseries 60} (2014) 94}.

\bibitem{borinsky_tropical_2023}
M.~Borinsky, H.J.~Munch and F.~Tellander, \emph{Tropical {{Feynman}}
  integration in the {{Minkowski}} regime},  Feb., 2023.
\newblock 10.48550/arXiv.2302.08955.

\bibitem{kompaniets_minimally_2017}
M.V.~Kompaniets and E.~Panzer, \emph{Minimally subtracted six loop
  renormalization of $o(n)$ -symmetric $\phi^4$ theory and critical exponents},
  \href{https://doi.org/10.1103/PhysRevD.96.036016}{\emph{Physical Review D}
  {\bfseries 96} (2017) 036016}
  [\href{https://arxiv.org/abs/1705.06483}{{\ttfamily 1705.06483}}].

\bibitem{nickel_compilation_1977}
B.G.~Nickel, D.I.~Meiron and G.A.~Baker, Jr., \emph{Compilation of 2-pt. and
  4-pt. graphs for continuous spin models}, {\emph{University of Guelph Report}
  (1977) }.

\bibitem{oeis}
N.J.A.~Sloane~(editor), ``The {{On-Line Encyclopedia}} of {{Integer
  Sequences}}.'' https://oeis.org, 2023.

\bibitem{meringer_fast_1999}
M.~Meringer, \emph{Fast generation of regular graphs and construction of
  cages},
  \href{https://doi.org/10.1002/(SICI)1097-0118(199902)30:2<137::AID-JGT7>3.0.CO;2-G}{\emph{Journal
  of Graph Theory} {\bfseries 30} (1999) 137}.

\bibitem{meringer_regular_2009}
M.~Meringer, ``Regular {{Graphs Page}}.''
  https://www.mathe2.uni-bayreuth.de/markus/reggraphs.html, June, 2009.

\bibitem{wormald_models_1999}
N.C.~Wormald, \emph{Models of {{Random Regular Graphs}}},  in \emph{Surveys in
  {{Combinatorics}}, 1999}, no.~267 in London {{Mathematical Society Lecture
  Note Series}}, pp.~239--298, {Cambridge University Press} (1999),
  \href{https://doi.org/10.1017/CBO9780511721335.010}{DOI}.

\bibitem{bollobas_random_2001}
B.~Bollob{\'a}s, \emph{Random {{Graphs}}}, Cambridge {{Studies}} in {{Advanced
  Mathematics}}, {Cambridge University Press}, {Cambridge}, 2~ed. (2001),
  \href{https://doi.org/10.1017/CBO9780511814068}{10.1017/CBO9780511814068}.

\bibitem{bollobas_modern_1998}
B.~Bollob{\'a}s, \emph{Modern {{Graph Theory}}}, {Springer}, {New York, NY}
  (1998).

\bibitem{wormald_asymptotic_1981}
N.C.~Wormald, \emph{The asymptotic connectivity of labelled regular graphs},
  \href{https://doi.org/10.1016/S0095-8956(81)80021-4}{\emph{Journal of
  Combinatorial Theory, Series B} {\bfseries 31} (1981) 156}.

\bibitem{bollobas_probabilistic_1980}
B.~Bollob{\'a}s, \emph{A {{Probabilistic Proof}} of an {{Asymptotic Formula}}
  for the {{Number}} of {{Labelled Regular Graphs}}},
  \href{https://doi.org/10.1016/S0195-6698(80)80030-8}{\emph{European Journal
  of Combinatorics} {\bfseries 1} (1980) 311}.

\bibitem{wormald_asymptotic_1981a}
N.C.~Wormald, \emph{The asymptotic distribution of short cycles in random
  regular graphs},
  \href{https://doi.org/10.1016/S0095-8956(81)80022-6}{\emph{Journal of
  Combinatorial Theory, Series B} {\bfseries 31} (1981) 168}.

\bibitem{bender_asymptotic_1978}
E.A.~Bender and E.R.~Canfield, \emph{The asymptotic number of labeled graphs
  with given degree sequences},
  \href{https://doi.org/10.1016/0097-3165(78)90059-6}{\emph{Journal of
  Combinatorial Theory, Series A} {\bfseries 24} (1978) 296}.

\bibitem{bollobas_asymptotic_1982}
B.~Bollob{\'a}s, \emph{The {{Asymptotic Number}} of {{Unlabelled Regular
  Graphs}}}, \href{https://doi.org/10.1112/jlms/s2-26.2.201}{\emph{Journal of
  the London Mathematical Society} {\bfseries s2-26} (1982) 201}.

\bibitem{bonichon_planar_2006}
N.~Bonichon, C.~Gavoille, N.~Hanusse, D.~Poulalhon and G.~Schaeffer,
  \emph{Planar {{Graphs}}, via {{Well-Orderly Maps}} and {{Trees}}},
  \href{https://doi.org/10.1007/s00373-006-0647-2}{\emph{Graphs and
  Combinatorics} {\bfseries 22} (2006) 185}.

\bibitem{cvitanovic_number_1978}
P.~Cvitanovi{\'c}, B.~Lautrup and R.B.~Pearson, \emph{Number and weights of
  {{Feynman}} diagrams},
  \href{https://doi.org/10.1103/PhysRevD.18.1939}{\emph{Physical Review D}
  {\bfseries 18} (1978) 1939}.

\bibitem{borinsky_renormalized_2017}
M.~Borinsky, \emph{Renormalized asymptotic enumeration of {{Feynman}}
  diagrams}, \href{https://doi.org/10.1016/j.aop.2017.07.009}{\emph{Annals of
  Physics} {\bfseries 385} (2017) 95}.

\bibitem{mckay_automorphisms_1984}
B.D.~McKay and N.C.~Wormald, \emph{Automorphisms of random graphs with
  specified vertices},
  \href{https://doi.org/10.1007/BF02579144}{\emph{Combinatorica} {\bfseries 4}
  (1984) 325}.

\bibitem{bender_number_2002}
E.A.~Bender, Z.~Gao and N.C.~Wormald, \emph{The {{Number}} of {{Labeled}}
  2-{{Connected Planar Graphs}}},
  \href{https://doi.org/10.37236/1659}{\emph{The Electronic Journal of
  Combinatorics} (2002) R43}.

\bibitem{panzer_hepp_2022}
E.~Panzer, \emph{Hepp's bound for {{Feynman}} graphs and matroids},
  \href{https://doi.org/10.4171/aihpd/126}{\emph{Annales de l'Institut Henri
  Poincar\'e D} {\bfseries 10} (2022) 31}.

\bibitem{broadhurst_exploiting_1986}
D.J.~Broadhurst, \emph{Exploiting the 1, 440-fold symmetry of the master
  two-loop diagram},
  \href{https://doi.org/10.1007/BF01552503}{\emph{Zeitschrift f\"ur Physik C
  Particles and Fields} {\bfseries 32} (1986) 249}.

\bibitem{golz_graphical_2017}
M.~Golz, E.~Panzer and O.~Schnetz, \emph{Graphical functions in parametric
  space}, \href{https://doi.org/10.1007/s11005-016-0935-6}{\emph{Letters in
  Mathematical Physics} {\bfseries 107} (2017) 1177}.

\bibitem{siek_boost_2001}
J.~Siek, L.-Q.~Lee and A.~Lumsdaine, ``The {{Boost Graph Library}} - 1.82.0.''
  https://www.boost.org/doc/libs/1\_82\_0/libs/graph/doc/index.html, 2001.

\bibitem{schnetz_quantum_2011}
O.~Schnetz, \emph{{Quantum field theory over $F_q$}},
  \href{https://doi.org/10.48550/arXiv.0909.0905}{\emph{The Electronic Journal
  of Combinatorics} {\bfseries 18} (2011) 102}
  [\href{https://arxiv.org/abs/0909.0905}{{\ttfamily 0909.0905}}].

\bibitem{panzer_feynman_2023}
E.~Panzer and K.~Yeats, \emph{Feynman symmetries of the {{Martin}} and $c_2$
  invariants of regular graphs},  Apr., 2023.
\newblock 10.48550/arXiv.2304.05299.

\bibitem{pelissetto_critical_2002}
A.~Pelissetto and E.~Vicari, \emph{Critical phenomena and renormalization-group
  theory}, \href{https://doi.org/10.1016/S0370-1573(02)00219-3}{\emph{Physics
  Reports} {\bfseries 368} (2002) 549}.

\bibitem{balduf_dyson_2023}
P.-H.~Balduf, \emph{Dyson\textendash{{Schwinger}} equations in minimal
  subtraction}, \href{https://doi.org/10.4171/aihpd/169}{\emph{Annales de
  l'Institut Henri Poincar\'e D} (2023) }.

\bibitem{schnetz_phi_2023}
O.~Schnetz, \emph{$\phi^{4}$ theory at seven loops},
  \href{https://doi.org/10.1103/PhysRevD.107.036002}{\emph{Physical Review D}
  {\bfseries 107} (2023) 036002}.

\bibitem{martin_enumerations_1977}
P.~Martin, \emph{{Enum\'erations eul\'eriennes dans les multigraphes et
  invariants de Tutte-Grothendieck}}, {thesis}, Institut National Polytechnique
  de Grenoble - INPG ; Universit\'e Joseph-Fourier - Grenoble I, Mar., 1977.

\bibitem{ellis-monaghan_new_1998}
J.A.~{Ellis-Monaghan}, \emph{New {{Results}} for the {{Martin}} polynomial},
  {\emph{Journal of Combinatorial Theory, Series B} {\bfseries 74} (1998) 326}.

\bibitem{bollobas_evaluations_2002}
B.~Bollob{\'a}s, \emph{Evaluations of the {{Circuit Partition Polynomial}}},
  \href{https://doi.org/10.1006/jctb.2001.2102}{\emph{Journal of Combinatorial
  Theory, Series B} {\bfseries 85} (2002) 261}.

\bibitem{bouchet_connectivity_1996}
A.~Bouchet and L.~Ghier, \emph{Connectivity and {$\beta$}-invariants of
  isotropic systems and 4-regular graphs},
  \href{https://doi.org/10.1016/0012-365X(95)00219-M}{\emph{Discrete
  Mathematics} {\bfseries 161} (1996) 25}.

\bibitem{kleinert_critical_2001}
H.~Kleinert and V.~{Schulte-Frohlinde}, \emph{Critical {{Properties}} of
  $\phi^4$-{{Theories}}}, {World Scientific} (2001).

\bibitem{mckane_nonperturbative_1984}
A.J.~McKane, D.J.~Wallace and O.F.d.A.~Bonfim, \emph{Non-perturbative
  renormalisation using dimensional regularisation: Applications to the
  {$\epsilon$} expansion},
  \href{https://doi.org/10.1088/0305-4470/17/9/021}{\emph{Journal of Physics A:
  Mathematical and General} {\bfseries 17} (1984) 1861}.

\bibitem{mckane_perturbation_2019}
A.J.~McKane, \emph{Perturbation expansions at large order: {{Results}} for
  scalar field theories revisited},
  \href{https://doi.org/10.1088/1751-8121/aaf768}{\emph{Journal of Physics A:
  Mathematical and Theoretical} {\bfseries 52} (2019) 055401}
  [\href{https://arxiv.org/abs/1807.00656}{{\ttfamily 1807.00656}}].

\bibitem{kinkelin_ueber_1860}
H.~Kinkelin, \emph{{Ueber eine mit der Gammafunction verwandte Transcendente
  und deren Anwendung auf die Integralrechnung.}},
  \href{https://doi.org/10.1515/crll.1860.57.122}{\emph{Journal f\"ur die reine
  und angewandte Mathematik} {\bfseries 57} (1860) 122}.

\bibitem{richardson_approximate_1911}
L.F.~Richardson and R.T.~Glazebrook, \emph{The approximate arithmetical
  solution by finite differences of physical problems involving differential
  equations, with an application to the stresses in a masonry dam},
  \href{https://doi.org/10.1098/rsta.1911.0009}{\emph{Philosophical
  Transactions of the Royal Society of London. Series A} {\bfseries 210} (1911)
  307}.

\bibitem{aniceto_primer_2019}
I.~Aniceto, G.~Ba{\c s}ar and R.~Schiappa, \emph{A primer on resurgent
  transseries and their asymptotics},
  \href{https://doi.org/10.1016/j.physrep.2019.02.003}{\emph{Physics Reports}
  {\bfseries 809} (2019) 1} [\href{https://arxiv.org/abs/1802.10441}{{\ttfamily
  1802.10441}}].

\bibitem{komarova_asymptotic_2001}
M.V.~Komarova and {M. Yu. Nalimov}, \emph{Asymptotic {{Behavior}} of
  {{Renormalization Constants}} in {{Higher Orders}} of the {{Perturbation
  Expansion}} for the $(4-\epsilon)$-{{Dimensionally Regularized}} $o(n)$
  symmetric $\phi^4$ theory},
  \href{https://doi.org/10.1023/A:1010367917876}{\emph{Theoretical and
  Mathematical Physics} {\bfseries 126} (2001) 339}.

\bibitem{hooft_planar_1974}
G.~'t~Hooft, \emph{A planar diagram theory for strong interactions},
  \href{https://doi.org/10.1016/0550-3213(74)90154-0}{\emph{Nuclear Physics B}
  {\bfseries 72} (1974) 461}.

\bibitem{gurau_colored_2012}
R.~Gurau and J.P.~Ryan, \emph{Colored {{Tensor Models}} - a {{Review}}},
  \href{https://doi.org/10.3842/SIGMA.2012.020}{\emph{SIGMA. Symmetry,
  Integrability and Geometry: Methods and Applications} {\bfseries 8} (2012)
  020}.

\bibitem{bollobas_diameter_1982}
B.~Bollob{\'a}s and W.~{Fernandez de la Vega}, \emph{The diameter of random
  regular graphs},
  \href{https://doi.org/10.1007/BF02579310}{\emph{Combinatorica} {\bfseries 2}
  (1982) 125}.

\bibitem{shenton_development_1975}
L.R.~Shenton and K.O.~Bowman, \emph{The {{Development}} of {{Techniques}} for
  the {{Evaluation}} of {{Sampling Moments}}},
  \href{https://doi.org/10.2307/1403115}{\emph{International Statistical Review
  / Revue Internationale de Statistique} {\bfseries 43} (1975) 317}
  [\href{https://arxiv.org/abs/1403115}{{\ttfamily 1403115}}].

\bibitem{davison_bootstrap_1997}
A.C.~Davison and D.V.~Hinkley, \emph{Bootstrap {{Methods}} and Their
  {{Application}}}, Cambridge {{Series}} in {{Statistical}} and {{Probabilistic
  Mathematics}}, {Cambridge University Press}, {Cambridge} (1997),
  \href{https://doi.org/10.1017/CBO9780511802843}{10.1017/CBO9780511802843}.

\bibitem{tille_sampling_2020}
Y.~Till{\'e}, \emph{Sampling and {{Estimation}} from {{Finite Populations}}},
  {John Wiley \& Sons, Ltd} (2020),
  \href{https://doi.org/10.1002/9781119071259.ch15}{10.1002/9781119071259.ch15}.

\bibitem{evans_probability_2023}
M.J.~Evans and J.S.~Rosenthal, ``Probability and {{Statistics}}: {{The
  Science}} of {{Uncertainty}}.'' 2023.

\bibitem{johnson_curious_2002}
W.P.~Johnson, \emph{The {{Curious History}} of {{Faa}} du {{Bruno}}'s
  {{Formula}}}, {\emph{The American Mathematical Monthly} {\bfseries 109}
  (2002) 217}.

\bibitem{bell_exponential_1934}
E.T.~Bell, \emph{Exponential {{Polynomials}}},
  \href{https://doi.org/10.2307/1968431}{\emph{Annals of Mathematics}
  {\bfseries 35} (1934) 258} [\href{https://arxiv.org/abs/1968431}{{\ttfamily
  1968431}}].

\bibitem{holtzman_unbiased_1950}
W.H.~Holtzman, \emph{The unbiased estimate of the population variance and
  standard deviation}, {\emph{The American Journal of Psychology} {\bfseries
  63} (1950) 615}.

\bibitem{angelova_moments_2012}
J.A.~Angelova, \emph{On moments of sample mean and variance},
  {\emph{International Journal of Pure and Applied Mathematics} {\bfseries 79}
  (2012) 67}.

\end{thebibliography}\endgroup

\end{document}